# Animal-associated marine Acidobacteria with a rich natural product repertoire

Stefan Leopold-Messer[1,4], Clara Chepkirui[1,4], Mathijs F.J. Mabesoone[1], Joshua Meyer[1], Lucas Paoli[1], Shinichi Sunagawa[1], Agustinus R. Uria[2], Toshiyuki Wakimoto[2], Jörn Piel[1,3*]

[1]Institute of Microbiology, Eidgenössische Technische Hochschule (ETH) Zürich, Vladimir-Prelog-Weg 4, 8093 Zürich, Switzerland

[2]Laboratory of Natural Product Chemistry, Faculty of Pharmaceutical Sciences, Hokkaido University, Japan

[3]Lead contact

[4]These authors contributed equally

*Correspondence: jpiel@micro.biol.ethz.ch

## SUMMARY

Sponges are well-known as rich sources of bioactive natural products. Various studies suggest that many of these compounds are produced by symbiotic bacteria. However, substance supplies and functional insights about the producers remain limited because cultivation remains unsuccessful. To identify alternative, sustainable sources of sponge-derived polyketides, we computationally analyzed 5289 characterized and orphan *trans*-acyltransferase polyketide synthases, enzymes with widespread roles in polyketide biosynthesis by bacterial symbionts. The workflow predicted marine animal-derived Acidobacteria of the family Acanthopleuribacteraceae with large sets of biosynthetic gene clusters to be enriched in sponge-type chemistry. Targeted compound isolation from a chiton-associated strain yielded new congeners of the phorboxazoles and calyculins, potent and scarce cytotoxins exclusively known from sponges, as well as the novel depsipeptides acidobactamides. These first natural products of Acidobacteria and new coral metagenomic data on a third family member suggest animal-associated Acanthopleuribacteraceae as a rich source of sponge-type as well as novel metabolites.

## INTRODUCTION

Marine invertebrates like sponges, tunicates, and bryozoans are prominent sources of bioactive natural products with therapeutic potential.[1,2] Commonly, the substances share structural characteristics of microbial natural products classes, including many compounds resembling complex polyketides (Figure 1). Examples are phorboxazole A (**1**) and spongistatin 1 (**2**) from sponges, exceedingly scarce and synthetically demanding polyketides that belong to the most potent antiproliferative agents reported to date.[3,4] In a growing number of cases, symbiotic bacteria have been demonstrated as the actual origin of invertebrate-derived natural products. Among the previously identified producers are members of the candidate genus 'Entotheonella' that synthesize numerous compounds in sponges, including calyculin A (**3**) and onnamide A (**4**) (Figure 1), 'Endolissoclinum faulkneri' producing patellazole B (**5**) in tunicates, and the bryostatin (**6**) producer 'Endobugula sertula' in bryozoans.[5-8] With the exception of a shipworm-associated bacterium,[9] none of the symbionts has, to our knowledge, been cultivated, but had to be identified by localizing the biosynthetic genes in cultivation-independent approaches, highlighting the difficulty to access these scaffolds with high novelty.

The growing evidence for widespread biosynthetic roles of symbionts offers major potential to identify and establish renewable bacterial production systems for rare and structurally distinct marine drug candidates. Besides cultivating producers, promising alternative avenues for biotechnological production are the expression of biosynthetic genes in surrogate hosts or the search for similar genes in cultivated bacteria.[10] However, none of these approaches are straightforward, with challenges including a limited access to invertebrates, the demanding and unpredictable heterologous expression of large polyketide synthase (PKS) and non-ribosomal peptide synthetase (NRPS) genes from non-standard bacteria, and biosynthetic genes that are often elusive or not clustered in biosynthetic gene clusters (BGCs) in symbionts.[8,11,12] As such, no broadly applicable method to facilitate access to symbiont-derived marine drug candidates has been established.

To address the challenge of sustainable production, we here asked whether alternative, cultivated producers for invertebrate natural products can be identified by *de novo* compound prediction from sequenced bacterial genomes, i.e., without prior knowledge of the symbiont genes. We focused on complex polyketides, which in non-actinomycete symbionts are predominantly generated by PKSs of the *trans*-acyltransferase (*trans*-AT) family.[13] A global bioinformatic analysis of orphan *trans*-AT PKSs in sequenced bacterial genomes revealed a metabolically rich acidobacterial taxon with a remarkably sponge-like chemical profile. We verified the predicted acidobacterial chemistry by isolating congeners of two rare families of complex marine polyketides as well as distinct depsipeptides from the bacterial cultures. With this, we also chemically validate Acidobacteria as an untapped, promising drug discovery source, and we highlight the use of global bioinformatic analyses of BGCs to identify and prioritize producers of novel scaffolds.[14]

## RESULTS

### Identification of phorboxazole-type polyketides

*Trans*-AT PKSs are multimodular enzymes containing unusually diverse biochemical features.[13,15-17] Even though they are widespread among bacteria and known to produce compounds of therapeutic value, many representatives occur in chemically poorly studied bacterial groups and are therefore underexplored.[13,18,19] Particularly striking is their prevalence in non-actinomycete symbiotic bacteria, in which they account for almost all complex polyketides with assigned biosynthetic genes.[11,13,20] Each module in a *trans*-AT PKS minimally contains, as integrated domains, an acyl carrier protein (ACP) that covalently binds a polyketide intermediate via a thioester moiety and a ketosynthase (KS) that performs intermediate chain elongation by adding an extension unit. This extension unit is usually derived from malonyl-CoA that is supplied by a *trans*-acting AT enzyme. Depending on a variable set of additional *cis*- and/or *trans*-acting biochemical components present in the upstream module, the structure of this incoming substrate can differ greatly in the α,β-region. We previously showed that these moieties tightly correlate with sequence features in the accepting downstream KS domains and can therefore be predicted with high confidence.[21] Concatenation of the chemical units predicted from all KSs in a *trans*-AT PKS then suggests the overall polyketide core structure, thus facilitating targeted polyketide discovery by genome mining.

Here, we tested a strategy to predict the production of sponge-type complex polyketides in cultivated bacteria. We first applied a computational analysis to identify architecturally unique *trans*-AT PKSs lacking assigned functions. Among these, we hoped to identify candidates for marine invertebrate-type polyketides by automated structural prediction. In this initial analysis, we processed all bacterial genomes published in the NCBI GenBank database between January 2021 and June 2022 through the BGC identification software antiSMASH.[22,23] Those BGCs classified as *trans*-AT type were combined with all *trans*-AT PKS BGCs deposited in the antiSMASH database (covering BGCs up to the year 2020), resulting in 5289 PKS pathways.[24] Each cluster was subsequently analyzed with an updated version of *trans*PACT,[25] which predicts KS substrates of *trans*-AT PKSs and infers a phylogenetic tree based on the KS patterns (Figure 2A). Mapping information on the genome source and on previously characterized PKSs onto these data then revealed functionally unassigned BGCs with unique architectures from cultivated bacteria for subsequent analysis by TransATor,[26] a computational tool that predicts chemical structures for *trans*-AT PKS products. Finally, the suggested structures of these orphan BGCs were subjected to substructure searches in the Dictionary of Natural Products,[27] PubChem,[28] and Reaxys.[29] In addition, structures were visually compared to a manually curated collection of polyketides reported from sponges and tunicates.[30]

Inspection of the phylogenetic tree inferred from the PKS domain sequences revealed multiple clades comprising PKSs of similar or identical architecture and assigned metabolites (Figure 2A). Disregarding clades with small PKSs comprised of 12 or fewer modules, the largest assigned clades were for bacillaenes, difficidins, rhizoxins, and oocydins (315, 134, 79, and 47 members, respectively). Conversely, several tree regions existed that mainly comprised metabolically unassigned and unique or rare *trans*-AT PKSs. To detect candidates for sponge-type pathways in free-living bacteria, we focused on such regions (Figure 2B) and PKSs from cultivated strains that were preferably available from strain collections. As first strong candidates based on TransATor predictions (see next section), we noticed two architecturally related hybrid *trans*-AT PKS-NRPS (*pho* and *phb*, Figure 2C) encoded in genomes of cultivated marine Acidobacteria of the family *Acanthopleuribacteraceae* (Figure 2C).[30,31] This was intriguing, since Acidobacteria are a diverse and ubiquitous, but functionally poorly characterized bacterial phylum with few cultivated members and no reported natural products. The candidate producers *Acanthopleuribacter pedis* FYK2218 and *Sulfidibacter corallicola* M133 have been isolated from a marine chiton[30] and a coral[31], respectively, and both contain large (>10 Mb) genomes.[31] The two strains comprise all presently available *Acanthopleuribacteraceae* genomes. AntiSMASH analyses indicated striking biosynthetic potential in both bacteria with at least 39 and 42 predicted BGCs, respectively (Figure 2D). In *A. pedis*, three of these BGCs belong to *trans*-AT PKS-NRPS hybrid systems, and a further PKS-NRPS BGC encodes *trans*-AT as well as *cis*-AT modules. For the initial chemical studies, we focused on the *pho* BGC (GenBank accession JAFREP010000003, Tables S1 and S2) from *A. pedis*.

## Discovery of phorbactazole-type complex polyketides

The *pho* BGC from *A. pedis* encodes five PKS or PKS-NRPS proteins with a total of 26 modules (Table S1). TransATor-based structural predictions suggested a compound with two oxazole rings and three (partially) saturated pyran rings (Figure 3, Figure S1, and Table S2). Manual inspection did not suggest further structural changes except for a potential additional hemiacetal ring formed in the C-39 to C-43 region. Substructure searches in natural product databases using various portions of the prediction retrieved phorboxazoles as convincingly close hits (Figure 3A). Phorboxazole A (**1**) and congeners were first reported in 1995 by Searle and Molinski from a marine *Phorbas* sp. sponge collected off the Muiron Islands.[3] In 2004, Capon and coworkers also reported phorboxazoles from an Australian *Raspailia* sp. sponge.[32] With activity against all 60 tested cell lines of the National Cancer Institute (NCI) panel and mean growth inhibition ($GI_{50}$) values in the single-digit nanomolar range or lower, phorboxazoles are among the most potent known cytotoxins. The isolation from two taxonomically distinct sponges led to the speculation of a potential microbial phorboxazole producer, which to date remains unknown.

The bioinformatic prediction suggested a compound with a molecular formula of approximately $C_{55}H_{84}N_2O_{18}$ (calculated mass *m/z* 1060.5719 $[M+H]^+$), which we used as guidance for its targeted isolation from *A. pedis*. In contrast to most other Acidobacteria that are either fastidious or uncultivated, bacterial cultures grew rapidly. Following pilot cultivation in 20 mL of six different liquid media for three days, we extracted the supernatants with ethyl acetate and the cell pellets with a mixture of methanol and acetone (1:1) and subjected the residues to high-performance liquid chromatography-heated electrospray ionization-high resolution mass spectrometry (HPLC-HESI-HRMS, Figure S2). Analysis of the supernatant extracts revealed a candidate ion at *m/z* 957.4961 $[M+H]^+$, with a suggested molecular formula of $C_{50}H_{73}N_2O_{16}$ (Δ +0.64 mmu, Figure S3), close to the prediction. Furthermore, molecular network analysis[33] of the tandem MS (MS/MS) data showed that the detected ion was part of a larger cluster of ions (Figure S3). This cluster also contained ions at *m/z* 1000.5023 $[M+H]^+$ ($C_{51}H_{74}N_3O_{17}$; Δ +1.03 mmu), *m/z* 1119.5496 $[M+H]^+$ ($C_{56}H_{83}N_2O_{21}$; Δ +1.32 mmu) and *m/z* 1162.5554 $[M+H]^+$ ($C_{57}H_{84}N_3O_{22}$; Δ +1.30 mmu). The mass difference of 43 Da between the ions 957.5/1000.5 and 1119.5/1162.6 suggested congeners differing by one C, H, N, and O atom. Other ion pairs featured a 162 Da mass difference, indicating glycosylation by a hexose ($\Delta C_6H_{10}O_5$).

Cultivation of 20 L of *A. pedis* and repetitive MS-guided reverse-phase HPLC purification from organic extracts yielded six compounds, named phorbactazole A (**7**, 1.2 mg), B (**8**, 0.1 mg), C (**9**, 1.2 mg), D (**10**, 0.2 mg), E (**11**, 0.7 mg), and F (**12**, 0.9 mg) (Figure 3B). Compound **7** has the molecular formula $C_{51}H_{73}N_3O_{17}$ and 17 degrees of unsaturation deduced from the HRMS data (Figure S4). Analysis of the one- and two-dimensional nuclear magnetic resonance spectroscopy (NMR) data (Figure S5-S10 and Table S3) revealed three spin systems **I-III** (Figure S10) that were connected based on heteronuclear multiple bond correlations (HMBC) (see Supplemental Information). The relative configuration (marked by an asterisk) of four tetrahydropyran-type rings was assigned from NOESY data: ring A (5*R**, 6*S**, 7*R**, 9*S**), ring B (17*S**, *18R**, 19*R**, 21*S** or 17*S**, *18S**, 19*R**, 21*S** for compounds **8** and **10**), ring C (28*R**, 30*S**, 31*R**, 32*R**), and ring D (40*S**, 41*R**, 43*S**). The relative configuration of C12 was inferred from the nuclear Overhauser- effect (NOE) correlations along the spin system **I** as (12*S**). C-13, C-15, and the hemiacetal C-39 could not be stereochemically assigned by NMR (Figure S11). To obtain insights into the absolute configuration, sequence motifs of the ketoreductase (KR) domains[26] were analyzed, which in modular PKSs correlate well with the configuration of hydroxyl-bearing carbons.[34] The *pho* KRs (Table S2) suggested the following absolute configurations of oxygen-bearing carbons: (7*R*, 9*S*, 13*R*, 15*R*, 19*R*, 21*S*, 30*S*, 32*S*, 41*R*, 43*S*). A complementary method to predict the absolute configuration is based on KS clades of *trans*-AT PKSs, and indicated (7*R*, 9*S*, 13*R*, 15*R*, 19*R*, 21*S*, 30*S*, 32*R*, 41*R*, 43*S*) (Table S2). The KR- and KS-based predictions agreed on all carbon centers of rings A, B, and D assignable by NMR, however for ring C only the KS-based prediction matched one of the NMR-elucidated enantiomers, with three axial protons (Figure S11C), rendering the KS-based prediction of C-32 more probable than the KR-based prediction. The combination of NMR and bioinformatic methods therefore suggest the absolute configuration of **7** as (5*R*, 6*S*, 7*R*, 9*S*, 12*S*, 13*R*, 15*R*, 17*S*, 18*S*, 19*R*, 21*S*, 28*R*, 30*S*, 31*R*, 32*R*, 40*S*, 41*R*, 43*S*) with only the hemiacetal C-39 unassigned (Figure 3B, Figure S10, Figure S11 and Table S2). [26,34]As predicted, compound **7** shares large molecular portions with the sponge-derived phorboxazoles.[3] The main differences to phorboxazoles are an expanded macrolide ring around C-10-17, a modified Western region that lacks the vinyl bromide moiety, and the presence of an *O*-carbamoyl substituent. Notably, our analyses also suggest an inverted absolute configuration of rings A and B, while the absolute configuration of rings C and D fits that reported for phorboxazole (Figure 3). However, these rings also carry different substituents, and the reported isolation and synthetic data suggest a correct assignment of the sponge compounds.[35]

NMR data for phorbactazole B (**8**) with the molecular formula $C_{50}H_{72}N_2O_{16}$ were similar to those of **7** (Figures S12-S16 and Table S3) but lacked signals for the carbamoyl unit (Figure 3B). Phorbactazole C (**9**, $C_{57}H_{83}N_3O_{22}$, Figure S17) yielded NMR spectra (Figures S18-S22 and Table S4) suggesting a glucose moiety attached to C-12 and carbamoylation at C-19 (Figure 3B, Figure S23). Phorbactazole D (**10**, $C_{56}H_{83}N_2O_{21}$, Figure S24) is a non-carbamoylated but glycosylated variant (Figures S25-S28 and Table S4). Two dehydration products of phorbactazole C, namely phorbactazole E (**11**, $C_{57}H_{81}N_3O_{21}$,, Figure S29-S33 and Table S5) and phorbactazole F (**12,** $C_{56}H_{77}N_3O_{20}$,, Figure S34-S38 and Table S5) were also isolated. Even though the two congeners were detected in the crude extracts, we could not rule out the possibility of them being degradation products from extraction process or the products of the isomerization observed in ring D (Figure S39). A comparison of the circular dichroism (CD) spectra of **7** with those of **8**-**12** suggested a similar configuration of all compounds, as the electronic circular dichroism (ECD) curves of these compounds were comparable in the range of 190-290 nm (Figure S40).

## Biosynthetic model of phorbactazoles

A proposed model for phorbactazole biosynthesis is shown in Figure 4. The KS-based polyketide prediction and other components of the *trans*-AT PKS-NRPS proteins PhoA-E largely agree with the phorbactazole structures. The two NRPS modules and three PKS modules with pyran synthase (PS) domains positionally match the presence of two oxazoles and three pyran-type rings in phorbactazoles (rings A-C), with another six-membered ring (ring D) likely arising from spontaneous hemiketal formation. A discrepancy of the PKS architecture to the phorbactazole structures lies in the first PKS module, in which the KS domain seems to lack a function. Inspection of the KS protein sequence showed overall conserved active-site residues except for a Cys-to-Ser mutation of a site critical for chain elongation. Together with the phorbactazole structure, this suggests the KS decarboxylates a malonyl unit on the first module to generate an acetyl starter. In analogy to functionally related domains in many *cis*-AT PKSs named $KS_Q$ due to a Gln mutation,[36] we termed the phorbactazole domain $KS_S$. Further, all phorbactazoles contain two off-pattern hydroxy groups at C-12 and C-18. In agreement, bimodules 15+16 and 19+20 of the *pho* PKS as well as the free-standing oxygenase PhoL match regarding architecture and protein homologies counterparts in the oocydin PKS (OocM) that we recently showed to install α-hydroxylations during chain elongation.[15] At the terminus of the assembly line, PhoE features an unusual architecture with internal TE and C domains. Internal TEs, for which we earlier proposed the term $TE_B$ for branching thioesterase, are also present in several other *trans*-AT PKSs,[15] and for a $TE_B$ of the oocydin PKS we showed that it installs an *O*-acetyl group.[17] A similar reaction in phorbactazole biosynthesis might facilitate elimination to generate the *Z* -configured double bond present at the biosynthetic polyketide terminus.[17] The *pho* BGC also contains the non-PKS genes *phoF* and *phoM* encoding a putative glycosyl and carbamoyl transferase, respectively, which match the glycosylation and carbamoylation substituents in phorbactazoles **7** and **9**-**12**. Although the *pho* architecture and phorbactazole structures are in good agreement, trials to genetically modify *A. pedis* for BGC inactivation have not yet been successful, and the biosynthetic model is therefore putative.

## A second *trans*-AT PKS in *A. pedis* produces sponge-type calyculinamides

Further analysis of the TransPACT output revealed another gene cluster in *A. pedis*, termed *cly* BGC, that was strikingly similar to the calyculin (*cal*) BGC previously identified in an uncultivated

'Entotheonella' sp. symbiont of the sponge *Discodermia calyx* (Figure 2A, 5 and Figure S41).[6] Calyculins are highly potent phosphatase inhibitors[37] with picomolar inhibitory activity against several tumor cell lines.[38] Calyculin A (**3**) is employed as a cellular probe[39] and currently sourced from the sponge for commercial supply. The *cly* PKS architecture perfectly agrees with that of the calyculin PKS (*cal*) except for the additional presence of an *O*-methyl transferase domain between KS14 and 15 (Figure S41). HPLC-HESI-HRMS data of a 20 mL pilot extraction revealed a candidate ion at *m/z* 1041.5776 $[M+H]^+$, which had a suggested molecular formula of $C_{51}H_{86}N_4O_{16}P$ (Δ +0.55 mmu), differing from **3** by $CH_4O$ (Figures S42 and S43). MS/MS fragmentation patterns and a GNPS library search supported the compound being related to the calyculin family (Figure S42).[33] For further insights into the identity of these compounds, we compared the bacterial extracts to those of the sponge *D. calyx* using molecular network analysis.[33] This showed a large cluster of ions shared between the two samples (Figure S42). Ions corresponding to reported calyculins A-J with nitrile moieties were only detected in the sponge. However, ions of similar *m/z* values and with matching MS/MS fragmentation patterns were located in both samples, suggesting closely related compounds.

To isolate the main compound (**13**), the supernatants of 12 L combined *A. pedis* cultures were extracted with ethyl acetate and subsequently separated by repetitive HPLC. MS-guided fractionation followed by structure elucidation (see Supplementary Information) yielded 0.2 mg of 21-methoxy calyculinamide A (**13**) (Figure 5, Figures S44-S50, Table S6-S8). While the NOESY data suggested the relative configuration of only the spirocyclic moiety, stereochemically informative features of the *cly* KS and KR domains perfectly matched those of the calyculin (**3**) PKS except for the KR between KS13 and 14. The combined bioinformatic data therefore suggested almost identical absolute configurations of **3** and **13** with an epimerized center at C-21 in **13** that is supported by the NMR data. In addition to this configurational difference, **13** but not **3** contains a methoxy group at C-21 that agrees with a unique *O*-methyltransferase domain located before KS15 in the *cly* PKS (Figure S41). A similar moiety at that position is found in the calyculin congeners geometricin[40] and swinhoeiamide[41] from sponges, for which the PKSs are unknown. These also harbor the same relative configuration of C-21 as **13**.

Further analysis of *A. pedis* cell pellets also revealed the ion *m/z* 1121.5449 $[M+H]^+$ with a suggested molecular formula of $C_{51}H_{87}N_4O_{19}P_2$ (Δ +0.94 mmu) corresponding to a variant of **13** with an additional phosphate group (Figure S51). Characteristic fragment ions are shared between this ion and compound **13**, corroborating related structures. Since this suggested phospho-21-methoxy-calyculinamide was present at minute amounts and highly unstable, isolation was not attempted.

### Acidobactamides A-C (15-17), peptide-polyketide hybrids from *A. pedis*

To gather additional insights into the biosynthetic potential of Acanthopleuribacteraceae, we aimed to isolate and characterize further natural products of *A. pedis*. A multimodular biosynthetic system comprising nine NRPS modules interspaced by three PKS modules, termed *aba* (Figure S52, Table S9), was suggested by antiSMASH[42] and PRISM analysis[43] to synthesize a peptide containing a threonine unit (module 6 in Figure S52) that is extended by one PKS module (module 7 in Figure S52), resulting in a 4-amino-3,5-dihydroxy-2-methylhexanoic acid (Admha) motif (Figure 5, Figure S52-S53). This caught our attention, since a SciFinder[44] search revealed depsipeptides from marine sponges as the only other known natural products with a closely related, non-methylated variant of this moiety, a 4-amino-3,5-dihydroxyhexanoic acid (Adha) residue (Figure 5, Figure S53). These compounds are the cyclolithistides isolated from *Discodermia japonica*[45] and *Theonella swinhoei*[46], and phoriospongins from *Phoriospongia* sp. and *Callyspongia bilamellata*.[47] To isolate the predicted compound, NMR-guided isolation was adopted. A close look into the $^1$H and HSQC NMR spectra (Figure S54-S55) of the crude extracts HPLC fraction F12 revealed the presence of proton signals attributed to the α-methines of amino acid residues ($\delta_H$ 3.9–5.0 ppm) and aromatic proton signals attributed to 1,4-disubstituted and benzene moieties ($\delta_H$ 6.9-7.31 ppm). MS analysis of this fraction revealed an ion at *m/z* 1343.6689 $[M+H]^+$ with a suggested molecular formula of $C_{62}H_{94}N_{12}O_{21}$ (Δ -3.244 mmu, Figure S56) Several rounds of purification and NMR-based structure elucidation yielded acidobactamide A (**15**, Figure 5, Figures S57-S81, Table S10-S12), a depsipeptide featuring several unusual residues: 3,6-diamino-5,7-dihydroxy-7-(4-methoxyphenyl)heptanoic acid (Ddmpha), the predicted Admha residue, 2-hydroxy-3-phenylpropanoic acid (Hppa), and *N*-methyl isoleucine (*N*-Me-Ile). Acidobactamide A (**15**) shares a central *N*-branched Ile-COO-Admha-Gly portion with the sponge-derived cyclolithistides (Figure 5) but differs in other molecular regions. A substructure search in natural product databases for the unusual Ddmpha residue yielded no results, and hence to our knowledge **15** is the first natural product containing this moiety.

Two additional congeners of **15**, acidobactamide B (**16**) and acidobactamide C (**17**) were characterized. The MS and NMR data of **16** (Figures S82-S90, Table S13), with the molecular formula $C_{63}H_{96}N_{12}O_{21}$, suggested a similar structure to **15** with an additional ß-methylation on the aspartic acid residue. Responsible for this modification might be one of the two cluster-encoded $B_{12}$-dependent radical *S*-adenosyl methionine methyltransferases, AbaH and AbaL (Figure S52, Table S9).[48] The MS and NMR data of acidobactamide C (**17**) (Figures S91-S99, Table S14), with the molecular formula $C_{61}H_{92}N_{12}O_{21}$, indicated that the central isoleucine of **15** is replaced by a valine in **17**. To identify candidate genes responsible for the β-hydroxylation of tyrosine in **15**-**17** we analyzed the protein AbaK using Phyre2.[49] Although the protein was genome-annotated as a metallohydrolase, two high-confidence Phyre2 hits were the amino acid β-hydroxylases FrsH and CmlA, which have been shown to act *in trans* on assembly line-bound substrates, suggesting a similar function for AbaK in tyrosine hydroxylation.[50,51] In summary, the acidobactamide (**15**-**17**) structures largely agree with the adenylation domain substrate selectivity and the PKS domain architecture of antiSMASH and PRISM predictions (Figures S52 and S53). For the two NRPS modules that had ambiguous predictions, Asp and Gln residues were elucidated from NMR data. Additionally, Marfey's analysis confirmed two D-configurated amino acids matching the presence of two epimerization (E) domains (Figure S52, Figure S100).

## Bioactivity of phorbactazoles, 21-methoxy calyculinamide, and acidobactamides

Assays with phorbactazoles A-F (**7-12**) and Henrietta Lacks (HeLa) cervical cancer cells showed activity in the micromolar range for phorbactazoles A-F ($IC_{50}$ 2.4, 10.2, 2.4, 3.2, 2.0, and 2.6 µM for **7**-**12**, respectively, Figure S101). These $IC_{50}$ values are higher than those reported for phorboxazole A (**1**, $IC_{50}$ < 1 nM) and comparable to hemi-phorboxazole A ($IC_{50}$ > 10 µM), a shortened phorboxazole congener.[53] Structure-activity relationship studies on phorboxazoles have shown that the vinyl bromide and the diene group present in **1** but not in **7-12** are crucial for cytotoxicity in the low nM range.[54] Of specific interest are the C-1 to C-32 and C-1 to C-38 phorboxazole analogs (Figure 3A) synthesized by Uckun and Forsyth.[54] The $IC_{50}$ values of these more phorbactazole-like analogs were outside the tested concentrations ($IC_{50}$ > 2 µM) against various cancer cell lines. The carbamoylation of phorbactazole A (**7**) increases its cytotoxicity about 4-fold compared to phorbactazole B (**8**). Carbamoylation has been reported to increase potency of other natural products.[55,56] Assay data of phorbactazole A (**7**) and C (**9**) indicate that the glycosylation does not influence activity. We also tested 21-methoxy calyculinamide (**13**) in this assay, which revealed potent activity at an $IC_{50}$ of 0.3 nM, similar to calyculins and calyculinamides.[57,58] Acidobactamides A-C (**15-17**) showed moderate cytotoxicity ($IC_{50}$ 15.0, 21.4, and 13.9 µM for **15-17**, respectively, Figure S101). Compounds **7-10, 13** and **15**-**17** were also evaluated for antimicrobial activity against a panel of bacteria (Table S15). Compounds **7**-**10** and **13** did not exhibit activities at concentrations ≤100 mg/mL, while **15**-**17** were inactive at any of the tested concentrations (≤50 mg/mL).

## Natural product richness and animal-associated lifestyle of *Acanthopleuribacteraceae*

The BGC-rich *Acanthopleuribacteraceae* members *A. pedis* and *S. corallicola* have been isolated from the chiton *Acanthopleura japonica* and the coral *Porites lutea*, respectively.[31] Based on genome features of these strains and on the detection of sequence reads from uncultivated members of this family in further animal microbiomes, an animal-associated lifestyle was suggested for this acidobacterial family.[31] To evaluate whether *A. pedis* may be more consistently associated with the chiton rather than being a transitory colonizer, we recollected the chiton *A. japonica* at the same location reported in the 2008 isolation study.[30] The presence of *A. pedis* in the chiton was assessed through the workflow described in Figure S102. We isolated high-molecular weight metagenomic DNA from six collected chiton samples. Using primers designed based on the 16S rRNA gene sequence of *A. pedis* (GenBank accession number NR_041599.1), we performed PCR experiments with each purified DNA sample as template. DNA electrophoresis showed correct-sized amplicons of ~1.4 kb in five of the six different metagenome samples, but not in negative controls (Figure S102 and Tables S16 and S17). Sequencing of cloned PCR products from two samples showed 100% sequence identity with the 16S rRNA gene sequence of *A. pedis* for one sample and a single point mutation for the other (Table S16, 17). This suggests that *A. pedis* is more commonly and over larger time scales associated with specimens of the chiton collected at Chiba (Figure S102 and Tables S16 and S17). We were not able to detect compounds **7**-**13** and **15**-**17** in the animal extracts. Further studies are required to characterize the nature of this association.

The results on *A. pedis* and *S. corallicola*, the only two *Acanthopleuribacteraceae* members for which genomic data are available, motivated us to perform a targeted search for additional representatives in metagenomic datasets as described before.[59] For a metagenome from the coral *Millepora platyphylla* collected at the *Tara Pacific* expedition (Figure 6A, TARA_CO-0003230_METAG), we were able to reconstruct a metagenome-assembled genome (MAG) of a further member that belong to an unassigned *Acanthopleuribacteraceae* genus according to phylogenetic analyses (Supplementary Information, Figure 6A). The genome of approximately 7.3 Mb was estimated as 91.88% complete with 0.85% contamination (Supplementary Information) and good integrity (79 scaffolds, longest scaffold >370 kb, N50 ~150 kb). Using antiSMASH we identified 39 BGC-containing scaffolds of which at least 7 belonged to *trans*-AT PKSs (Figure 6C). In total, 69 KS domains of modular PKSs and 156 adenylation domains of NRPSs were detected (Figure 6). A TransPACT analysis was performed to estimate the diversity of the *trans*-AT PKSs in the three genomes (Figure 6D). None of the *trans*-AT PKS BGCs are architecturally identical between the three bacteria. Qualitative analysis of the *Millepora* MAG BGCs suggest complex biosynthetic pathways, of which two architecturally unusual *trans*-AT PKS-NRPS hybrids are shown in Figure 6E. These data corroborate the impressive BGC richness and common animal-associated lifestyle of this acidobacterial family.

## DISCUSSION

Acidobacteria are a large and abundant bacterial phylum with members detected in diverse habitats.[12,14,30,31,61] In studies on different soil environments, Acidobacteria comprised on average ca. 20% of the total bacterial communities.[62,63] Despite their ubiquitous occurrence, only few members of this phylum have been cultivated, with *A. pedis* reported in 2008 as only the fourth cultured representative.[30,64] Genome analyses of cultivated strains as well as metagenomic studies of soil communities have suggested rich biosynthetic potential for some Acidobacteria based on the presence of multiple NRPS, PKS, and further gene clusters.[14,65] However, to the best of our knowledge, no natural product has been reported to date from this phylum. In this work, we provide functional evidence for Acidobacteria as a rich and untapped natural product resource by isolating bioactive, complex polyketides as well as depsipeptides from three different *A. pedis* pathways. Our analysis suggests that most *trans*-AT PKSs present in currently available acidobacterial genomes are found in the talented natural product family of *Acanthopleuribacteraceae*. All three available *Acanthopleuribacteraceae* genomes, one a MAG reconstructed in this study, feature large numbers of BGCs suggesting a metabolic richness comparable to that of the industrially relevant streptomycetes.[66] In agreement, HPLC-MS profiling (Figure S2) and MS network analysis (Figure S3) revealed further metabolites unrelated to the compounds isolated here. These data suggest that targeted cultivation efforts to expand the number of available strains could be rewarded with a substantial number of bioactive compounds. Particularly remarkable is the resemblance of the isolated *A. pedis* compounds to those from currently uncultivated sponge symbionts, as well as the presence of a further variant of a phorboxazole-type BGC in *S. corallicola*. Such high frequency of sponge-type pathways in closely related bacteria is, to our knowledge, unprecedented and raises intriguing questions on the broader chemical diversity of this taxon, the ecological roles of the substances in host-bacterial interactions, and the evolutionary origin of the natural product pathways in sponge symbionts. In a previous study[31], GenBank entries with sequence reads with >92% identity to the 16S rRNA gene of *S. corallicola* and *A. pedis* were identified for microbiomes of corals (*P. lutea*, *Acropora eurystoma*) and fish (*Seriola rivoliana*). Our updated analysis that includes 16S rRNA genes above 90% identity retrieved further sequences from microbiomes of sponge, corals and the coccolithophore *Emiliania huxleyi* (Table S18). In addition, chiton recollection and reconstruction of a third *Acanthopleuribacteraceae* genome from the coral *M. platyphylla* corroborate a close association with marine animals.[31] However, the frequency and roles of symbiotic relationships in this Acidobacterial family and function of the metabolites in host-bacterial interactions remains to be studied.

The identification of alternative, cultivated bacterial sources has been a successful method to obtain polyketides and modified peptides from marine macroorganisms. Examples are swinholides[67], oocydins[68], lobatamides[69,70], didemnins[71], microsclerodermins[72], bengamides[73], and polytheonamides[10]. With the exception of the ribosomally synthesized polytheonamides[74], these previous cases were serendipitous discoveries. In this work we show that the TransPACT tool can pinpoint sustainable supplies of rare polyketides in a systematic, targeted fashion. This is feasible for substances with known biosynthetic genes, as demonstrated for calyculinamides, as well as through *de-novo* prediction from orphan BGCs aided by TransATor, as applied to phorboxazol-type polyketides. While the nonribosomal peptides in this study were isolated by individual BGC mining, the same large-scale methods as applied for polyketides might also be feasible for nonribosomal peptides, which are also generated by multimodular assembly lines well-suited for prediction. With the rapid increase of publicly available bacterial genomes, we anticipate that the mining strategy presented here will reveal sustainable sources for many further natural products.

## EXPERIMENTAL PROCEDURES

### Resource availability

#### *Lead Contact*

Further information and requests for resources and reagents should be directed to the lead contact, Prof. Jörn Piel (jpiel@micro.biol.ethz.ch).

#### *Materials availability*

Compounds generated in this study will be made available on request. All other materials are commercially available or can be prepared as described.

#### *Data and code availability*

All data supporting the findings of this study are included within the article and its Supplemental Information and are also available from the authors upon request. Scripts and sequencing data files used for bioinformatic analysis are available at *https://github.com/mathijs-m/transPACT_wrapper*. The phorbactazole, 21-methoxy calyculinamide, and acidobactamide BGCs have been deposited in MIBiG under accession no. BGC0002813, BGC0002814, and BGC0002815, respectively. Data underlying figures 2 and 6, and genomic information of the *M. platyphylla*-associated *Acantipleuribacteraceae* member is available in the Zenodo repository https://doi.org/10.5281/zenodo.8255904.

## SUPPLEMENTAL INFORMATION

Supplemental Information can be found online at ###

## ACKNOWLEDGEMENTS

We thank Kefu Yu and Guanghua Wang for insightful discussions and Hans-Joachim Ruscheweyh as well as the ETH IT team for their computational support. We also thank Dr. Kensuke Yanagi (Coastal Branch of Natural History Museum and Institute, Chiba) for his help in collecting the chitons. M.M. has received funding from the European Union's Horizon 2020 research and innovation programme under the Marie Skłodowska-Curie grant agreement No. 897571. J.P. acknowledges funding by the European Research Council (ERC) under the European Union's Horizon 2020 research and innovation program (grant agreement No 742739), the Gordon and Betty Moore Foundation (#9204, https://doi.org/10.37807/GBMF9204), the Swiss National Science Foundation (SNSF; 185077, 197245), and the ETH (Research Grant ETH-21 18-2). T.W. acknowledges funding by the JSPS KAKENHI (Grant numbers JP21H02635 and JP22H05128). A.U has received funding from the JSPS KAKENHI (Grant number JP 21K06612). J.P, T.W. and A.U. acknowledges the support of Global Station for Biosurfaces and Drug Discovery, a project of Global Institution for Collaborative Research and Education in Hokkaido University. S.S. acknowledges funding by the SNSF (205321_184955). We would like to thank the expedition participants and consortium partners of Tara Pacific (https://zenodo.org/record/3777760#.YfEEsfXMLjB), Eric Rottinger, Ryan McMinds, Cecile Rottier, and Claudia Pogoreutz for collecting samples, and Stéphane Pesant for curating the data of the coral-associated microbial genome presented in this study.

## AUTHOR CONTRIBUTIONS

M.M. and S.L.-M. performed bioinformatic studies on polyketide predictions. S.L.-M. and C.C. isolated and characterized natural products. S.L.-M. and C.C. performed bioactivity assays. A.R.U. and T.W. sampled chitons and performed metagenomic analyses. L.P. and S.S. performed analyses of the Tara Pacific dataset and metagenomic genome reconstruction. S.L.-M and J.P. wrote the paper with contributions of all authors.

## DECLARATION OF INTERESTS

The authors declare no competing interests.

## INCLUSION AND DIVERSITY

One or more of the authors of this paper self-identifies as an underrepresented ethnic minority in science

.

## Figure Captions

### Figure 1. Examples of complex polyketides from marine invertebrates.

Some of these compounds (**3**-**6**) have been linked to symbiotic bacteria while for others (**1**, **2**) the producers and biosynthetic genes remain elusive.

### Figure 2. Global analysis of bacterial *trans*-AT PKS gene clusters and identification of a talented producer taxon.

(A) *Trans*-AT PKSs gene clusters in genomes published in GenBank[23] were analyzed with the TransPACT tool.[25] TransPACT predicts for each PKS a series of polyketide moieties as substrates of all KSs and represents them as color-coded patterns. The phylogenetic tree indicates the relatedness of PKS BGCs based on their KS patterns[25] (see Supplemental Information and GitHub for details on visualization). This visualization allows rapid dereplication of PKSs similar to already characterized ones (annotated for some examples, e.g., bacillaene), identification of rare orphan PKSs, and identification of alternative producers for BGCs of interest (e.g., a cultivated calyculin producer). Red branches indicate BGCs found in Acidobacteria, blue branches indicate the *pho* BGC in *A. pedis* and the *phb* BGC in *S. corallicola*, and the orange branch indicates the *cly* cluster in *A. pedis* as well as the *cal* BGC in '*Ca.* Entotheonella sp.'. The legend indicates the predicted KS substrates. KSs individually assignable to pyran units, oxazole rings, chain branching, and oxygen insertion are summarized in the KS legend as "specialized" (Supplemental Information).

(B) Expanded view of a tree section containing the *pho*, *phb*, *cly*, and *cal* BGCs.

(C) KS domain sequences of the Acidobacterial BGCs.

(D) AntiSMASH analysis of the two bacteria harboring the *pho*, *phb*, *cly*, and *cal* PKSs. Both strains belong to the family *Acanthopleuribacteraceae*, have large genomes, and contain around 40 BGCs, suggesting a talented producer taxon within the phylum Acidobacteria. RiPP: Ribosomally synthesized and post-translationally modified peptide.

### Figure 3. Products of the *pho trans*-AT PKS isolated from *A. pedis*.

(A) Top: TransATor-based structural prediction of the *pho* BGC product. Bottom: structure of sponge-derived phorboxazole A (**1**). Moieties shared by both structures are highlighted in blue.

(B) Phorbactazoles A-F (**7-12**) isolated in this study. The relative configuration (marked by an asterisk) of rings A-D in **7** was assigned by NOESY correlations (Figure S11). The absolute configuration of the remaining hydroxy group-bearing carbon centers was suggested by analysis of KR and KS domains (Table S2), with a relative configuration matching the NMR data. The only exception was C-32, for which bioinformatic data were contradictory but for which the configuration could be determined by NMR. Due to different atom priorities of C-17 and C-19 the assignment of C-18 in **8** and **10** (18*S*) is inverted to that in **7**, **9**, **11**, **12** (18*R*).

### Figure 4. Proposed biosynthesis of phorbactazoles by the *pho trans*-AT PKS.

(A) Map of the *pho* biosynthetic locus.

(B) Biosynthetic model of the phorbactazole assembly line. The predicted flavin-dependent oxidoreductase PhoL might be responsible for α-hydroxylation in modules 17 and 21. The glycosyltransferase PhoF and the *O*-carbamoyltransferase PhoM would generate the final products. Suggested configurations of hydroxy groups based on bioinformatic prediction, are shown above KR domains. The KR-based prediction upstream of KS9 is opposite of what the KS and NOESY data suggest. OX: oxygenase, Cy: cyclization domain, A: adenylation domain, DH: dehydratase domain. Filled black circles: acyl or peptidyl carrier protein.

### Figure 5. Additional polyketides and depsipeptides isolated from *A. pedis* and related compounds from sponges.

(A) 21-Methoxy calyculinamide A (**13**) of the *cly trans*-AT PKS in *A. pedis* (left) and the structurally related previously reported sponge-derived calyculinamide A (**14**) and calyculin (**3**) (right). The relative configuration (indicated by an asterisk) of the spiro system in **13** was determined by NMR (Supplemental Information), the absolute configuration is suggested by PKS domain analyses and matches that of **14** with the exception of C-21. Several stereocenters of **13** could not be determined by NMR or predicted from the PKS sequence.

(B) The depsipeptides acidobactamide A-C (**15**-**17**) are putatively biosynthesized by the *aba* NRPS-PKS hybrid (Figure S52) and were isolated from the cell pellet of *A. pedis*. The structure features a macrocycle and multiple unusual building blocks. A nonmethylated Admha residue, Adha, is also present in sponge derived depsipeptides such as cyclolithistide A (**18**), embedded in a shared three-residue moiety highlighted in purple. Ddmpha: 6-diamino-5,7-dihydroxy-7-(4-methoxyphenyl)heptanoic acid; Hppa: 2-hydroxy-3-phenylpropanoic acid; Admha: 4-amino-3,5-dihydroxy-2-methylhexanoic acid. The absolute configuration was assigned using the modified Mosher method (Supplemental Information).[52] Chemical shift differences were only observed on one site of the plane for the hydroxy group of carbon C-45, possibly because of the close proximity of MTPA substituents.

### Figure 6. Characterized and uncharacterized *Acanthopleuribacteraceae* are talented and diverse natural product producers.

(A) Collection and metagenome sequencing of a *Millepora* coral collected by the *Tara Pacific* expedition near Tobi island (2°59'54.5"N 131°07'31.4"E; Republic of Palau; sample ID: TARA_CO-0003230). Inset: Photo of *Millepora platyphylla.*

(B) Phylogenetic analysis based on a single-copy marker gene (COG0012; Supplementary Information) resolves the evolutionary relationships between *A. pedis*, *S. corallicola*, and a previously unreported coral-associated Acanthopleuribacteraceae strain for which a metagenome assembled genome was reconstructed. GTDBTk annotations (Supplementary Information) indicate that all three genomes belong to different genera of the Acanthopleuribacteraceae family.

(C) AntiSMASH analysis reveals that the *M. platyphylla*-associated *Acantipleuribacteraceae* member is as chemically rich as *A. pedis* and *S. corallicola* and features an impressive number of biosynthetic gene clusters, including multiple *trans*-AT PKS BGCs (only BGCs longer than 5 kb were included in the analysis).

(D) *Trans*PACT analysis indicates that the *trans*-AT PKSs found in *Acantipleuribacteraceae* are diverse and distinct when compared to the BGCs deposited in MIBiG[60]. For this analysis, clusters that contained one or more PKSs with at least two KS domains and no *cis*-AT domains were extracted from the MIBiG database. The legend refers to the inner tracks. The legend referring to the outer KS phylogenies can be found in Figure 2.
(E) Two examples of architecturally unusual *trans*-AT PKS-NRPS BGCs from the *Millepora*-associated strain. The assembly lines comprise 10 and 24 modules, respectively, with a repeated occurrence of domains of unknown function (DUF) adjacent to pyridoxal phosphate-(PLP-)dependent domains.

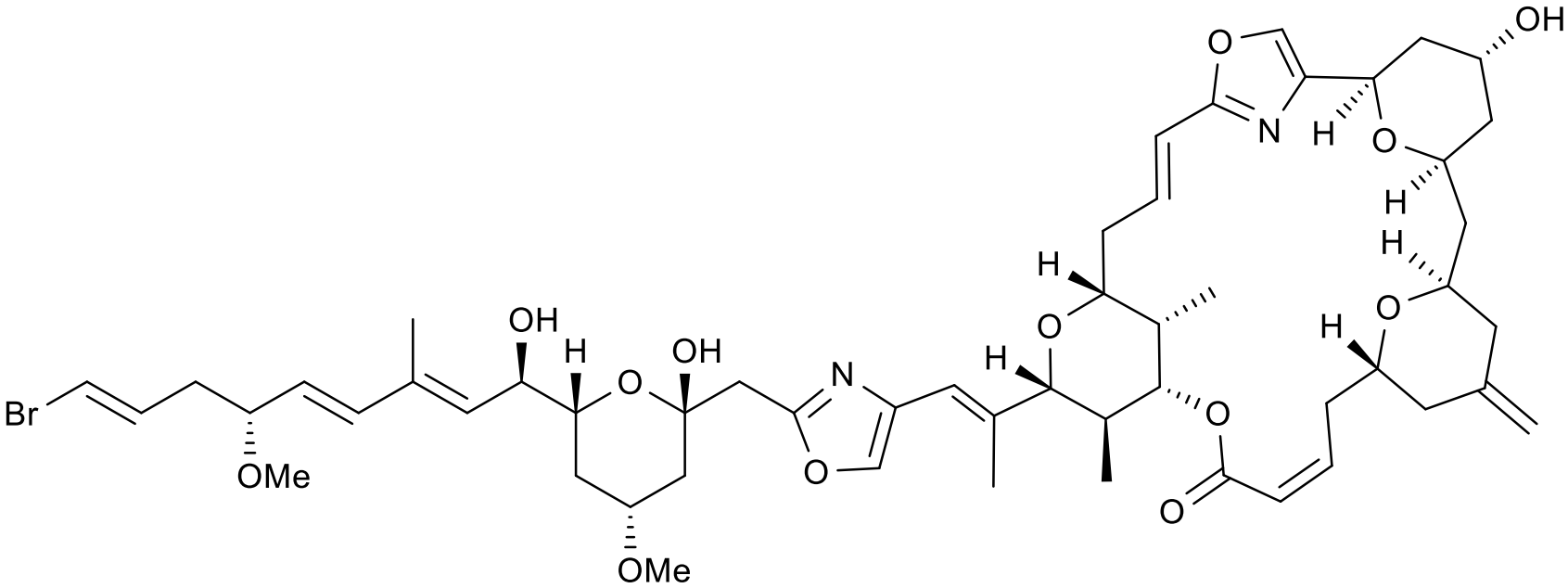

phorboxazole A (**1**)
producer unkown; isolated from a *Phorbas* sp. sponge

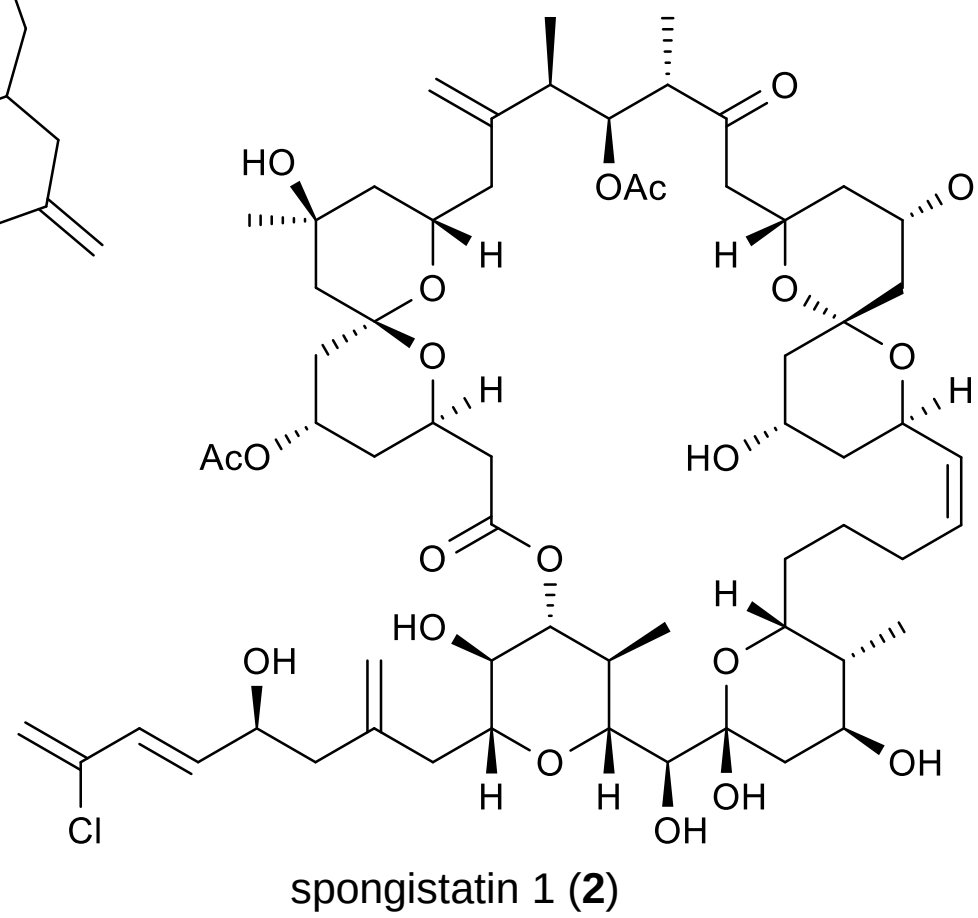

spongistatin 1 (**2**)
producer unknown; isolated from a *Spongia* sp. sponge

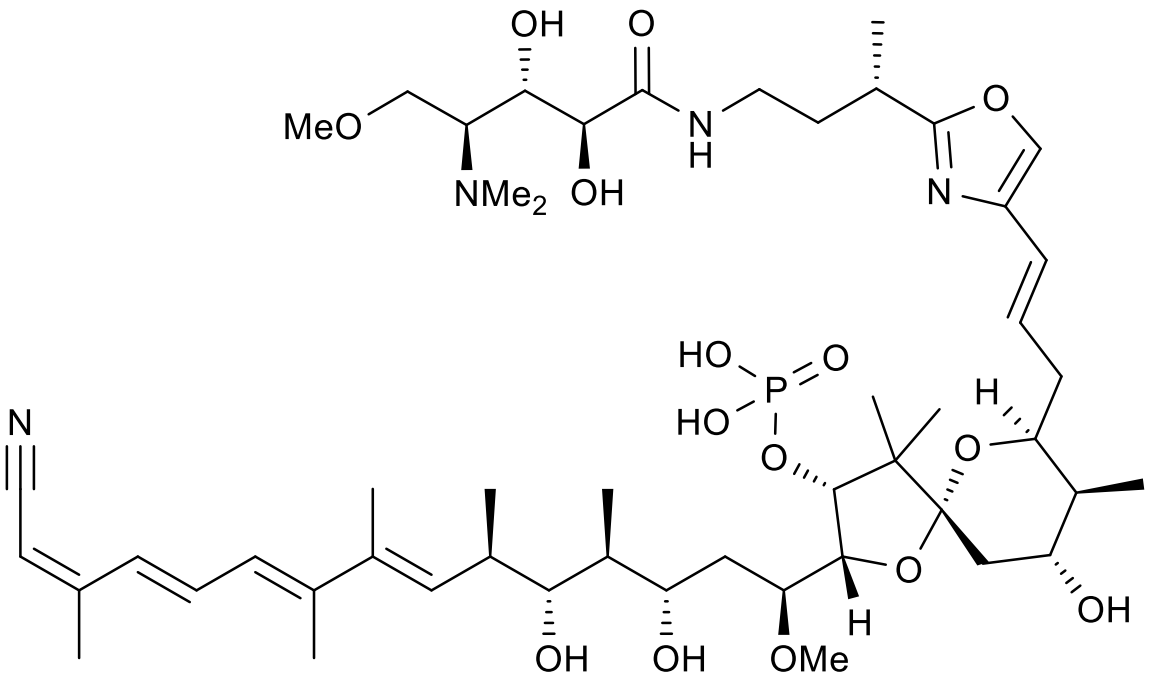

calyculin A (**3**)
*Candidatus* Entotheonella sp.; symbiont of the sponge *Discodermia calyx*

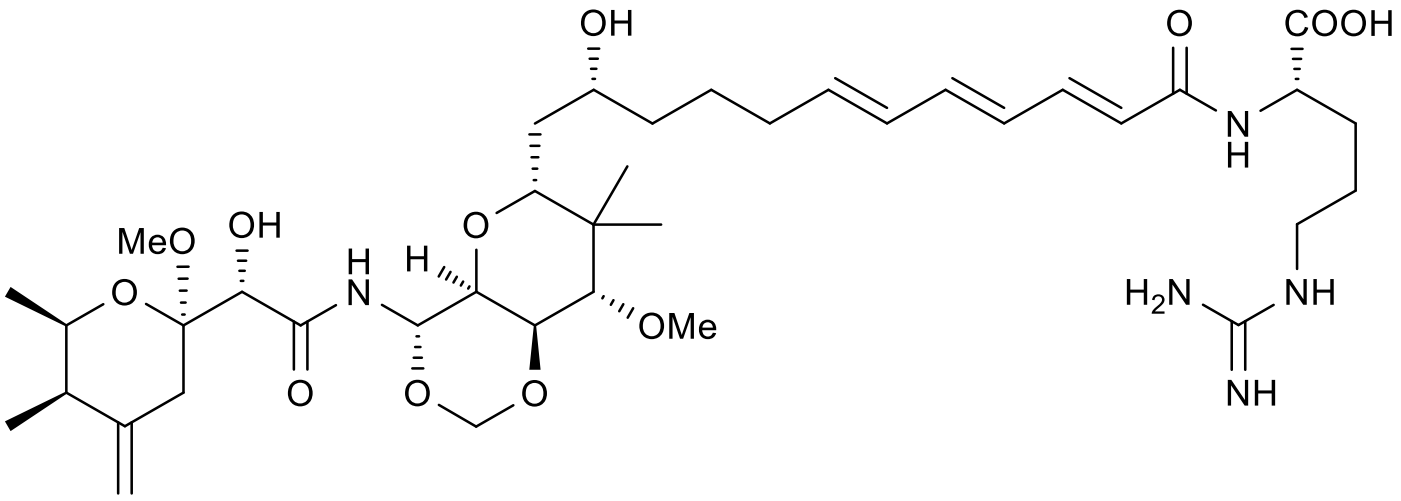

onnamide A (**4**)
*Candidatus* Entotheonella sp.; symbiont of the sponge *Theonella* sp.

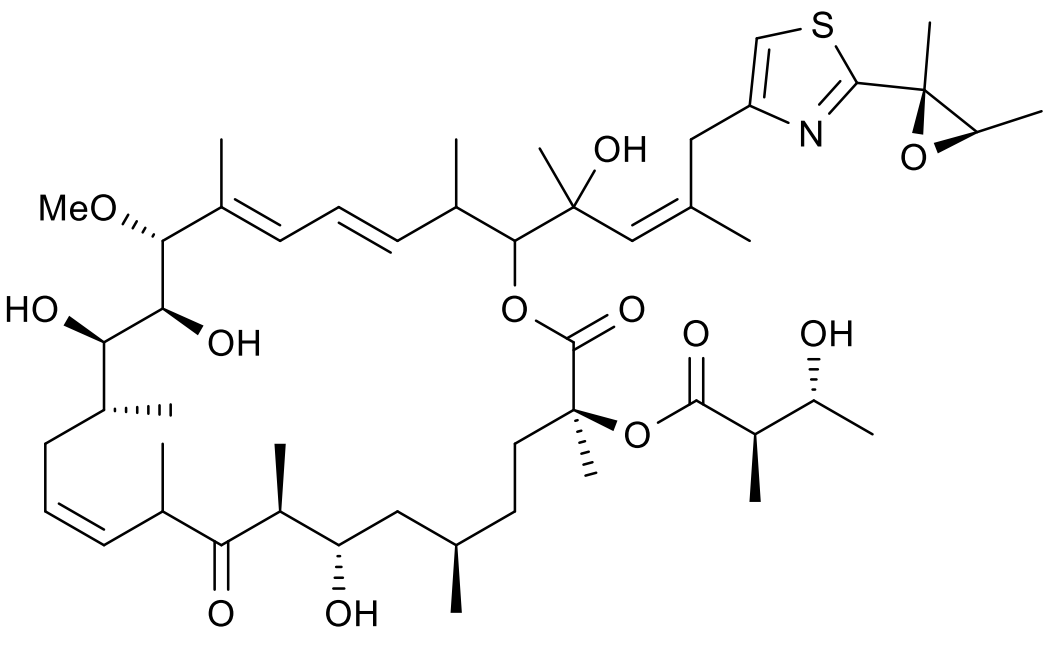

patellazole B (**5**)
*Candidatus* Endolissoclinum faulkneri; symbiont of the tunicate *Lissoclinum patella*

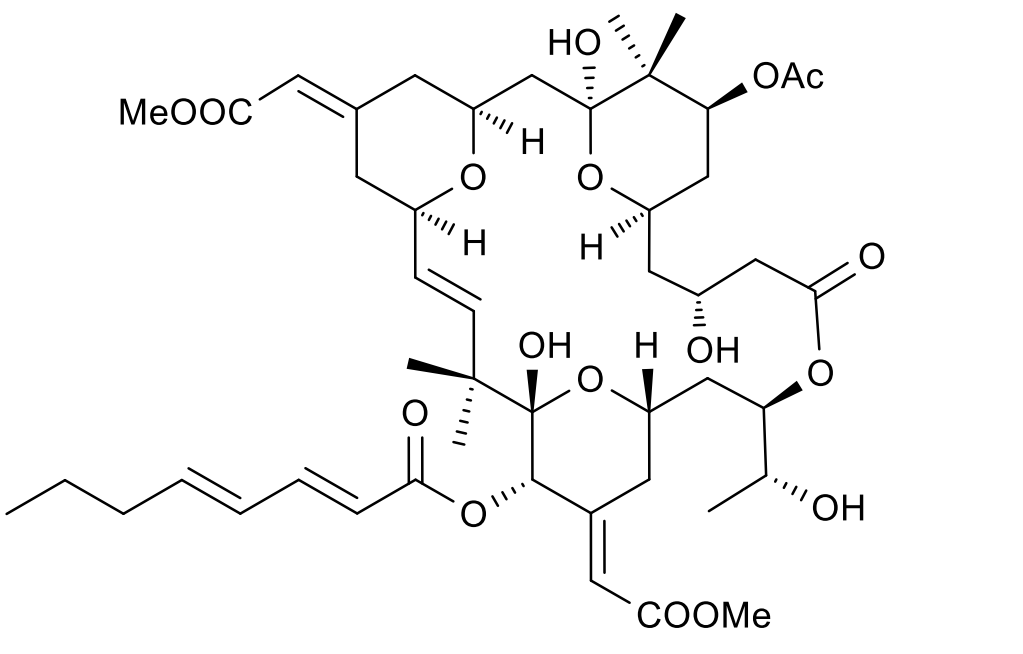

bryostatin 1 (**6**)
*Candidatus* Endobugula sertula; symbiont of the bryozoan *Bugula neritina*

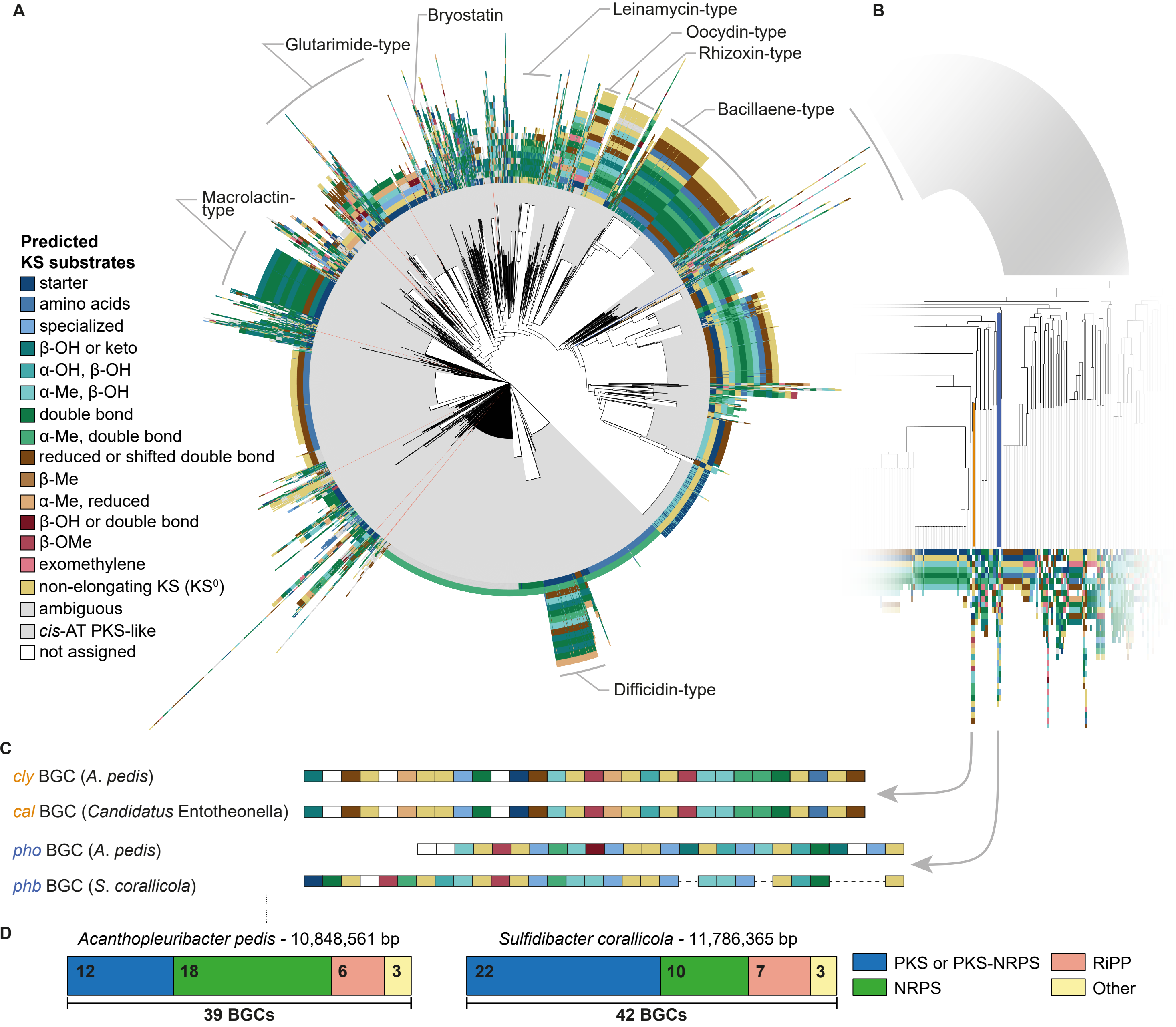

**A**

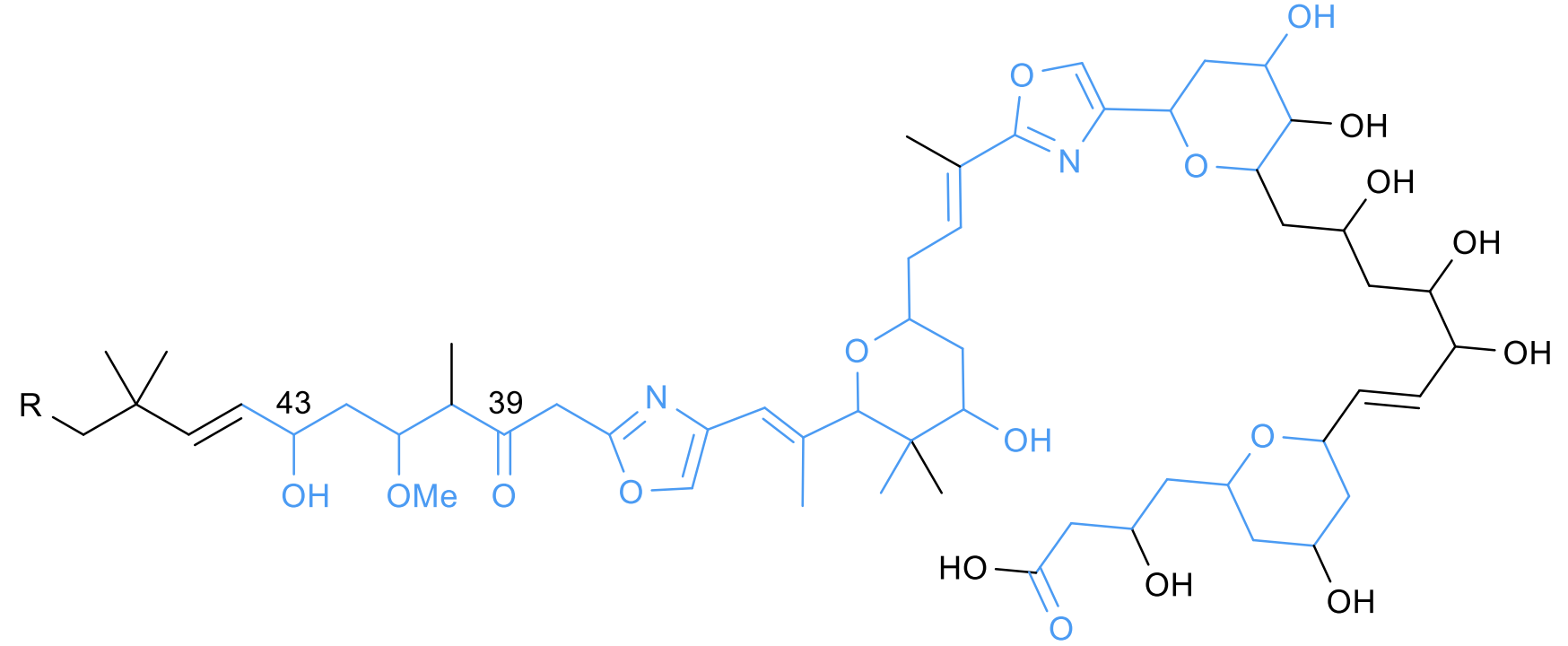

automated TransATor prediction of PhoA-E

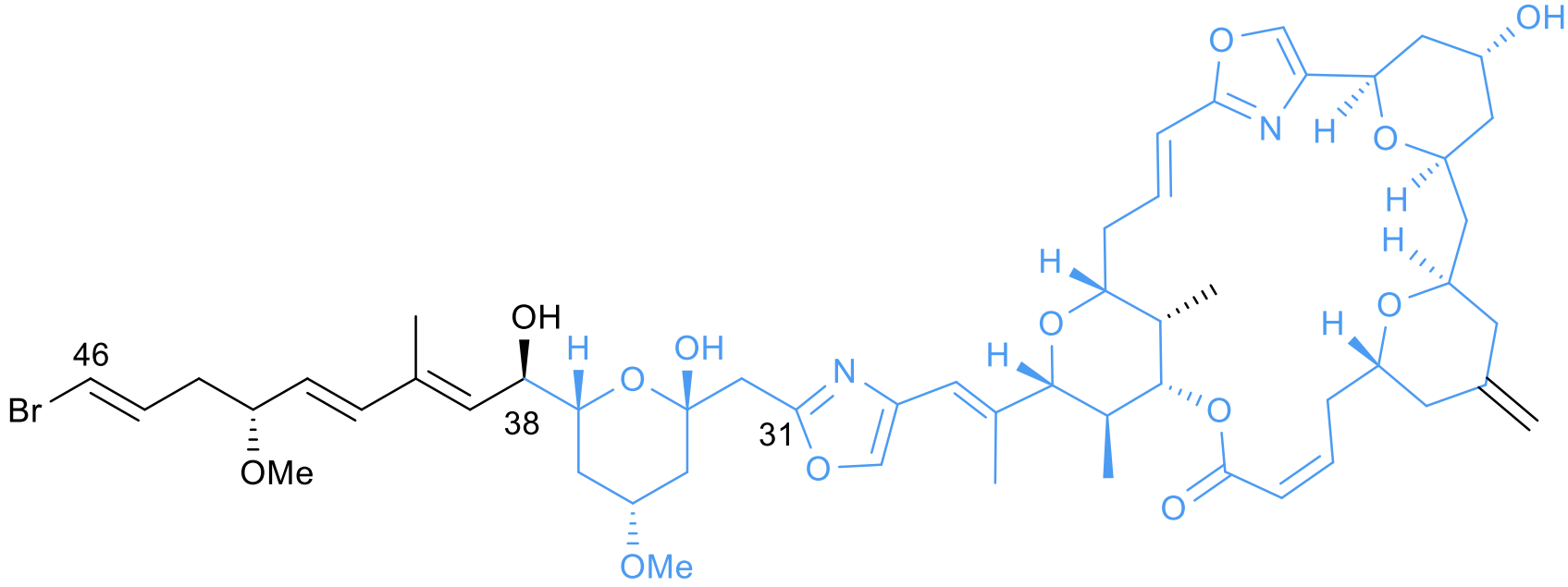

phorboxazole A (**1**)

**B**

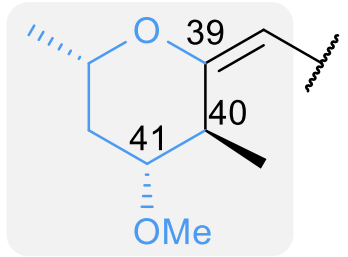

phorbactazole E (**11**)

R¹ = CONH$_2$ R² = D-glucose

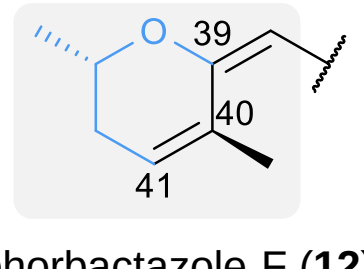

phorbactazole F (**12**)

R¹ = CONH$_2$ R² = D-glucose

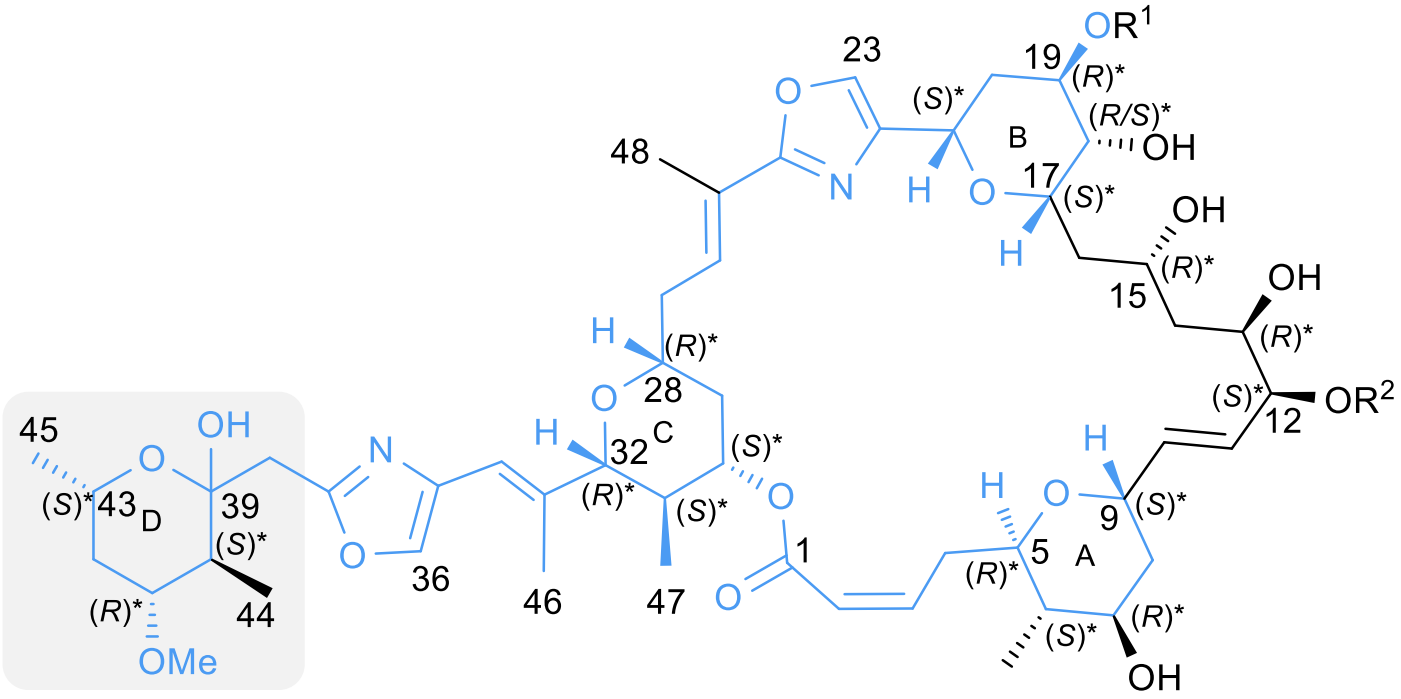

| | | |
|---|---|---|
| phorbactazole A (**7**) | R¹ = CONH$_2$ | R² = H |
| phorbactazole B (**8**) | R¹ = H | R² = H |
| phorbactazole C (**9**) | R¹ = CONH$_2$ | R² = D-glucose |
| phorbactazole D (**10**) | R¹ = H | R² = D-glucose |

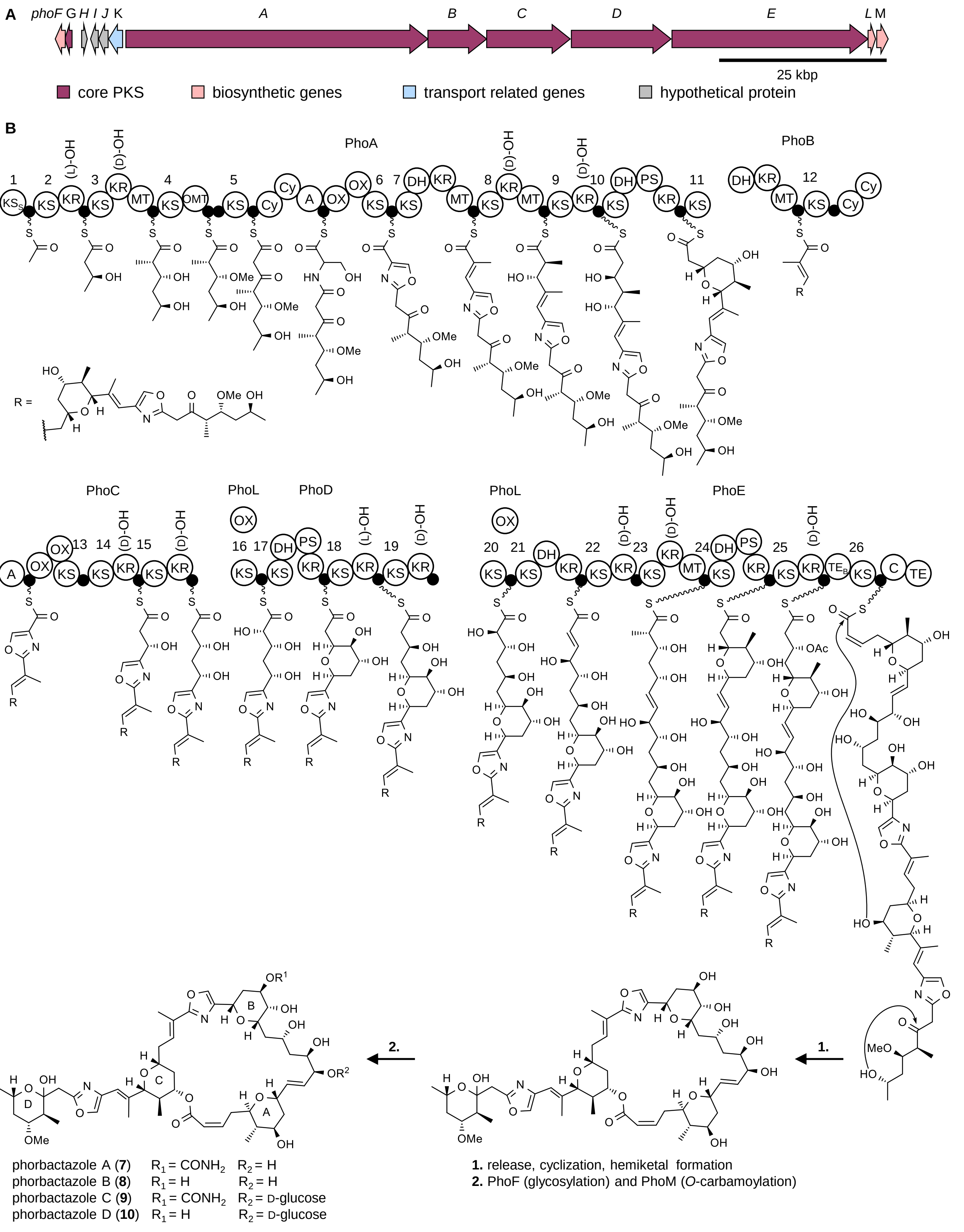
A
phoF G H I J K
A
B
C
D
E
L M
25 kbp
core PKS
biosynthetic genes
transport related genes
hypothetical protein
B
PhoA
PhoB
PhoC
PhoL
PhoD
PhoL
PhoE
R =
1. release, cyclization, hemiketal formation
2. PhoF (glycosylation) and PhoM (O-carbamoylation)
phorbactazole A (7) R1 = CONH2 R2 = H
phorbactazole B (8) R1 = H R2 = H
phorbactazole C (9) R1 = CONH2 R2 = D-glucose
phorbactazole D (10) R1 = H R2 = D-glucose

**A**

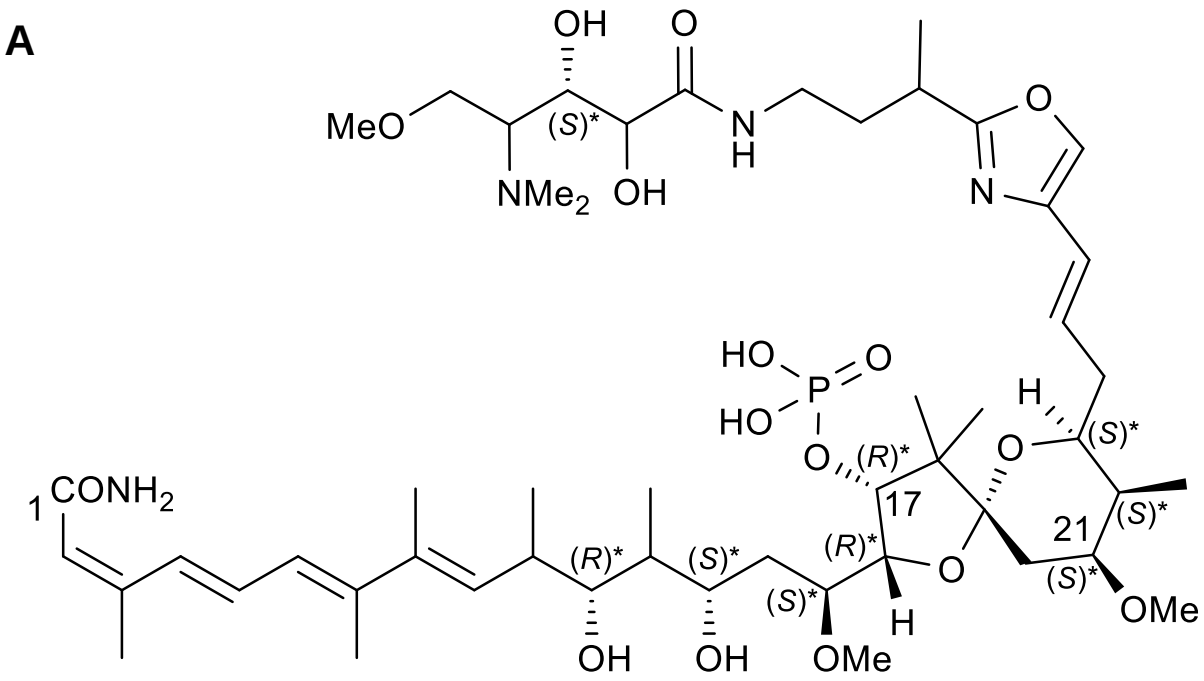

21-methoxy calyculinamide A (**13**)
isolated from *Acanthopleuribacter pedis*

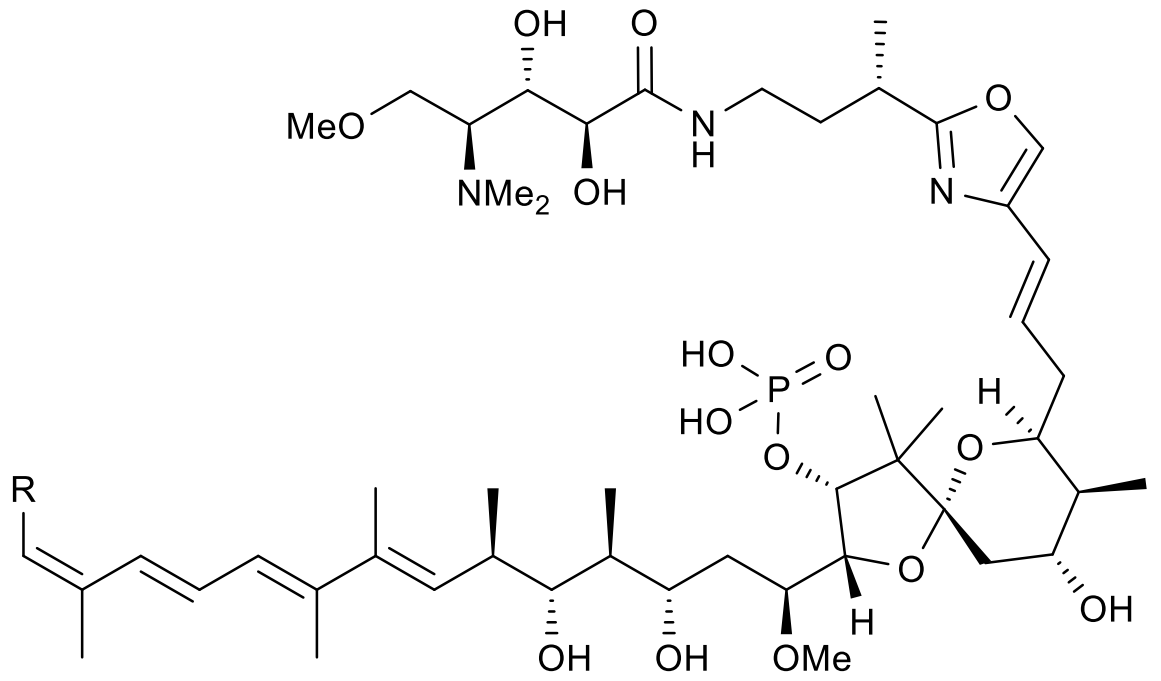

calyculin A (**3**) R = CN
calyculinamide A (**14**) R = $CONH_2$
isolated from the sponge *Discodermia calyx*

**B**

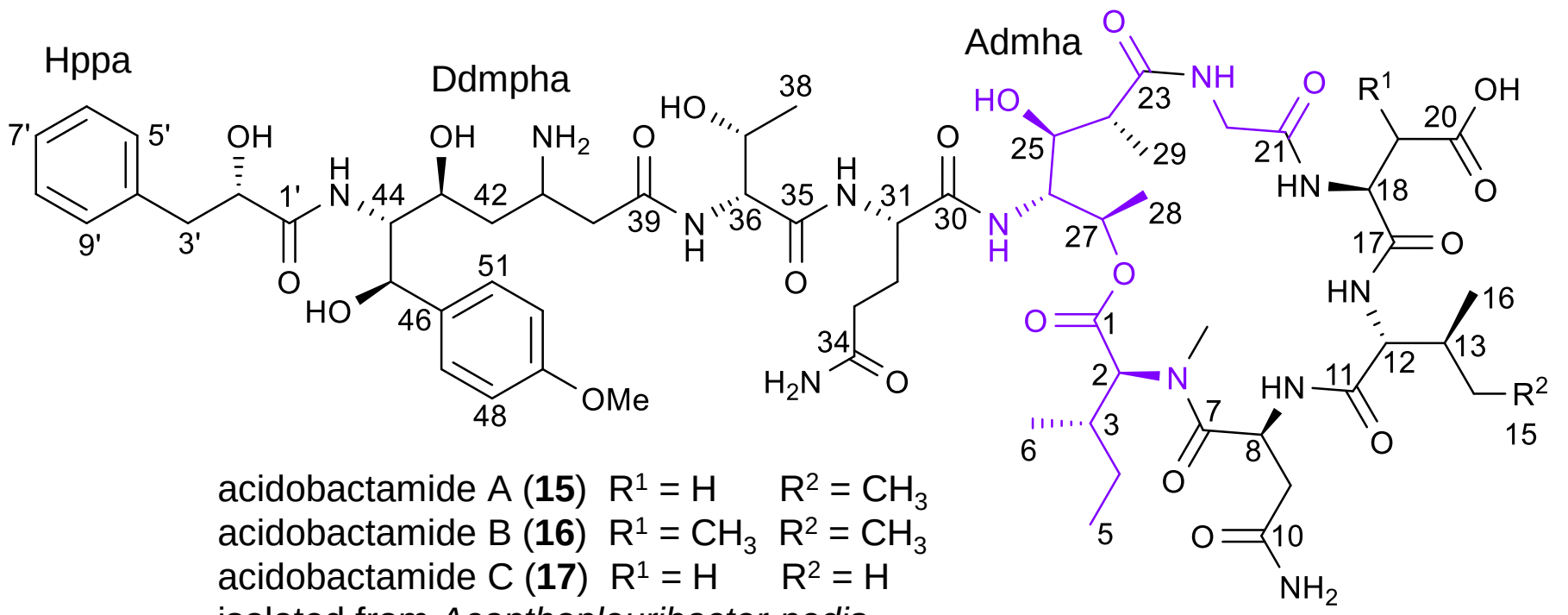

acidobactamide A (**15**) $R^1$ = H $R^2$ = $CH_3$
acidobactamide B (**16**) $R^1$ = $CH_3$ $R^2$ = $CH_3$
acidobactamide C (**17**) $R^1$ = H $R^2$ = H
isolated from *Acanthopleuribacter pedis*

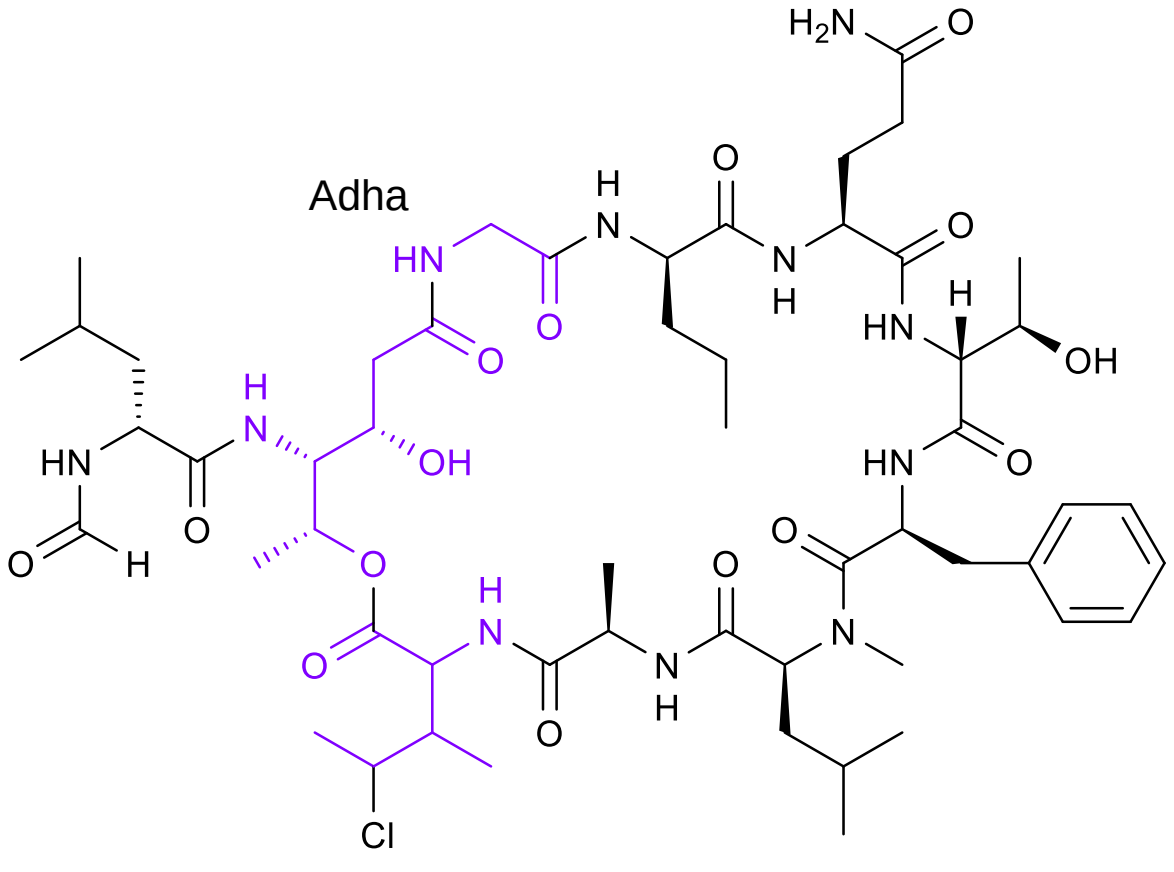

cyclolithistide A (**18**)
isolated from the sponges *Discodermia japonica* and *Theonella swinhoei*

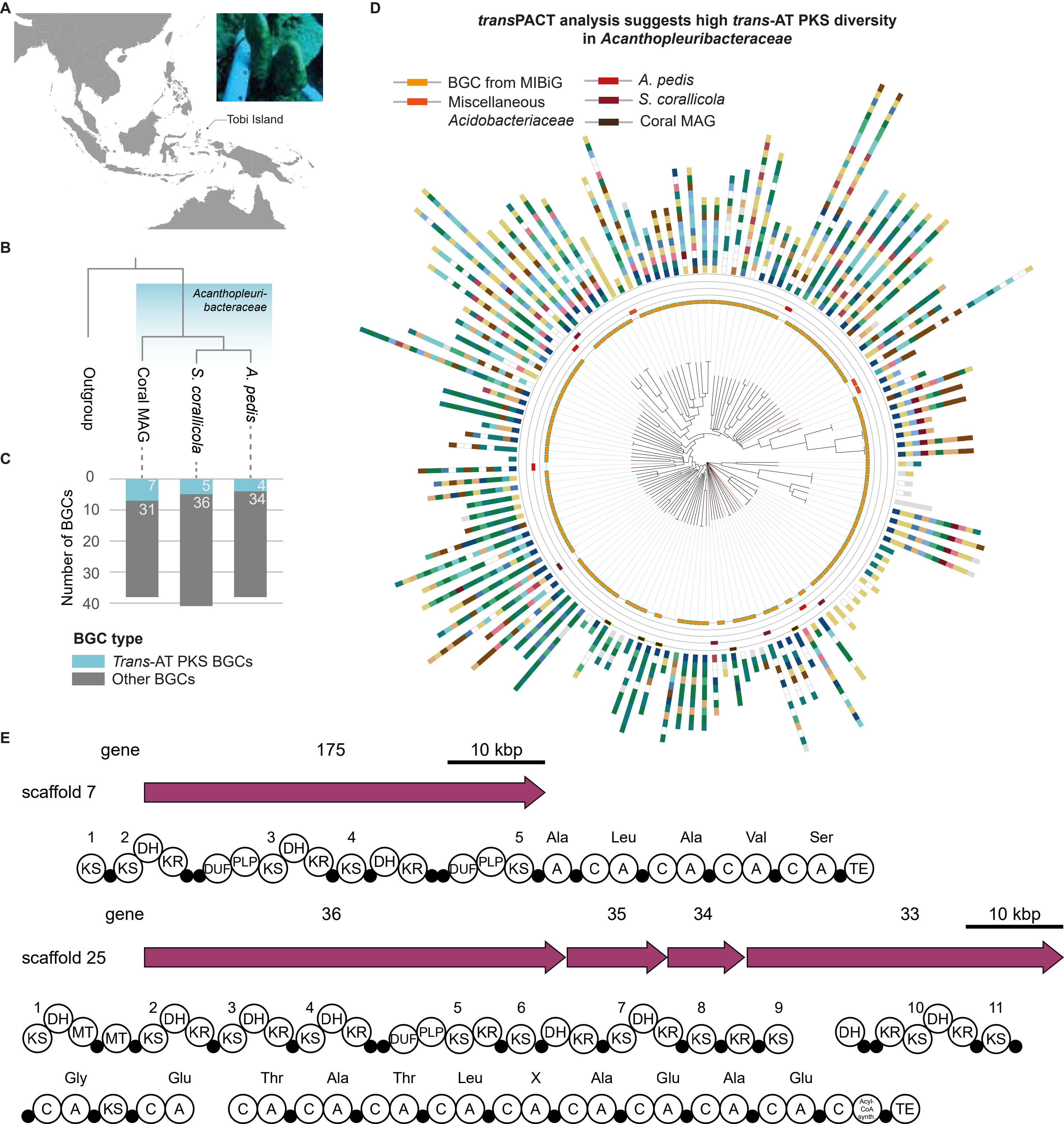
A
Tobi Island
B
Acanthopleuri-bacteraceae
Outgroup
Coral MAG
S. corallicola
A. pedis
C
Number of BGCs
0
10
20
30
40
7
31
5
36
4
34
BGC type
Trans-AT PKS BGCs
Other BGCs
D
transPACT analysis suggests high trans-AT PKS diversity in Acanthopleuribacteraceae
BGC from MIBiG
Miscellaneous Acidobacteriaceae
A. pedis
S. corallicola
Coral MAG
E
gene
175
10 kbp
scaffold 7
1 2 3 4 5 Ala Leu Ala Val Ser
KS KS DH KR DUF PLP KS DH KR KS DH KR DUF PLP KS A C A C A C A C A TE
gene
36
35
34
33
10 kbp
scaffold 25
1 2 3 4 5 6 7 8 9 10 11
KS DH MT MT KS DH KR KS DH KR KS DH KR DUF PLP KS KR KS DH KR KS DH KR KS KR KS DH KR KS DH KR KS
Gly Glu Thr Ala Thr Leu X Ala Glu Ala Glu
C A KS C A C A C A C A C A C A C A C A C A C A C Acyl-CoA synth. TE

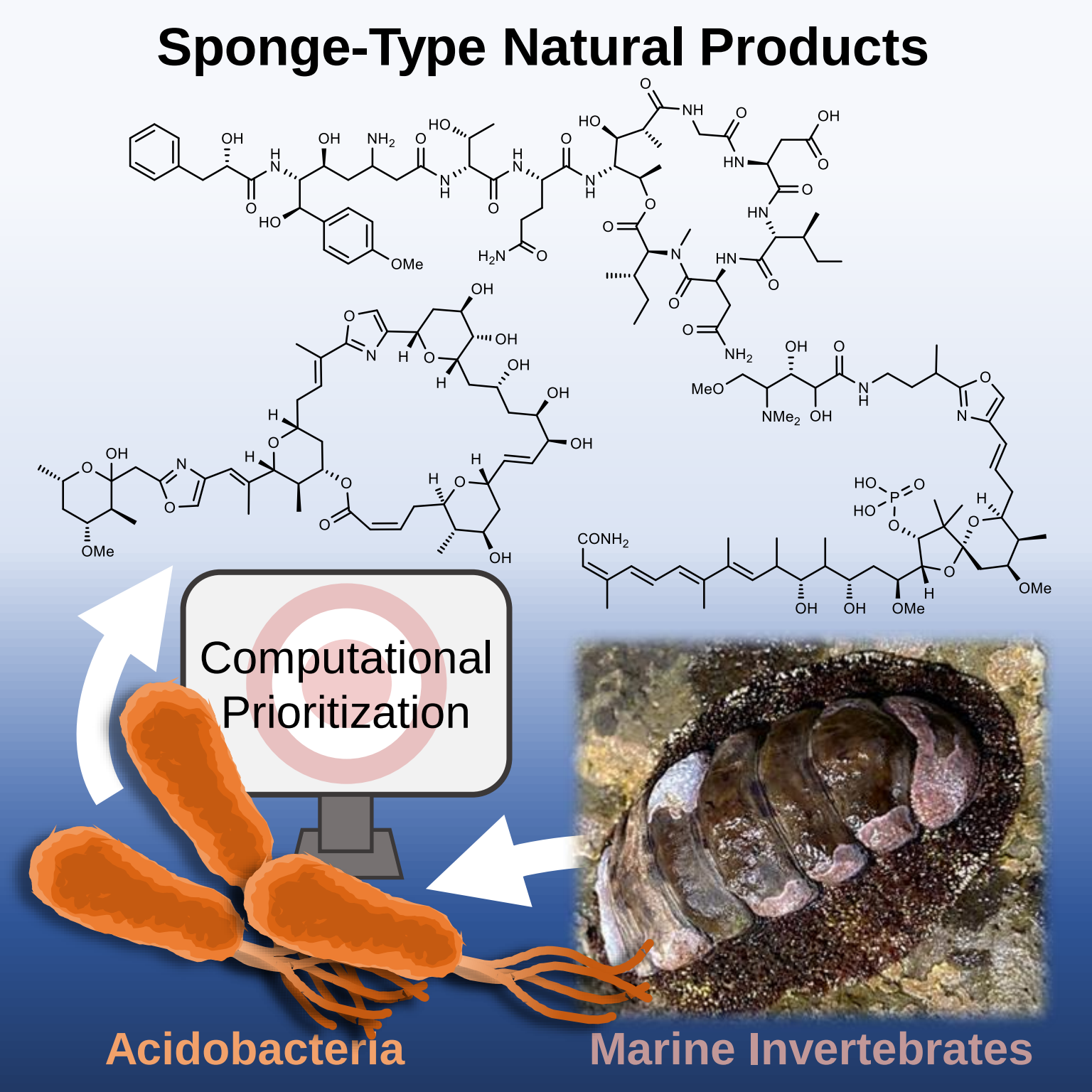
Sponge-Type Natural Products
Computational Prioritization
Acidobacteria
Marine Invertebrates

# Supplemental Information

## Contents

## List of Tables

## List of Figures

## Supplemental Experimental Procedures

### General materials and methods

Heated electrospray ionization high resolution tandem mass spectrometry (HESI-HRMS/MS) was performed on a Thermo Scientific Q Exactive or LTQ Orbitrap XL mass spectrometer coupled to a Dionex Ultimate 3000 HPLC system operated by Xcalibur 4.1 and Chromeleon Xpress 7.2 (Thermo Scientific) respectively (column: Kinetex 2.6 µm XB-C18 100 Å 150x4.6 nm; solvent A ($H_2O$ + 0.1% formic acid); solvent B (MeCN + 0.1% formic acid); gradient, 5% B for 0.5 min increasing to 100% B in 17 min and then maintaining 100% B for 2 min; flow rate 0.7 mL/min; 27 °C UV-Vis detection 200–600 nm combined with *m/z* detection (scan range 100–2500 *m/z*, capillary voltage 4500 V, dry temperature 200 °C). NMR spectra were recorded on a Bruker Avance III spectrometer equipped with a cryoprobe at 500 MHz and 600 MHz for $^{1}H$ NMR and 125 MHz and 150 MHz for $^{13}C$ NMR at 298 K operated by TopSpin 3.5/4.1 (Bruker). Chemical shifts were referenced to the solvent peaks at $\delta_H$ 3.31 ppm and $\delta_C$ 49.15 ppm for methanol-$d_4$, and $\delta_H$ 1.94 ppm and $\delta_C$ 118.2 ppm for acetonitrile-$d_3$. Data were analyzed using Xcalibur 4.1, TopSpin 4.1 and MestReNova (Mestrelab Research). Purification was achieved on Agilent Infinity 1260 HPLCs operated by Open Lab CDS 2.2 (Agilent). Eukaryotic cell lines were not authenticated as they were used to test cytotoxic effects of compounds.

### Global analysis of *trans*-AT PKSs and BGC tree generation

For the global analysis of *trans*-AT PKS BGC architectures, all bacterial genomes with a publication date between January 1st 2021 and August 31st 2021 were downloaded (query: "bct"[Division]AND("2021/01/08"[Publication Date] : "2021/08/01"[Publication Date], 930,534 entries, of which the whole genome shotgun sequencing projects were downloaded on the fly). For BGCs deposited before 2021, all clusters in the MiBIG[1] database (2,516 clusters) and all entries from the antiSMASH database with either a *trans*-AT PKS or *trans*-AT PKS-like BGC type (antiSMASH categories, 2,410 clusters) were downloaded and analyzed with antiSMASH 5.1[1,2] using the prodigal gene-finding tool.[3] Regions in which a *trans*-AT PKS was predicted to occur were saved and used for later analysis.

The bacterial genomes published between August 31st 2021 and June 1st 2022 were obtained by downloading newly published genomes from Genbank on a daily basis. These records were filtered to include only records of the BCT, UNA, GSS, HTG, HTC, and ENV divisions, have a length of more than 1,000 bases and a non-eukaryotic or non-viral taxonomy. After filtration, the genomes were similarly analyzed with antiSMASH 5.1 using the prodigal gene-finding tool.

After antiSMASH analysis, all clusters predicted to contain *trans*-AT PKS BGCs were submitted to a *trans*PACT analysis.[4] In this analysis, the KS sequences were extracted from PKSs that did not contain domains annotated as PKS_AT by antiSMASH and had two or more KS domains, and their substrate specificity was predicted using the KS specificity prediction pipeline of transPACT[4] (substrate_from_faa.py). In total, we analyzed 36,131 KS sequences from 5,289 BGCs. TransPACT's substrate_from_faa.py was run on batches of ten KS sequences at the time. We noted that changing the batch size or composition might alter the alignments constructed by transPACT and result in KSs being placed in slightly different clades. From the 5289 clusters parsed, 5256 clusters contained at least one assigned KS. These 5256 cluster architectures were then used to construct the phylogenetic tree with ete3[5] of the *trans*-AT PKS BGCs using the dendrogram pipeline in transPACT (generate_dendrogram_userweights.py). A wrapper script to streamline and connect the two processes is provided on Github (https://github.com/mathijs-m/transPACT_wrapper). Besides the two Acanthopleuribacteraceae genomes, four additional acidobacterial genomes (NZ_LMUD01000039, JADCJL010000352.1, JADCJL010000127.1, JADCJL010000354.1), all of them belonging to members of the family Acidobacteriaceae, contain *trans*-AT PKS BGCs (Supplementary Material). With the additional *de novo*-constructed MAG (see further below) this adds up to a total of seven acidobacterial genomes containing *trans*-AT PKS BGCs, from a total of 40 identified Acidobacteria amongst the >$10^6$ analyzed bacterial genomes.

To improve visibility in Figure 2 multiple clades, as reported by Helfrich *et. al.*[6], were combined ("starter": clades combined 6, 7, 8, 9, 10, 12, 41, 42, 45, 92, 104; "amino acid": clades combined 1, 2, 20, 30; "bOH|keto": clades combined 3, 4, 22, 29, 32, 37, 50, 51, 54, 55, 58, 59, 71, 78, 89, 96, 100, 111, 112; "specialized": clades combined 11, 23, 24, 25, 39, 44; "reduced|shifted_double_bonds": clades combined

5, 12, 15, 16, 18, 26, 38; “ambigious”: clades combined 13, 14, 17, 67, 120, no clade; “aMe_red”: clades combined 27, 90, 93, 95; “exometh”: clades combined 87, 88; “double bond”: clades combined 28, 31, 43, 62, 63, 64, 65, 66, 68, 69, 70, 91, 105, 106, 107; “aMe_double_bond”: clades combined 46, 47, 48, 82, 84; “bMe”: clades combined 97, 98; “aMe_red”: 27, 90, 93, 95; “bOMe”: 81, 83; “non_elongating”: clades combined 33, 34, 35, 36, 49, 60, 61, 72, 73, 74, 75, 79, 86, 98, 108).

### *Trans*-AT PKS-based prediction of phorbactazoles

For a first structural proposal, the concatenated sequences of PhoA-E were submitted to the automated prediction tool TransATor (transator.ethz.ch, Figure S1)[4]. For manual improvement including new PKS modules not yet implemented in TransATor[4,7,8], extracted amino acid sequences of all KS domains from *trans*-AT PKSs from *Acanthopleuribacter pedis* were aligned with 1,069 previously annotated KS domain sequences (Helfrich *et al*.[9] and in house data). Alignments were performed using the MUSCLE algorithm with default settings[10]. A phylogenetic tree was constructed using the default settings of FastTree version: 2.1.10 +SSE3 +OpenMP (16 threads)[11]. The types of intermediates accepted by KS domains in the PKS proteins were inferred from functionally assigned KSs in the same or phylogenetically close clades (Table S1).

### Optimization of growth conditions for *Acantopleuribacter pedis*

*A*. *pedis* FYK2218 was purchased from the German Collection of Microorganisms and Cell Cultures GmbH (DSMZ 28897). Bacteria were cultivated at 28 °C in 100 mL Erlenmeyer flasks containing 20 mL of one of the following liquid media: self-made marine broth (MB) made of 5 g bacteriological peptone, 1 g yeast extract, 33 g instant ocean (Aquarium Systems), 1 L water; self-made marine broth with only 16.5 g instant ocean per L water (½ MB); PH-103[12] medium made of 20 g tryptone, 5 g yeast extract, 4 g glucose, 4 g maltose, 4 g $CaCO_3$, 27 g instant ocean; and purchased marine broth (DIFCO 2216). MB and PH-103 were also tested at 30 °C, resulting in a total of six tested growth conditions (Figure S2). Following cultivation, ethyl acetate extracts of the culture supernatants were analyzed by HPLC-HESI-HRMS. Cultures grown in MB at 28 °C showed the most diverse total ion chromatogram (Figure S2).

### Large-scale cultivation of *A. pedis* and isolation of phorbactazoles

50 liquid cultures of *A*. *pedis* were grown in 400 mL self-made MB in 1 L flasks (total volume 20 L) at 28 °C for three days at 150 rpm. The pellet and supernatant from the 20 L-scale cultivation were separated via centrifugation. The pellet was extracted with 4 × 500 mL of acetone in an ultrasonic bath for 30 min. The extracts were combined and concentrated by rotary evaporation. The remaining water phase of the extract was suspended in 500 mL of distilled water, extracted with 3 × 500 mL of ethyl acetate, and filtered through anhydrous sodium sulfate. The resulting ethyl acetate extracts were evaporated to dryness, leaving a brown solid. The supernatant was extracted with an equal volume of ethyl acetate. The organic phase was filtered through anhydrous sodium sulfate and evaporated to dryness.

The supernatant and pellet crude extracts were fractionated using a preparative reverse phase (RP)-HPLC (Agilent 1260) Infinity system, equipped with a Phenomenex Luna 5µ C18, φ 20 x 250 mm column. Deionized water (Milli-Q, Millipore) (solvent A) and acetonitrile (solvent B) were used as the mobile phase. An elution gradient of 5–100% solvent B in 40 min followed by isocratic conditions at 100% solvent B for 10 min was applied. UV detection was carried out at 210, 254, and 280 nm. HPLC-HESI-HRMS analysis of the fractions collected allowed the identification of the fractions containing the compounds of interest.

To further purify the fractions containing phorbactazoles, MS-guided repetitive HPLC purification was performed by semi-preparative RP-HPLC (Agilent 1260 Infinity system, equipped with a Phenomenex Kinetex 5 µm C18 10 x 250 mm column). Phorbactazoles A (**7**) (1.2 mg) and B (**8**) (0.1 mg) were obtained by purification of fraction F-20 by reverse phase LC (solvent A/solvent B), elution gradient 30–65% solvent B for 30 min followed by gradient shift from 60 to 100% in 5 min, and finally isocratic condition at 100% solvent B for 5 min.

Fractions 17 and 18 which contained phorbactazoles C-F (**9-12**) were purified applying the same principle as the two compounds mentioned above. Elution gradient, 40–65% solvent B for 30 min was applied, followed by gradient shift from 65 to 100% in 5 min and isocratic condition at 100% solvent B for 5 min to afford 1.2 mg, 0.2 mg, 0.7 mg, and 0.9 mg of the four molecules respectively.

## Acid hydrolysis of phorbactazole C (9)

Phorbactazole C (**9**) (1 mg) was hydrolyzed with 10% aqueous HCl (1 mL) at 90 °C for 12 h. The reaction mixture was then diluted with 10 mL of $H_2O$ and extracted with equal volume of ethyl acetate. The aqueous layer was concentrated under vacuum (0.3 mg). The CD spectra of the aqueous phase was compared with that of D-glucose standard.

## Structure elucidation of phorbactazoles (7-10)

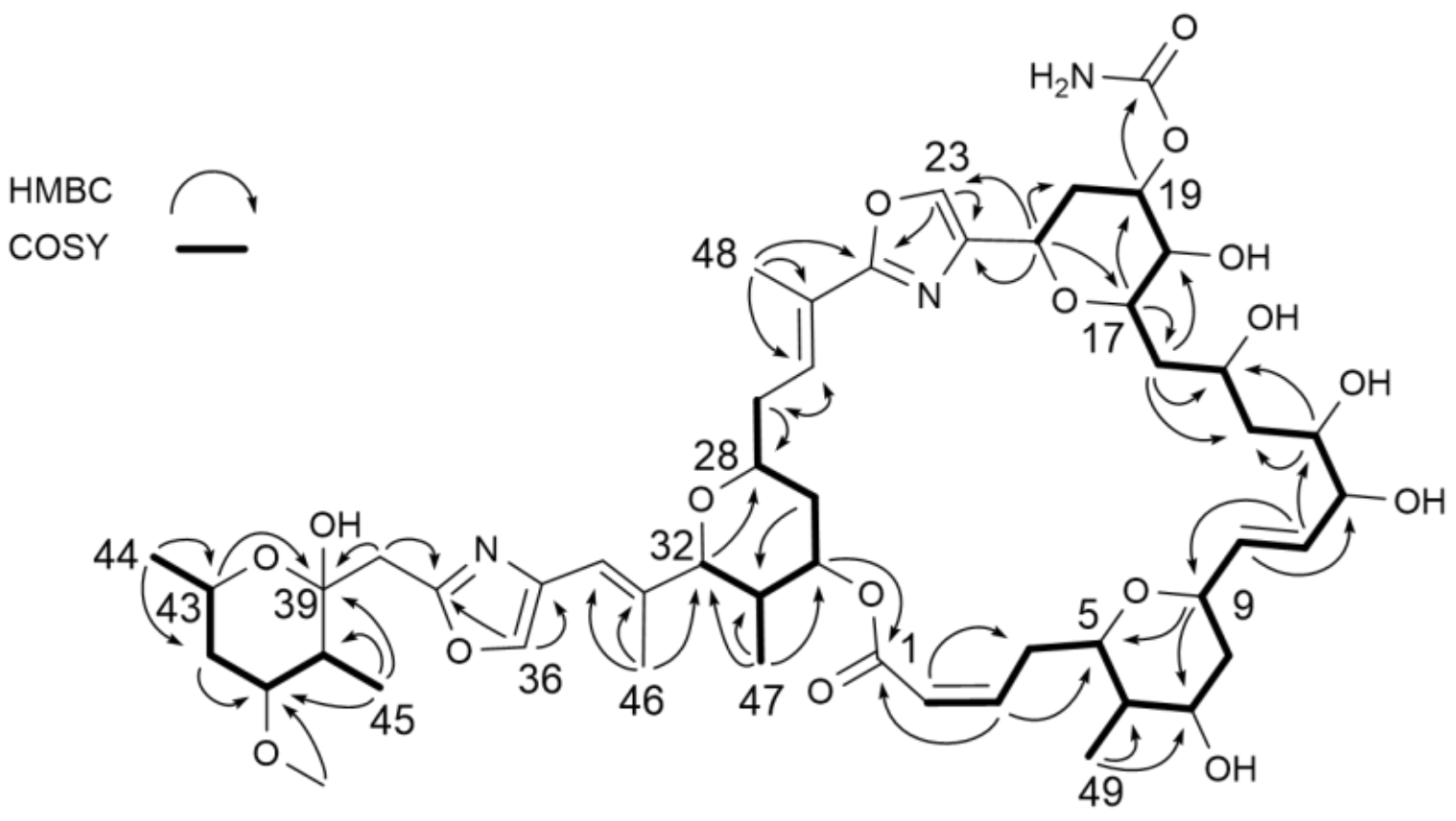

phorbactazole (**7**)

Compound **7** (phorbactazole A) with the molecular formula $C_{51}H_{73}N_3O_{17}$ and 17 degrees of unsaturation deduced from the HR-MS data (Figure S4) was isolated from both the supernatant and bacterial pellet extracts. Four methyl doublets with resonance at δ 0.77 ($H_3$-49), 0.83 ($H_3$-47), 1.11 ($H_3$-44), 1.14 ($H_3$-45) together with two singlets at δ 1.95 ($H_3$-46) and 2.07 ($H_3$-48) were recorded in the $^1$H NMR spectrum. In addition, a methoxy singlet at δ 3.34 (-$OCH_3$), two signals at δ 7.79 (H-23) and 7.84 (H-36) characteristic for oxazoles, and six olefinic proton signals were observed in the $^1$H-NMR spectrum (Figure S5). Analysis of the HSQC data revealed the presence of 26 methines, nine methylene, and seven methyl groups (Figure S6, Table S3).

A total of three spin systems were elucidated from the COSY and HMBC NMR data (Figure S7-Figure S8). The first and the longest spin system in the molecule which was established from the COSY data spanned from H-2 to H-21. These COSY correlations along with the HMBC correlations of H-2 to C-1/C-4, $H_3$-49 to C-5/C-6/C-7, H-10 to C-9/C-11/C-12, H-12 to C-11/C-13/C-14/, H-17 to C15/C-16/C-21 and H-18 to C-19/C-20 established the substructure C-1 to C-21 of the molecule. One of the tetrahydropyran rings in this fragment was established based on HMBC correlations of H-9 to C-5 and a downfield shift of H-9 (δ 4.58). A second tetrahydropyran ring was elucidated by HMBC correlations of H-17 to C-21. The downfield shifts of H-17 (δ 4.12) and H-21 (δ 4.77) further supported these assignments. A carbamoyl moiety attached to C-19 was deduced from HMBC correlations of H-19 to the carbamoyl carbon (δ 160.7) and a downfield shift of H-19 (δ 4.89). The configuration of the olefinic protons H-2 and H-3 was assigned as *Z* based on the small coupling constant recorded ($J$ = 10.9 Hz) between the two protons. The configuration at C-10 to C-11 was assigned as *E* because of a large coupling constant between H-10 and H-11 ($J$ = 16.2 Hz). The second spin system was deduced from COSY correlations of H-28 to $H_2$-27/$H_2$-29, H-30 to $H_2$-29/H-31 and H-31 to H-30/H32/$H_3$-47. Further, cross peaks on the HMBC spectrum were observed for correlations of $H_2$-26 to C-25/C-26/C-28 and of H-31 to C-30/C-32/C-47. Downfield shifts of C-32 (δ 89.1) along with the HMBC correlations of H-28 to C-32 established a tetrahydropyran ring in this spin system. An oxazole ring was elucidated from HMBC correlations of H-23 to C-22 (δ 142.6)/C-24 (δ 164.8). The position of H-23 in the oxazole ring was confirmed from NOESY correlations of H-23 to H-21. The two spin systems were connected through the oxazole ring because of HMBC correlations of $H_3$-48 (δ 2.07) to C-24/C-25/-C-26 and of H-21 to C-22/C-23. Furthermore, the proton H-30 (δ 4.68) showed HMBC correlations to C-1 (δ 167.4), thus establishing the 30-carbon macrocyclic ring. Cross peaks in the HMBC spectrum between H-36 (δ 7.84) and C-35/C-37 established the presence of the second oxazole moiety in the molecule. This moiety was connected to the rest of the molecule through C-33 (δ139.0) and C-34 (δ 119.8) because of HMBC correlations of $H_3$-46 (δ 1.95) to C-32/C-33/C-34 and H-34 to C-32/C-36/C-46. NOESY correlations of H-34 to $H_3$-46 ascertained the position of C-34 in the oxazole ring. The upfield shifts of C-48 (δ 13.2) and C-47 (δ 14.2) indicated that the two olefinic bonds were *E*-configured. Furthermore, cross peaks in the NOESY spectrum were recorded between $H_3$-46 and H-36, H-34, and H-32, and between $H_3$-48 and $H_2$-27 (Figures 3B and Figure S9). The third spin system was elucidated by COSY correlations of H-40 to $H_3$-45/H-41, of $H_2$-42 to H-41/H-43, and of H-43 to H-42/$H_3$-44. These assignments were further supported by HMBC correlations of $H_3$-44 to C-42/C-43, of $H_3$-45 to C-39/C-40/C-41, and of $H_2$-42 to C-41/C-43/C-44. A methoxy group (δ 56.4) was attached to C-41 (δ 79.9) based on HMBC correlation of the methoxy protons (δ 3.34) to this carbon. This moiety of the molecule was connected to the rest of the molecule through C-38 (δ 39.5) as HMBC correlations of the C-38 diastereotopic protons to C-37/C-39/C-40 were recorded, hence completing the structure of **7** (Figure S10).

A total of 18 stereocenters were identified in phorbactazole A (**7**). The relative configuration (Figure 3) was assigned based on both the NOESY and genomic data (Figure S11, Table S1). A network of NOESY correlations recorded of H-30 to H-32/$H_3$-47/ $H_2$-29 (δ 2.12), H-32 to H-30/H-34/ $H_3$-47 and H-28 to H-30/ $H_2$-29 (δ 2.12) indicated the orientation of these protons on the same plane. $H_3$-46 on the other hand exhibited NOESY correlations to H-31/H-36/ $H_2$-29 (δ 1.31), and H-31 to $H_3$-46/$H_2$-29 (δ 1.31), and hence was concluded to be on the opposite plane. The large coupling constants of H-32 (*J*=10.3 Hz) and H-30 (*J*=10.6 Hz) indicated the axial orientation of these protons with reference to H-31 (Figure S11C) and hence 28*R**, 30*S**, 31*R**, 32*R** was assigned as the relative configuration (noted with *).

In ring B, NOESY correlations were observed for H-21 to $H_2$-20 (δ 1.76)/H-17 while H-19 showed correlations to H-18 and the H-20 diasterotopic protons (Figure S11B). Axial arrangement of H-21 relative to $H_2$-20 (δ 2.37) was assigned based on the large coupling constant (*J*=12.3 Hz) recorded between these protons and consequently, the relative configuration at C-21 was assigned as *S**. The small coupling constant recorded between H-18 and H-17/H-19 (3.4 Hz) together with the NOESY correlation of H-18 to both H-17 and H-19 placed H-19 and H-18 on the equatorial position. Therefore, the relative configuration in ring B of **7** was assigned as 17*S**, 18*R**, 19*R**, 21*S**.

A network of NOESY cross peaks correlations of H-10 to H-12, H-13 to $H_2$-14 (δ 1.76) / H-12/$H_2$-16 (δ 1.51) and H-17 to $H_2$-16 (1.51) placed these protons on the same side of the plane. Conversely, NOESY correlations of H-15 to $H_2$-14 (δ 1.47)/$H_2$-16 (δ 1.85) were recorded. The idea was to use the stereochemistry at C-17 as a reference to this spin system. This will infer to the relative stereochemistry along this spin system as 12*S**, 13*R** and 15*R**. However, artifact signals cannot be ruled out for NOESY cross peaks observed between H-15 (δ 4.02) and H-17 (δ 4.12) because of the small difference between these two signals' chemical shifts and the proximity to the diagonal. Also, unambiguous assignment could not be deduced because the resonances of three signals were overlapping in the 3.99-4.05 ppm range. Thus, the stereochemistry at C-12, C-13, and C-15 could not be assigned based on the available NOESY data. The stereochemistry on these three centers was therefore assigned based on the bioinformatic KR domain-based prediction[4] of the D-hydroxy at C-13 and L-hydroxy group for C-15 (Table S1). The C-12 hydroxy group is installed by hydroxylation and could therefore not be predicted using the KR method. However, Since NOESY correlations were recorded between H-12 and H-13, the 12*S** was inferred for the C-12 stereocenter.

For ring A, correlations of $H_3$-49 to H-5/H-6, H-5/H-7/$H_3$-49/ $H_2$-4 (δ 3.05) and $H_2$-8 (δ 2.07) to H-7 were observed in the NOESY spectrum (Figure S9 and Figure S11). Conversely, NOESY correlations of $H_2$-8 (δ 1.70) to H-6/H-9 and H-9 to H-10/ $H_2$-8 (δ 1.70)/ $H_2$-8 (δ 2.07) were observed. Furthermore, D-hydroxy groups predicted for C-7 and C-9 from the genomic data together with the with the large coupling constant of H-7 (J=10.2 Hz) and H-9 resonating as a broad complex multiplet led to the 5*R**, 6*S**, 7*R** and 9*S** assignment as the relative configuration. In ring D, coupling constants of 6.4 Hz between $H_3$-45 and H-40 and 5.9 Hz between $H_3$-44 and H-43 placed the two methyl groups on equatorial position and consequently the axial positions for H-40 and H-43 (Figure S11). The chair conformation completely agrees with the restrictions set by the NOEs between H-43 and H-41 and $H_3$-44 to $H_2$-42 (δ 2.09). Therefore, the relative configuration in ring D was assigned as 39*R**40*S**, 41*R**, 43*S**.

Compound **8** (phorbactazole B) was isolated from the supernatant as a white solid with the molecular formula $C_{50}H_{72}N_2O_{16}$ obtained from the HRMS data (Figure S12). The NMR data of **8** were similar to **7** (Figure S13-Figure S16, Table S3) with the difference being the absence of the carbamoyl moiety (Figure 2B) This was established from the absence of the HMBC correlations of H-19 to the carbamoyl carbon and the upfield shift of H-19 from δ 4.89 in **7** to δ 4.00 in **8**. Also, the mass difference of 43 Da recorded between these two molecules corresponding to carbamoyl confirmed this assignment.

Another congener, phorbactazole C (**9**) (Figure 2B) with the molecular formula $C_{57}H_{83}N_3O_{22}$ and 18 degrees of unsaturation established from the HRMS data was isolated from both the supernatant and the pellet (Figure S17). The NMR data for **9** (Figure S18-Figure S22, Table S4) were similar to compound **7** with the difference being the presence of a glucose moiety in **9**. The glucoside moiety was elucidated from the COSY correlations of H-2' to H-1'/H3', H-4' to H-3'/H-5' and $H_2$-6' to H-5' (Figure S11F). HMBC correlations of $H_2$-6' to C-5', H-3' to C-2'/C-4' and H-5' to C-1' were recorded. The glycoside moiety was attached to the core structure of the molecule through C-12 because of HMBC correlations of H-12 (δ 4.29) to C-1' (δ 101.4) (Figure S21) and the downfield shift of C-12 (δ 82.0) (Figure S19). The relative configuration of the glycoside was assigned by analysis of NOESY correlations (Figure S22) and the coupling constants of this moiety. H-1' signal resonance at δ 4.26 with $J_{1',2'}$ = 7.9 Hz suggested a β-glycoside. According to Giner *et*. *al*. 2016[13], the typical 1'-H signal for a β-glycoside shows a resonance around δ ~4.29 with a coupling constant of ~7.6 Hz, while a signal at δ ~5.2 with coupling constant of ~3.6 Hz is expected in the case of an α-glycoside. A chain of vicinal *trans* couplings of $J_{2',3'}$ = 9.1 Hz, $J_{3',4'}$= 8.9 Hz and $J_{4',5'}$= 9.0 Hz indicated that the methine protons H-1′ to H-5′ all occupy axial positions (Figure S11E and F). To determine the configuration of the glycoside moiety, circular dichroism (CD) of the aqueous phase of the acid hydrolysate of compound **9** was recorded. Comparison of the electronic circular dichroism (ECD) curves of the aqueous phase of the acid hydrolysate with that of a D-glucose gave similar cotton effect in the region 180-192 nm (Figure S23).

Phorbactazole D (**10**), with a molecular formula $C_{56}H_{82}N_2O_{21}$ deduced from the MS data was also isolated from both the supernatant and the pellet (Figure S24). The NMR data of **10** (Figure S25-Figure S28,Table S4) revealed a similar structure as **8** with an additional glycoside moiety attached to C-12 based on the HMBC correlations of 1′-H to C-12.

Phorbactazole E (**11**) with the molecular formula $C_{57}H_{81}N_3O_{21}$ and phorbactazole F (**12**) with molecular formula $C_{56}H_{77}N_3O_{20}$ established from the HRMS data were isolated from the pellet extracts (Figure S29 and Figure S34). The NMR data for **11** (Figure S30-Figure S33,Table S5) were similar to those of **9** with a difference being the olefinic bond between C-38 and C-39. The C-38 methylene signal was missing in the HSQC data and instead a methine peak with $^{13}$C resonance at δ 94.5 and $^{1}$H resonance peak at δ 5.64 was recorded. The HMBC correlations of the H-38 to C-36/ C-39/C-40 confirm the olefinic bond at C-38-C-39. This was further supported by the downfield shift of C-39 (δ 167.2) and the HMBC correlations of $H_3$-45 to this carbon. The structure elucidated for Phorbactazole F (**12**) was similar to that of **11** with the difference being an additional double bond at C-40 to C-41 (Figure S35-Figure S38, Table S5). The methine signals for these two carbons δ 41.8 and δ 81.2 together with the methoxy signal were missing and instead two signals at δ 129.3 (C-40) and δ 132.4 (C-41) were identified in the HSQC spectrum (Figure S36). The structure of ring D in **12** was unambiguously confirmed by the COSY correlations of H-43 to $H_3$-44/$H_2$-42 and H-41 to $H_2$-42 (Figure S37). HMBC correlations of $H_3$-44 to C-33 (o C-43/C-42, H-1 to C-43/C- 45/C-39, $H_3$-45 to C-39/C-40/C-41 and H-43 to C-39 supported the assigned structure for ring D (Figure S38).

The $^{13}$C signals for C-33 (δ141.3), C-34(δ 117.7) and C-35 (δ 141.5) in **12** were slightly shifted compared to their corresponding carbon shifts in **9** and **11.** This could be attributed to the new conjugated spin system (C-33 to C-41) established by the introduction of two additional double bonds. It is also worth to mention the instability of the ring D in phorbactazoles under acidic conditions. When isolation was carried out with solvents with 0.05 % TFA, additional peaks which could have resulted from isomerization in ring D were recorded. Downfield shift of methyl protons for $H_3$-44 (from δ 1.11 to δ 1.17) and the upfield shift of $H_3$-45 (from δ 1.14 to δ 1.08) characterized this isomerization. The quaternary carbon C-39 was also shifted downfield (from δ 99.5 to δ 102.5-106.2) (Figure S39)

The relative configuration of **7-12** was assigned using **7** and **9** as the model compounds. The comparison of the CD spectra of **7** with those of **8-12** suggested similar relative configuration for the stereocenters discussed for **7** as the ECD curves of these compounds were comparable in the range of 190–290 nm (Figure S40).

### Large-scale cultivation of *A. pedis* and 21-methoxy-calyculinamide (13) isolation

For a total volume of 12 L, 36 cultures of *A. pedis* were grown in 330 mL homemade MB in 1 L baffled flasks at 28 °C for three days at 150 rpm. Cells were harvested by centrifugation. The supernatant was extracted three times with an equal volume of ethyl acetate and the solvent was removed from the organic phase by evaporation. The crude extract was dissolved in 5 mL methanol and fractionated by preparative RP-HPLC (Phenomenex Luna 5µ C18, φ 20 x 250 mm, 15.0 mL/min, 364 nm, 280 nm, 254 nm) using solvents, A ($H_2O$ + 0.1% formic acid) and B (acetonitrile + 0.1% formic acid) with a gradient elution: 5 min 5% B, 37 min 95% B, 47 min 95% B, 47.1 min 5% B. Fractions collected between minute 34 and 36 were further purified to yield 21-methoxy-calyculinamide (**13**). **13** was further purified by RP-HPLC (Phenomenex Luna 5µ C18, φ 10 x 250 mm, 2.0 mL/min, 340 nm) with 70% MeCN in $H_2O$ + 0.05% TFA. From the fractions around 10 min 0.2 mg of **13** was isolated and analyzed by NMR.

### Structure elucidation of 21-methoxy-calyculinamide (13)

21-methoxy calyculinamide (**13**) had the predicted molecular formula of $C_{51}H_{86}N_4O_{16}P$ based on HRMS (*m*/*z* 1041.5809 $[M+H]^+$, Δ +3.65 mmu) (Figure S42 and Figure S43). $^1$H NMR in conjunction with HSQC data suggested four doublet methyls, two singlets aliphatic methyls, three vinylic methyls, three methoxy groups, two methyls connected to a amines, six methylene groups, eight protons connected to $sp^2$ carbons, ten methanetriyl groups connected to hetero atoms, and four aliphatic methanetriyl groups (Figure S45 and Figure S46 and Table S8). The HMBC spectrum revealed nine quaternary carbons accounting for all 51 carbons of the suggested molecular formula (Figure S47). From the COSY spectrum eight spin systems were identified (Figure S46). Five were connected by HMBC correlations to assemble fragment **I** (Figure S50). Based on HMBC correlations, the singlet C-2 was connected to a carbonyl which was suggested to be a terminal amide based on the molecular formula. HMBC correlations from the methyls $H_3$-51 to C-2/C-3/C-4, $H_3$-50 to C-6/C-7/C-8, $H_3$-49 to C-7/C-8/C-9, $H_3$-48 to C-9/C-10/C-11 and $H_3$-47 to C-11/C-12/C-13 in conjunction with the COSY spin system from H-4 over H-5 and H-6 established a tetraene moiety. The COSY spin system from C-9 to C-10 was connected to that of C-11 to C-17 because of HMBC correlations from H-48 to C-9/C-10/C-11 and weak HMBC correlations from H-11 to C-9/C-18. In the COSY spin system from H-11 to H-17 hydroxy groups were attached to C-11 and C-13 based on their chemical shifts. A methoxy group was placed on C-15 based on HMBC interactions with $H_3$-46, and a phosphate group was placed on C-17 based on $^1$H-$^{31}$P COSY correlations (Figure S49). HMBC correlations from H-16, H-17, and H-20 to C-19 (δ 109.3) suggested a ketal at C-19. Correlations from the two singlet methyl groups $H_3$-44 and $H_3$-45 to C-17/C-18/C-19 established a quaternary carbon at C-18. The COSY spin system H-20 to H-26 was connected to the ketal at C-19. No HMBC correlation from H-16 or H-23 to C-19 were observed, however the downfield carbon shift of C-19 (δ 109.3) together with the proton shifts of H-16 and H-23 (both δ 3.93) indicate a spiroketal on C-19. Two dehydrative ring closures are in agreement with the suggested molecular formula suggesting C-19 as a spiro center connecting a tetrahydrofuran ring with a tetrahydropyran ring. Based on HMBC correlations, a methoxy group was placed on C-21. Differentiation between the methoxy groups C-43 and C-46 was difficult because of the similar chemical shifts of C-21 and C-15. However, NOESY correlations between H-42 and H-43 suggested the methoxy group at δ 3.27 ppm as C-43. In the spin system H-20 to H-26, the large coupling constant of the doublet of doublets H-26 ($J$ = 16.0 Hz) indicated 25*E* configuration. HMBC correlations from H-25 and H-26 to the quaternary carbon C-27. NOESY correlation of the singlet H-28 to H-26 placed it next to C-28. A strong HMBC correlation from H-27 to C-29 and a weak correlation from H-27 to C-27 suggested an oxazole ring. HMBC correlations from H-41 and H-30 to C-29 connected the oxazole ring with the spin system $H_3$-41 to H-31. Due to overlapping signals H-32 could not be connected to the spin system of H-31. H-32 showed COSY correlations to a nitrogen proton, however no HMBC correlations to a carbonyl center were observed. Hence, the amide connected to the methylene H-32 was assigned as fragment **II**. Two COSY spin systems between H-34 and H-35 and H-36 and H-37 were connected by HMBC correlations from H-34 to H-37 to assemble fragment **III**. A methoxy group was attached to C-37 based on HMBC correlations from $H_3$-38 to C-37. The two methyl groups were attached to the amine based on HMBC correlations from $H_3$-39/ $H_3$-40 to C-36. H-34/H-35 showed correlations to a carbonyl group. The fragments **I**, **II** and **II** could not be connected by HMBC correlations due to lacking signals. However, the suggested molecular formula of the parent ion (*m*/*z* 1041.5809) indicates connection of **I**, **II**, and **III**. MS/MS analysis revealed an ion at *m/z* 190.1077 $[M+H]^+$ ($C_8H_{16}N_1O_4$; Δ +0.32 mmu). This ion can only arise from fragment **III**. Furthermore, fragment ions at *m/z* 353.1707 $[M+H]^+$ ($C_{17}H_{25}N_2O_6$; Δ -0.01 mmu), *m/z* 398.2290 $[M+H]^+$ ($C_{19}H_{32}N_3O_6$; Δ +0.44 mmu), and *m/z* 506.3227 $[M+H]^+$ ($C_{27}H_{44}N_3O_6$; Δ +0.24 mmu) were observed (Figure S43). The observation of ions with three nitrogen atoms suggest that the oxazole ring of **I**, the amide of **II**, and the tertiary amine of **III** are connected and exclude a connection of **I**, **II**, and **III** through C-1.The relative configuration of the spiroketal was determined based on NOESY correlations of $H_3$-45 to H-17/ and H-16 in the tetrahydrofuran ring, and correlations from $H_3$-45 to H-20a and H-20b as well as $H_3$-44 to H-20a suggesting (15*R**,16*R**,19*R**) configuration (Figure S48). NOESY correlations from H-20a to $H_3$-42, from H-21 to H-23 as well as H-22 to H-24 suggested (15*R**,16*R**,19*R**, 21*S**, 22*S**, 23*S**) configuration (Figure S48). The configuration of the tetraene moiety was suggested to be (*2Z*, *4E*, *6E*, *8E*) because of NOESY correlations between $H_3$-51 and H-2/H-5, H-4 and H-6, $H_3$-49 and H-6 as well as $H_3$-50 and H-9. The coupling constants between H-4/H-5 ($J$ = 15.4 Hz), H-5/H-6 ($J$ = 11.1 Hz) further confirmed the assigned *4E*, *6E* configuration. The phosphate group was placed on C-17 based on the upfield chemical shift of H-17 and $^{31}$P-COSY correlations between H-17 and the phosphorous atom (Figure S49).

### *In silico* analysis of the proposed acidobactamide biosynthetic gene cluster

AntiSMASH predicted the adenylation (A) domain substrate specificity for Tyr, Thr, Thr, Gly, Asn, and *N*-Me-Ile residues for the NRPS modules.[2] The substrate predictions for two further A domains were not conclusive. PRISM predicted the incorporation of Admha and Ddmpha residues in the peptide (Figure 5, Figure S53).[14] Epimerization (E) domains were present in module 4 and 8 suggesting the incorporation of D-configured Thr and Ile (Figure S52). The presence of *N*-methyl transferase domain in AbaI indicated incorporation of *N*-Me-Ile residue. Two $B_{12}$-dependent radical *S*-adenosyl methionine methyltransferases, AbaH and AbaL, were encoded in the *aba* BGC indicating additional methylations (Figure S52).

### Large-scale cultivation of *A. pedis* and isolation of acidobactamides

16 liquid cultures of *A. pedis* were grown at 30 °C for three days at 150 rpm in individual 2 L flasks, each containing 1000 mL marine broth (Difco™) for a total volume of 16 L. The pellet and supernatant from the 16 L-scale cultivation were separated via centrifugation. The pellet was extracted with 4 × 800 mL of acetone in an ultrasonic bath for 30 min. The extracts were combined and concentrated by rotary evaporation to dryness leaving a brown solid.

The crude pellet extracts were fractionated using a preparative reverse phase (RP)-HPLC (Agilent 1260) Infinity system, equipped with a Phenomenex Luna 5μ C18, φ 20 x 250 mm column. Deionized water (Milli-Q, Millipore) (solvent A) and acetonitrile (solvent B) were used as the mobile phase. An elution gradient of 5–100% solvent B in 40 min followed by isocratic conditions at 100% solvent B for 10 min was applied. UV detection was carried out at 210, 254, and 280 nm. Fractions were collected at two-minute intervals. Acidobactamides (**15**-**17**) were identified by HPLC-HESI-HRMS analysis in fractions F12 and F13.

To further purify acidobactamides, MS-guided repetitive HPLC purification was performed by semi-preparative RP-HPLC (Agilent 1260 Infinity system, equipped with a Phenomenex Kinetex 5 µm C18 10 x 250 mm column). The fractions F12 and F13 were purified by reverse phase LC (solvent A/solvent B), elution gradient 25–65% solvent B for 30 min, followed by gradient shift from 65 to 100% in 5 min, and finally isocratic condition at 100% solvent B for 5 min. Acidobactamide A (**15**) (10 mg), acidobactamide B (**16**) (3 mg), and acidobactamide C (**17**) (2.5 mg) were obtained.

### Acid hydrolysis and Marfey's analysis of acidobactamides A (15) and C (17)

Hydrolysis and Marfey analysis were carried out according to Fujii et al.[15] with slight modification Briefly,1 mL 6 M HCl was added separately to 1 mg of acidobactamide A (**15**) and 0.5 mg of and acidobactamide C (**17**) and the reaction mixtures were heated to 100 °C with stirring for 24 h. The reaction mixture was dried under reduced pressure and the resulting residues were dissolved in 250 µL of $H_2O$. To 50 µL of the hydrolysate were added 20 µL of sodium bicarbonate (1 M) and 100 µL of Nα-(2,4-dinitro-5-fluorophenyl)-L-valinamide (FDVA) (4 mg/mL in acetone). The reaction was incubated for 1 h at 40 °C, then quenched with 20 µL 1 N HCl. The solution was then diluted with 200 µL of $H_2O$ for HPLC-HESI-HRMS analysis. Analysis was performed on a Thermo Scientific Q Exactive or LTQ Orbitrap XL mass spectrometer coupled to a Dionex Ultimate 3000 HPLC system operated by Xcalibur 4.1 and Chromeleon Xpress 7.2 (Thermo Scientific) software, respectively (column: Kinetex 2.6 µm XB-C18 100 Å 150x4.6 nm; solvent A ($H_2O$ + 0.1% formic acid); solvent B (MeCN + 0.1% formic acid), flow; 0.8 mL/min. A gradient from: 5% solvent B from 0-2min followed by a gradient increase to 100% solvent B in 20 min and linear gradient at 100% solvent B for 1 min. For Ile and *N*-Me-Ile isomers, a gradient from 5-100 % in 55 minutes was applied.

Amino acid standards Thr, Glu, Asp, Ile, *N*-Me-Ile and Val (1 mg/mL) were dissolved in water and derivatized with FDVA in the same manner described above for the acid hydrolysate for retention time comparison. For 2-hydroxy-3-phenlylpropanoic acid (Hppa), its corresponding L and D standards were used for the analysis. Retention times for amino acids derived from **15** and **17** (Val), with amino acids standards are summarized in Figure S100.

**Partial hydrolysis of acidobactamides A-C (15-17)**

0.2 mg of **15-17** were dissolved in 0.02 N NaOH (1 mL) and stirred at room temperature for 22 h. The reaction mixture was evaporated to dryness and resuspended in 500 µL of methanol and MS/MS data acquired by HPLC-HESI-HRMS. (Figure S66-Figure S68, Figure S87, Figure S89, Figure S96-Figure S98).

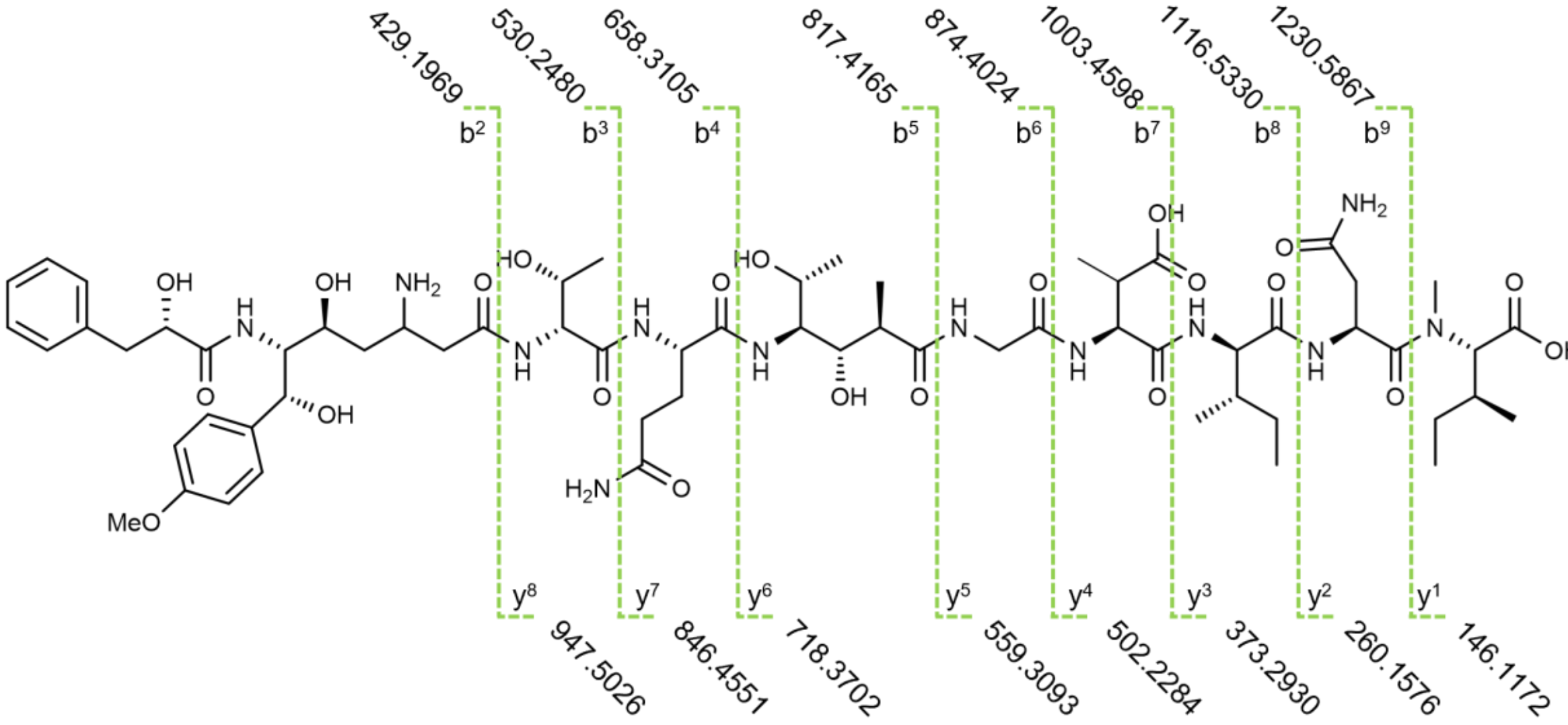

**Determination of the configuration for the Admha and Ddmpha residues**

5 mg of **15** were hydrolysed as described above and the resulting hydrolysate was extracted with EtOAc and concentrated under reduced pressure. The hydrolysate was suspended in 500 µL of 1,4 dioxane and treated with 4 mg of pyridinium *p*-toluenesulfonate and 20 µ of 2,2-dimethoxypropane. The mixture was stirred at 40 °C for 12 h followed by purification by reverse phase HPLC (Kinetex 5 µm C18 10 x 250 mm column, flow 2 mL/min). A gradient of 20%-80% MeCN in 40 mins was applied. The fractions containing **20** were combined and dried under reduced to afford 1.2 mg of **20** and analyzed by HPLC-HESI-HRMS and NMR (Figure S72-Figure S77).

The absolute configuration was determined by Mosher method with slight modification.[16] Briefly, two portions of **15** (1 mg each) was suspended in 200 µl of pyridine and treated with 2 µL of *S* and *R*-MTPA-Cl respectively. The solutions were stirred for 6 h, dried under reduced pressure and analyzed by NMR (Figure S78-Figure S81).

**Synthesis of Boc-Ile-OH** and **Synthesis of *N*-Me-Ile**

Boc-Ile-OH and Boc-D-Ile-OH were prepared according to Umezawa et al. [17] The *N*-Me-L-Ile and *N*-Me-D-Ile were synthesized according to Tajima et al.[18] with slight modification. To 30 mg of either Boc-L-Leu or Boc-D-Ile-OH dissolved in THF (200 µL), sodium hydride (31.2 mg) was added at 0 °C. After stirring for 1 h, iodomethane (80 µL) was added and the reaction mixture stirred at 0 °C for 2 h and held at rt for 18 h. The reaction mixture was quenched with $H_2O$ and then acidified by HCl to pH 1–2. The mixture was extracted with EtOAc, dried over sodium sulfate and concentrated under reduce*d* pressure. The resulting white solid was treated with TFA (500 µL) at rt for 1 h and then freeze-dried. *N*-Me-L-Ile was obtained as a white solid (16.7 mg) and *N*-Me-D-Ile (12.3 mg) (Figure S103-Figure S111).

**Structure elucidation of acidobactamides A-C (15-17)**

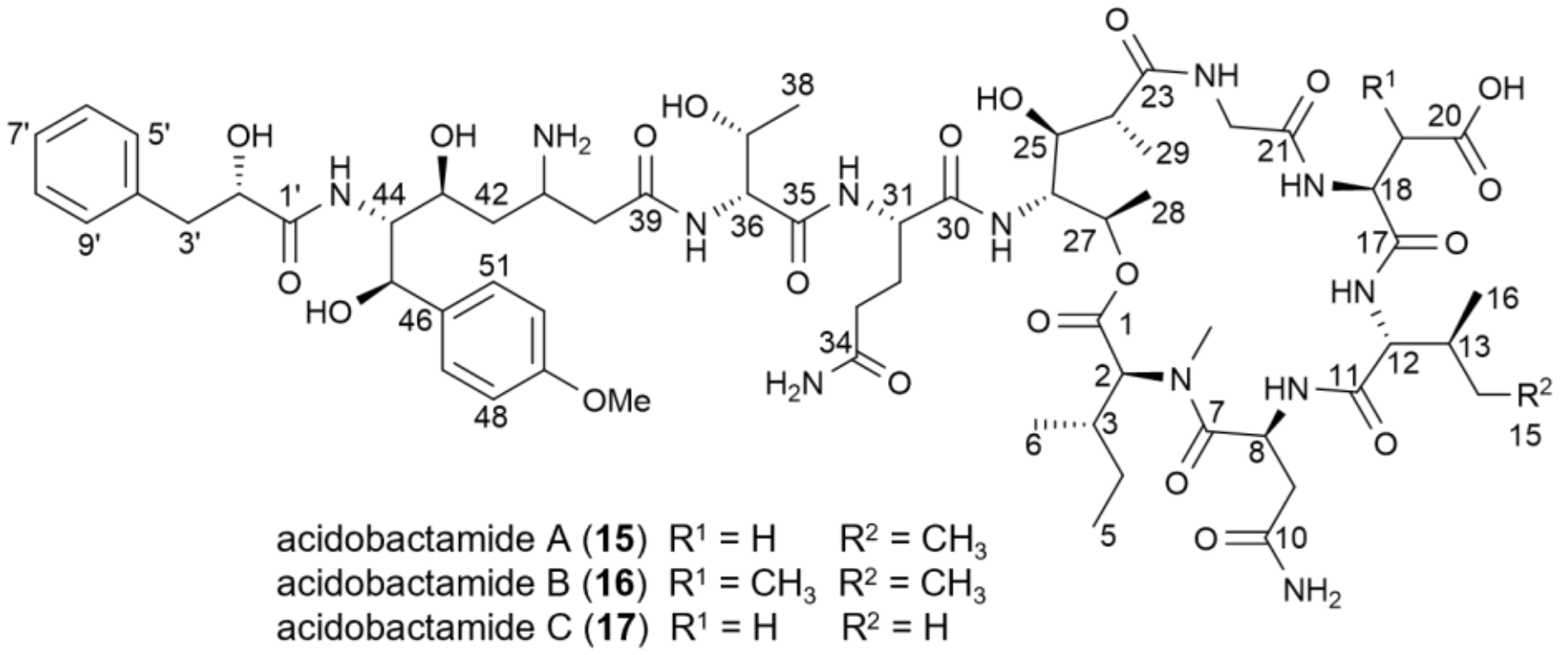

acidobactamide A (**15**) $R^1$ = H $R^2$ = $CH_3$
acidobactamide B (**16**) $R^1$ = $CH_3$ $R^2$ = $CH_3$
acidobactamide C (**17**) $R^1$ = H $R^2$ = H

The structure of acidobactamide A (**15**) was elucidated by interpreting the $^1$H NMR, COSY, TOCSY, HSQC, HMBC, and NOESY spectra (Figure S57-Figure S61 and Figure S62-Figure S65, Table S10). HESI-HRMS of **15** suggested a molecular formula of $C_{62}H_{94}N_{12}O_{21}$ (*m/z* 1343.6689 for [M + H]$^+$, Δ -3.244 mmu, Figure S56). Signals corresponding to the α-methine of the amino-acid residues (δ 4.2–5.0 ppm) were recorded in the $^1$H NMR (Figure S54). Furthermore, characteristic aromatic proton signals (δ 6.90 and δ 7.31 ppm) corresponding to a 1,4-disubstitued benzene, unsubstituted benzene (δ 7.12, δ 7.18 and δ 7.24 ppm) and four methyl doublets were recorded in the $^1$H NMR spectra (Figure S57) Analysis of the COSY, TOCSY, and HMBC spectra (Figure S58, Figure S60 and Figure S61) revealed the presence of nine amino acid residues including threonine (Thr), glutamine (Gln), glycine (Gly), aspartic acid (Asp), isoleucine (Ile), asparagine (Asn), *N*-methyl isoleucine (*N*-Me-Ile), the unusual 4-amino-3,5-dihydroxy-2-methylhexanoic acid (Admha), and 3,6-diamino-5,7-dihydroxy-7-(4-methoxyphenyl)heptanoic acid (Ddmpha) residue. The first spin system in (Ddmpha) was elucidated from the COSY correlations of H-41 to $H_2$-40/ $H_2$-42, H-43 to $H_2$-42/H-44 and H-44 to H-43/H-45 (Figure S58). These COSY correlations were supported by the HMBC correlations of $H_2$-40 to C-39/C-41/C-42, $H_2$-42 to C-40/C-41/C-43/C/-44, H-44 to C-42/C-43/C-45/C-46.The 1,2-substituted phenyl moiety was established from the COSY correlations of H-47/H-51 to H-48/H-50 and the HMBC correlations of H-48/H-50 to C-46 and H-47/H/51 to C-49. Cross peaks were recorded between the methoxy protons (δ 3.76 ppm) and C-49 (δ 160.7 ppm). This moiety was connected to the first spin system through the HMBC correlations of H-46/51 to C-45 and H-45 to H-C-46/C-47/C-51. The connection of the Ddmpha moiety to Thr was confirmed by the HMBC correlations of H -41/H-40 to C-39 and Thr α-methine (δ 4.07 ppm) to C-39 (δ 172.4 ppm). The 2-hydroxy-3-phenylpropanoic acid (Hppa) moiety was established from the COSY correlations of $H_2$-2'- to H-3', H-6'/H-8' to H-7'/H-5'/H-9' and HMBC correlations of H-3' to C-4'/C-5'/C-9'. The Hppa unit was connected to the Ddmpha moiety based on the HMBC correlations of H-44 (δ 3.91) to C-1' (δ 176.6 ppm). The Admha was elucidated from the COSY correlations of H-24 to $H_3$-29/H-25, H-26 to H-25/H-27 and H-27 to H-26/$H_3$-28. Admha was connected to Gly because of the HMBC correlations of $H_3$-29 to C-23/C-C-24/C-25 and Gly diasterotopic protons (δ 3.60 and δ 4.25 ppm) to C-23 (δ 179.8 ppm). The connection to Gln was established from the HMBC correlation of H-26 to C-30 (δ 175.2 ppm). The HMBC cross peaks between the methyl protons (δ 3.01 ppm) and C-2/C-7 confirm the *N*-Me-Ile moiety. The cyclic hexapeptide was concluded from the HMBC correlation of the Admha H-27 (δ 5.27 ppm) to *N*-Me-Ile C-1 (δ 170.6 ppm). The amino acid sequence in **15** was confirmed by the inter-residue HMBC correlations of the α-methine to the next amino acid carbonyl carbon and the NOESY correlations of NH to the α-methine protons (Figure S65). The elucidated structure of **15** was further supported by the MS/MS data (Figure S66-Figure S68)**.**

Acidobactamide B (**16**), a congener of **15** with the molecular formula $C_{63}H_{96}N_{12}O_{21}$ established from HESI-HRMS data (*m/z* 1357.6854 for [M + H]$^+$, Δ -3.154 mmu) (), was isolated. The 1D and 2D NMR data of **16** (Figure S83-Figure S86, Table S13) were similar to those of **15** with an additional *C*-methylation on the Asp residue. The structure of the β-methyl aspartic acid (β-Me-Asp) was deduced from COSY correlations of the β-methyl group protons $H_3$-20 (δ 1.17 ppm) to the β-methine proton H-19 (δ 2.69 ppm) and HMBC correlation of $H_3$-20 to C-18 (δ 58.0 ppm)/C-19 (δ 41.4 ppm)/C-21 (δ 178.0 ppm). The third congener, acidobactamide C (**17**), had the molecular formula $C_{61}H_{92}N_{12}O_{21}$ deduced from the HESI-HRMS data (*m/z* 1329.6538 for [M + H]$^+$, Δ -2.464 mmu) (Figure S91). The 1D and 2D NMR data of **17** (Figure S92-Figure S95, Table S14) suggested the presence of valine (Val) in **17** replacing the Ile in **15**.

To assign the stereochemistry of the constituent amino acids, acid hydrolysis of acidobactamide A (**15**) and B (**17**) was carried out. The resulting hydrolysate was derivatized with $N_\alpha$-(2,4-dinitro-5-fluorophenyl)-L-valinamide (FDVA) and analyzed by HPLC-HESI-HRMS. The results were compared to the FDVA-derivatized Thr, Glu, Asp, Ile, Val, *N*-Me-Ile, and Hppa standards. Hydrolysis of Asn and Gln results in Asp and Glu, respectively. Therefore, Asp and Glu were used as their corresponding standards. The HESI-HRMS analysis confirmed the presence of D-*allo*-Thr, L-Gln, L-Asp, D-*allo*-Ile, L-Asn, *N*-Me-L-Ile and L-Hppa in **15** (Figure S100). The retention time of Val in **17** matched that of the D-Val standard, therefore confirming the presence of D-Val in **17.** To assign the stereochemistry at the Ddmpha and Admha residues, **15** was hydrolyzed and the resulting hydrolysate was chemically transformed to the acetonide at C-25-27 and C-43-C-45, resulting in **20**. The transformation was confirmed by the HMBC correlations of H-45 (δ 5.26) and H-25 (δ 4.09) to the quaternary carbon of the *O*-isopropylidene moiety (δ 103.3). The NOESY correlations of the isopropylidene methyl protons $H_3$-57 (δ 1.57) and $H_3$-54 (δ 1.58) to H-27 and H-43 respectively suggested the axial-equatorial arrangement of these protons. The small coupling constant of 4.7 Hz between H-26 and H-25/H-27 and 5 Hz between H-44 and H-43/H-45 supported the assigned axial-equatorial arrangement of the protons (Figure S100). These proton arrangements in the *O*-isopropylidene moieties suggested the relative configuration as either (25*S*, 26*R*, 27*R*) and (43*S*, 44*R*, 45*R*) or (25*R*, 26*S*, 27*S*) and (43*R*, 44*S*, 45*S*). The absolute stereochemistry was assigned as (25*S*, 26*R*, 27*R*) and (43*S*, 44*R*, 45*R*) by derivatizing **15** with (*R*) and (*S*)-α-methoxy-α-(trifluoromethyl)phenylacetyl chloride (MTPA-Cl). The configuration at C-24 was inferred as *R* from the NOESY correlations of $H_3$-29 to H-26. A clear difference in the chemical shift between the *S* and the *R* ester in the Ddmpha moiety was not observed, possibly due to the close proximity of the MTPA substituents (Figure S71). Because of the clear positive values recorded for H-47 and H-48, C-45 was assigned as *R*.

**Chemical data**

**Phorbactazole A (1):** $[\alpha]_D^{27}$ -3.9 (c 0.04, MeOH); UV (MeOH, c= 0.00005 mM) $\lambda_{max}$ (log ε) 262 (5.774), 210 (6.775), 198 (.6.757); HRMS *m/z* 1000.5018 [M+H]$^+$, calcd $C_{51}H_{74}N_3O_{17}$ , 1000.5018.

**Phorbactazole B (8):** $[\alpha]_D^{27}$ -4.7 (c 0.03, MeOH). UV (MeOH, c= 0.125) $\lambda_{max}$ (log ε) 258 (6.171), 210 (7.762), 198 (.7.128); HRMS *m/z* 957.4977 [M+H]$^+$, calcd $C_{50}H_{73}N_2O_{16}$ , 957.4960.

**Phorbactazole C (9):** $[\alpha]_D^{27}$ -10.8 (c 0.03, MeOH); UV (MeOH, c= 0.0003 mM) $\lambda_{max}$ (log ε) 218 (6.944); HRMS *m/z* 1162.5560 [M+H]$^+$, calcd $C_{57}H_{84}N_3O_{22}$, 1162.5547.

**Phorbactazole D (10):** $[\alpha]_D^{27}$ -7.3 (c 0.03, MeOH); UV (MeOH, c= 0.0003 mM) $\lambda_{max}$ (log ε) 218 (7.042); HRMS *m/z* 1119.5518 [M+H]$^+$, calcd $C_{56}H_{83}N_2O_{21}$, 1119.5488.

**Phorbactazole E (11):** $[\alpha]_D^{27}$ -14.1 (c 0.03, MeOH); UV (MeOH, c= 0.0004 mM) $\lambda_{max}$ (log ε) 262 (6.653), 218 (6.954); HRMS *m/z* 1144.5414 [M+H]$^+$, calcd $C_{57}H_{82}N_3O_{21}$, 1144.5440.

**Phorbactazole F (12):** $[\alpha]_D^{27}$ -11.8 (c 0.04, MeOH); UV (MeOH, c= 0.0005mM) $\lambda_{max}$ (log ε) 318 (6.441), 218 (6.744) HRMS *m/z* 1112.5309 [M+H]$^+$, calcd $C_{56}H_{78}N_3O_{20}$, 1112.5179.

**21-Methoxy-calyculinamide (13)**: $[\alpha]_D^{27}$ -12.8 (c 0.03, MeOH). UV (MeOH, c= 0.0005 mM) $\lambda_{max}$ (log ε) 230 (6.767). HRMS *m/z* 1041.5809 [M+H]$^+$, calcd $C_{51}H_{86}N_4O_{16}P$ 1041.5776.

**Acidobactamide A (15)**: $[\alpha]_D^{27}$ -12.6 (c 0.2, MeOH). UV (MeOH, c= 0.00372 mM) $\lambda_{max}$ (log ε) 274 (4.207),198 (5.894), 194 (5.887). HRMS *m/z* 1343.6689 [M+H ]$^+$, calcd $C_{62}H_{95}N_{12}O_{21}$ 1343.6735.

**Acidobactamide B (16)**: $[\alpha]_D^{27}$ -5.9 (c 0.03, MeOH). UV (MeOH, c= 0.00369 mM) $\lambda_{max}$ (log ε) 274 (4.132), 198 (5.901), 194 (5.892). HRMS *m/z* 1357.6854 [M+H]$^+$, calcd $C_{63}H_{97}N_{12}O_{21}$ 1357.6891.

**Acidobactamide C (17)**: $[\alpha]_D^{27}$ -1.8 (c 0.05, MeOH). UV (MeOH, c= 0.00376 mM) $\lambda_{max}$ (log ε) (MeOH, c= 0.00368 mM) $\lambda_{max}$ (log ε) 274 (4.132), 198 (5.888), 194 (5.873). HRMS *m/z* 1329.6538 [M+H]$^+$, calcd $C_{61}H_{93}N_{12}O_{21}$ 1329.6578.

**Boc-L-Ile-OH**: $^1$H NMR (600 MHz, $CD_3OD$): δ 4.05 (d, *J* = 5.7 Hz, 1H, CH), 1.84 (m, 1H, CH) 1.49 (m, 1H, $CH_2$), 1.45 (s, 9H, 3 x $CH_3$), 1.23 (m, 1H, $CH_2$), 0.95 (d, J=7.3 Hz, 3H, $CH_3$), 0.92 (t, *J*=7.3 Hz, 3H $CH_3$); $^{13}$C NMR (150 MHz, $CD_3OD$): δ 175.7, 158.3, 80.6, 59.6, 38.6, 28.9, 26.3, 16.2, 12.0. HRMS *m/z* 232.1532 [M+H]$^+$.

**Boc-D-Ile-OH**: $^1$H NMR (600 MHz, $CD_3OD$): δ 4.05 (d, *J* = 5.6 Hz, 1H, CH), 1.84 (m, 1H, CH) 1.49 (m, 1H, $CH_2$), 1.45 (s, 9H, 3 x $CH_3$), 1.23 (m, 1H, $CH_2$), 0.95 (d, *J*=7.2 Hz, 3H, $CH_3$), 0.92 (t, J=7.2 Hz, 3H, $CH_3$); $^{13}$C NMR (150 MHz, $CD_3OD$): δ 175.8, 158.4, 80.6, 59.7, 38.6, 28.9, 26.3, 16.2, 12.0. HRMS *m/z* 232.1531 [M+H]$^+$.

***N*-Me-L-Ile**: $^1$H NMR (600 MHz, $CD_3OD$): δ 3.80 (d, *J* = 3.8 Hz, 1H, CH), 2.77 (s, 3H, $CH_3$), 1.56 (m, 1H, CH), 1.37 (m, 2H, $CH_2$), 1.01 (d, *J*=7.3 Hz, 3H, $CH_3$), 0.97 (t, J=7.3 Hz, 3H, $CH_3$); $^{13}$C NMR (150 MHz, $CD_3OD$): δ 171.2, 66.6, 35.6, 32.4, 25.5, 13.9, 11.0. HRMS *m/z* 232.1531 [M+H]$^+$.

***N*-Me-D-Ile**: $^1$H NMR (600 MHz, $CD_3OD$): δ 3.80 (d, *J* = 3.8 Hz, 1H, CH), 2.77 (s, 3H, $CH_3$), 1.56 (m, 1H, CH), 1.37 (m, 2H, $CH_2$), 1.01 (d, *J*=7.3 Hz, 3H, $CH_3$), 0.97 (t, J=7.3 Hz, 3H, $CH_3$); $^{13}$C NMR (150 MHz, $CD_3OD$): δ 171.2, 66.6, 35.6, 32.4, 25.5, 13.9, 11.0. HRMS *m/z* 232.1531 $[M+H]^+$.

### Bioactivity tests of isolated compounds 7-13 and 15-17 against HeLa cells

HeLa cells purchased from ATCC were cultivated at 37 °C, 5% (v/v) $CO_2$ for 3-4 days. Cells were washed with PBS buffer (Sigma D8537), 0.05 % trypsin-EDTA solution (Thermo 25300-054) was added, and the plate was incubated for 5 min at 37 °C. The cells were resuspended in 10 mL medium (DMEM-GlutaMAX) supplemented with 10% FCS (Eurobio CVFSVF00-01), and 50 µg/mL penicillin-streptomycin (Corning). After counting the cells under a ZEISS Axiovert 25 microscope using a Neubauer hemocytometer, a 13,000 cells/mL suspension was prepared and 200 µL were transferred into each well of a 96-well plate. After one day of cultivation 2 µL of compounds (**7-13**, and **15-17**) were added and a five-fold serial dilution was performed. After four days of cultivation (three days for **15-17**), 50 µL of 3-(4,5-dimethylthiazol-2-yl)-2,5-diphenyltetrazolium bromide (MTT, 1 mg/mL in sterile $H_2O$) were added and the cells incubated for 3 h at 37 °C. The supernatant was discarded, the plate washed with PBS buffer and then 150 µL of DMSO were added to the wells. The absorbance at 570 nm was measured on a spectraMAXplus spectrometer (Molecular Devices LLC). Results are shown in Figure S101.

### Sampling of chitons and metagenomic DNA

Ten specimens of the Chiton *Acanthopleura japonica*[19] were collected at the Pacific Ocean coast near the Coastal Branch of Natural History Museum and Institute, Katsuura, Boso peninsula, Chiba, Japan (coordinates approximately 35°08'05.6"N 140°17'05.1"E, Figure S102). Chiton specimens were placed in 50-mL Falcon tubes containing seawater and brought to the Faculty of Pharmaceutical Sciences, Hokkaido University for analyses. Metagenomic DNA was prepared from each chiton individual using conventional cetyl trimethylammonium bromide (CTAB) method with slight modification[20]. The chiton body was initially cut vertically from top to bottom using a sterile blade to allow obtaining representative DNA from different parts. The cut body part was separated from the mantle/shell, and subsequently ground in liquid nitrogen. To the homogenized sample, 10 mL lysis buffer (50 mM Tris-Cl pH 7.5, 50 mM urea, 700 mM NaCl, 2% sarkosyl, 8 M EDTA) were added and incubated at 50 °C for 20 min with gentle mixing every two minutes. Lysed sample was extracted two times with phenol:chloroform:isoamyl alcohol (1:24:25) mixture, and one time with chloroform. The extracted DNA was precipitated with 0.7 volume isopropanol, washed two times with cold 70% ethanol, dried on air, and dissolved with 100-200 µL $dH_2O$. The DNA extract obtained from each sample was separated on a 1% agarose gel. The high-molecular-weight (HMW) DNA bands were sliced and purified from the gel using Zymoclean large fragment DNA recovery kit (Zymo Research). The concentration and quality of purified metagenomic DNA were checked by Nanodrop.

To test whether the target bacterium *A. pedis* is detectable in chitons, the following PCR primers were designed and used to amplify a 1400-bp 16S rRNA gene of *A. pedis*: 16SAPedis-F1 (5'-TAC GCG TTG TTT TGG CTT CGG CTG AGG CAA C-3'), APedis16S-F2 (5'-AGC ATC TTG GTG TGT AAA GCC CTG-3'), and 16SAPedis-R (5'-CAC CAA TCA CAC CGT AAA CAG CTA CCT CCC CAA GG-3'). Every PCR reaction was carried out in total 25 µL volume using KOD One™ PCR Master Mix (TOYOBO) with the following composition: 12.5 µL of KOD One™ PCR Master Mix, 10.5 µL of $dH_2O$, and 0.5 µL of 50 mM forward primer 16SAPedis-F1 or APedis16S-F2, 0.5 µL of 50 mM reverse primer 16SAPedis-R, and 0.5 µL of purified chiton metagenomic DNA. Five PCR reactions without DNA template were also prepared as negative controls. This PCR experiment was repeated three times for validation. PCR product of the 16S DNA genes were used as DNA templates for positive controls. The PCR program was set up at 40 cycles consisting of denaturation at 98 °C for 10 seconds, annealing at 68 °C for 5 seconds, and elongation at 68 °C for 10 seconds. Each correct-sized PCR band on the gel was transferred into PCR master mix (20 µL in total) consisting of 10-µL of 2X GoTaq® Green Master Mix, 10 µL $ddH_20$, 0.5 µL of 10 µM each primer above, and 1 µL DMSO. The PCR program was set up at 35 cycles consisting of pre-denaturation at 95 °C for 2 min, denaturation at 95°C for 30 s, annealing at 57 °C for 30 s, elongation at 72 °C for 1 h 20 min, and final elongation at 72 °C for 5 min. Resulting target PCR products were individually extracted from the gel using GSCN buffer (5.5 M guanidinium thiocyanate in 20 mM Tris-Cl pH 6.5) with silica membrane mini spin columns (EconoSpin®), and subsequently cloned into *E. coli* DH5α using the pT7Blue™ T-Vector (Novagen). The transformed cells were plated on Luria Bertani (LB) agar plates containing 100 µg/ml ampicilin, 40 µg/ml X-Gal (5-bromo-4-chloro-3-indolyl-β-D-galactoside), and 1 mM IPTG (isopropylthio-β-galactoside) and incubated overnight at 37 °C. The resulting white colonies were randomly picked and screened by colony PCR using GoTaq® Green Master Mix (Promega) with M13 primers. Recombinant plasmids were prepared from overnight LB-cultures of positive clones, sequenced, and analyzed. Recombinant plasmids were prepared from overnight LB cultures of positive clones, sequenced, and analyzed.

### Recovery of a new Acanthopleuribacteraceae genome from a coral metagenome

To extend the genomic representation of the Acanthopleuribacteraceae family, we explored the *Tara Pacific* metagenomic data collection[21-23] for Acanthopleuribacteraceae genomes. The Illumina raw metagenomic reads were processed and assembled as described previously[24]. The resulting assemblies were binned with MetaBAT2 v2.15 (with parameters --minContig 2000 --maxEdges 500 -x 1 --minClsSize 200000)[25]. The bins were then taxonomically annotated using GTDB-Tk v2.1.0 and GTDB r207[26]. From these, we identified a genome representing a candidate new Acanthopleuribacteraceae genus recovered from the metagenome of a fire coral (*Millepora* sp.) colony (Figure 6A). The corresponding biological sample was collected by the *Tara Pacific* Expedition near Tobi island (2°59'54.5"N 131°07'31.4"E; Republic of Palau; sample ID: TARA_CO-0003230). This Acanthopleuribacteraceae MAG is of high quality (91.88% completeness and 0.85% contamination) as estimated by CheckM v1.2.0[27] and good integrity (79 scaffolds, longest >370kb, N50 ~150kb). We then predicted the biosynthetic potential of this genome using antiSMASH v5.1.1[2] and identified 38 BGCs, including seven *trans*-AT PKS, of which one contains only a single PKS with one KS, which is not included in Figure 6D. The MAG sequence is publicly available and deposited at Zenodo under the DOI https://doi.org/10.5281/zenodo.8013902.

### Phylogeny of Acanthopleuribacteraceae genomes

To better understand the diversity of the Acanthopleuribacteraceae family, we sought to establish a phylogeny based on the three available genomes. We therefore used the *A. pedis* and *S. corallicola* genomes (accessions GCF_017377855.1 and GCF_017498545.1, respectively), the coral MAG (https://doi.org/10.5281/zenodo.8013902) as well as a *Bryobacter* genome as an outgroup (accession GCA_000702445.1). We predicted single-copy universal marker genes using fetchMGs (https://github.com/motu-tool/fetchMGs) and used the COG0012 gene to build the phylogeny. We then aligned the genes using MUSCLE v3.8[10], processed the alignment with trimAl v1.4 (with parameters -gt 0.75)[28], and built the tree with IQ-TREE v2.0.3[29]. The biosynthetic potential of these genomes was identified as described above.

**Table S1. ORFs of the *pho* genes and their putative functions.**

| ORF, GenBank accession number | Protein size [aa] | Proposed function | Closest homolog in GenBank, Source organism based on E value | Identity [%] | GenBank accession number |
|---|---|---|---|---|---|
| PhoA WP_207857246.1 | 15100 | PKS-NRPS (KSS ACP KS KR ACP KS KR MT ACP KS OMT ACP ACP KS ACP Cy Cy A ACP OX OX KS ACP KS DH KR MT ACP KS KR MT ACP KS KR ACP KS DH PS KR ACP KS) | *Sulfidibacter corallicola* | 48 | QTD53419.1 |
| PhoB WP_207857247.1 | 2914 | PKS-NRPS (DH KR MT ACP KS ACP Cy Cy) | *Sulfidibacter corallicola* | 51 | QTD53420.1 |
| PhoC WP_207857248.1 | 4251 | PKS-NRPS (A ACP OX OX KS ACP KS KR ACP KS KR ACP) | *Sulfidibacter corallicola* | 44 | QTD53421.1 |
| PhoD WP_207857249.1 | 5009 | PKS (KS ACP KS DH PS KR ACP KS KR ACP KS KR ACP) | *Sulfidibacter corallicola* | 37 | QTD53421.1 |
| PhoE WP_207857250.1 | 9819 | PKS-NRPS (KS ACP KS DH KR ACP KS KR ACP KS KR MT ACP KS DH PS KR ACP KS KR ACP TEB KS C TE) | symbiont bacterium of *Paederus fuscipes* | 39 | AAS47564.1 |
| PhoF WP_207857233.1 | 439 | glycosyltransferase | *Sulfidibacter corallicola* | 48 | QTD53425.1 |
| PhoG WP_207857234.1 | 380 | AT | *Sulfidibacter corallicola* | 58 | QTD53413.1 |
| PhoH WP_207857235.1 | 326 | 6-bladed beta-propeller | Acidobacteria bacterium | 24 | PIE01935.1 |
| PhoI WP_207857236.1 | 372 | hypothetical protein | *Candidatus* Thiomargarita nelsonii | 36 | KHD07458.1 |
| PhoJ WP_207857237.1 | 500 | AAA family ATPase | *Candidatus* Viridilinea mediisalina | 40 | WP_097642444.1 |
| PhoK WP_207857238.1 | 710 | ABC transporter ATP-binding protein | *Sulfidibacter corallicola* | 55 | QTD53414.1 |
| PhoL WP_207857251.1 | 383 | LLM flavin-dependent oxidoreductase | *Sulfidibacter corallicola* | 70 | QTD53423.1 |
| PhoM WP_207857252.1 | 599 | *O*-carbamoyltransferase | *Sulfidibacter corallicola* | 66 | QTD53427.1 |

**Table S2. Biosynthetic product prediction for the *pho* cluster.**

Shown are the predicted structures of KS substrates (α,β-region of the thioester) based on the KS phylogeny and the TransATor prediction tool (Figure S1).[4]

| KS Domain | Predicted accepted intermediate type by TransATor | Predicted intermediate type accepted by the KS based on phylogeny (see SI) | Suggested KS intermediate type based on NMR data | Predicted KR product specificity |
|---|---|---|---|---|
| PhoA, KS1 | various | ambiguous | ? | |
| PhoA, KS2 | β-keto or double bonds | β-keto | acetyl | |
| PhoA, KS3 | β-(L)-OH | β-(L)-OH | β-OH | (L)-OH |
| PhoA, KS4 | non-elongating β-(D)-OH | non-elongating α-methyl-β-OH | non-elongating α-methyl-β-OH | (D)-OH |
| PhoA, KS5 | β-OMe | β-(D)-OMe | α-methyl-β-OMe | |
| PhoA, KS6 | non-elongating (oxazole/thiazole) | non-elongating (oxazole/thiazole) | non-elongating, oxazole | |
| PhoA, KS7 | amino acids (oxazole/thiazole) | amino acids (oxazole/thiazole) | oxazole | |
| PhoA, KS8 | α-methyl-double bond | α-methyl-(*E*)-double bond | α-methyl-(*E*)-double bond | (D)-OH |
| PhoA, KS9 | α-L-methyl-β-(L)-OH or α-dimethyl-β-(L)-OH | α-(L)-methyl-β-(L)-OH | α-methyl-β-OH | (D)-OH |
| PhoA, KS10 | β-(D)-OH | β-(D)-OH | β-OH | (D)-OH |
| PhoA, KS11 | pyran/furan | pyran/furan | tetrahydropyran | (D)-OH |
| PhoB, KS12 | non-elongating double bond | non-elongating (*E*) double bond | α-methyl-(*E*)-double bond | (D)-OH |
| PhoC, KS13 | non-elongating (oxazole/thiazole) | non-elongating (oxazole/thiazole) | non-elongating, oxazole | |
| PhoC, KS14 | amino acids (oxazole/thiazole) | amino acids (oxazole/thiazole) | oxazole | |
| PhoC, KS15 | β-(D)-OH | β-(D)-OH | β-OH | (D)-OH |
| PhoD, KS16 | non-elongating, β-OH | non-elongating, β-OH | non-elongating, β-OH | (D)-OH |
| PhoD, KS17 | α-(D)-OH-β-(D)-OH | α-(D)-OH-β-(D)-OH | α-OH-β-OH | |
| PhoD, KS18 | pyran/furan | pyran/furan | tetrahydropyran | (D)-OH |
| PhoD, KS19 | β-(L)-OH | β-(L)-OH | β-OH | (L)-OH |
| PhoE, KS20 | non-elongating, β-OH | non-elongating, β-OH | non-elongating, β-OH | (D)-OH |
| PhoE, KS21 | α-(D)-OH-β-(D)-OH | α-(D)-OH-β-(D)-OH | α-OH-β-OH | |
| PhoE, KS22 | double bond | (*E*)-double bond | (*E*)-double bond | (D)-OH |
| PhoE, KS23 | β-(D)-OH | β-(D)-OH | β-OH | (D)-OH |
| PhoE, KS24 | β-(D)-OH | α-methyl-β-(D)-OH | α-methyl-β-OH | (D)-OH |
| PhoE, KS25 | pyran/furan | pyran/furan | pyran | (D)-OH |
| PhoE, KS26 | non-elongating, β-branching | non-elongating, double bond formation via β-*O*-acetylation (new clade for which recognition is not yet implemented in TransATor) | non-elongating, (*Z*)-double bond | (D)-OH |

# *Trans*-AT PKS derived polyketide prediction results

The annotation of the different *trans*-AT PKS KS clades on the submitted sequences produces the following structure:

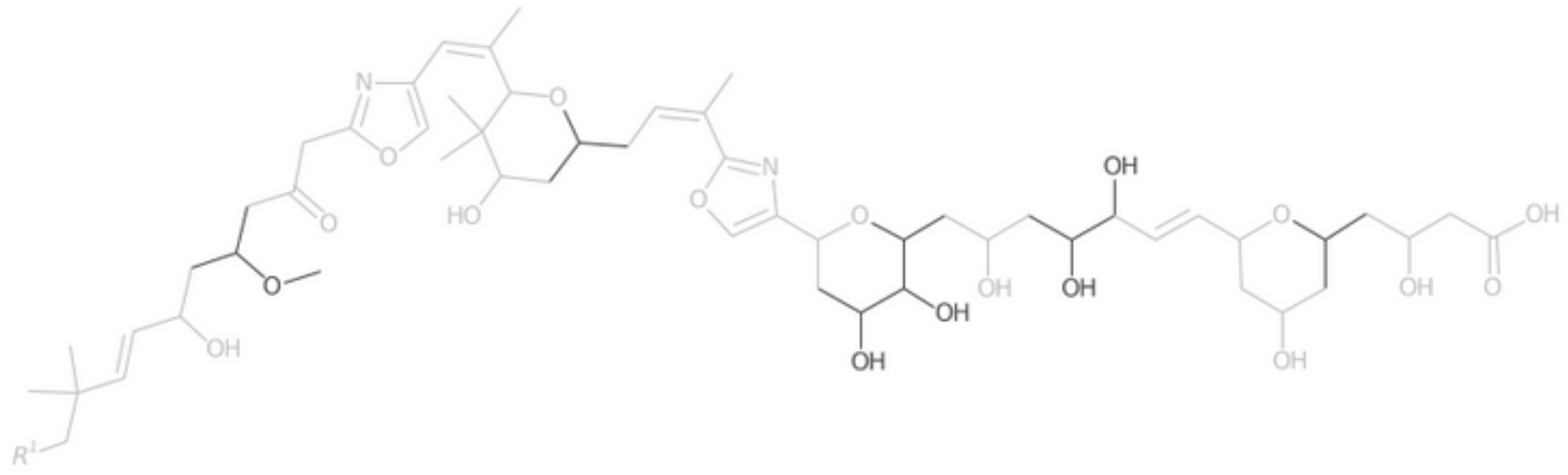

SMILES: C(*)C(C)(C)C=CC(CC(CC(CC1=NC(=CO1)C=C(C)C2C(C)(C)C(CC(CC=C(C)C3=NC(=CO3)C4CC(C(O)C(CC(CC(C(O)C=CC5CC(CC(CC(CC(O)=O)O)O5)O)O)O)O4)O)O2)O)=O)OC)O

The annotation for each sequence submitted can be seen in the sections below.

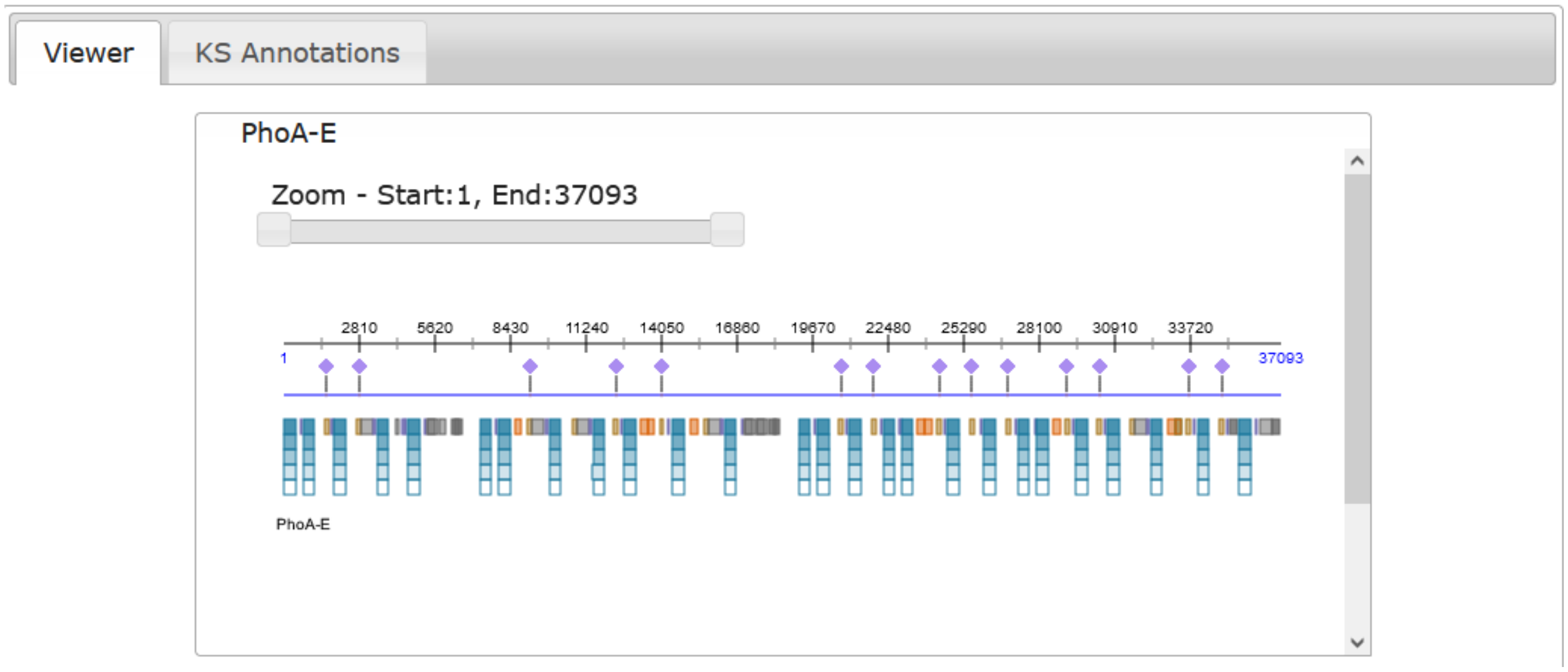

**Figure S1. Biosynthetic product prediction for the *pho* cluster showing a direct output of the TransATor web interface (https://transator.ethz.ch/).**

The assembly line proteins PhoA-E were concatenated as a query and submitted to TransATor[4].
Top: structural prediction.
Bottom: map of KS (blue boxes), ACP (purple boxes), MT/OMT (gray), KR (brown and purple), and GNAT, C, or A (gray) domains is displayed.

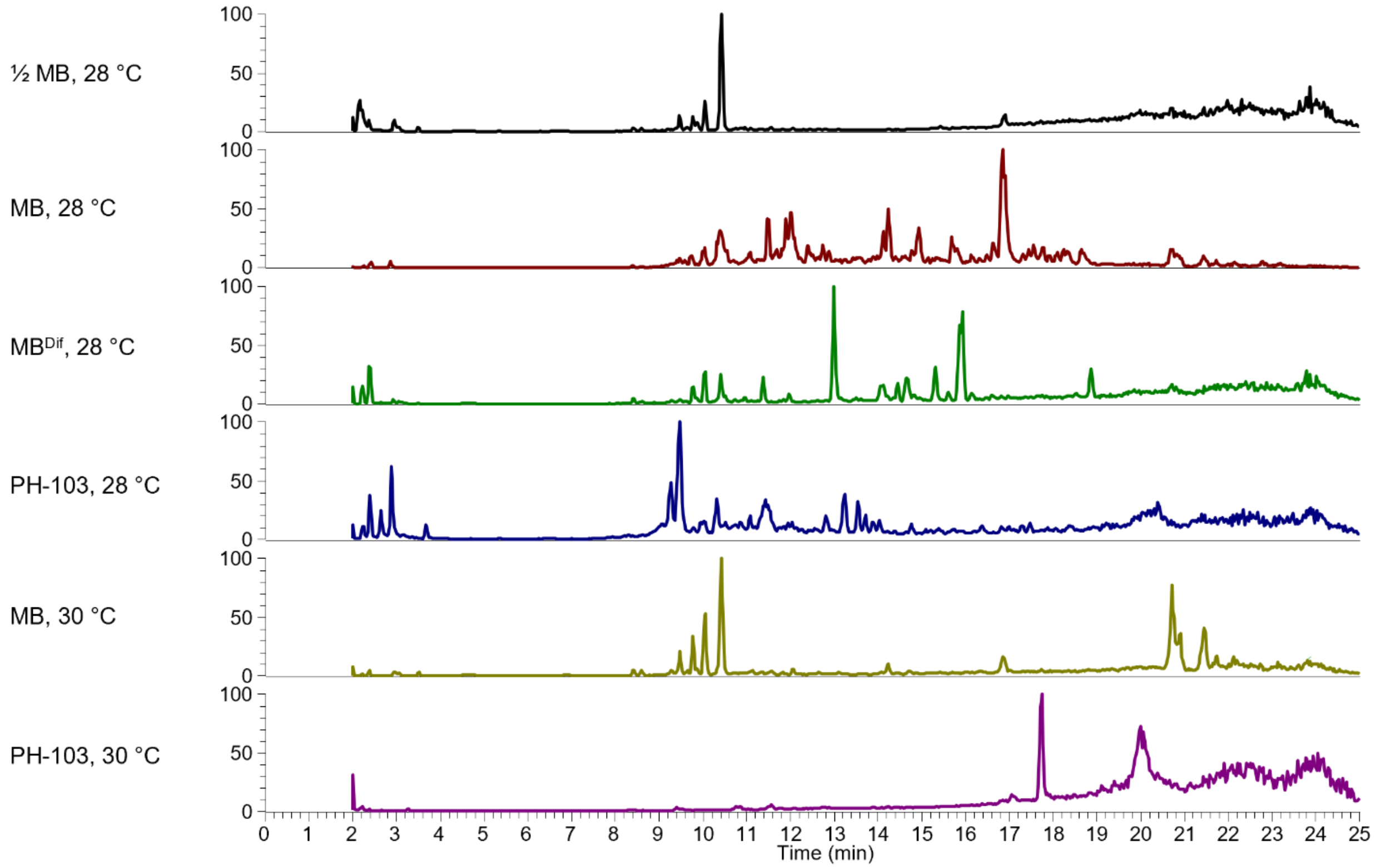

**Figure S2. Screening of six different culture conditions for *A. pedis*.**

After three days of growth in baffled 100 mL flasks containing different media at 28 or 30 °C and at 150 rpm, cultures were extracted with ethyl acetate. Cultures grown in MB at 28 °C showed the most diverse total ion chromatogram. ½ MB, self-made marine broth at half salt concentration; MB, self-made marine broth; $MB^{DIF}$, marine broth Difco 2216; PH-103 medium described by Acebal *et al*.[12]

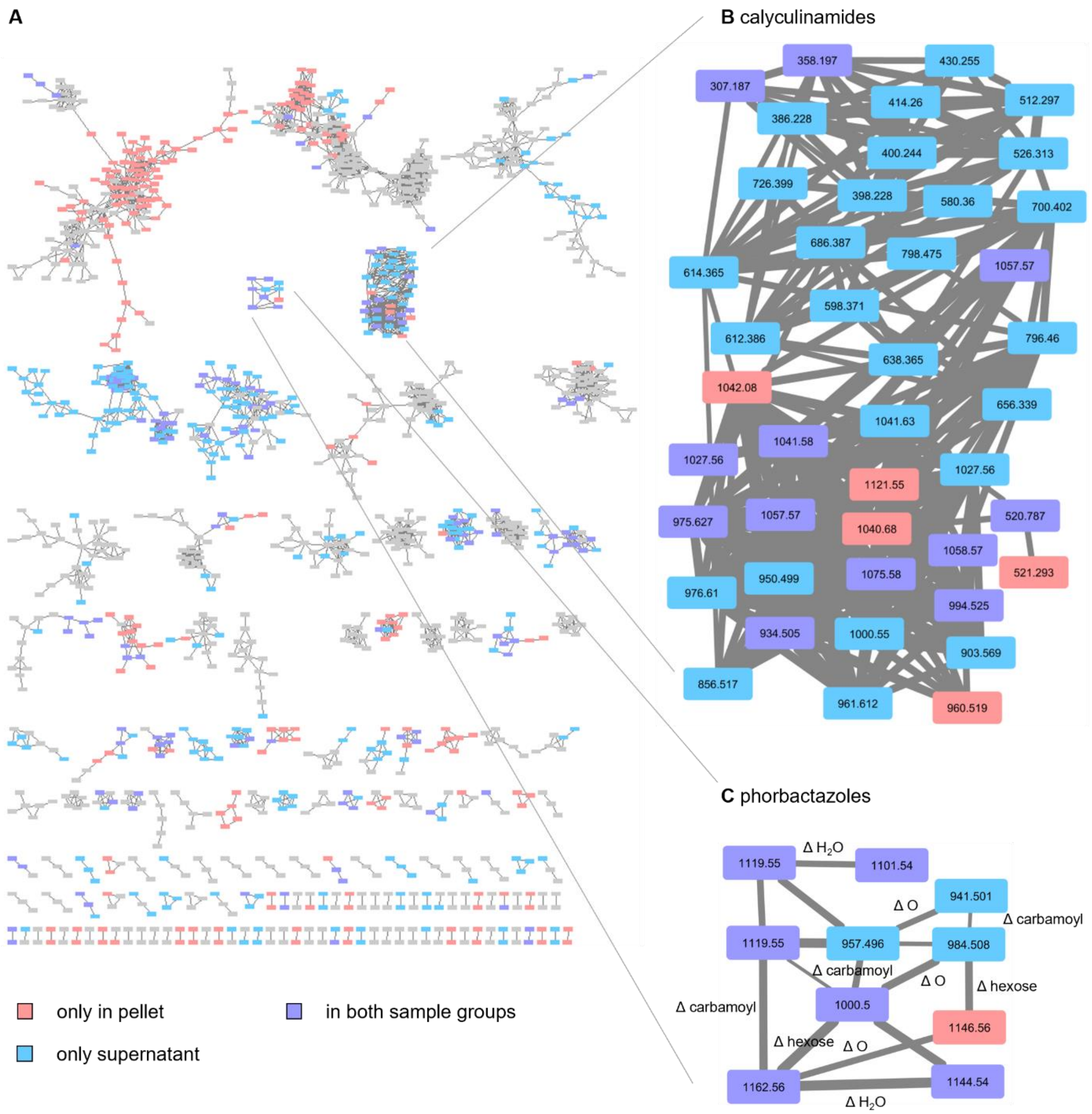

**Figure S3. MS molecular network of the extracted pellet (red) and supernatant (blue) of *A. pedis* cultivated in self-made marine broth.**

(A) A molecular network was constructed from HPLC-HESI-HRMS/MS data.[30]
(B) Ions belonging to the calyculin compound family. 21-methoxy calyculinamide (**11**, *m/z* 1041.58) is the most abundant peak in the extracts. Phospho-21-methoxy-calyculinamide (*m/z* 1121.55) was only detected in the pellet (Figure S51).
(C) Phorbactazole compound family. Differences of one oxygen (16.00 Da), water (18.01 Da), a carbamoyl group (43.01 Da) and a hexose (162.05 Da) are highlighted.

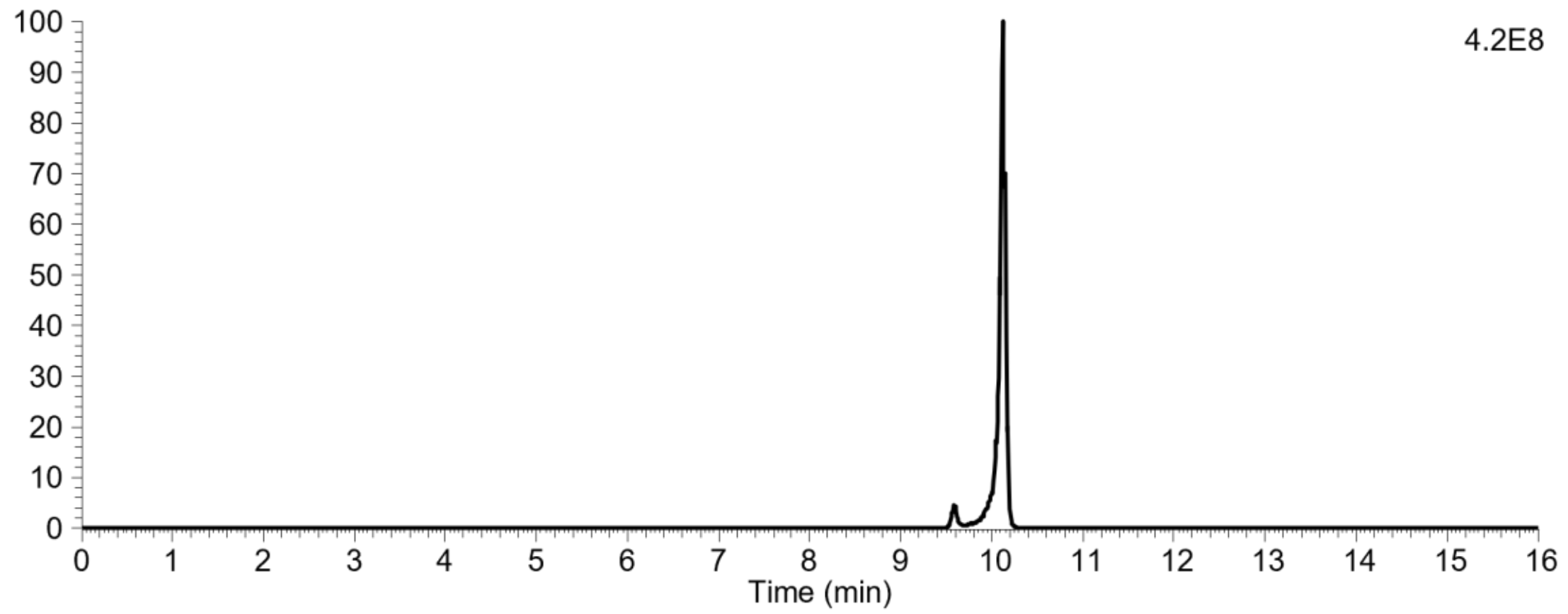

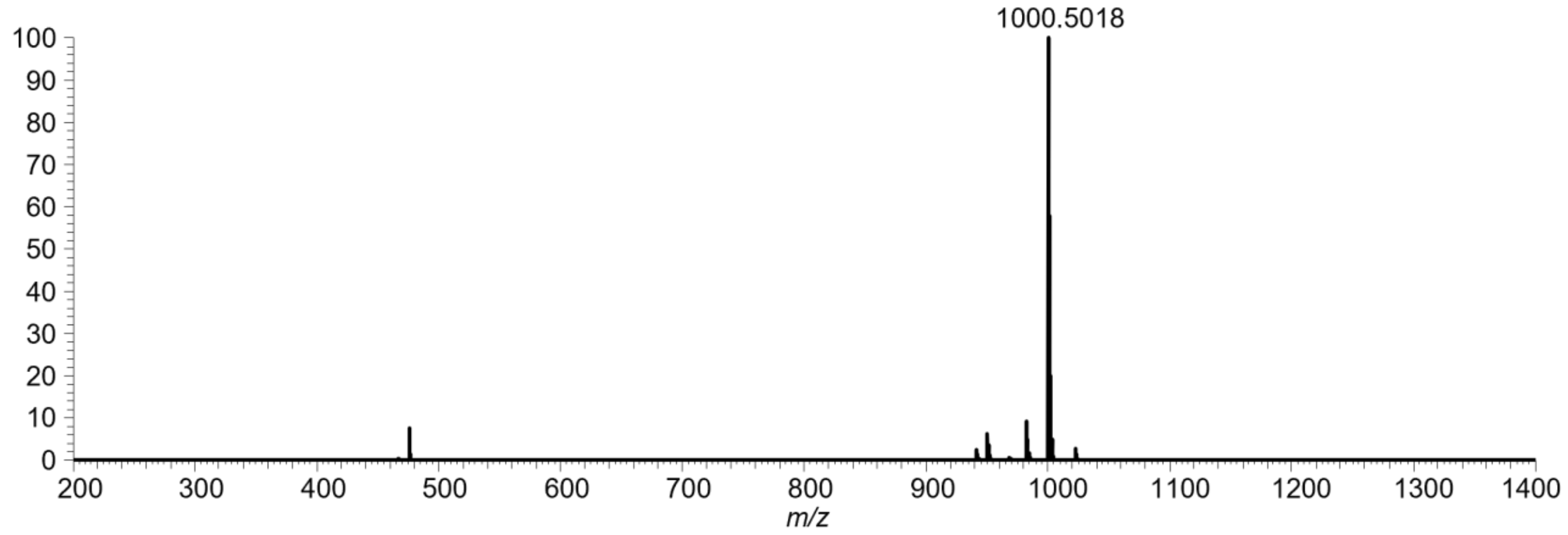

**Figure S4. HPLC-HESI-HRMS data of phorbactazole A (7).**

Top: Extracted ion chromatogram *m/z* 1000.5018.
Bottom: MS data of phorbactazole A (**7**) (*m/z* 1000.5018 $[M+H]^+$).

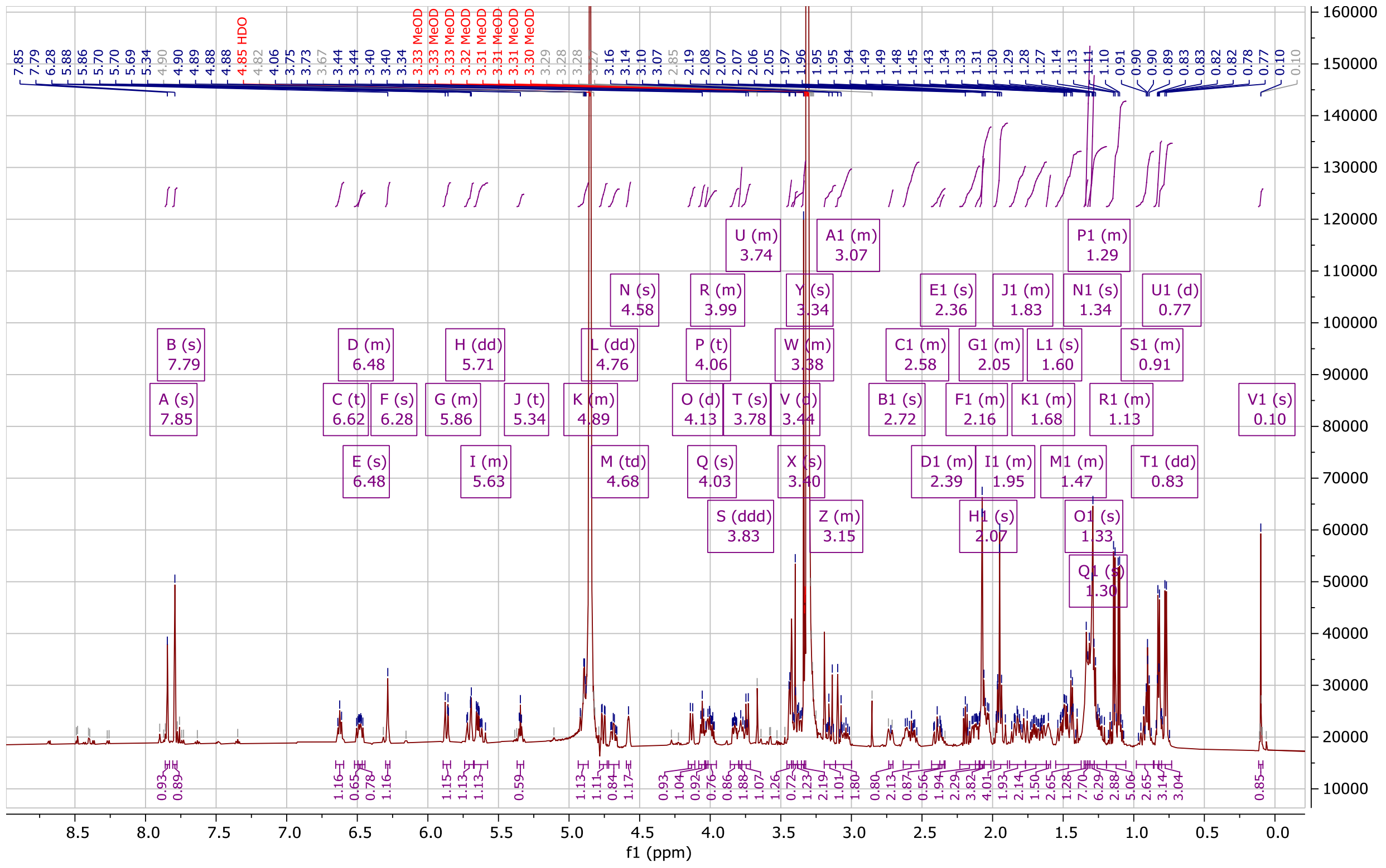

**Figure S5. $^{1}$H NMR spectrum of phorbactazole A (7) in methanol-$d_4$.**

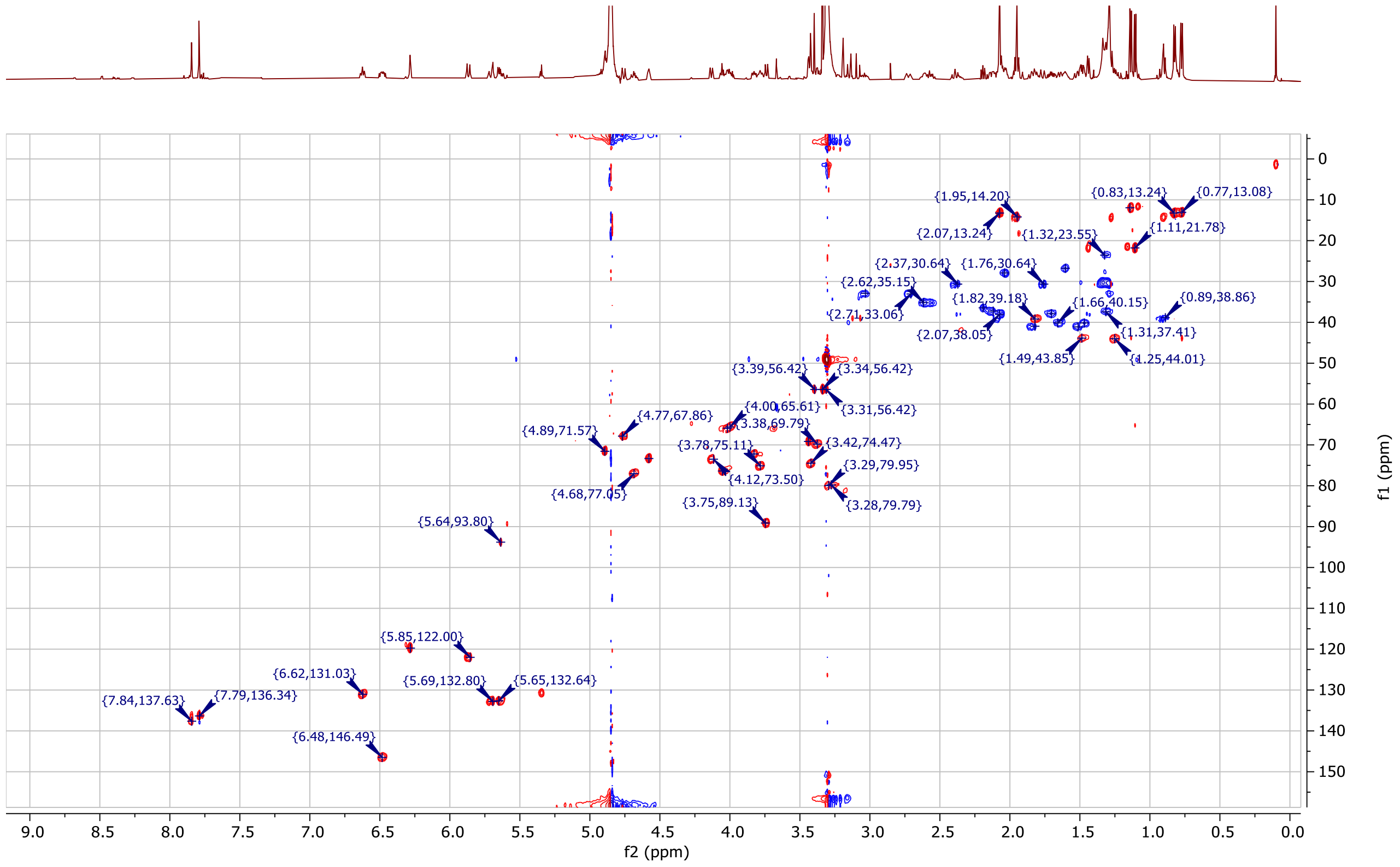

**Figure S6. HSQC spectrum of phorbactazole A (7) in methanol-$d_4$.**

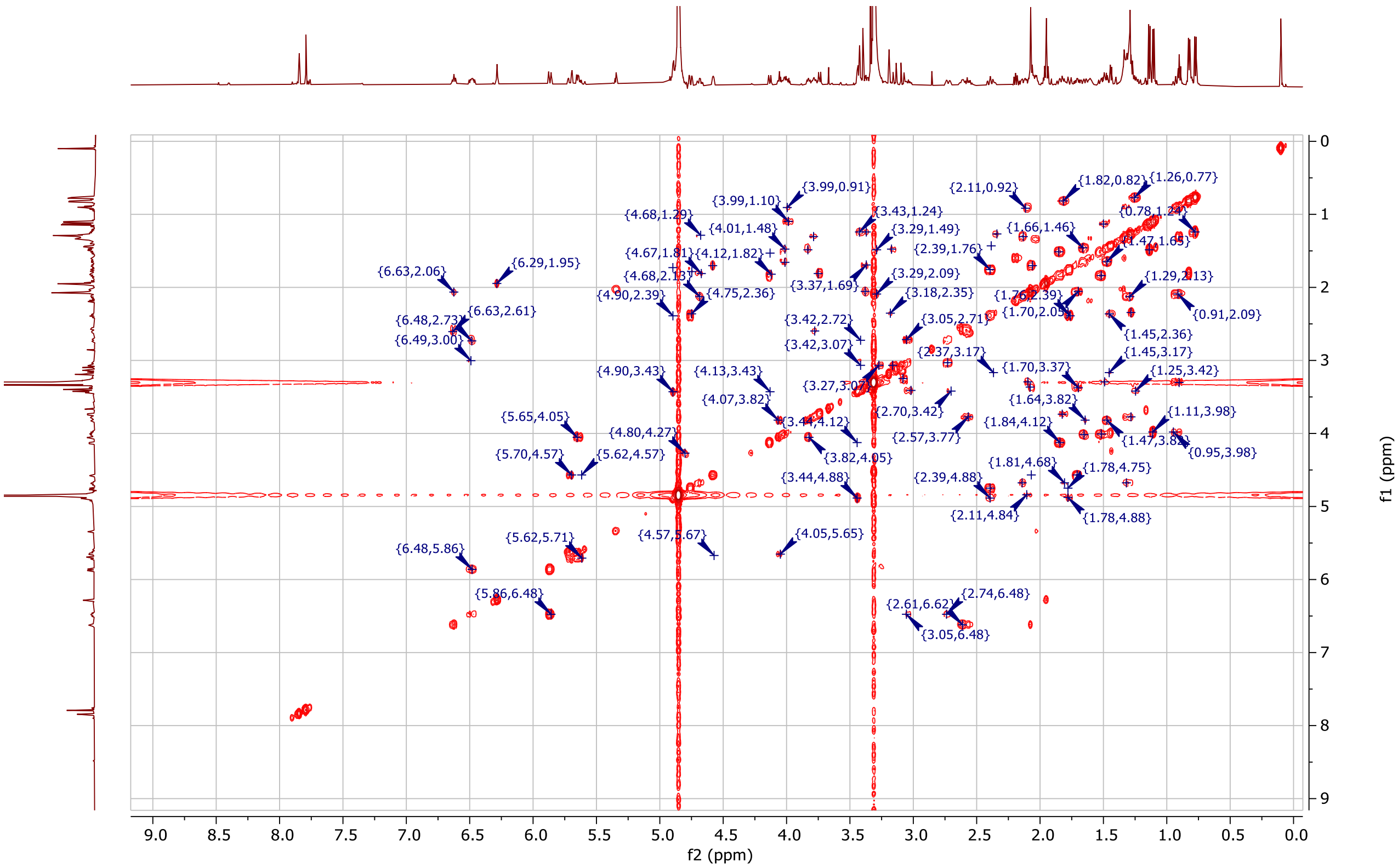

**Figure S7. COSY spectrum of phorbactazole A (7) in methanol-$d_4$.**

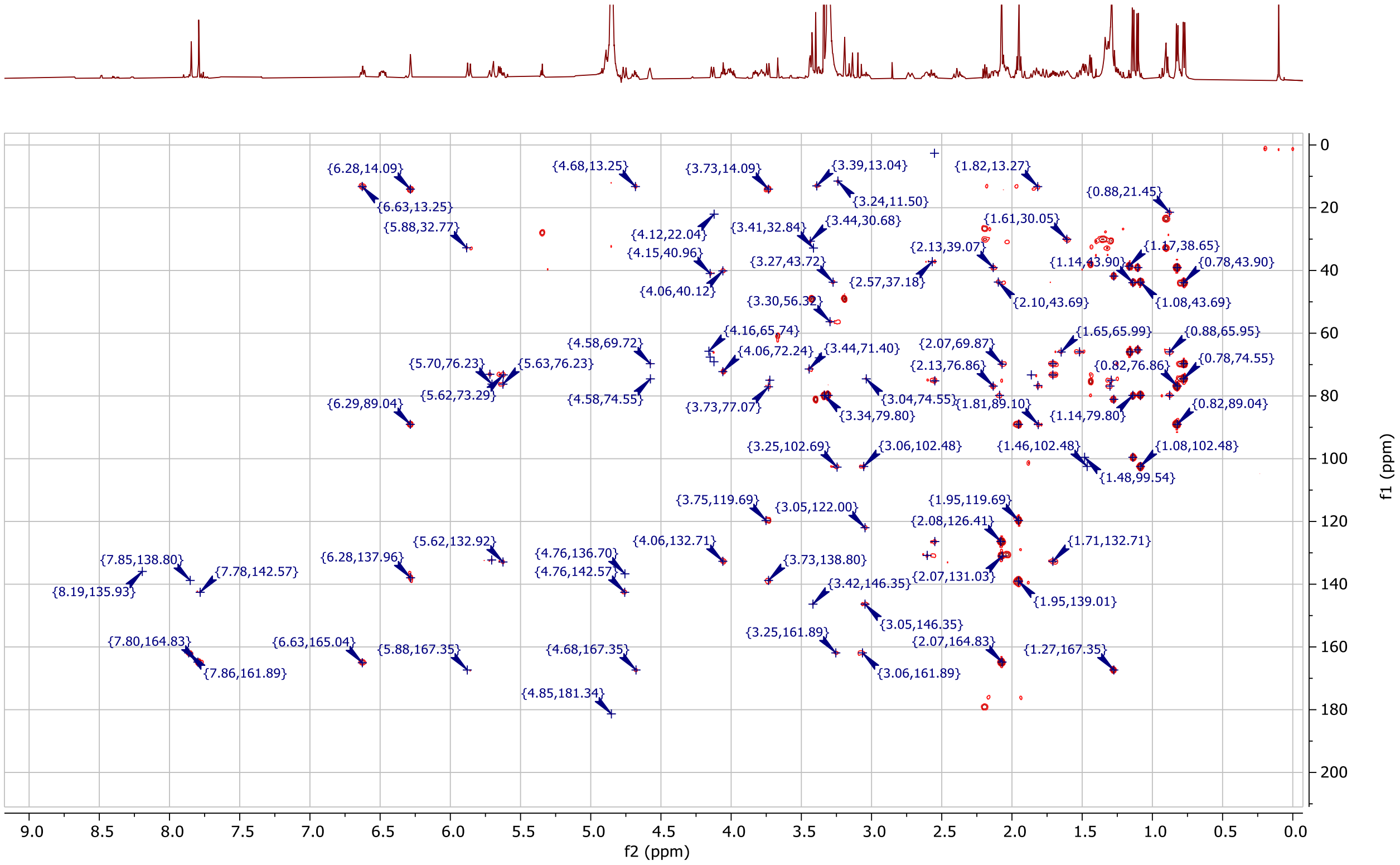

**Figure S8. HMBC spectrum of phorbactazole A (7) in methanol-$d_4$.**

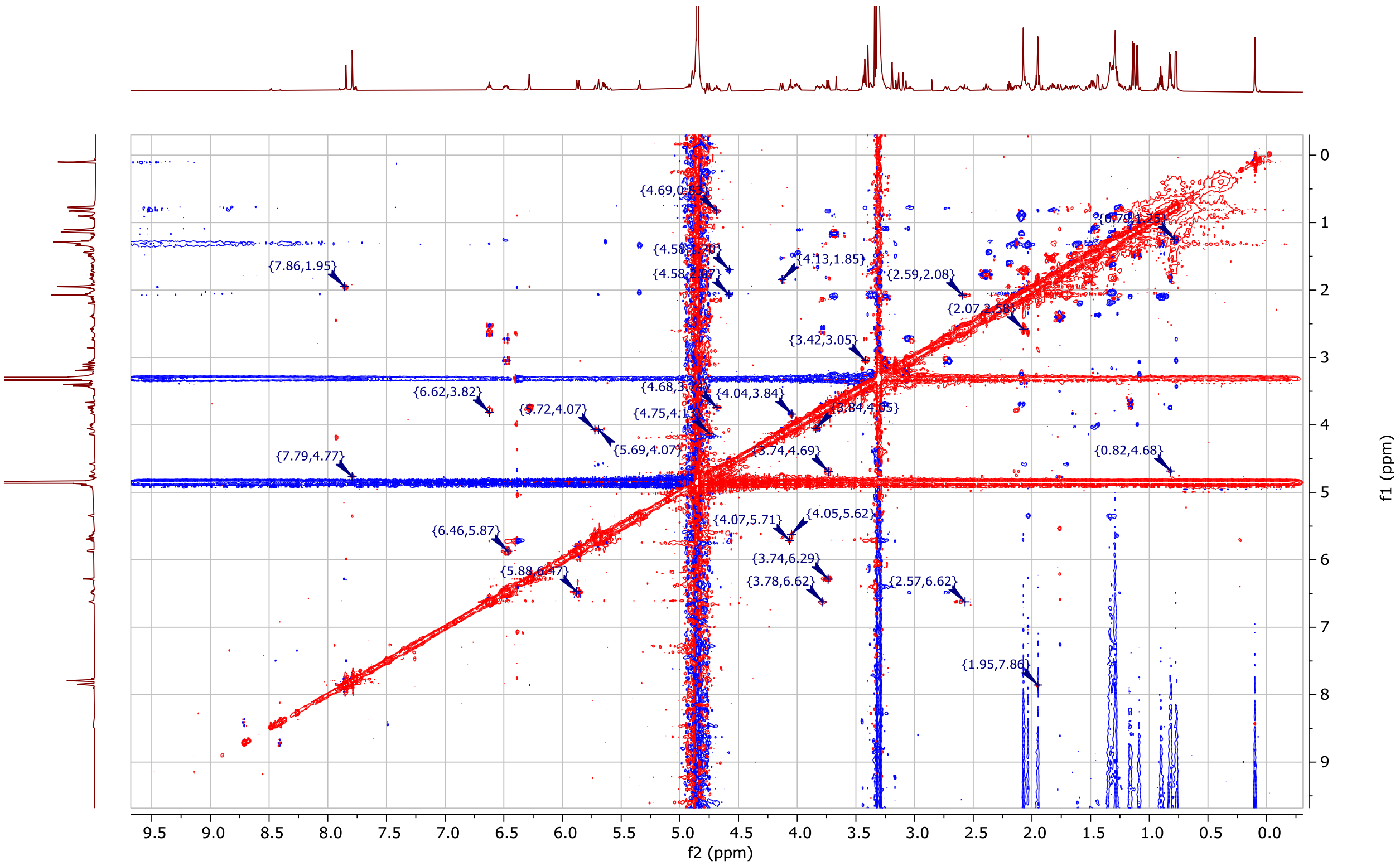

**Figure S9. NOESY spectrum of phorbactazole A (7) in methanol-$d_4$.**

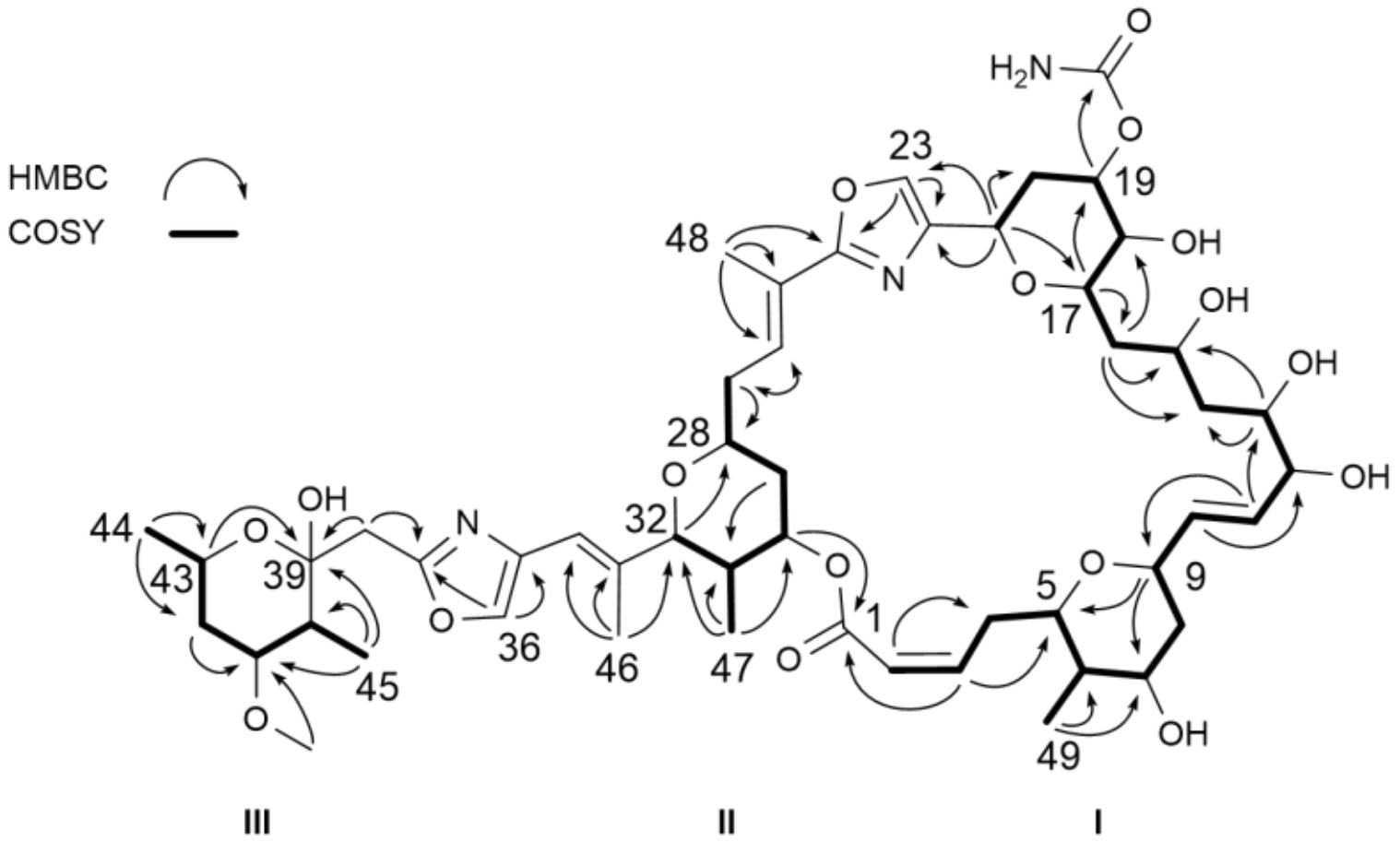

**Figure S10. COSY and HMBC correlations of phorbactazole A (7).**

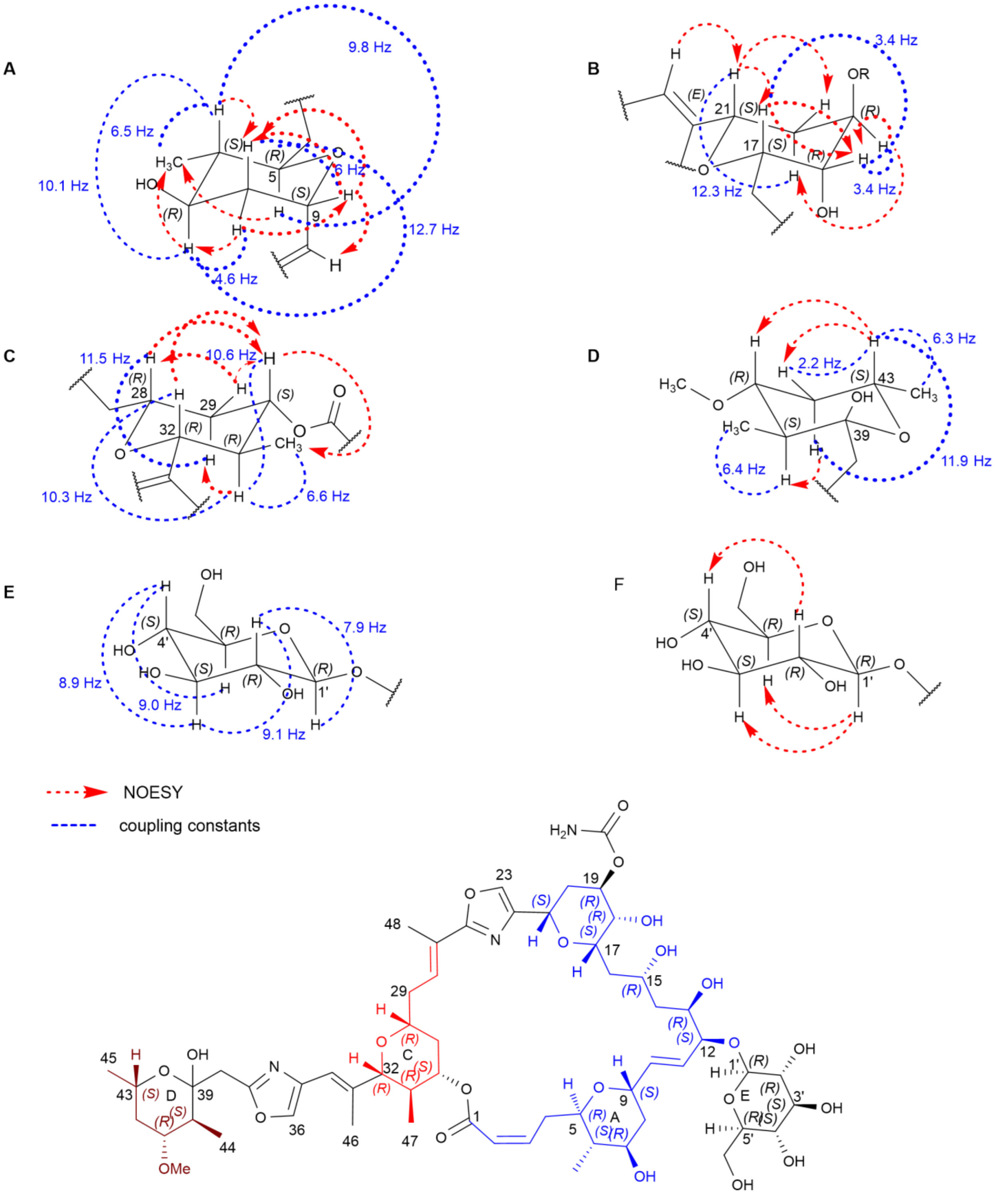

**Figure S11. Determination of the relative configuration of phorbactazoles by NOESY correlations.**

Rings A-D were determined for phorbactazole A (**7**). The hexose was elucidated for phorbactazole C (**9**). Regions of different color refer to spin systems.

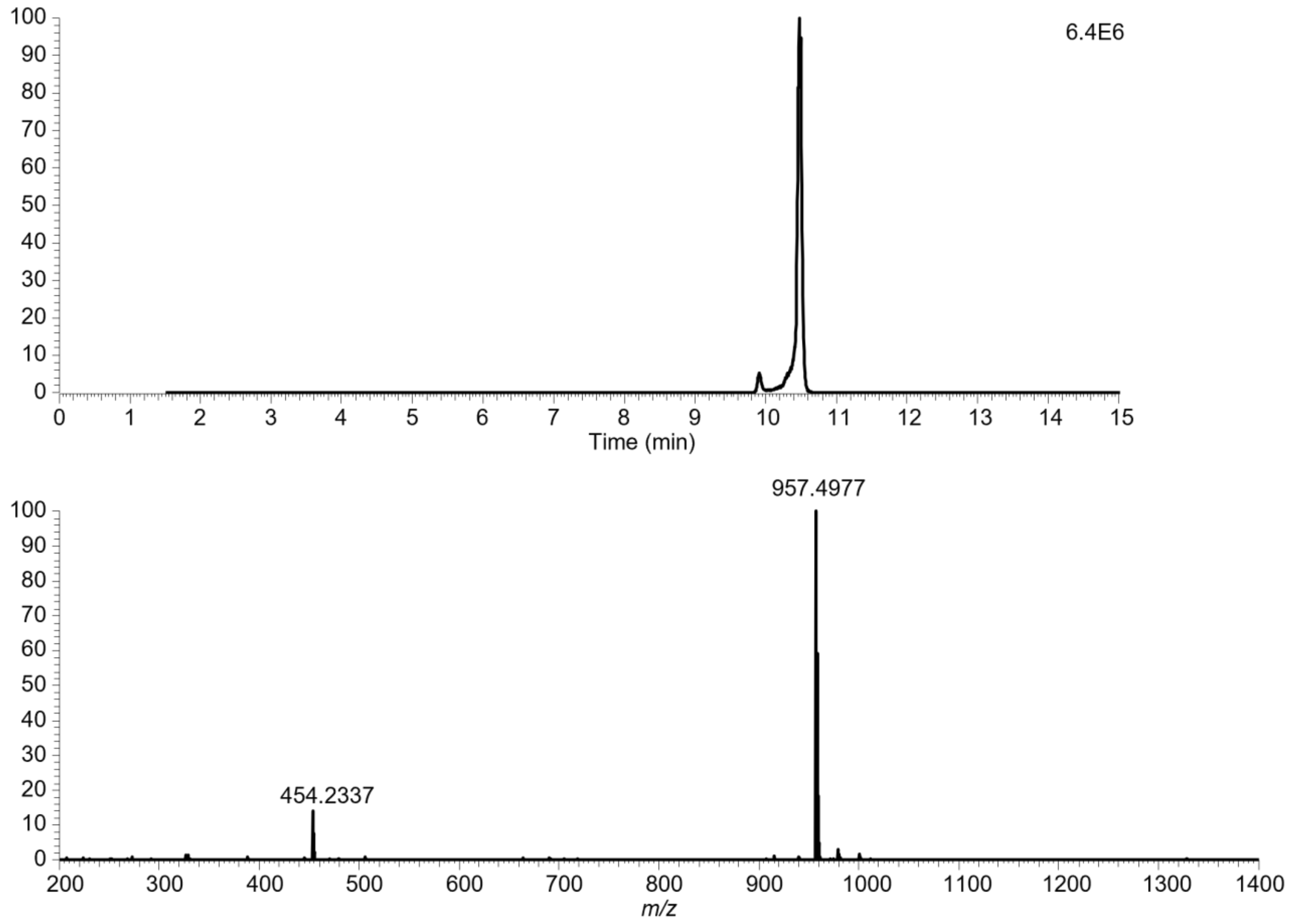

**Figure S12. HPLC-HESI-HRMS data of phorbactazole B (8).**

Top: Extracted ion chromatogram *m/z* 957.4961.
Bottom: MS data of phorbactazole B (**8**) (*m/z* 957.4977 $[M+H]^+$).

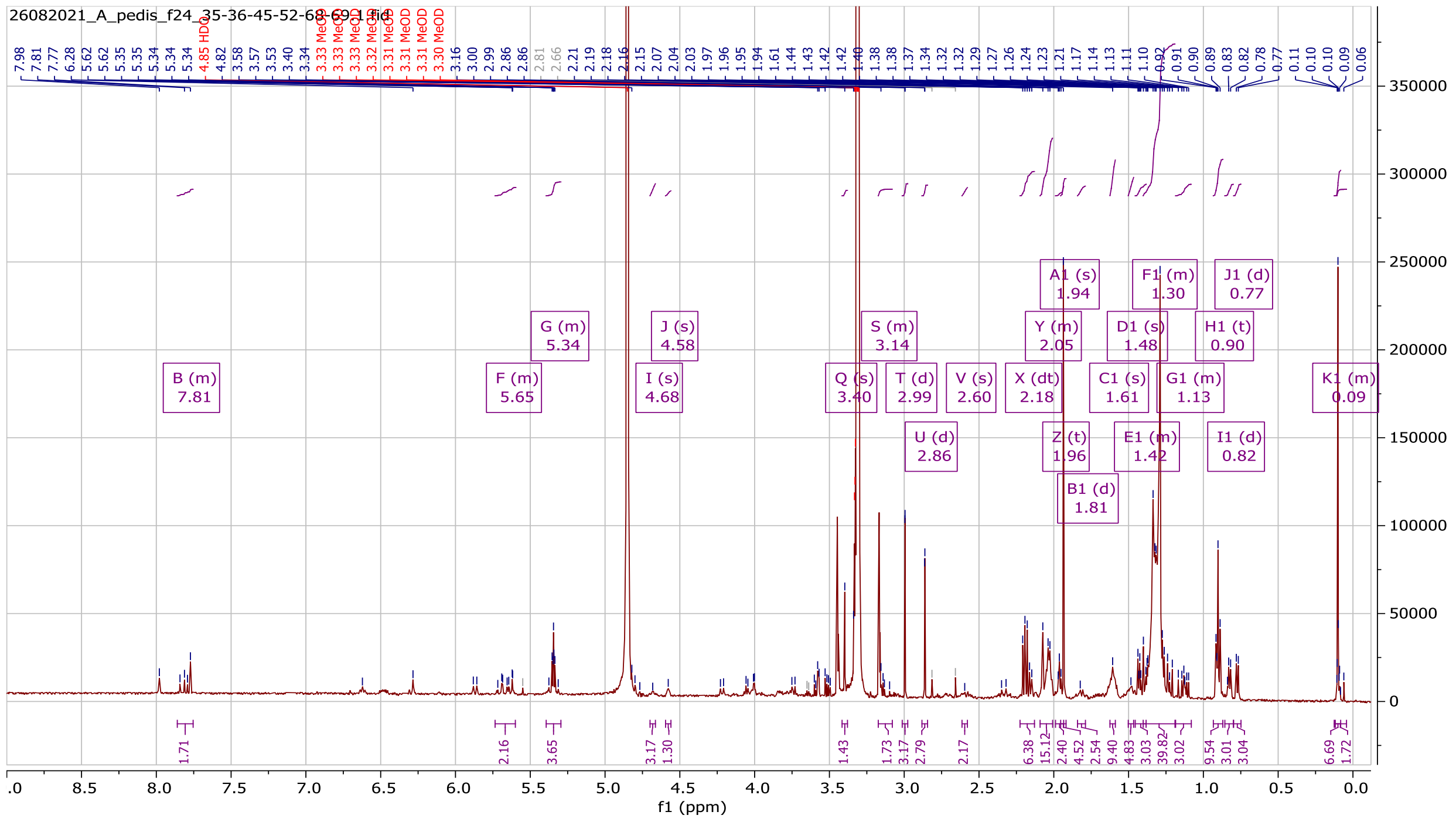

**Figure S13. $^{1}$H NMR spectrum of phorbactazole B (8) in methanol-$d_4$.**

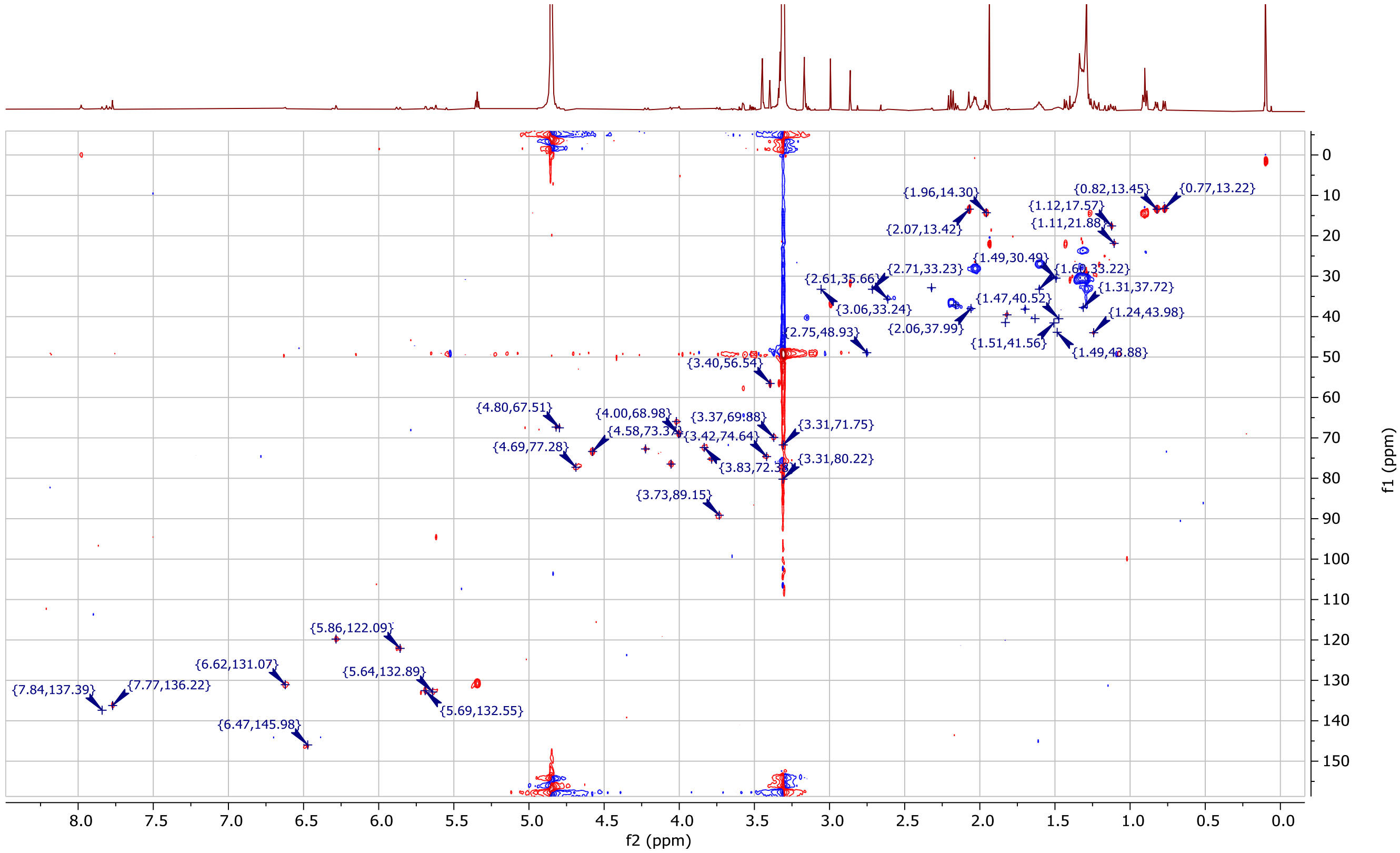

**Figure S14. HSQC spectrum of phorbactazole B (8) in methanol-$d_4$.**

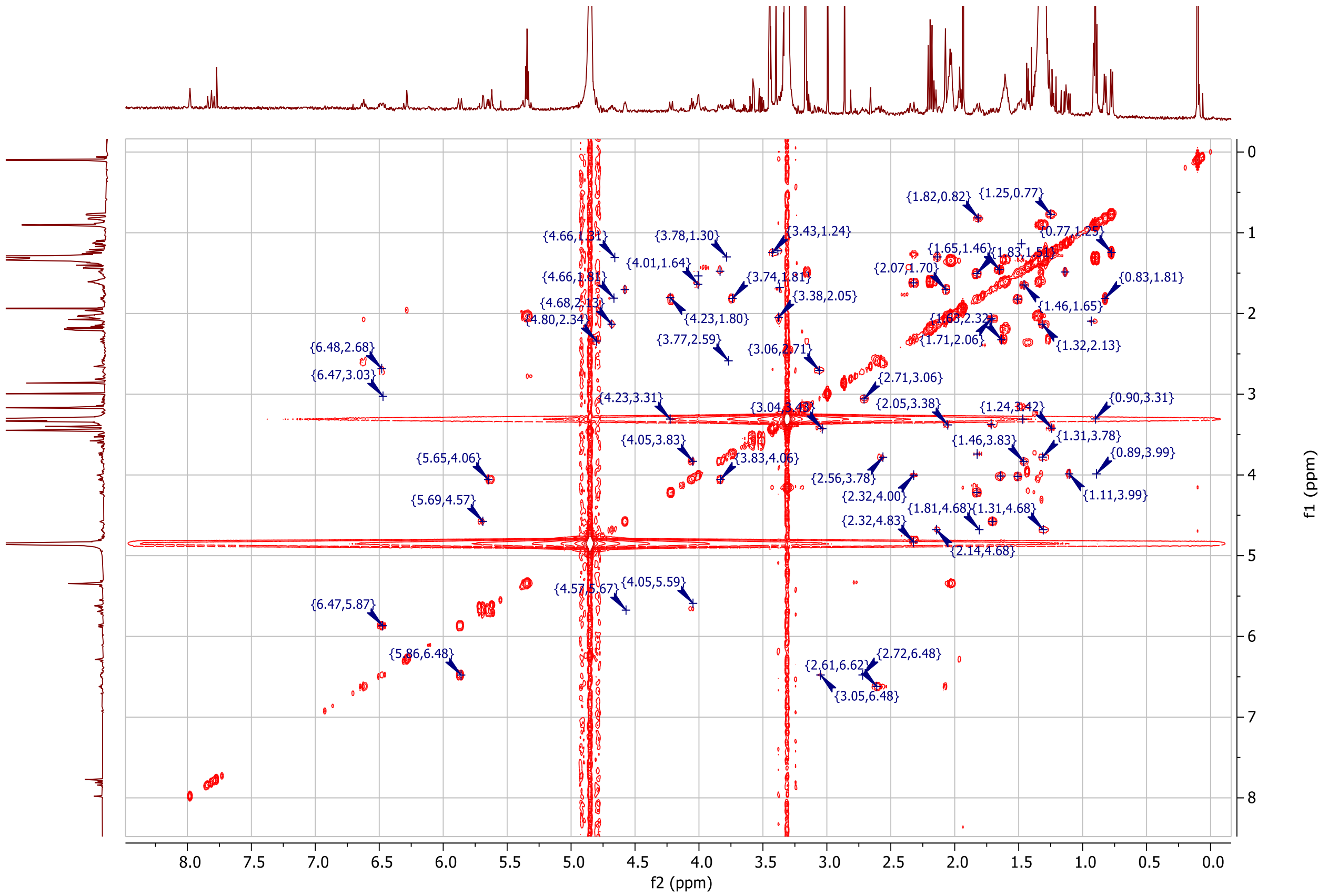

**Figure S15. COSY spectrum of phorbactazole B (8) in methanol-$d_4$.**

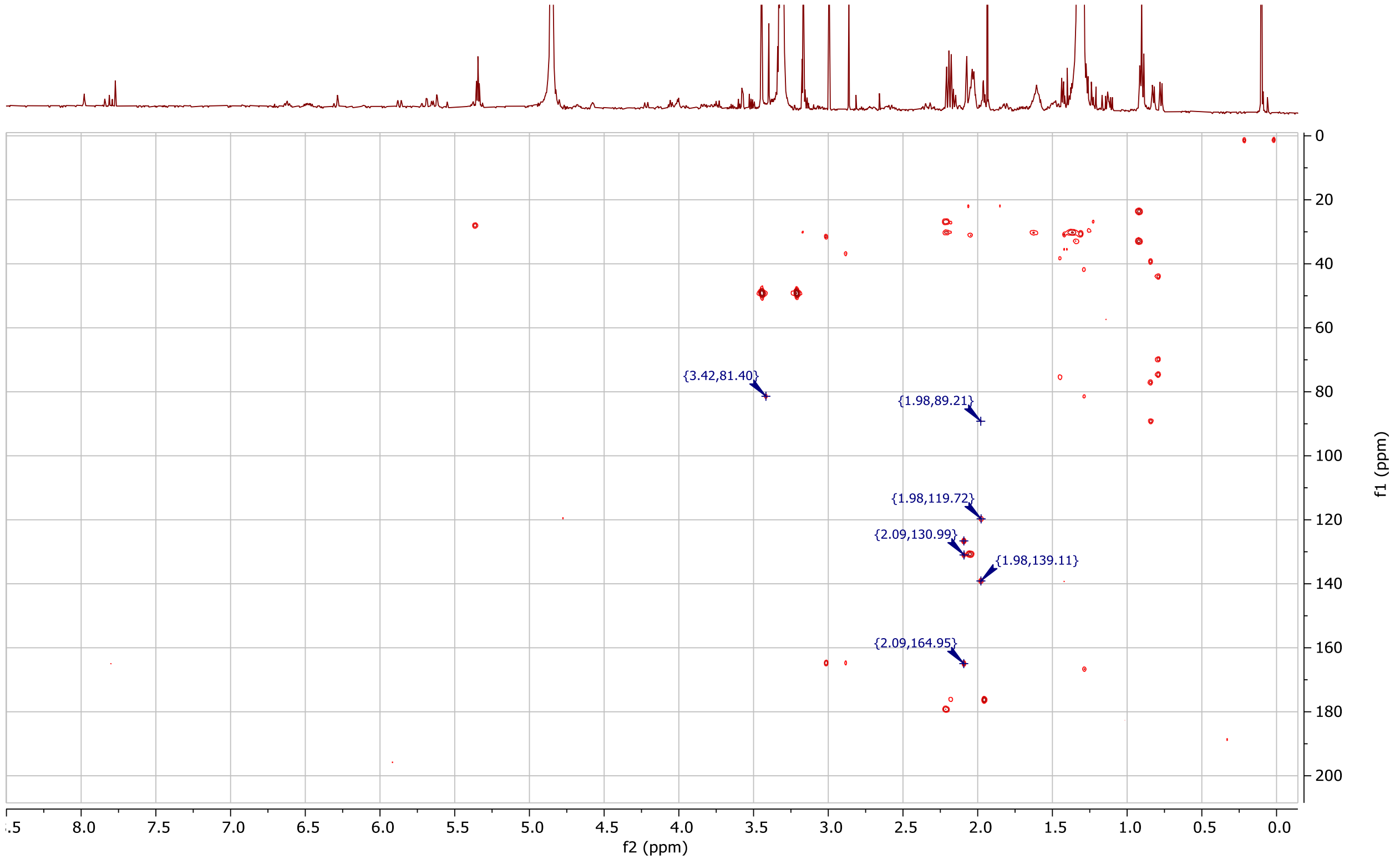

**Figure S16. HMBC spectrum of phorbactazole B (8) in methanol-$d_4$.**

**Table S3. NMR Data ($^1$H 600 MHz, $^{13}$C 150 MHz) for 7 and 8 in methanol-$d_4$.**

| | Phorbactazole A (7) | | Phorbactazole C (8) | |
|---|---|---|---|---|
| **No.** | **$^{13}$C/HSQC** | **$^1$H (multiplicity)** | **$^{13}$C/HSQC** | **$^1$H (multiplicity)** |
| 1 | 167.4, C | | 167.4, C | |
| 2 | 122.0, CH | 5.86 (d), *J* = 10.9 Hz | 122.1, CH | 5.86 (d), *J* = 10.84 Hz |
| 3 | 146.5, CH | 6.48 (m) | 146.3, CH | 6.48 (m) |
| 4 | 33.1, $CH_2$ | 2.72 (m), 3.03 (m) | 33.2, $CH_2$ | 2.71 (m), 3.04 (m) |
| 5 | 74.5, CH | 3.42 (m) | 74.6, CH | 3.42 (m) |
| 6 | 44.0, CH | 1.25 (m) | 44.0, CH | 1.24 (m) |
| 7 | 69.8, CH | 3.38 (td), *J* = 3.6, 10.2 Hz | 69.8, CH | 3.37 (m) |
| 8 | 37.9, $CH_2$ | 1.70 (m), 2.07 (m) | 38.0, $CH_2$ | 1.70 (m), 2.06 (m) |
| 9 | 73.3, CH | 4.58 (m) | 73.4, CH | 4.58 (m) |
| 10 | 132.8, CH | 5.70 (dd), *J* = 3.1, 16.2 Hz | 132.8, CH | 5.70 (dd), *J* = 3.1, 16.4 Hz |
| 11 | 132.7, CH | 5.65 (dd), *J* = 5.6, 16.2 Hz | 132.8, CH | 5.64 (dd), *J* = 5.44, 16.4 Hz |
| 12 | 76.4, CH | 4.05 (t), *J* = 5.63 Hz | 76.5, CH | 4.05 (t), *J* = 5.90 Hz |
| 13 | 72.2, CH | 3.83 (ddd), *J* = 2.8, 5.7. 9.0 Hz | 72.4, CH | 3.83 (ddd), *J* = 2.5, 5.8. 9.3 Hz |
| 14 | 40.2, $CH_2$ | 1.47 (m), 1.66 (m) | 40.3, $CH_2$ | 1.47 (m), 1.65 (m) |
| 15 | 65.9, CH | 4.02 (m) | 65.9, CH | 4.02 (m) |
| 16 | 41.1, $CH_2$ | 1.51 (m), 1.85 (m) | 41.4, $CH_2$ | 1.51 (m), 1.85 (m) |
| 17 | 73.5, CH | 4.14 (dt), *J* = 3.7, 10.1 Hz | 72.7, CH | 4.22 (m) |
| 18 | 69.2, CH | 3.44 (dd), *J* = 3.4 Hz | 71.8, CH | 3.31 (m) |
| 19 | 71.6, CH | 4.89 (q), *J* = 3.4 Hz | 69.0, CH | 4.00 (m) |
| 20 | 30.6, $CH_2$ | 1.76 (m), 2.39 (m) | 33.2, $CH_2$ | 1.60 (m), 2.32 (m) |
| 21 | 67.7, CH | 4.76 (dd), *J* = 2.4, 12.3 Hz | 67.5, CH | 4.81 (dd) *J* = 2.0, 12.1 Hz |
| 22 | 142.6, C | | 142.6, C | |
| 23 | 136.3, CH | 7.79 (s) | 136.2, CH | 7.77 (s) |
| 24 | 164.8, C | | 164.9, C | |
| 25 | 126.4, C | | 126.6, C | |
| 26 | 131.0, CH | 6.62 (t), *J* = 7.9 Hz | 131.1, CH | 6.62 (t), *J* = 8.2 Hz |
| 27 | 35.2, $CH_2$ | 2.59 (m) | 35.4, $CH_2$ | 2.59 (m) |
| 28 | 75.1, CH | 3.78 (m) | 75.3, CH | 3.78 (m) |
| 29 | 37.4, $CH_2$ | 1.30 (m), 2.12 (m) | 37.5, $CH_2$ | 1.31 (m), 2.12 (m) |
| 30 | 77.1, CH | 4.68 (td), *J* = 4.6, 10.6 Hz | 77.2, CH | 4.69 (td), *J* = 4.2, 10.7 Hz |
| 31 | 39.2, CH | 1.82 (m) | 39.6, CH | 1.82 (m) |
| 32 | 89.1, CH | 3.74 (d), *J* = 10.3 Hz | 89.2, CH | 3.73 (d), *J* = 10.4 Hz |
| 33 | 139.0, C | | 139.1, C | |
| 34 | 119.8, CH | 6.28 (s) | 119.8, CH | 6.28 (s) |
| 35 | 138.8, C | | 138.8, C | |
| 36 | 137.6, CH | 7.84 (s) | 137.5, CH | 7.84 (s) |
| 37 | 161.9, C | | 161.9, C | |
| 38 | 39.5, $CH_2$ | 3.07 (d), *J* = 14.5 Hz<br>3.14 (d), *J* = 14.5 Hz | 40.2, $CH_2$ | 3,06 (d), *J* = 14.6 Hz<br>3.23 (s), *J* = 14.6 Hz |
| 39 | 99.5, C | | 99.3, C | |
| 40 | 43.9, CH | 1.49 | 43.9, CH | 1.49 |
| 41 | 79.9, CH | 3.29, (m) | 80.0, CH | 3.30 (m) |
| 42 | 38.9, $CH_2$ | 0.89 (m), 2.09 (m) | 38.9, $CH_2$ | 0.89 (m), 2.09 (m) |
| 43 | 65.9, CH | 4.00 (ddd), *J* = 2.3 ,5.6, 11.8 Hz | 65.9, CH | 4.00 (m) |
| 44 | 21.8, $CH_3$ | 1.11 (d), *J* = 5.9 Hz | 21.9, $CH_3$ | 1.11 (d), *J* = 5.9 Hz |
| 45 | 11.9, $CH_3$ | 1.14 (d), *J* = 6.4 Hz | 12.0, $CH_3$ | 1.13 (d), *J* = 6.5 Hz |
| 46 | 14.2, $CH_3$ | 1.95 (s) | 14.2, $CH_3$ | 1.95 (s) |
| 47 | 13.2, $CH_3$ | 0.83 (d), *J* = 6.6 Hz | 13.4, $CH_3$ | 0.82 (d), *J* = 6.7 Hz |
| 48 | 13.2, $CH_3$ | 2.07 (s) | 13.4, $CH_3$ | 2.07 (s) |
| 49 | 13.1, $CH_3$ | 0.77 (d), *J* = 6.5 Hz | 13.2, $CH_3$ | 0.77 (d), *J* = 6.6 Hz |
| 41 $OCH_3$ | 56.4, $CH_3$ | 3.34 (s) | 56.5, $CH_3$ | 3.40 (s) |
| 19 -$CONH_2$ | 158.7, C | | | |

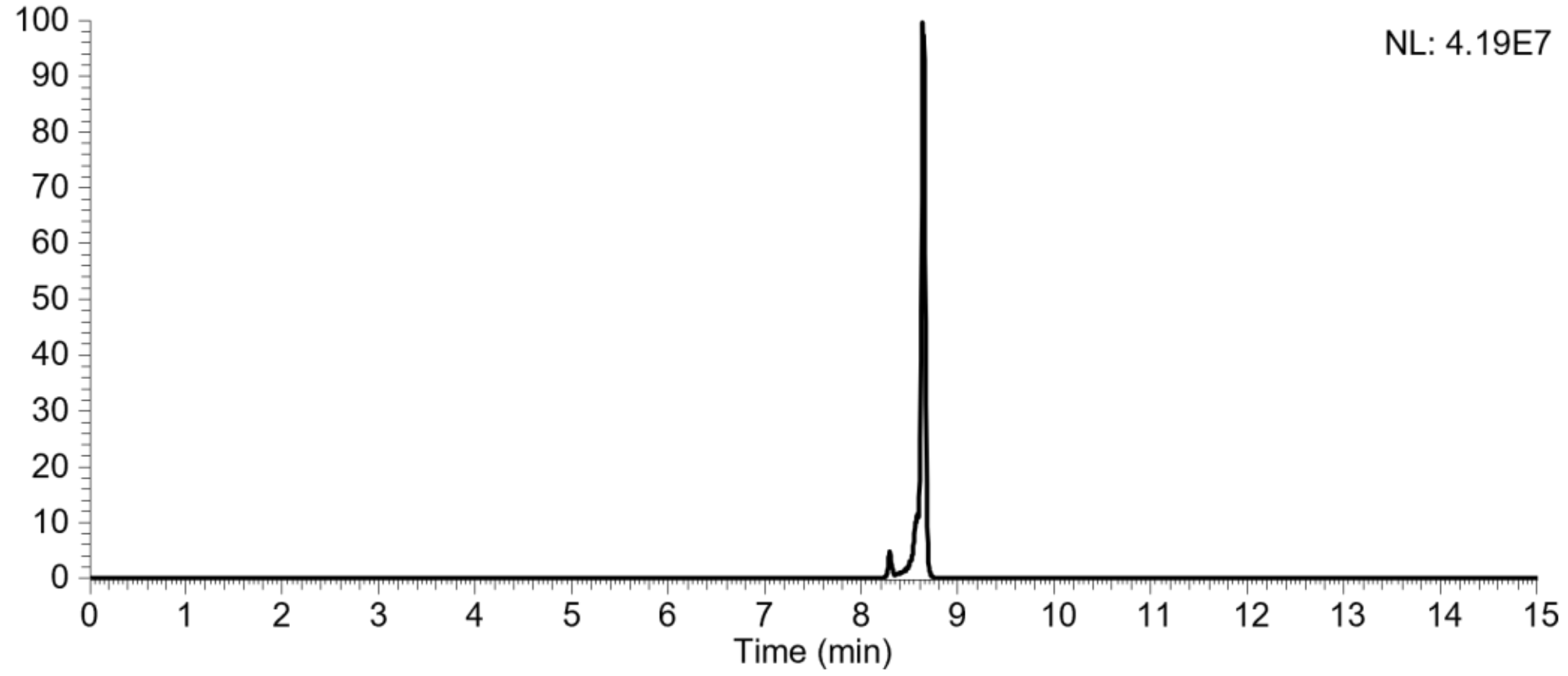

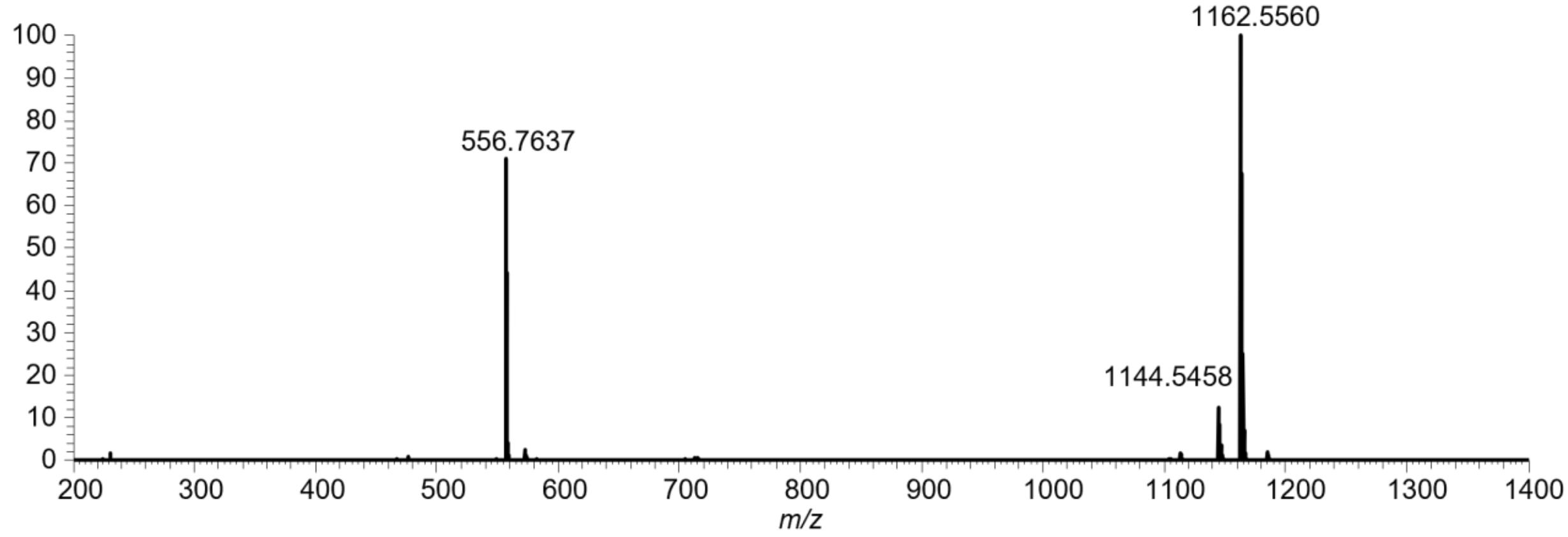

**Figure S17. HPLC-HESI-HRMS data of phorbactazole C (9).**

Top: Extracted ion chromatogram *m/z* 1162.5554.
Bottom: MS data of phorbactazole C (**9**) (*m/z* 1162.5560 [M+H]$^+$).

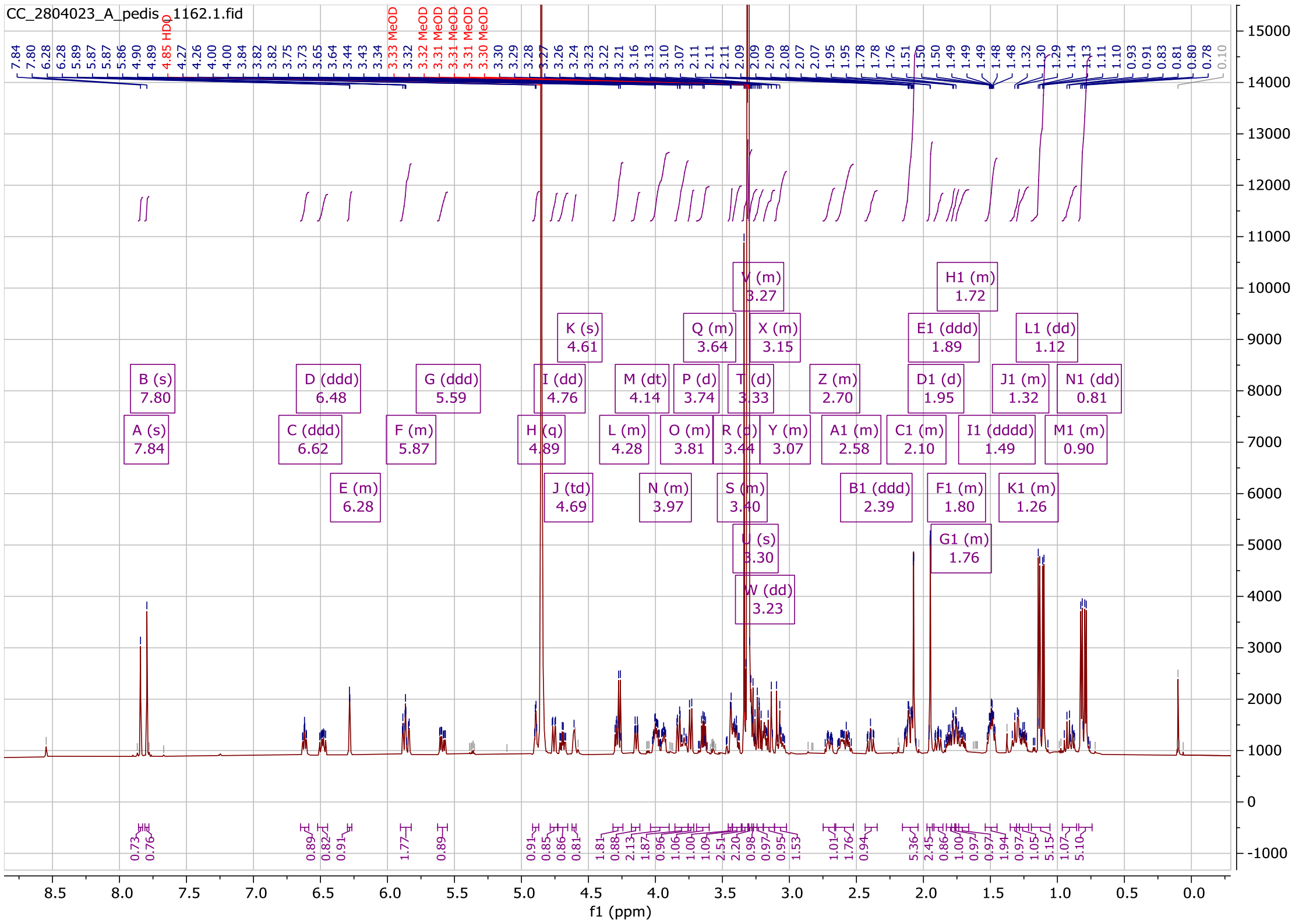

**Figure S18.** [1]**H NMR spectrum of phorbactazole C (9) in methanol-*d*$_4$.**

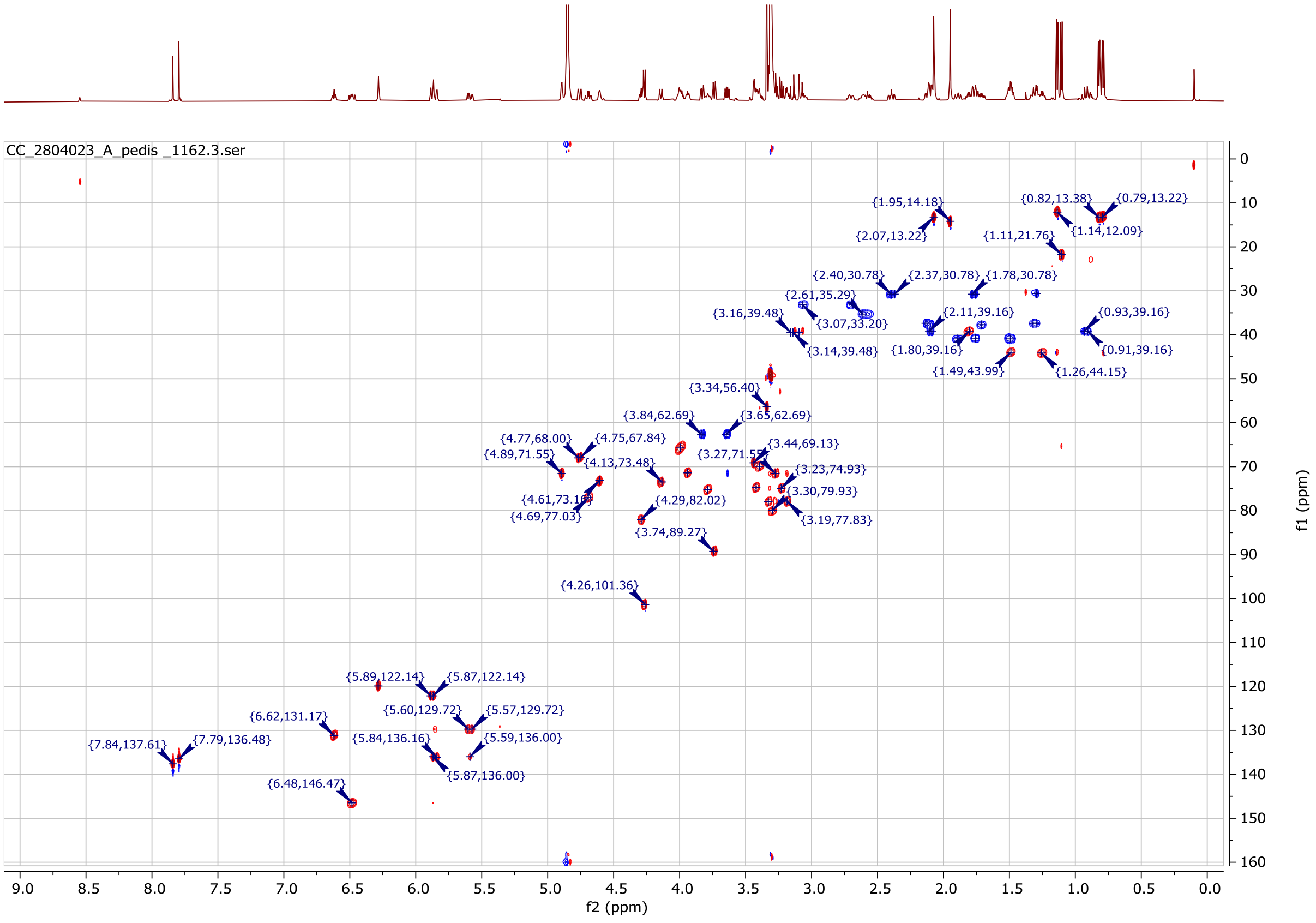

**Figure S19. HSQC spectrum of phorbactazole C (9) in methanol-*d*$_4$.**

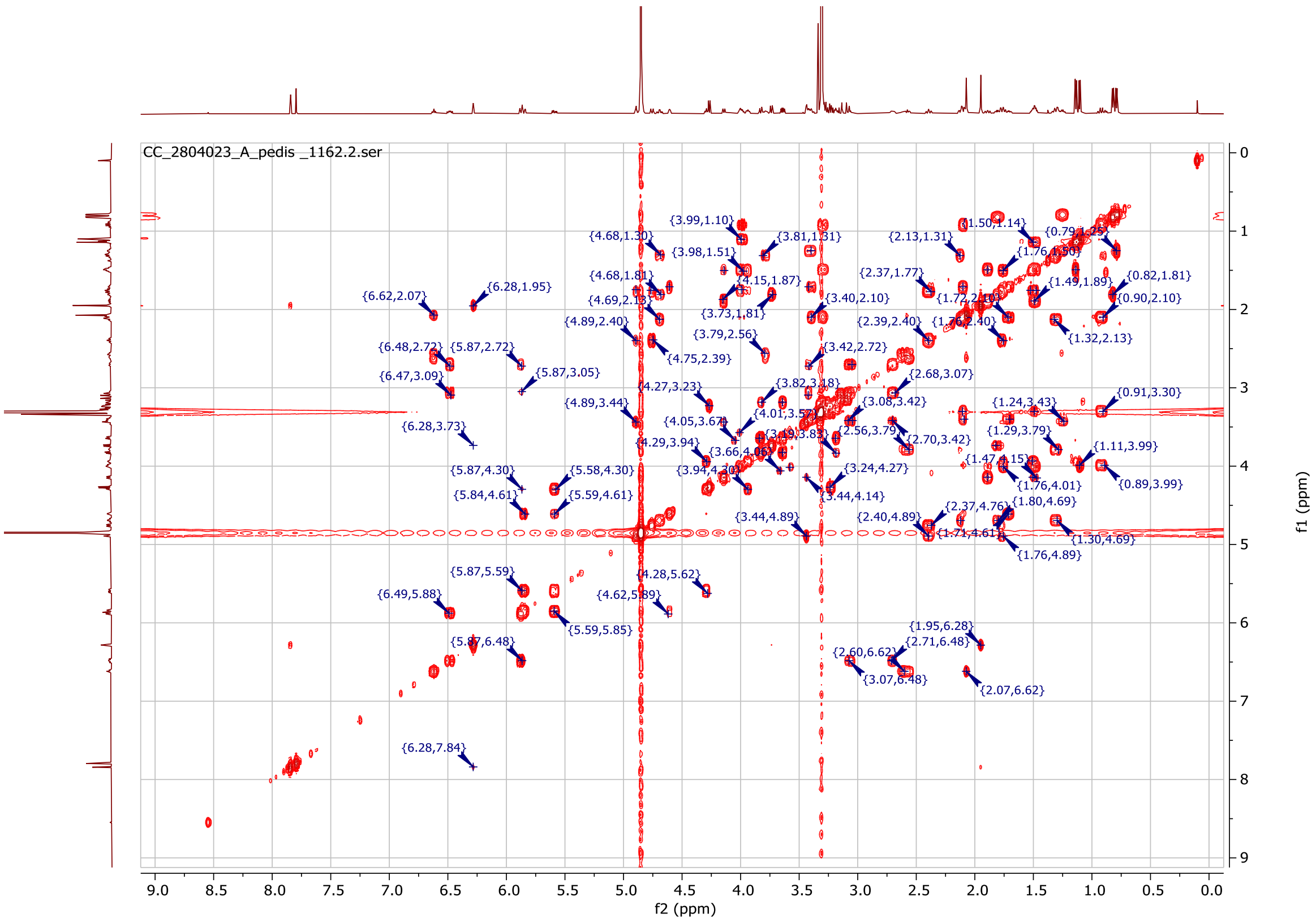

**Figure S20. COSY spectrum of phorbactazole C (9) in methanol-$d_4$.**

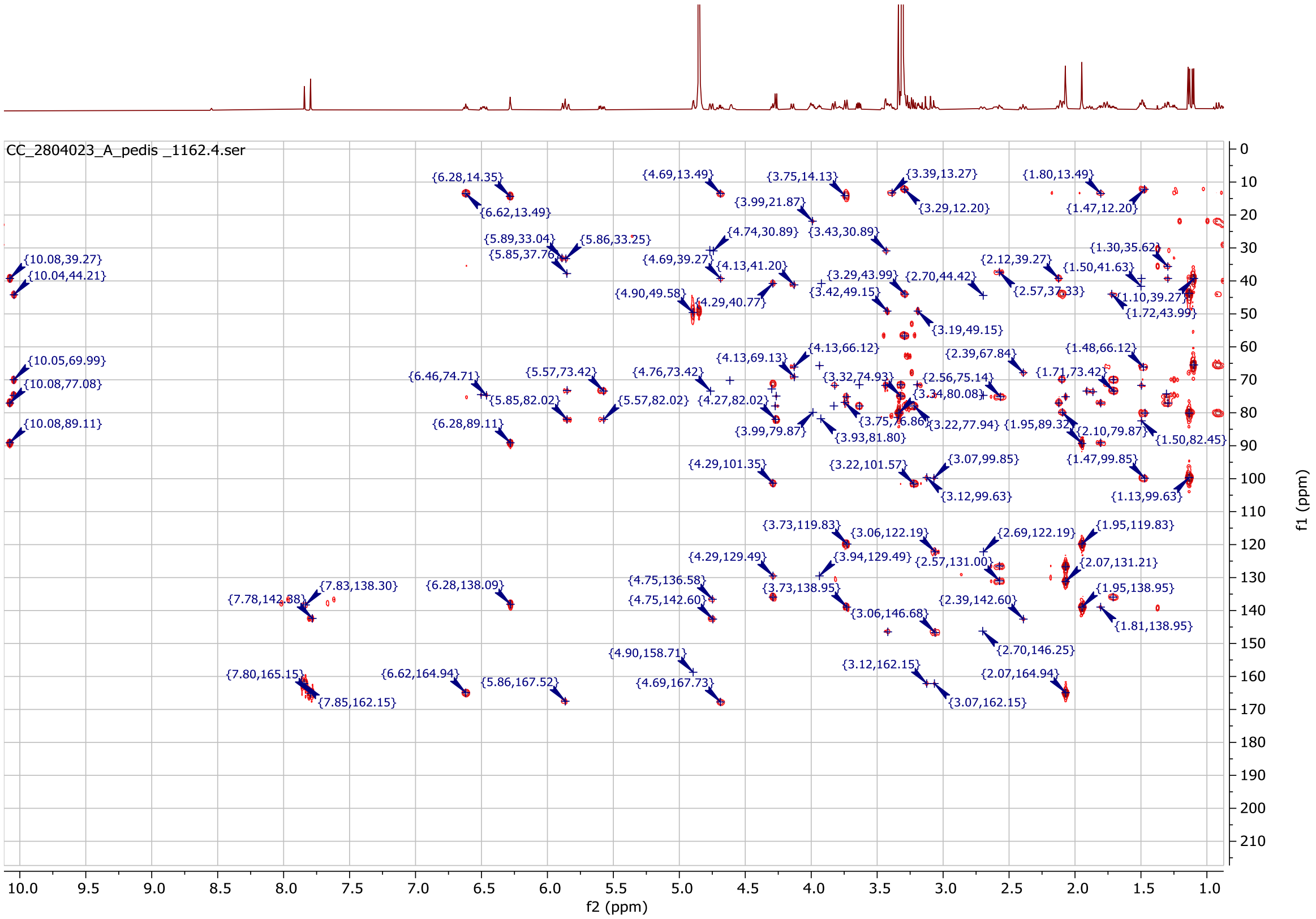

**Figure S21. HMBC spectrum of phorbactazole C (9) in methanol-$d_4$.**

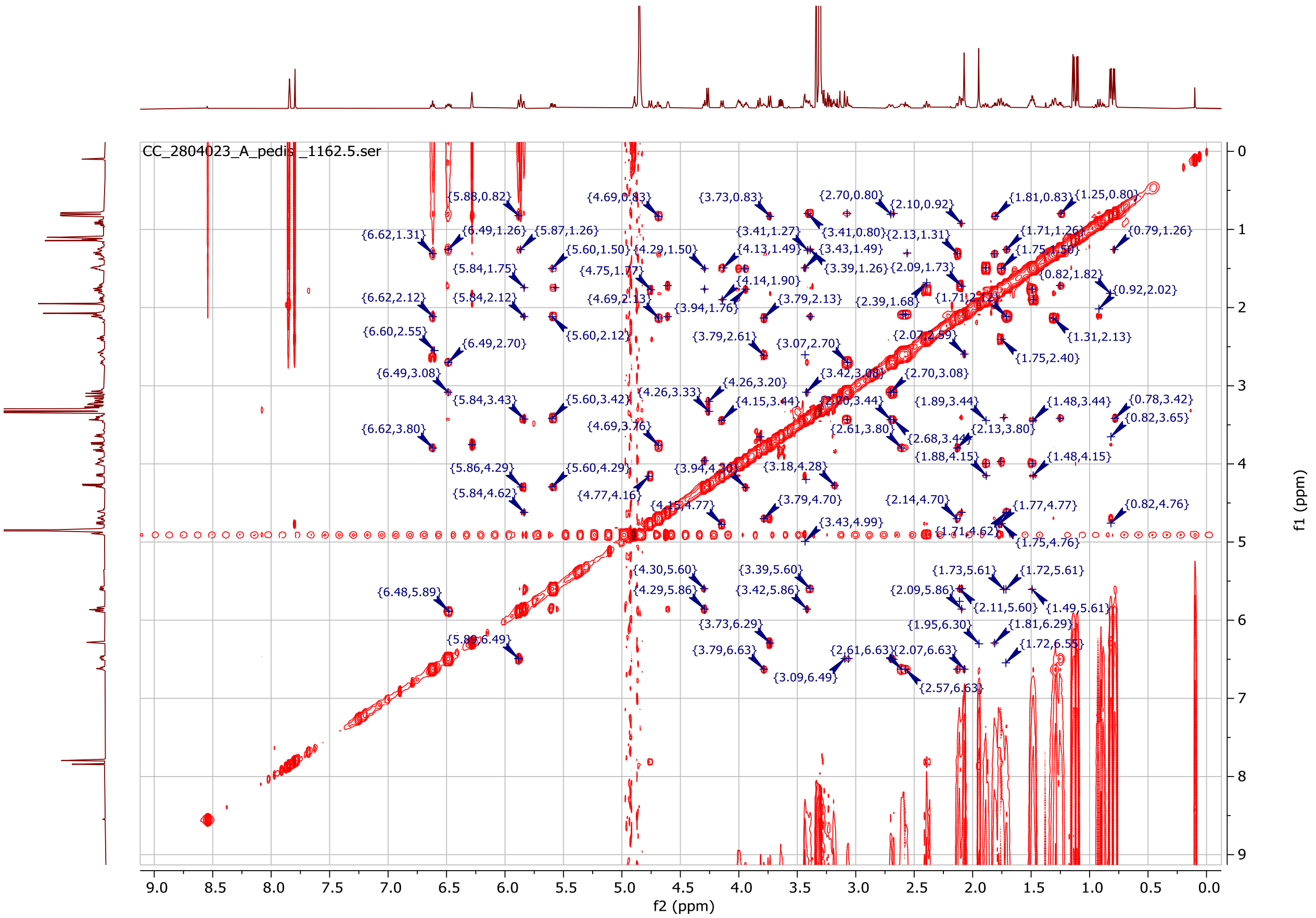

**Figure S22. NOESY spectrum of phorbactazole C (9) in methanol-*d*$_4$.**

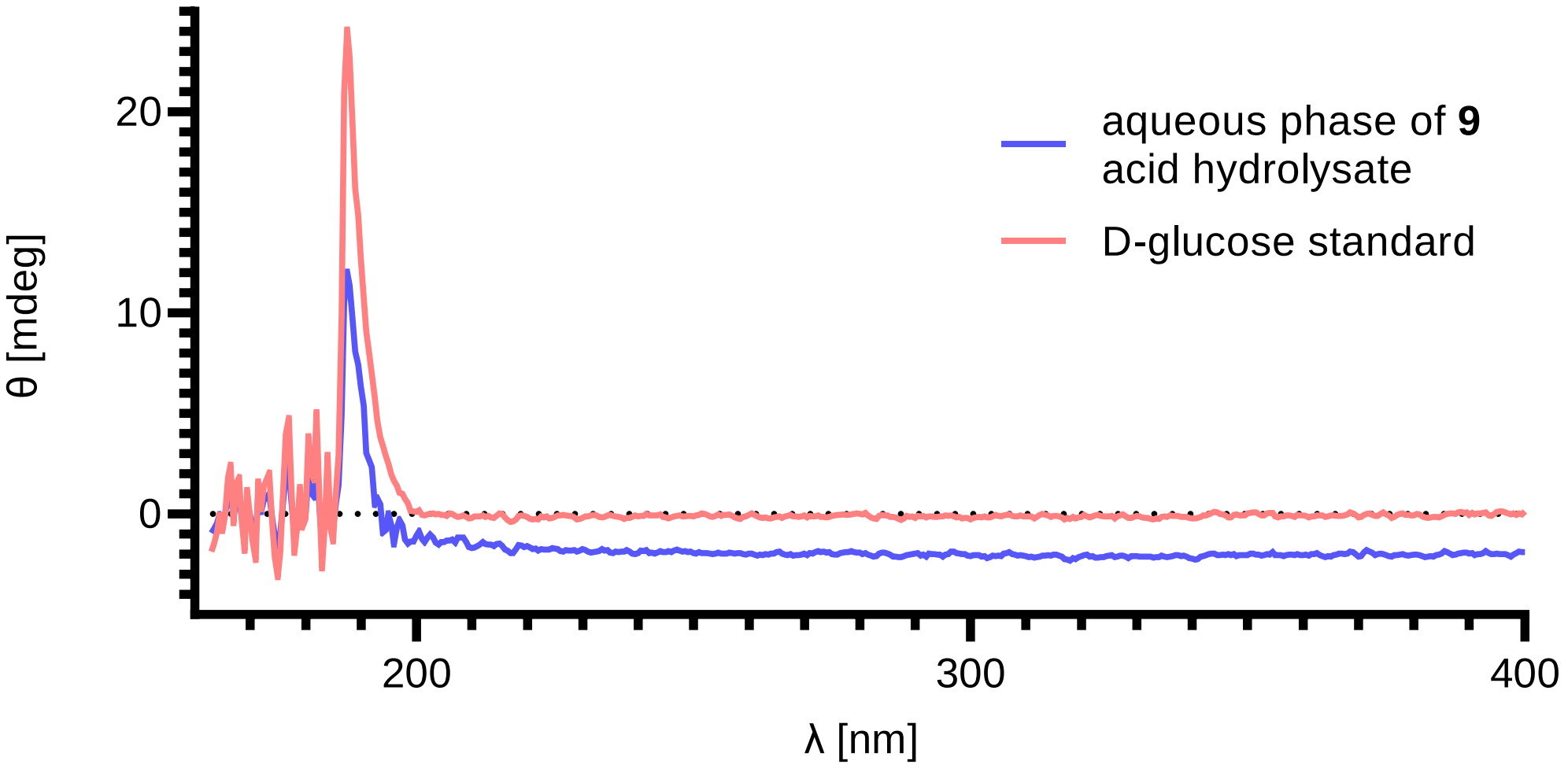

**Figure S23. ECD spectra of D-glucose standard and aqueous phase of phorbactazole C (9) acid hydrolysate.**

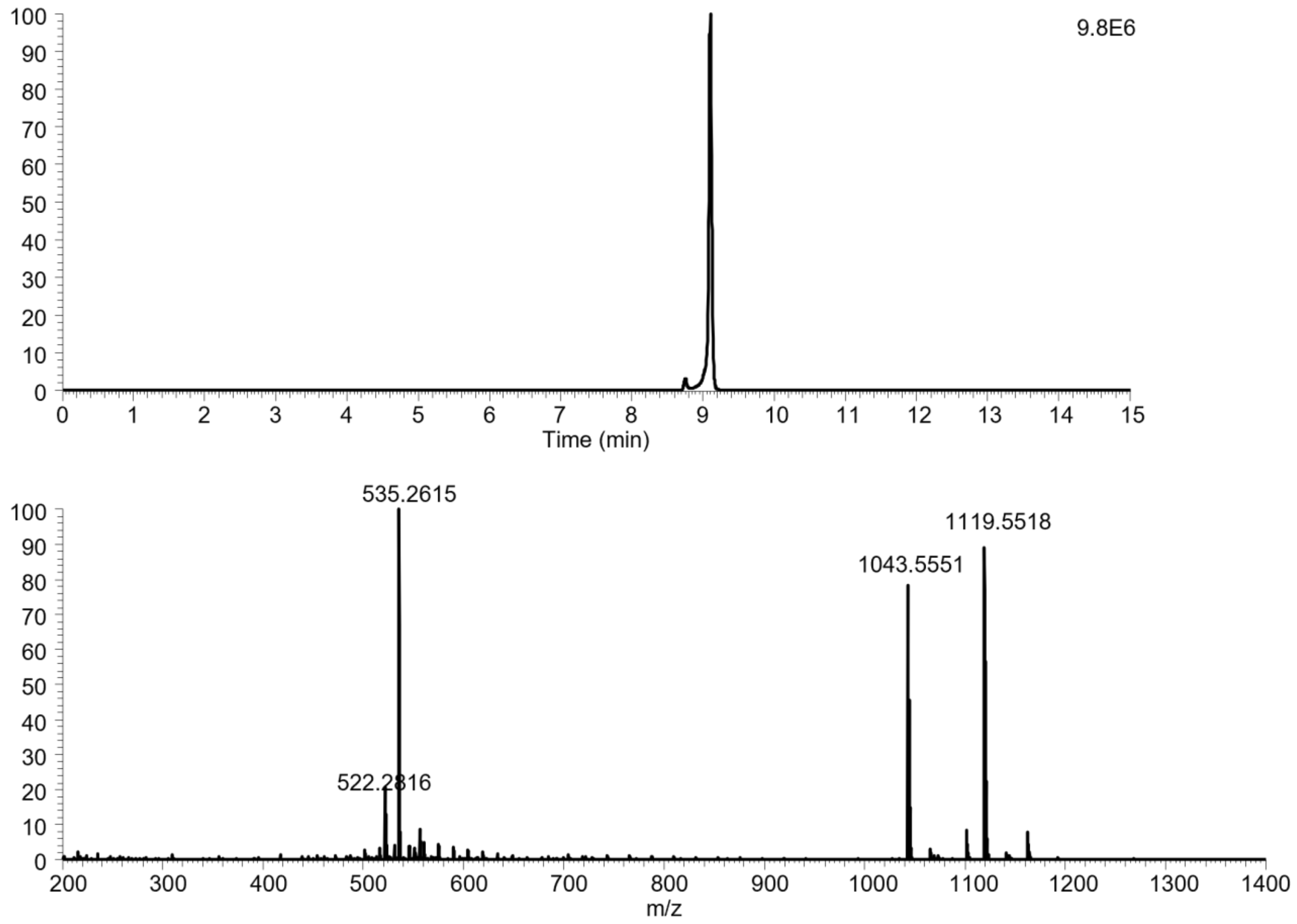

**Figure S24. HPLC-HESI-HRMS data of phorbactazole D (10).**

Top: Extracted ion chromatogram *m/z* 1119.5518.
Bottom: MS data of phorbactazole D (**10**) (*m/z* 1119.5518 $[M+H]^+$).

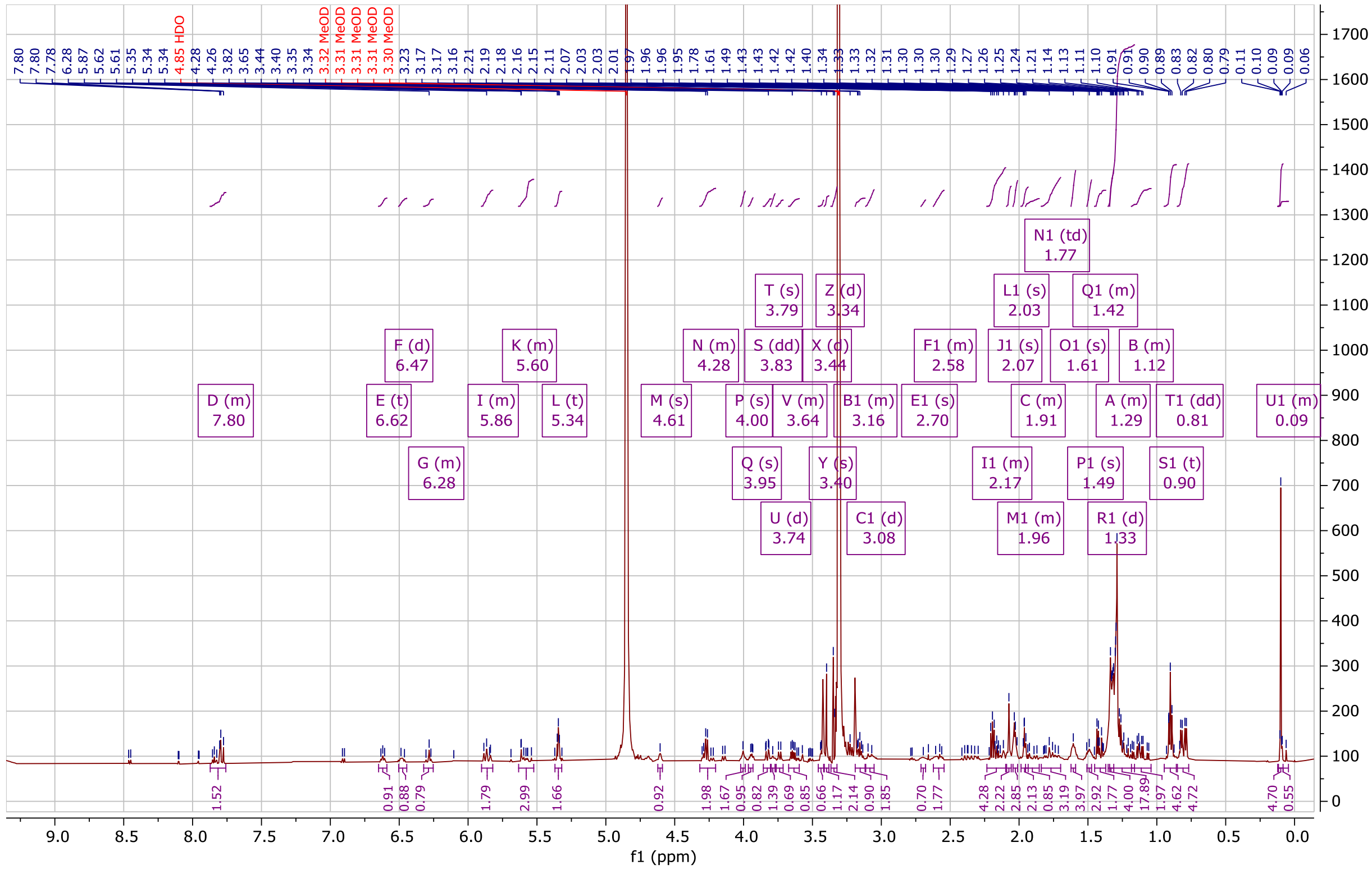

**Figure S25.** $^{1}$H NMR spectrum of phorbactazole D (10) in methanol-$d_4$.

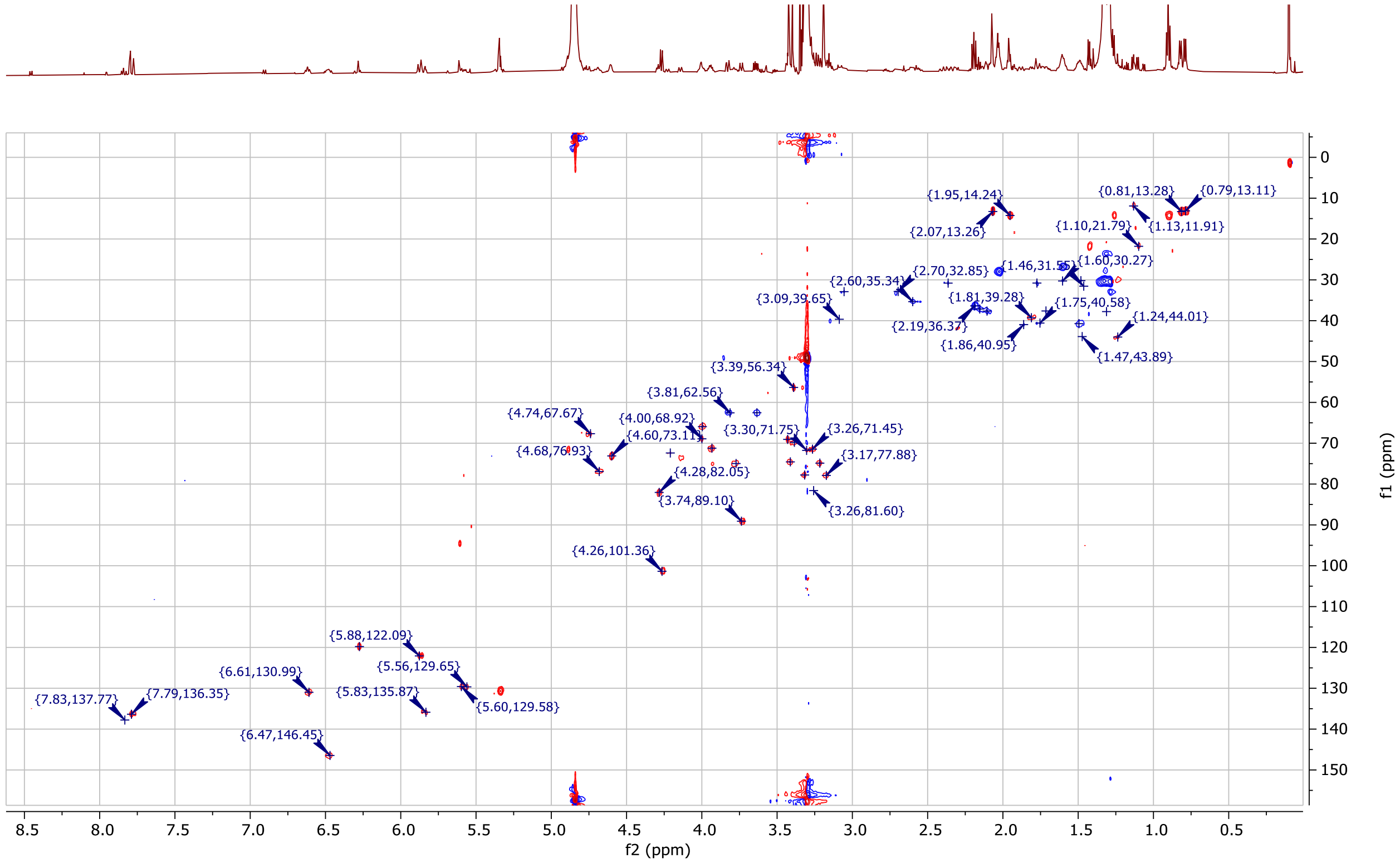

**Figure S26.** HSQC spectrum of phorbactazole D (10) in methanol-$d_4$.

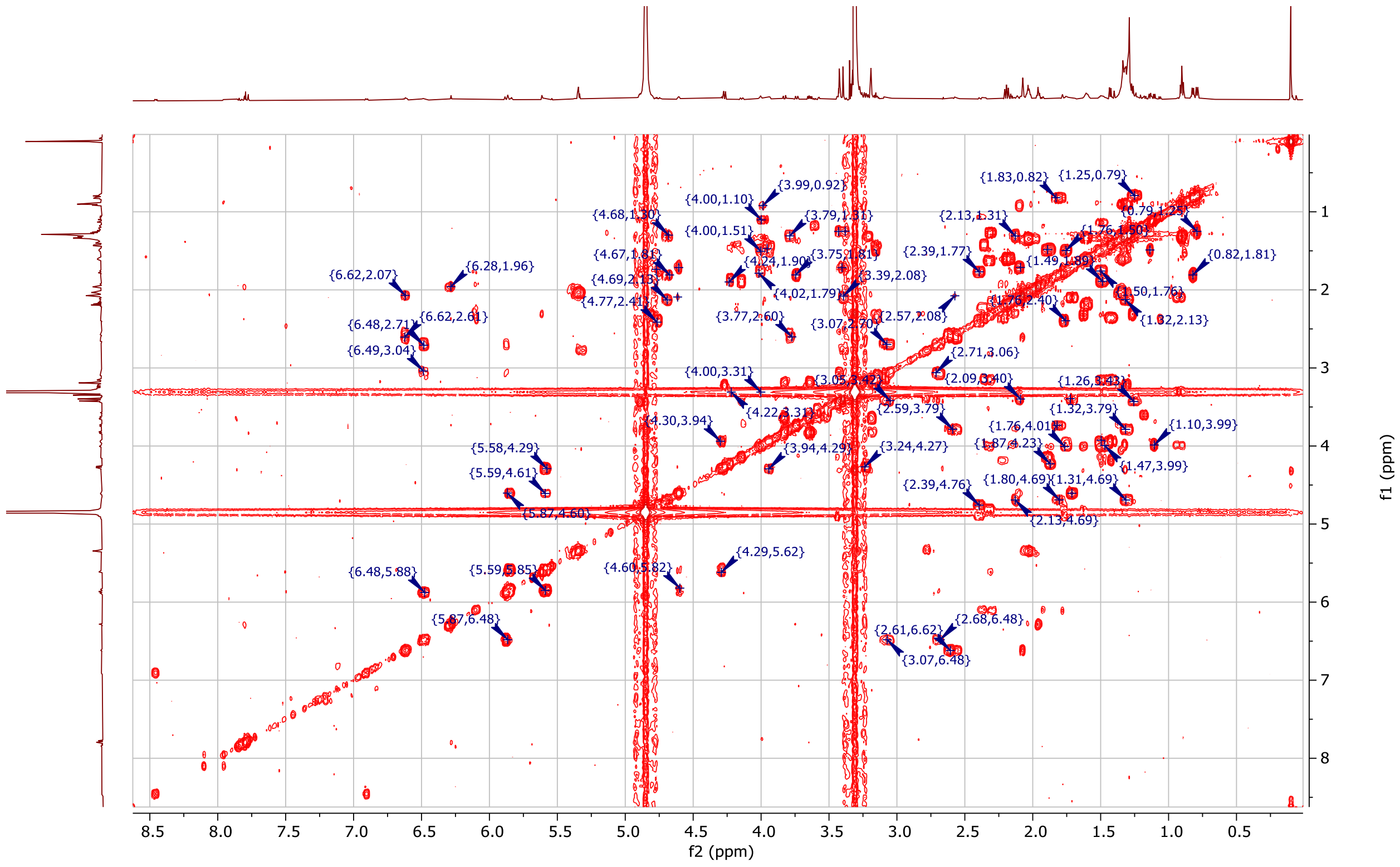

**Figure S27. COSY spectrum of phorbactazole D (10) in methanol-$d_4$.**

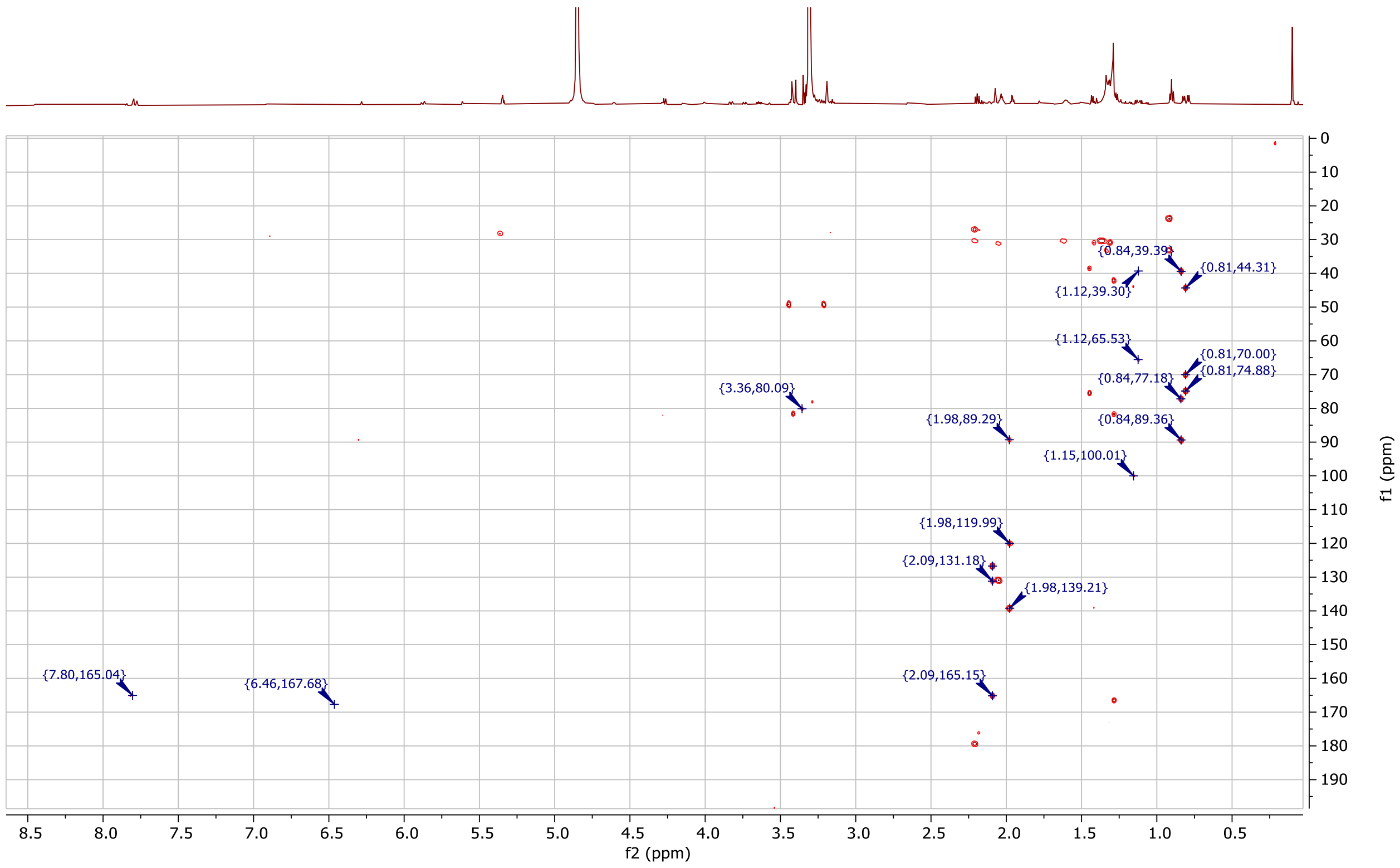

**Figure S28. HMBC spectrum of phorbactazole D (10) in methanol-$d_4$.**

**Table S4. NMR data ($^1$H 600 MHz, $^{13}$C 150 MHz) for 9 and 10 in methanol-$d_4$.**

| | Phorbactazole C (9) | | Phorbactazole C (10) | |
|---|---|---|---|---|
| No. | $^{13}$C/HSQC | $^1$H (multiplicity) | $^{13}$C/HSQC | $^1$H (multiplicity) |
| 1 | 167.4, C | | 167.7, C | |
| 2 | 122.0, CH | 5.85 (d), *J* = 10.9 Hz | 122.1, CH | 5.87 (d), *J* = 11.2 Hz |
| 3 | 146.5, CH | 6.48 (m) | 146.5, CH | 6.47 (m) |
| 4 | 33.0, $CH_2$ | 2.69 (m), 3.07 (m) | 32.9, $CH_2$ | 2.69 (m), 3.06 (m) |
| 5 | 74.6, CH | 3.42 (m) | 74.6, CH | 3.41 (m) |
| 6 | 44.2, CH | 1.26 (m) | 44.0, CH | 1.25 (m) |
| 7 | 69.8, CH | 3.39 (td), *J* = 3.8, 10.1 Hz | 69.8, CH | 3.39 *J* = 3.5, 10.3 Hz |
| 8 | 37.8, $CH_2$ | 1.71 (m), 2.09 (m) | 37.7, $CH_2$ | 1.72 (m), 2.11 (m) |
| 9 | 73.2, CH | 4.61 (m) | 73.1, CH | 4.60 (m) |
| 10 | 136.2, CH | 5.86 (dd), *J* = 3.2, 16.2 Hz | 135.9, CH | 5.83 (dd), *J* = 3.2, 16.1 Hz |
| 11 | 129.7, CH | 5.60 (dd), *J* = 7.4, 16.2 Hz | 129.7, CH | 5.60 (dd), *J* = 7.4, 16.2 Hz |
| 12 | 82.0, CH | 4.29 (t), *J* = 7.4 Hz | 82.1, CH | 4.29 (t), *J* = 7.5 Hz |
| 13 | 71.4, CH | 3.93 (m) | 71.2, CH | 3.93 (m) |
| 14 | 40.8, $CH_2$ | 1.49 (m), 1.76 (m) | 40.7, $CH_2$ | 1.49 (m), 1.76 (m) |
| 15 | 66.1, CH | 4.00 (m) | 66.0, CH | 4.00 (m) |
| 16 | 40.9, $CH_2$ | 1.49 (m),<br>1.89 (ddd), *J* = 2.5, 10.4, 14.6 Hz | 41.0, $CH_2$ | 1.49 (m), 1.87 (m) |
| 17 | 73.4, CH | 4.13 (dt), *J*=3.4, 10.3 Hz | 72.4, CH | 4.21 (dt), *J* = 3.1, 10.1 Hz |
| 18 | 69.1, CH | 3.44 (t), *J* = 3.6 Hz | 71.5, CH | 3.30 (t), *J* = 3.6 Hz |
| 19 | 71.6, CH | 4.89 (q), *J* = 3.3 Hz | 68.9, CH | 4.00 (q), *J* = 3.2 Hz |
| 20 | 30.7, $CH_2$ | 1.76 (m), 2.40 (m) | 30.8, $CH_2$ | 1.76 (m), 2.37 (m) |
| 21 | 67.7, CH | 4.75 (dd), *J* = 2.4, 12.1 Hz | 67.7, CH | 4.75 (dd), *J* = 2.4, 11.9 Hz |
| 22 | 142.6, C | | 142.6, C | |
| 23 | 136.5, CH | 7.79 (s) | 136.4, CH | 7.79 (s) |
| 24 | 164.9, C | | 165.2 C | |
| 25 | 126.7, C | | 126.7, C | |
| 26 | 131.0, CH | 6.62 (t), *J* = 7.7 Hz | 131.0, CH | 6.61 (t), *J* = 8.1 Hz |
| 27 | 35.3, $CH_2$ | 2.60 (m), 2.07 (m) | 35.4, $CH_2$ | 2.60 (m), 2.07 (m) |
| 28 | 75.3, CH | 3.79 (m) | 74.9, CH | 3.77 (m) |
| 29 | 37.4, $CH_2$ | 1.29 (m), 2.11 (m) | 37.6, $CH_2$ | 1.31 (m), 2.11 (m) |
| 30 | 77.0, CH | 4.69 (m) | 76.9, CH | 4.68 (tt), *J* = 5.0, 10.3 Hz |
| 31 | 39.2, CH | 1.81 (m) | 44.0, CH | 1.81 (m) |
| 32 | 89.1, CH | 3.73 (d), *J* = 10.7 Hz | 89.1, CH | 3.73 (d), *J* = 9.6 Hz |
| 33 | 139.0, C | | 139.2, C | |
| 34 | 119.7, CH | 6.28 (s) | 119.8, CH | 6.28 (s) |
| 35 | 138.8, C | | 138.7, C | |
| 36 | 137.8, CH | 7.86 (s) | 137.6, CH | 7.83 (s) |
| 37 | 162.1, C | | 162.0, C | |
| 38 | 39.5 $CH_2$ | 3.07 (d), *J* = 14.5 Hz<br>3.14 (d), *J* = 14.5 Hz | 39.8 $CH_2$ | 3.06 (d), *J* = 14.4 Hz<br>3.14 (d), *J* = 14.3 Hz |
| 39 | 99.9, C | | 100.0, C | |
| 40 | 44.0, CH | 1.49 | 43.9, CH | 1.47 (m) |
| 41 | 80.0, CH | 3.29, (m) | 80.1, CH | 3.26, (m) |
| 42 | 39.1, $CH_2$ | 0.90 (m), 2.09 (m) | 38.9, $CH_2$ | 0.90 (m), 2.10 (m) |
| 43 | 65.8, CH | 4.00 (qd), *J* = 2.2, 6.3, 11.9 Hz | 66.0, CH | 4.00 (m) |
| 44 | 21.8, $CH_3$ | 1.11 (d), *J* = 6.3 Hz | 21.8, $CH_3$ | 1.10 (d), *J* = 6.3 Hz |
| 45 | 12.1, $CH_3$ | 1.14 (d), *J* = 6.8 Hz | 11.9, $CH_3$ | 1.13 (d), *J* = 6.7 Hz |
| 46 | 14.2, $CH_3$ | 1.95 (s) | 14.2, $CH_3$ | 1.95 (s) |
| 47 | 13.2, $CH_3$ | 0.81 (d), *J* = 6.7 Hz | 13.3, $CH_3$ | 0.81 (d), *J* = 6.3 Hz |
| 48 | 13.2, $CH_3$ | 2.07 (s) | 13.3, $CH_3$ | 2.07 (s) |
| 49 | 13.1, $CH_3$ | 0.79 (d), *J* = 6.4 Hz | 13.1, $CH_3$ | 0.79 (d), *J* = 6.7 Hz |
| 1' | 101.4, CH | 4.26 (d), *J* = 7.9 Hz | 101.4, CH | 4.26 (d), *J* = 7.6 Hz |
| 2' | 74.9, CH | 3.23 (dd), *J* = 7.9, 9.1 Hz | 74.9, CH | 3.2 (dd), *J* = 7.3, 9.1 Hz |
| 3' | 77.8, CH | 3.33 (dd), 8.9, 9.1 Hz | 77.8, CH | 3.32 (m) |
| 4' | 71.6, CH | 3.27 (t), *J* = 8.9, 9.0Hz | 71.5, CH | 3.27 (dd), *J* = 9.0, 9.4 Hz |
| 5' | 77.8, CH | 3.18 (m) | 77.8, CH | 3.18 (m) |
| 6' | 62.5, $CH_2$ | 3.64 (dd), *J* = 5.4, 12.0 Hz<br>3.82 (dd), *J* = 1.9, 12.0 Hz | 62.6, $CH_2$ | 3.64 (dd), *J* = 5.8, 12.1 Hz<br>3.82 (dd), *J* = 12.3, 11.9 Hz |
| 41 $OCH_3$ | 56.6, $CH_3$ | 3.34 (s) | 56.3, $CH_3$ | 3.39 (s) |
| -$CONH_2$ | 158.5, C | | | |

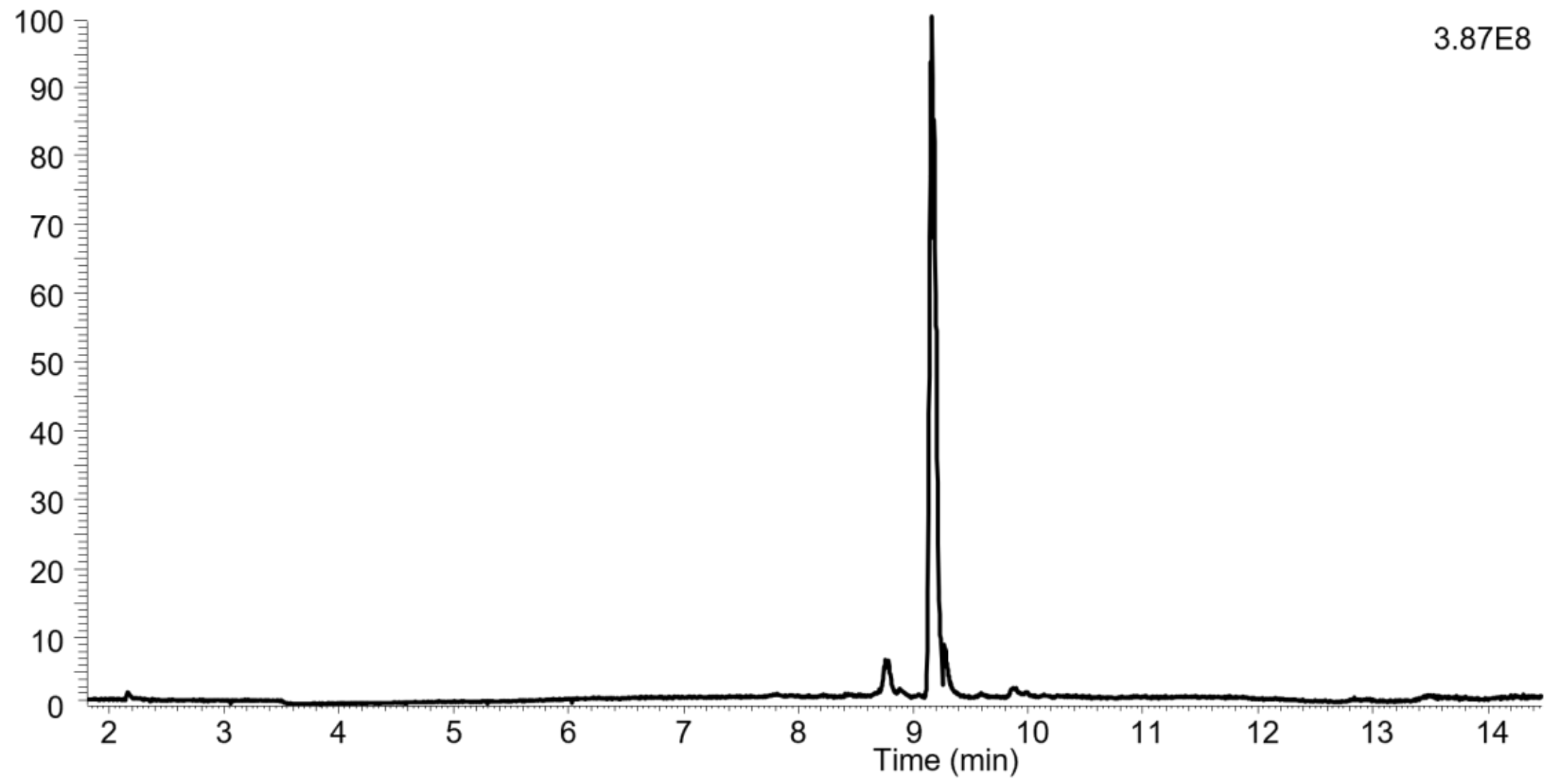

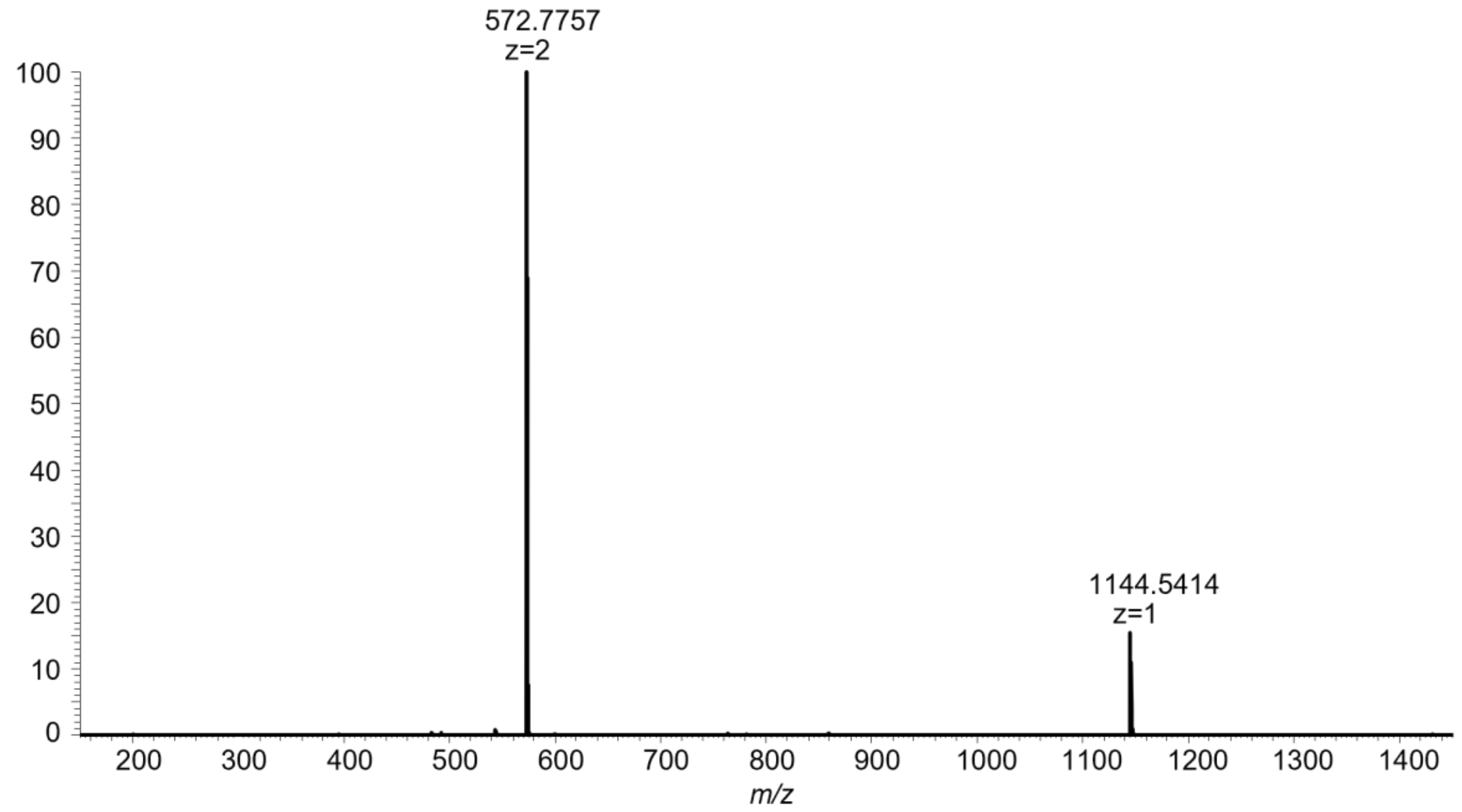

**Figure S29. HPLC-HESI-HRMS data of phorbactazole E (11).**

Top: Total ion chromatogram
Bottom: MS data of phorbactazole E (**11**) (*m/z* 1144.5414 $[M+H]^+$).

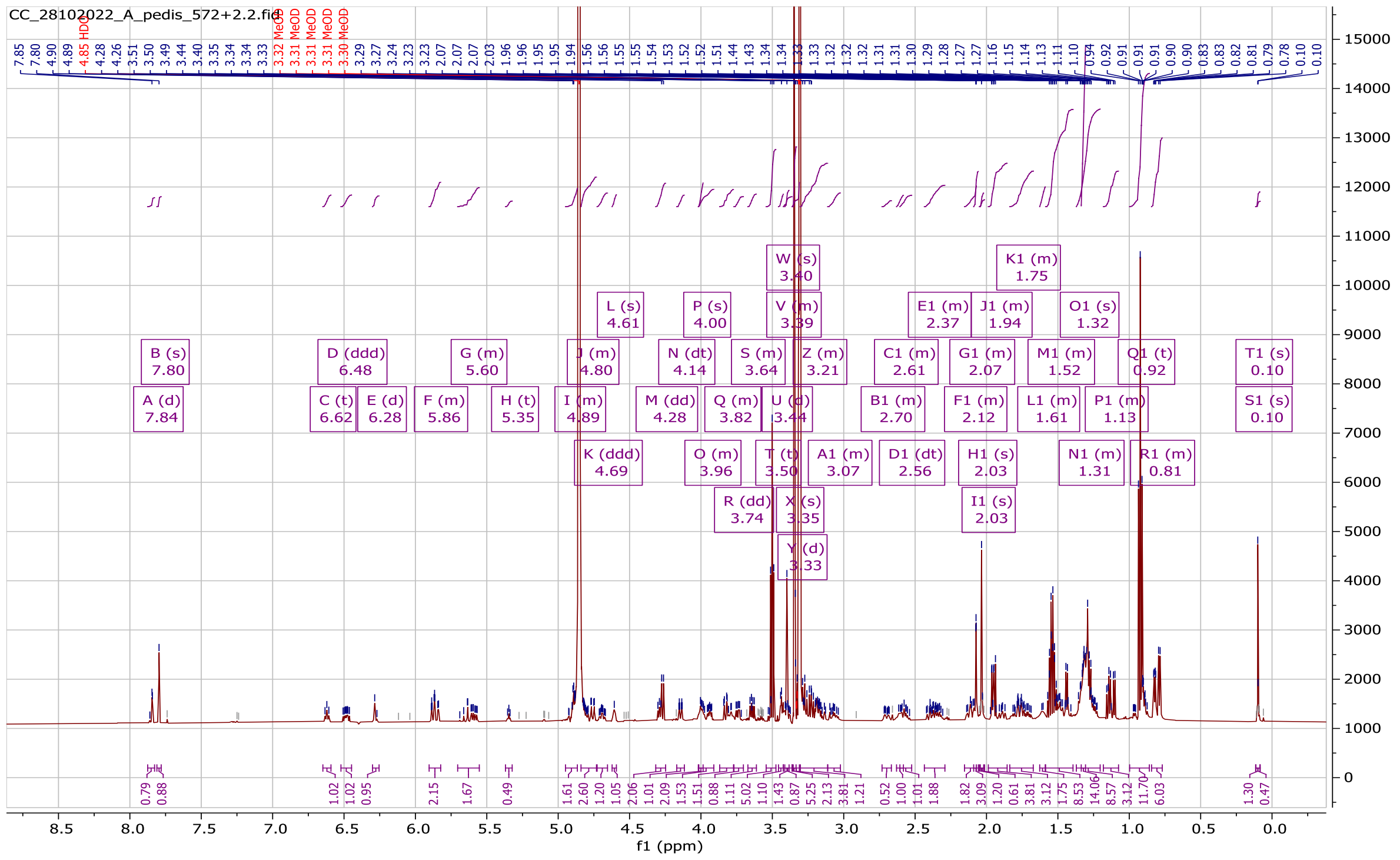

**Figure S30.** [1]**H NMR spectrum of phorbactazole E (11) in methanol-$d_4$.**

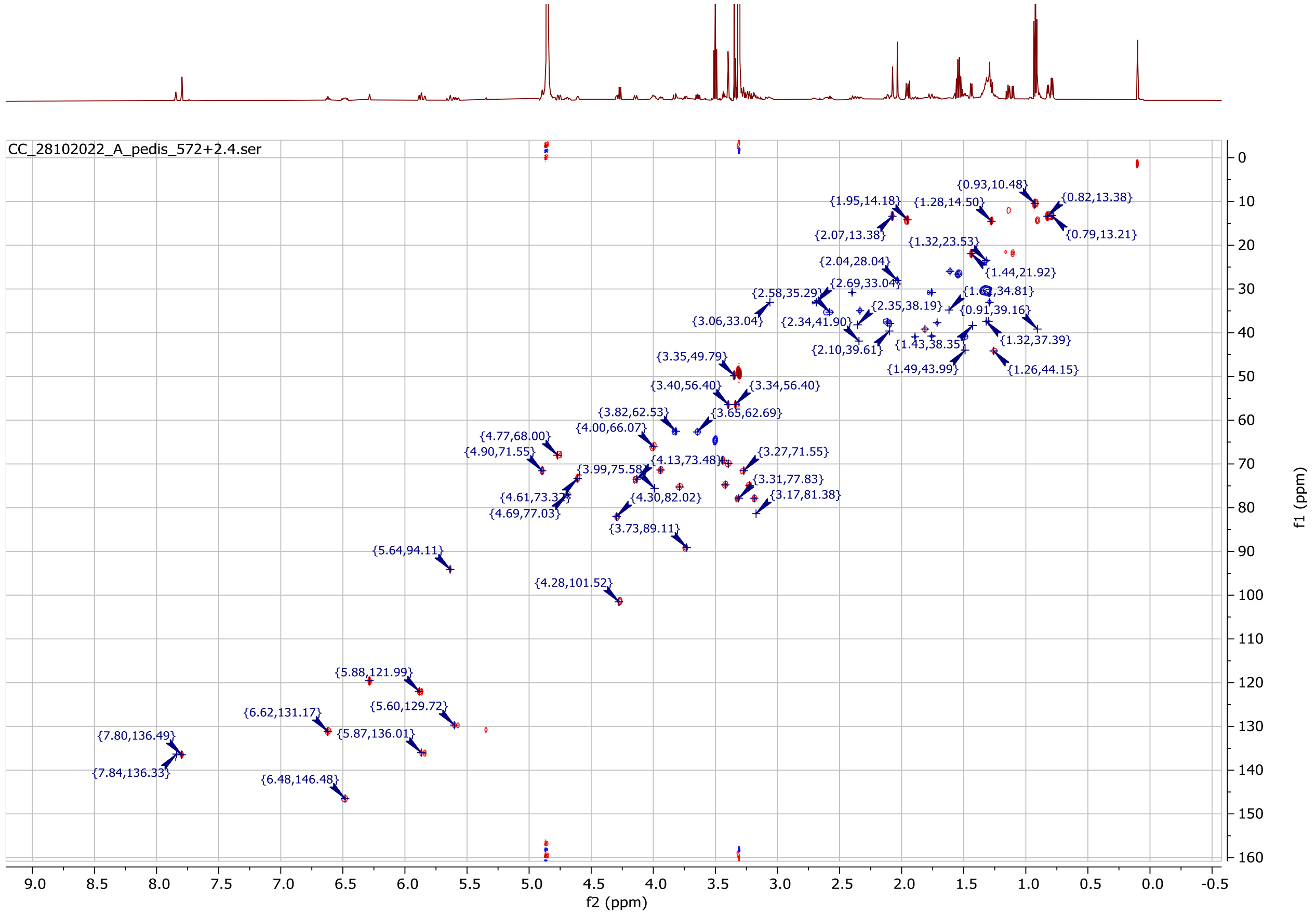

**Figure S31. HSQC spectrum of phorbactazole E (11) in methanol-$d_4$.**

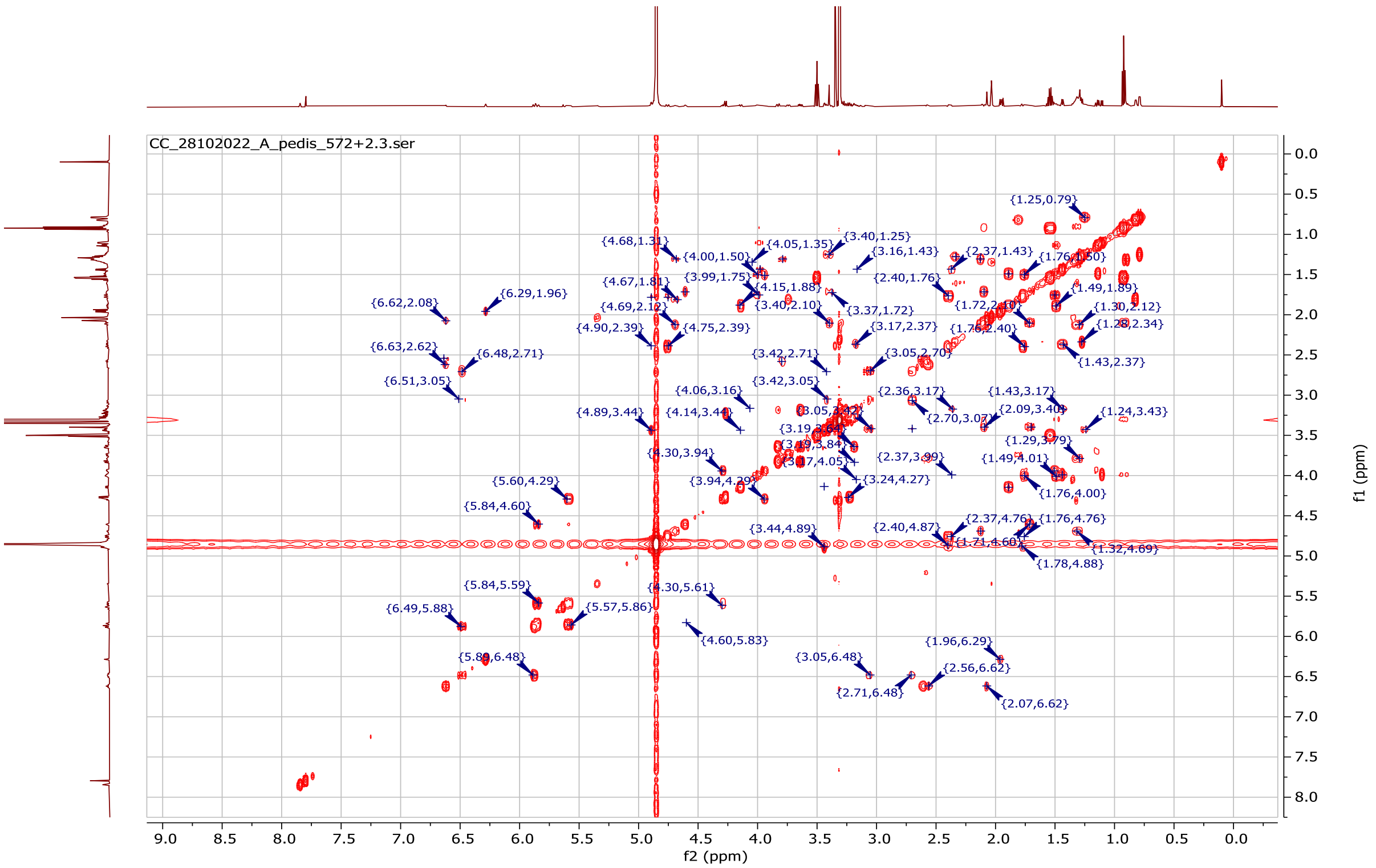

**Figure S32. COSY spectrum of phorbactazole E (11) in methanol-$d_4$.**

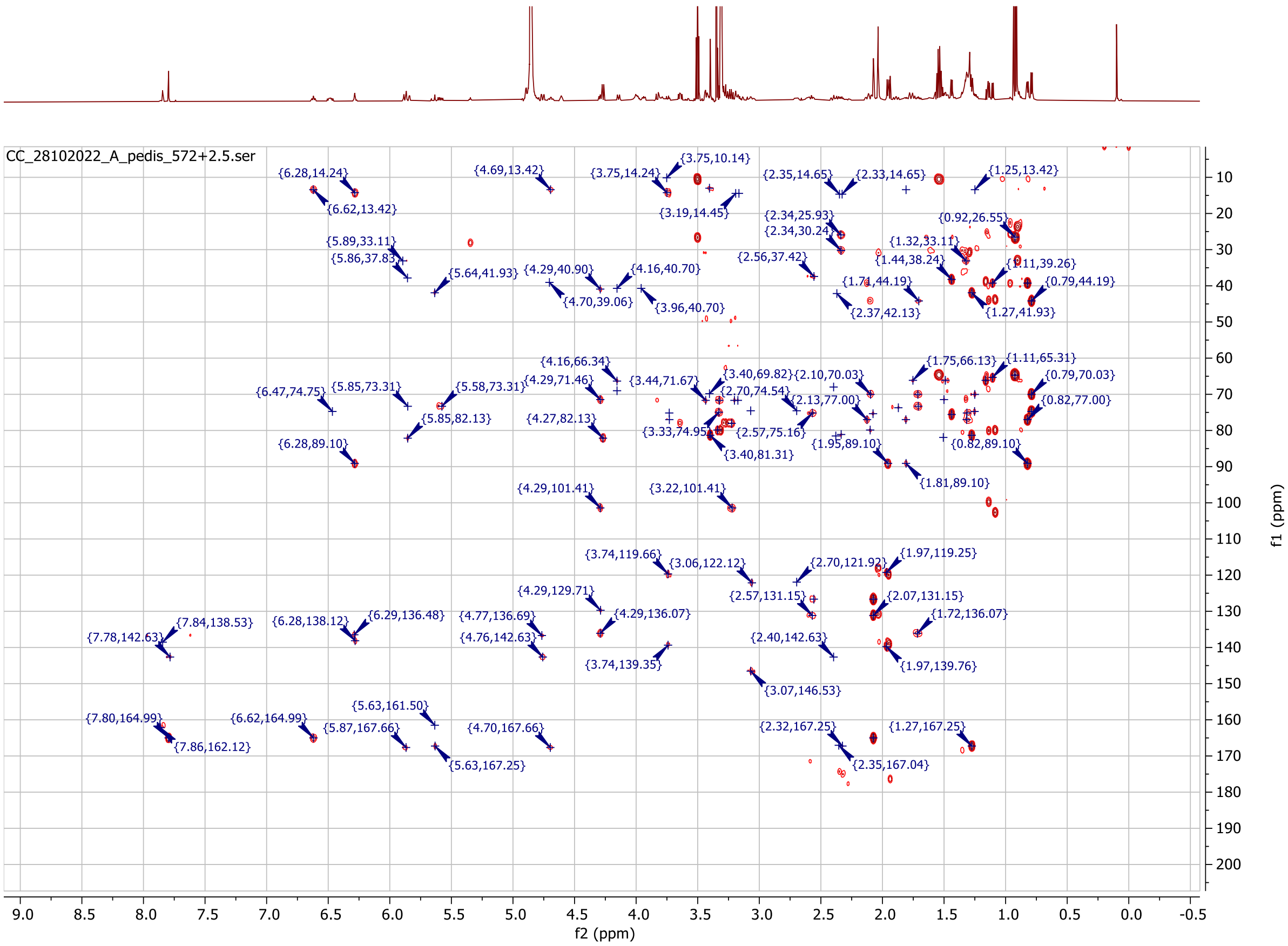

**Figure S33. HMBC spectrum of phorbactazole E (11) in methanol-$d_4$.**

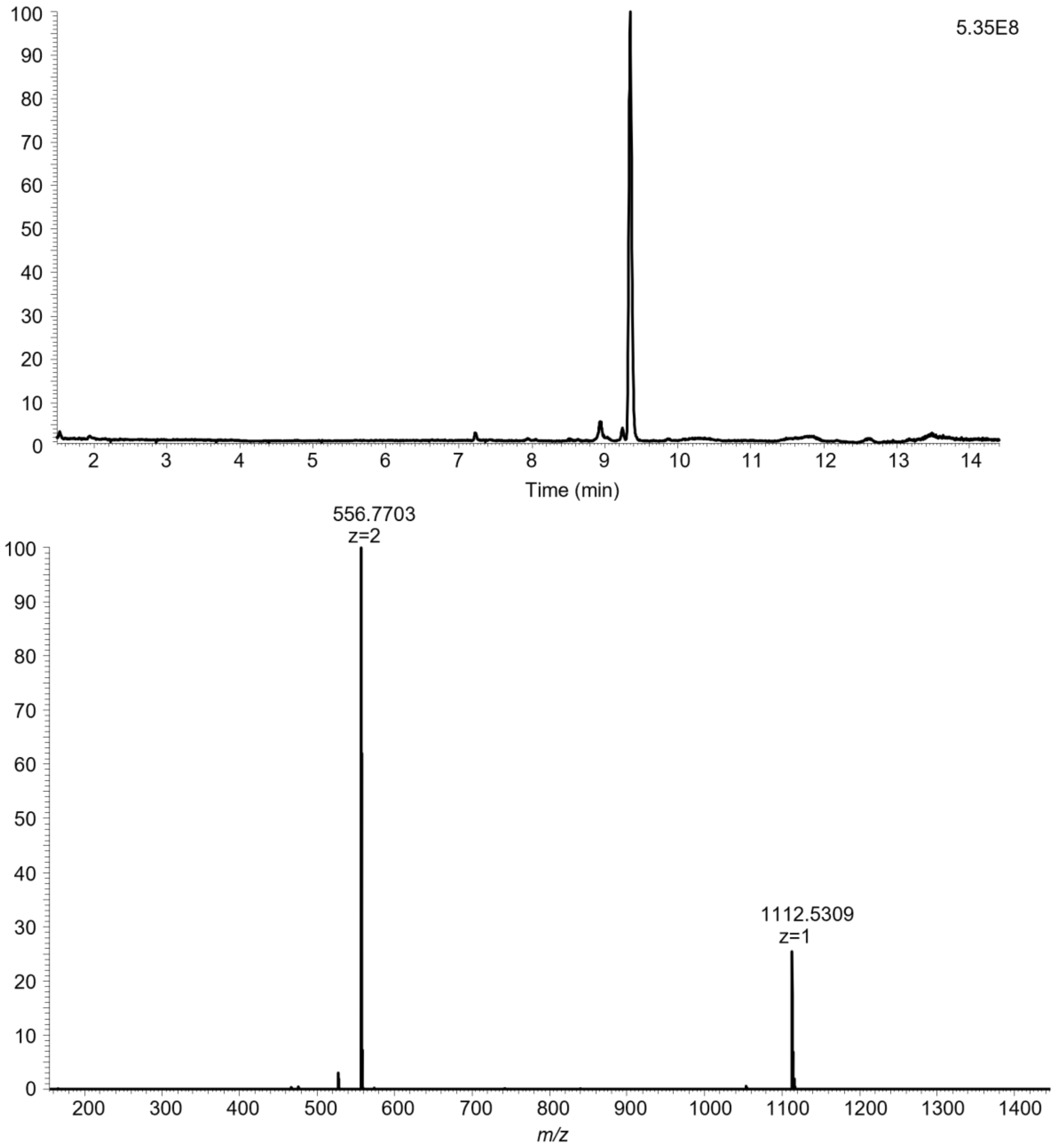

**Figure S34. HPLC-HESI-HRMS data of phorbactazole F (12).**

Top: Total ion chromatogram
Bottom: MS data of phorbactazole F (**12**) (*m*/*z* 1112.5309 $[M+H]^+$).

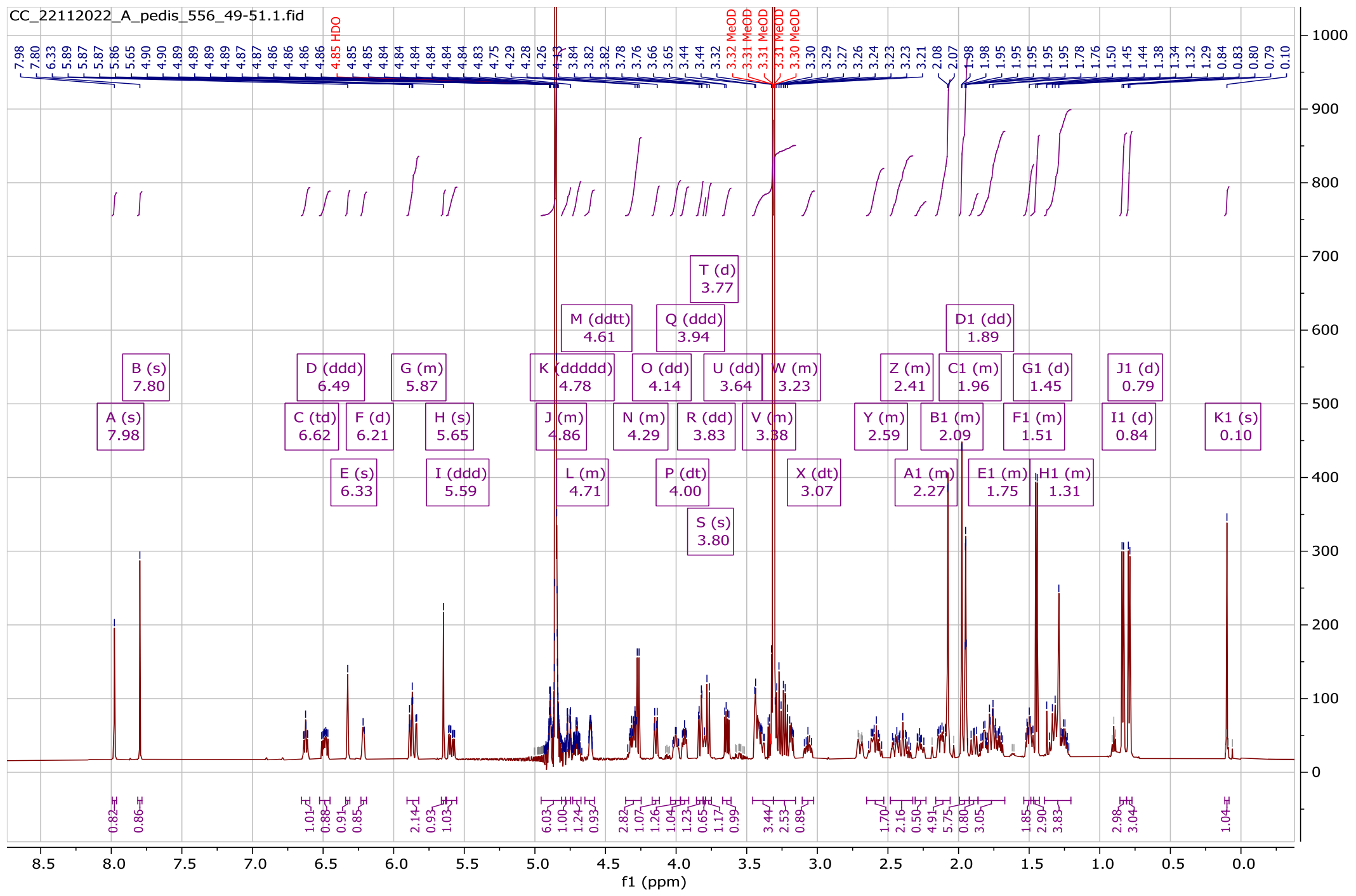

**Figure S35. ¹H NMR spectrum of phorbactazole F (12) in methanol-*d*$_4$.**

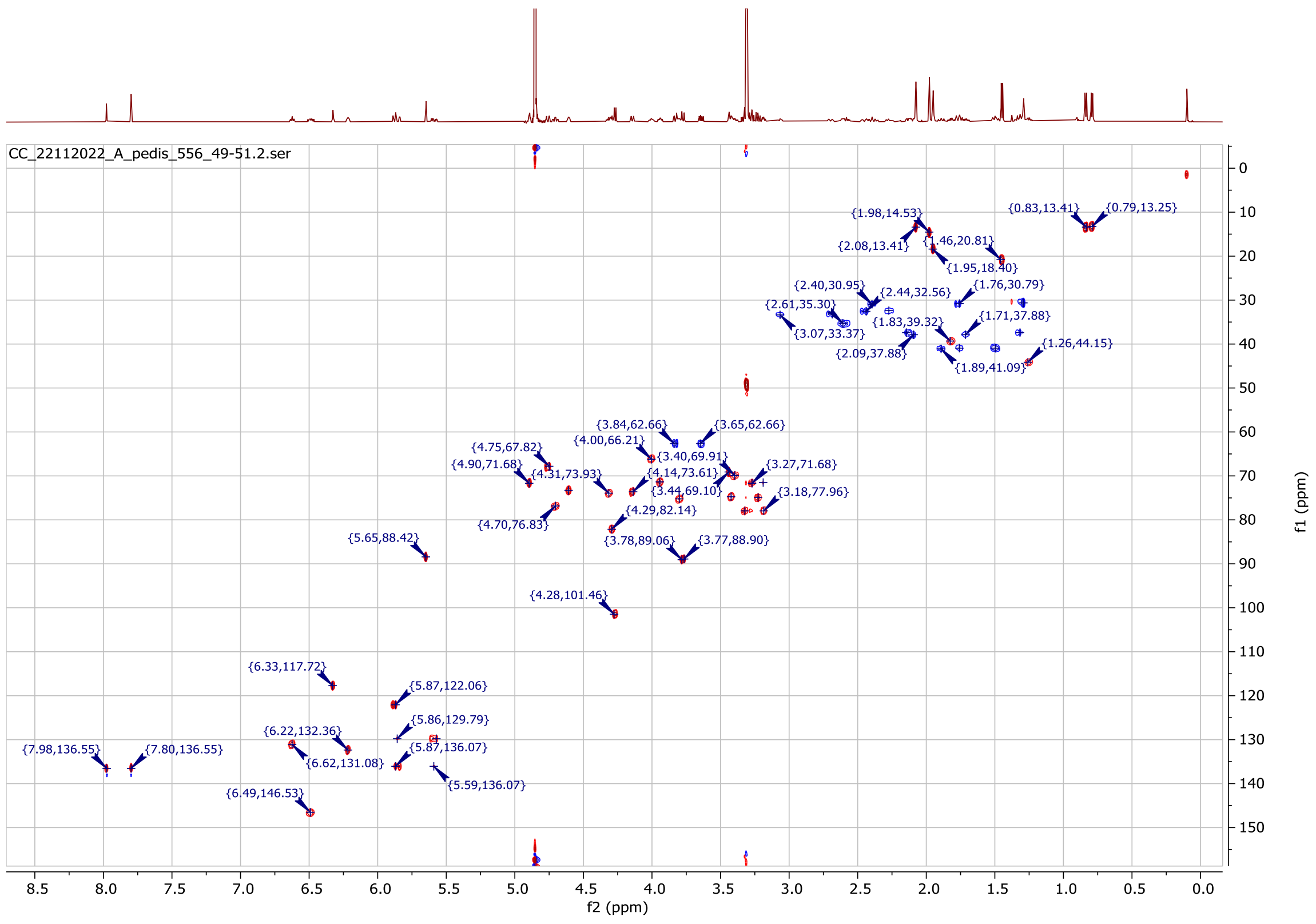

**Figure S36. HSQC spectrum of phorbactazole F (12) in methanol-*d*$_4$.**

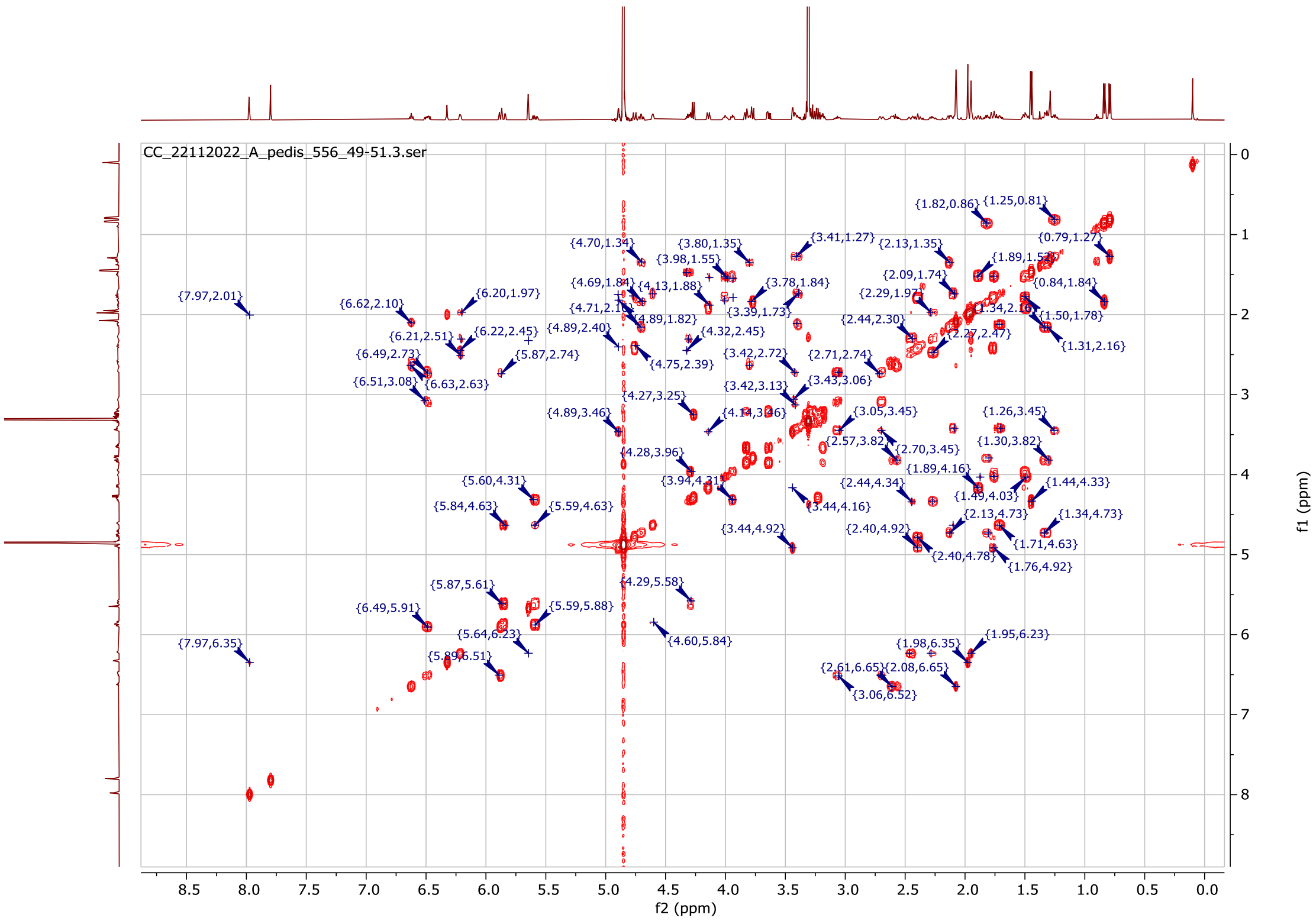

**Figure S37. COSY spectrum of phorbactazole F (12) in methanol-$d_4$.**

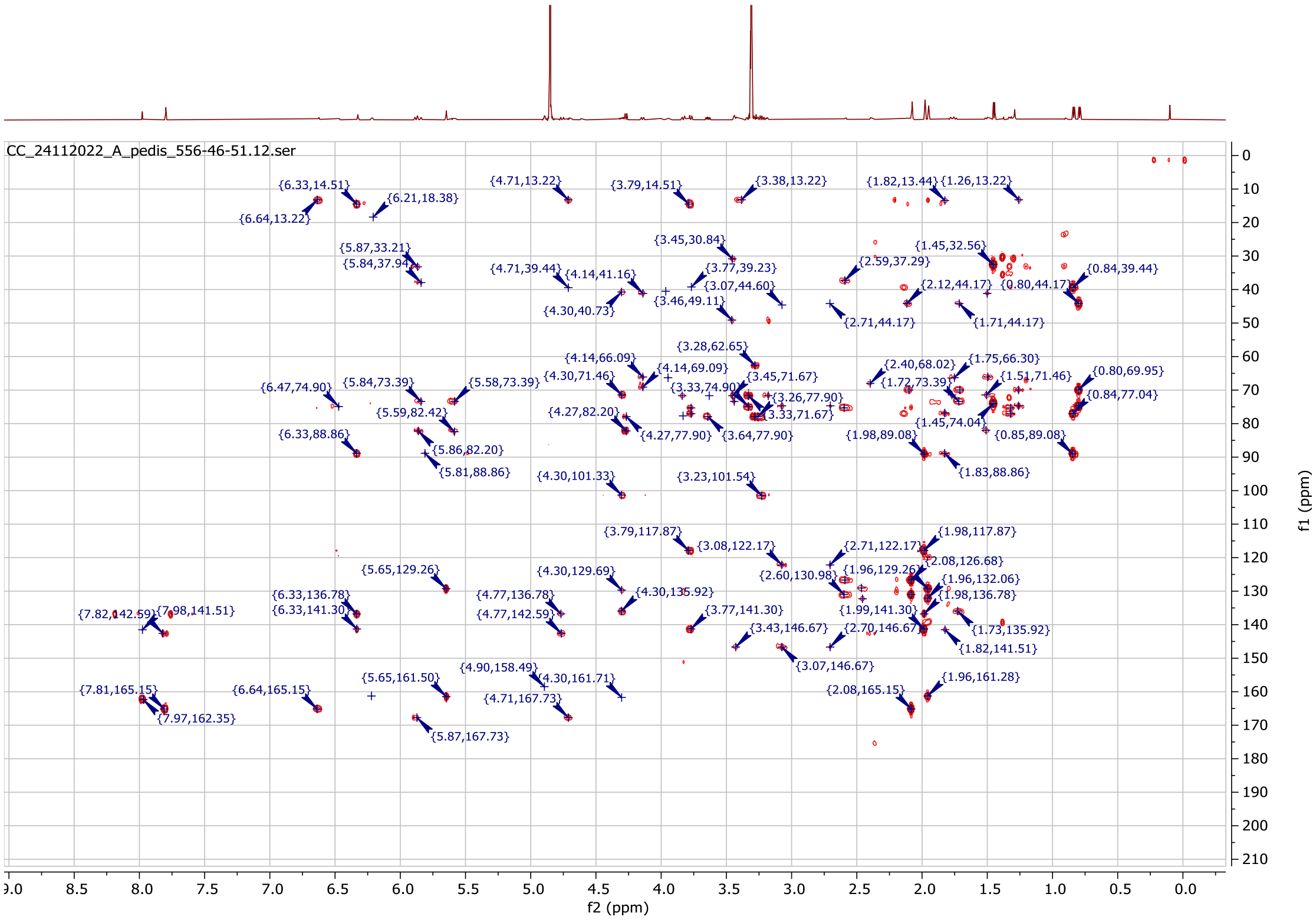

**Figure S38. HMBC spectrum of phorbactazole F (12) in methanol-$d_4$.**

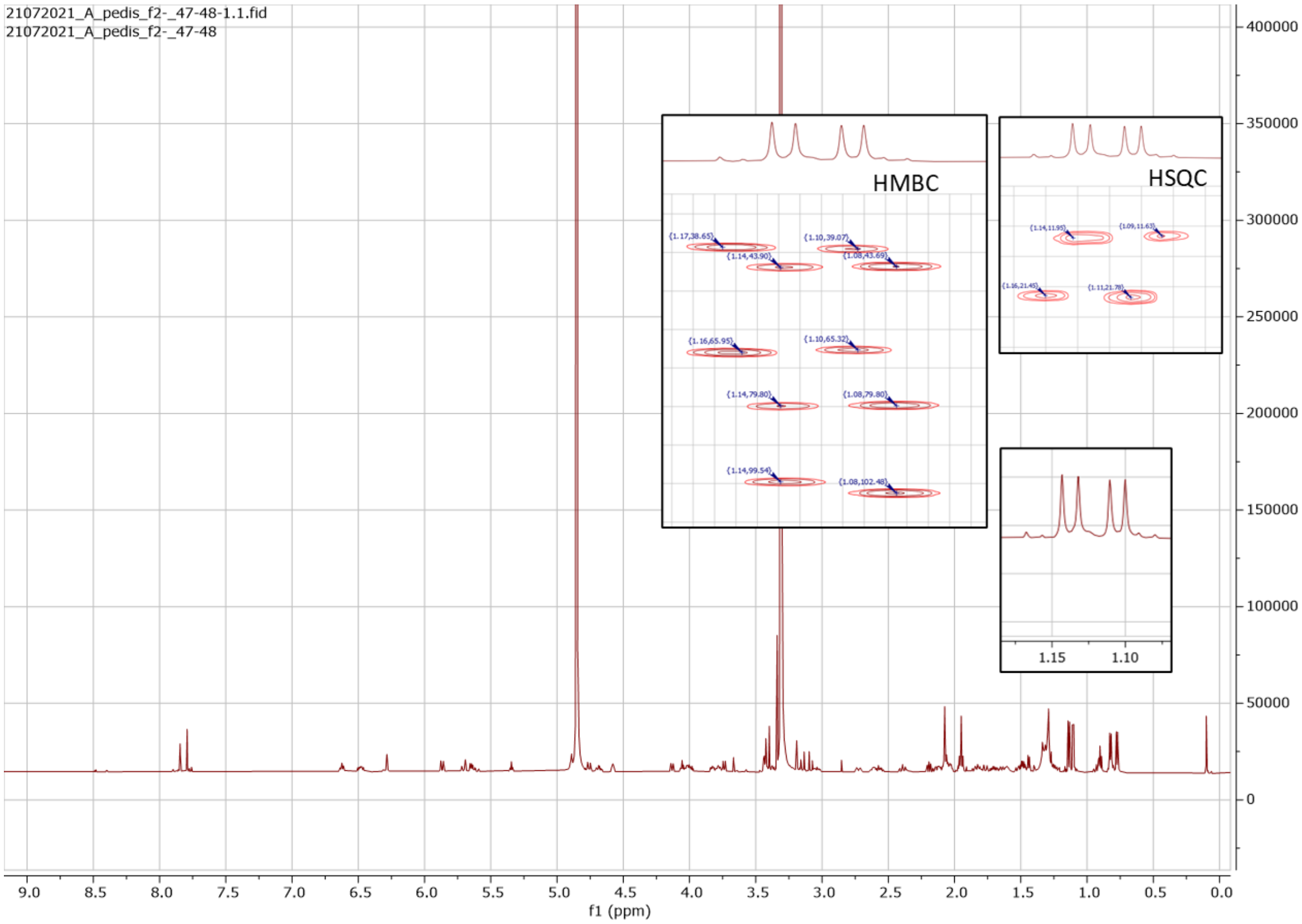

**Figure S39. Isomerization of phorbactazole A (7) in ring D.**

$^1$H NMR for phorbactazole A (**7**) showing the two methyl doublets in ring D and their corresponding HSQC and HMBC correlations signals in ring D.

**Table S5. NMR data ($^1$H 600 MHz, $^{13}$C 150 MHz) for 11 and 12 in methanol-$d_4$.**

| | Phorbactazole E (11) | | Phorbactazole F (12) | |
|---|---|---|---|---|
| No. | $^{13}$C/HSQC | $^1$H (multiplicity) | $^{13}$C/HSQC | $^1$H (multiplicity) |
| 1 | 167.7, C | | 167.7, C | |
| 2 | 122.0, CH | 5.88 (m) | 122.1, CH | 5.87 (m) |
| 3 | 146.5, CH | 6.48 (ddd), *J* = 5.8, 9.0, 11.7 Hz | 146.5, CH | 6.49(ddd), *J* = 5.8, 9.0, 11.7 Hz |
| 4 | 33.0, $CH_2$ | 2.69 (m), 3.06 (m) | 33.2, $CH_2$ | 2.69 (m), 3.07 (m) |
| 5 | 74.8, CH | 3.42 (m) | 74.7, CH | 3.42 (m) |
| 6 | 44.2, CH | 1.26 (m) | 44.2, CH | 1.26 (m) |
| 7 | 69.9, CH | 3.40 (m) | 69.9, CH | 3.40 (m) |
| 8 | 37.8, $CH_2$ | 1.71 (m)<br>2.09 (m) | 37.9, $CH_2$ | 1.71 (m)<br>2.09 (m) |
| 9 | 73.3, CH | 4.61 (m) | 73.3, CH | 4.61 (m) |
| 10 | 136.0, CH | 5.86 (m) | 136.1, CH | 5.86 (m) |
| 11 | 129.7, CH | 5.59 (ddd), *J* = 2.0, 7.2, 16.1 Hz | 129.8, CH | 5.59 (ddd), *J* = 2.0, 7.2, 16.1 Hz |
| 12 | 82.0, CH | 4.30 (m) | 82.0, CH | 4.29 (m) |
| 13 | 71.4, CH | 3.94 (m) | 71.4, CH | 3.94 (m) |
| 14 | 40.8, $CH_2$ | 1.50 (m)<br>1.76 (m) | 40.9, $CH_2$ | 1.50 (m), 1.76 (m) |
| 15 | 66.1, CH | 4.00 (m) | 66.2, CH | 4.00 (m) |
| 16 | 40.9, $CH_2$ | 1.50 (m)<br>1.89 (m) | 41.0, $CH_2$ | 1.49 (m)<br>1.89 (m) |
| 17 | 73.5, CH | 4.13 (dt), *J* = *J*=3.7, 10.0 Hz | 73.6, CH | 4.14 (dd), *J* =3.7, 10.1 Hz |
| 18 | 69.1, CH | 3.44 (m) | 69.1, CH | 3.44 (m) |
| 19 | 71.6, CH | 4.90 (m) | 71.7, CH | 4.90 (m) |
| 20 | 30.8, $CH_2$ | 1.76 (m)<br>2.40 (m) | 30.8, $CH_2$ | 1.76 (m), 2.40 (m) |
| 21 | 67.8, CH | 4.76 (dd), *J* = 1.9, 10.6 Hz | 67.8, CH | 4.76 (dd), *J* = 1.9, 10.6 Hz, 1 H |
| 22 | 142.6, C | | 142.6, C | |
| 23 | 136.5, CH | 7.80, (s) | 136.6, CH | 7.80, (s) |
| 24 | 165.0, C | | 165.1, C | |
| 25 | 126.6, C | | 126.7, C | |
| 26 | 131.1, CH | 6.62, (td), *J* = 1.5, 7.9 Hz | 131.1, CH | 6.62, (td), *J* = 1.5, 7.9 Hz |
| 27 | 35.3, $CH_2$ | 2.60 (m) | 35.3, $CH_2$ | 2.60 (m) |
| 28 | 75.3, CH | 3.79 (m) | 75.2, CH | 3.80 (m) |
| 29 | 37.4, $CH_2$ | 1.30 (m)<br>2.11 (m) | 37.4, $CH_2$ | 1.35 (t), *J* = 11.5 Hz<br>2.13 (m) |
| 30 | 77.0, CH | 4.69 (td), *J* = 5.2, 11.0 Hz | 76.8, CH | 4.70 (td), *J* = 5.2, 11.0 Hz |
| 31 | 39.2, CH | 1.81 (m) | 39.3, CH | 1.83 (m) |
| 32 | 89.1, CH | 3.73 (d), *J* = 10.0 Hz | 89.1, CH | 3.75 (d), *J* = 10.0 Hz |
| 33 | 139.0 C | | 141.3 C | |
| 34 | 119.6, CH | 6.29 (s) | 117.7, CH | 6.33 (s) |
| 35 | 138.5, C | | 141.5, C | |
| 36 | 136.3, CH | 7.84 (s) | 136.5, CH | 7.98 (s) |
| 37 | 162.1, C | | 162.4, C | |
| 38 | 94.1, CH | 5.64 (s) | 88.4, CH | 5.65 (s) |
| 39 | 167.2, C | | 161.5, C | |
| 40 | 41.9, CH | 2.34 (m) | 129.3, C | |
| 41 | 81.3, CH | 3.17, (m) | 132.4, CH | 6.22, (m) |
| 42 | 38.4, $CH_2$ | 1.44 (m)<br>2.36 (m) | 32.4, $CH_2$ | 2.27 (m), 2.45 (m) |
| 43 | 75.6, CH | 3.99 (m) | 73.9, CH | 4.31(m) |
| 44 | 21.9, $CH_3$ | 1.43 (d), $J$ = 6.4 Hz | 20.8, $CH_3$ | 1.46 (d), $J$ = 6.3 Hz |
| 45 | 14.5 $CH_3$ | 1.28 (d), $J$ = 6.7 Hz | 18.4 $CH_3$ | 1.95 (s) |
| 46 | 14.2, $CH_3$ | 1.95 (s) | 14.5, $CH_3$ | 1.98 (s) |
| 47 | 13.4, $CH_3$ | 0.82 (d), $J$ = 6.7 Hz | 13.4, $CH_3$ | 0.83 (d), $J$ = 6.6 Hz |
| 48 | 13.4, $CH_3$ | 2.07 (s) | 13.4, $CH_3$ | 2.08 (s) |
| 49 | 13.2, $CH_3$ | 0.79 (d), $J$ = 6.7 Hz | 13.3, $CH_3$ | 0.79 (d), $J$ = 6.5 Hz |
| 1' | 101.5, CH | 4.28 (d), $J$ = 7.8 Hz | 101.5, CH | 4.28 (d), $J$ = 7.8 Hz |
| 2' | 74.8, CH | 3.23 (t), $J$ = 8.9 Hz | 74.7, CH | 3.23 (t), $J$ = 9.0 Hz |
| 3' | 77.8, CH | 3.32 (m) | 77.8, CH | 3.32 (m) |
| 4' | 71.6, CH | 3.27 (t), $J$ = 9.0 Hz | 71.7, CH | 3.27 (t), $J$ = 9.0 Hz |
| 5' | 77.8, CH | 3.19 (m) | 77.9, CH | 3.18 (m) |
| 6' | 62.7, $CH_2$ | 3.65 (dd), $J$ = 5.7, 12.1 Hz<br>3.85 (dd), $J$ = 2.1, 12.1 Hz | 62.6, $CH_2$ | 3.65 (dd), $J$ = 5.8, 11.8 Hz<br>3.85 (dd), $J$ = 2.2, 12.02 Hz |
| $OCH_3$ | 56.4, $OCH_3$ | 3.40 (s) | | |
| -$CONH_2$ | 158.5, C | | 158.5, C | |

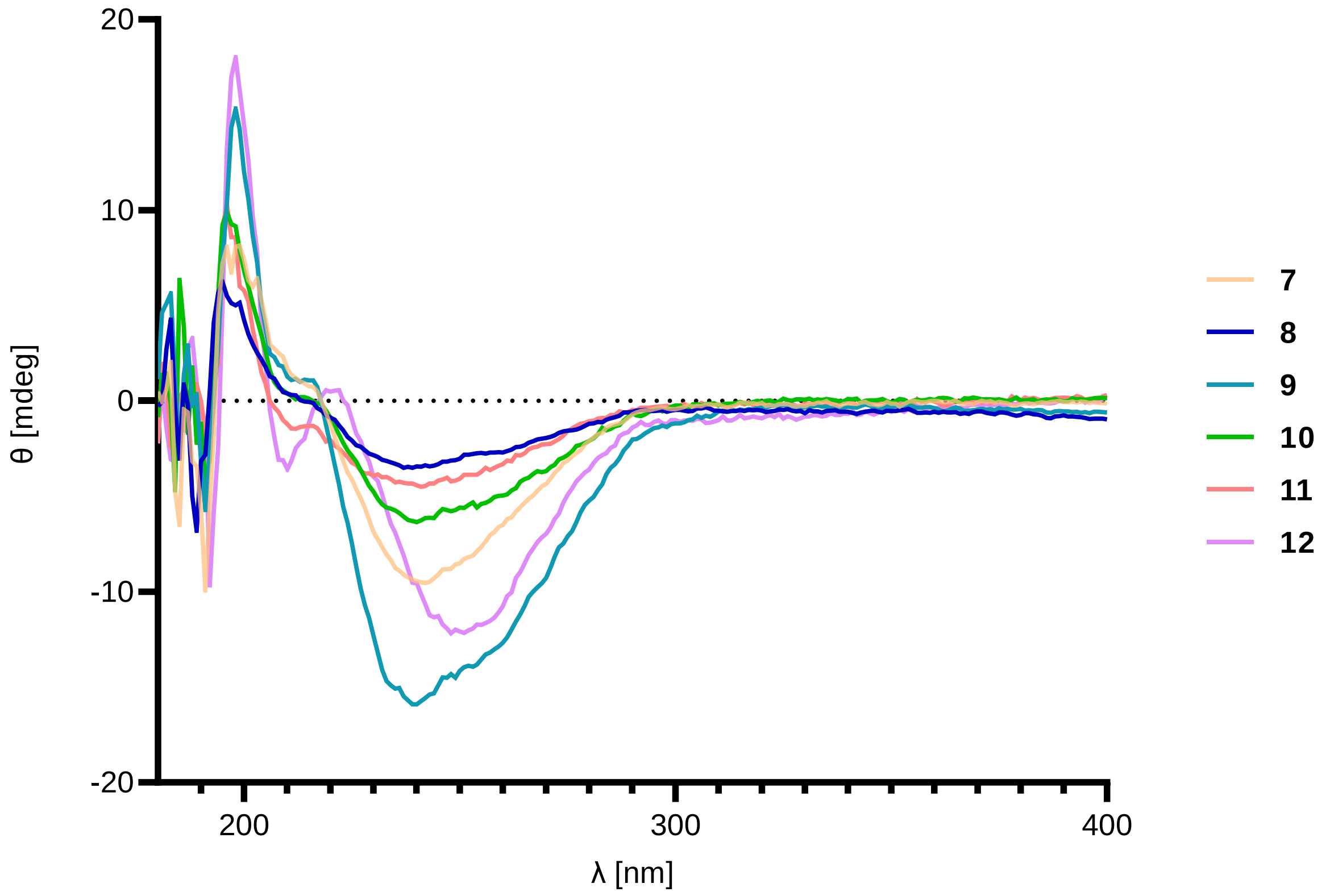

**Figure S40. ECD spectra of phorbactazoles A-F (7-12) in ethanol.**

**Table S6. ORFs of the *cly* genes and their putative functions.**

| ORF, GenBank accession number | Protein size [aa] | Proposed function | Closest homolog in GenBank, Source organism based on E value | Identity [%] | GenBank accession number |
|---|---|---|---|---|---|
| ClyA, WP_207859127.1 | 3902 | PKS-NRPS (A cMT ACP nMT KS ACP KS KR KR ACP) | CalA, *Candidatus* Entotheonella sp. | 64 | BAP05589.1 |
| ClyB WP_207859128.1 | 6493 | PKS-NRPS (KS ACP C A ACP KS DH KR cMT ACP ACP ER KS ACP KS ACP Cy Cy) | CalB, *Candidatus* Entotheonella sp. | 57 | BAP05590.1 |
| ClyC WP_207859129.1 | 4566 | PKS-NRPS (A ACP ACP OX OX KS ACP KS DH KR ACP KS KR ACP) | CalC, *Candidatus* Entotheonella sp. | 56 | BAP05591.1 |
| ClyD WP_207859130.1 | 384 | LLM class flavin-dependent oxidoreductase | CalD, *Candidatus* Entotheonella sp. | 80 | BAP05592.1 |
| ClyE WP_207859131.1 | 8170 | PKS (KS DH ACP KS ACP KS KR cMT ACP KS KR ACP KS oMT ACP KS cMT ACP ACP KS KR ACP) | CalE, *Candidatus* Entotheonella sp. | 59 | BAP05593.1 |
| ClyF WP_207859132.1 | 5801 | PKS (KS ACP KS KR ACP KS oMT ACP KS KR cMT ACP KS KR) | CalF, *Candidatus* Entotheonella sp. | 65 | BAP05594.1 |
| ClyG WP_207859133.1 | 5632 | PKS (cMT ACP KS DH KR cMT ACP KS ACP ACP KS ACP DH KR ACP KS) | CalG, *Candidatus* Entotheonella sp. | 64 | BAP05595.1 |
| ClyH WP_207859134.1 | 4745 | PKS-NRPS (ECH ECH ACP ACP KS ACP C A ACP KS DH KR ACP KS) | CalH, *Candidatus* Entotheonella sp. | 62 | BAP05596.1 |
| ClyI WP_207859135.1 | 2634 | PKS-NRPS (DH ACP KS ACP ACP C A ACP TE) | CalI, *Candidatus* Entotheonella sp. | 55 | BAP05597.1 |
| ClyL WP_207859125.1 | 442 | matallophosphoesterase | CalL, *Candidatus* Entotheonella sp. | 65 | BAP05586.1 |
| ClyM WP_207859115.1 | 232 | Hypothetical protein | CalM, *Candidatus* Entotheonella sp. | 78 | BAP05585.1 |
| ClyO WP_207859114.1 | 272 | VOC family protein | CalO, *Candidatus* Entotheonella sp. | 54 | BAP05583.1 |
| ClyP WP_207859116.1 | 231 | Hypothetical protein | CalP, *Candidatus* Entotheonella sp. | 77 | BAP05582.1 |
| ClyQ WP_207859117.1 | 373 | DUF1679 domain-containing protein | CalQ, *Candidatus* Entotheonella sp. | 68 | BAP05581.1 |
| ClyR WP_207859118.1 | 247 | ECH | CalR, *Candidatus* Entotheonella sp. | 80 | BAP05580.1 |
| ClyS WP_207859119.1 | 266 | ECH | CalS, *Candidatus* Entotheonella sp. | 75 | BAP05579.1 |
| ClyT<br>WP_207859120.1 | 423 | hydroxymethylglutaryl-CoA synthase family protein (HMG) | CalT, *Candidatus* Entotheonella sp. | 84 | BAP05578.1 |
| ClyU<br>WP_207859121.1 | 713 | ABC transporter ATP-binding protein/permease | CalU, *Candidatus* Entotheonella sp. | 66 | BAP05577.1 |
| ClyW<br>WP_207859124.1 | 410 | KS | Algicola sp.,<br>also similar to CalW | 60 | NQZ12188.1 |
| ClyX<br>WP_207859123.1 | 82 | ACP | CalX, *Candidatus* Entotheonella factor | 74 | ETX03756.1 |
| ClyY<br>WP_207859122.1 | 845 | AT | CalY, *Candidatus* Entotheonella sp. | 61 | BAP05573.1 |

**Table S7. Biosynthetic product prediction for the *cly* cluster.**

Shown are the predicted structures of KS substrates (α,β-region of the thioester) based on the KS phylogeny (Figure S41). To avoid bias, KSs from the almost identical *cal* BGC were ignored for this analysis.

| KS Domain | Predicted intermediate type accepted by the KS based on phylogeny (see SI) | Suggested KS intermediate type based on NMR data | Predicted KR product specificity |
|---|---|---|---|
| ClyA, KS1 | non-elongating<br>amino acids | non-elongating<br>amino acids | |
| ClyA, KS2 | amino acids | amino acids | |
| ClyB, KS3 | non-elongating<br>β-(L)-OH | non-elongating<br>β-OH | (L)-OH |
| ClyB, KS4 | amino acids (Gly) | amino acids (Gly) | |
| ClyB, KS5 | non-elongating<br>reduced α-L-methyl | non-elongating<br>reduced α-L-methyl | |
| ClyB, KS6 | non-elongating<br>reduced α-L-methyl | non-elongating<br>reduced α-L-methyl | |
| ClyC, KS7 | non-elongating<br>(oxazole/thiazole) | non-elongating<br>(oxazole/thiazole) | |
| ClyC, KS8 | amino acids<br>(oxazole/thiazole) | amino acids<br>(oxazole/thiazole) | |
| ClyC, KS9 | (*E*)-double bond | (*E*)-double bond | |
| ClyE, KS10 | non-elongating<br>ambiguous | oxygen insertion | |
| ClyE, KS11 | ambiguous | unclear | |
| ClyE, KS12 | ambiguous | unclear | |
| ClyE, KS13 | α-L-methyl-β-(L)-OH or<br>α-dimethyl-β-(L)-OH | α-methyl-OH | (L)-OH |
| ClyE, KS14 | β-(D)-OH | β-OH | (D)-OH |
| ClyE, KS15 | β-(D)-methoxy | β-methoxy | |
| ClyE, KS16 | α-dimethyl-β-keto | α-dimethyl-β-keto | |
| ClyF, KS17 | non-elongating<br>β-OH | non-elongating<br>β-OH | (L)-OH |
| ClyF, KS18 | α-(D)-OH-β-(D)-OH | α-OH-β-OH | |
| ClyF, KS19 | non-elongating<br>β-(D)-OH | non-elongating<br>β-OH | (D)-OH |
| ClyF, KS20 | β-(D)-methoxy | β-methoxy | |
| ClyF, KS21 | α-D-methyl-β-(L)-OH | α-methyl-β-OH | (L)-OH |
| ClyG, KS22 | α-L-methyl-β-(L)-OH | α-methyl-β-OH | (L)-OH |
| ClyG, KS23 | α-methyl-double bond | α-methyl-double bond | |
| ClyG, KS24 | β-methyl-double bond | β-methyl-double bond | |
| ClyG, KS25 | (*E*)-double bond | (*E*)-double bond | |
| ClyH, KS26 | non-elongating<br>ambiguous | non-elongating<br>ambiguous | |
| ClyH, KS27 | amino acids (Ala) | unknown | |
| ClyH, KS28 | non-elongating<br>double bond | unknown | |
| ClyI, KS29 | shifted double bond | unknown | |

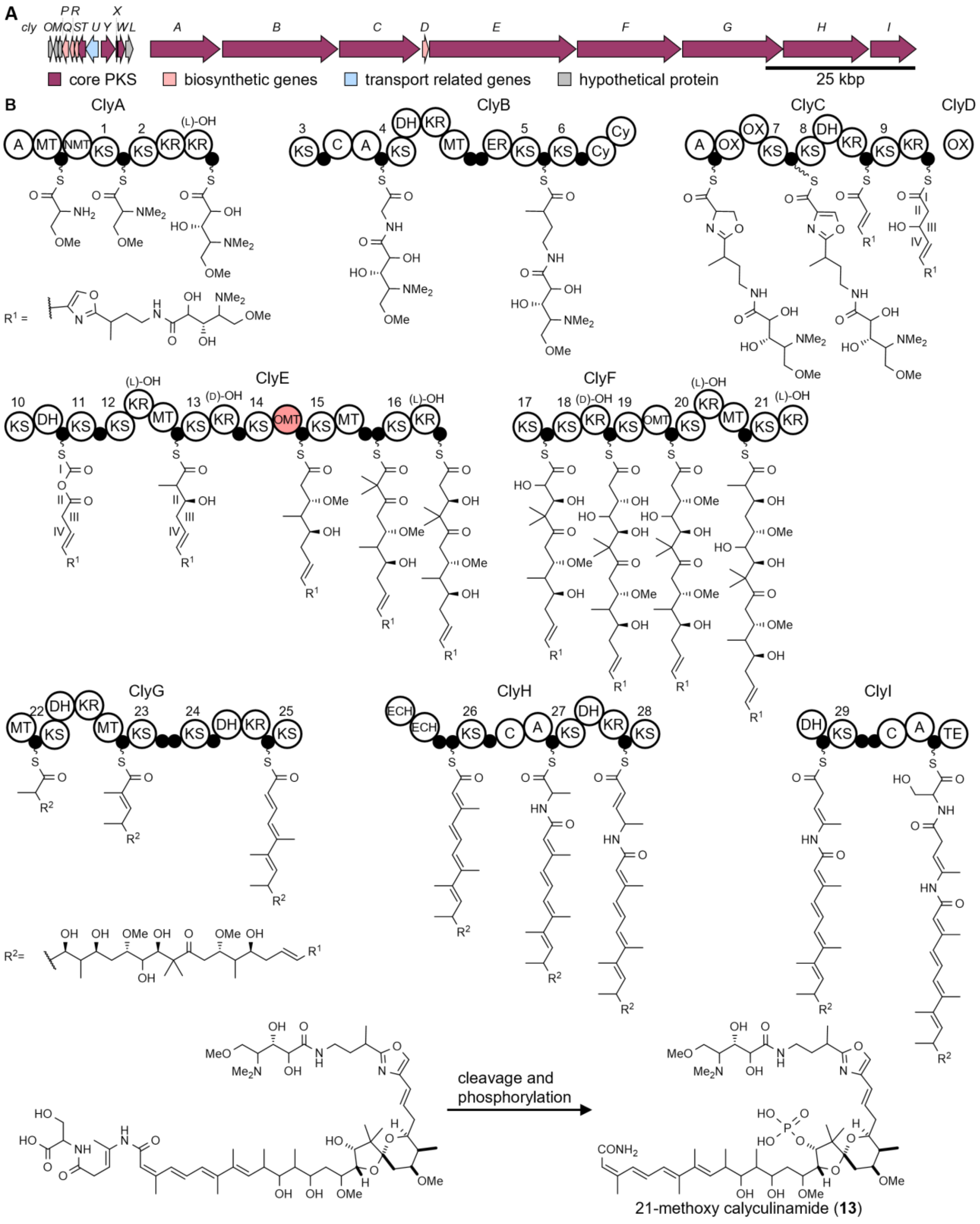

**Figure S41. Biosynthetic model of 21-methoxy calyculinamide (13) biosynthesis.**

(A) The *cly* biosynthetic gene cluster in *A. pedis*.
(B) The biosynthesis differs from the calyculin biosynthesis (*cal*) in "*Candidatus* Enthotheonella" only by an *O*-methyltransferase located upstream of KS 15 (red) and the product specificity of the KR in module 14. Full-length calyculinamide could not be detected. ECH: enoyl-CoA hydratase.

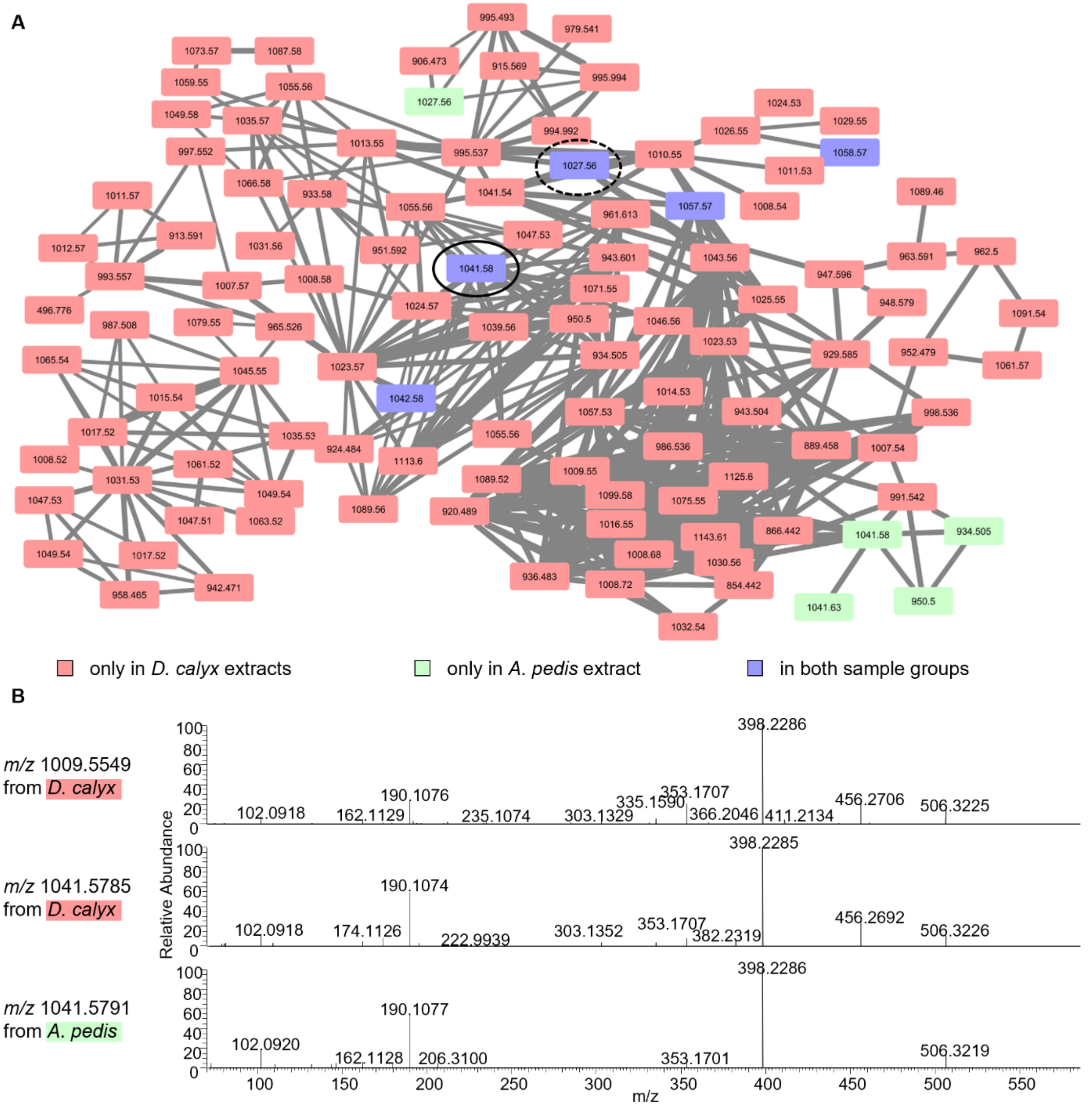

## Figure S42. Identification of calyculinamides in bacterial extracts.

(A) Extracts from the sponge *D. calyx* (red) and of *A. pedis* (green) were analyzed by HPLC-HESI-HRMS and in a molecular network.[30] It revealed ions that were shared between the sample groups (blue). Note that for all nodes 'only detected in the bacterial extract' (green) nodes with the same *m/z* were detected also in the sponge extract (red/blue). The most abundant ion of the bacterial extracts in this cluster is circled in black. One ion detected in the bacterial extracts yielded the GNPS library hit calyculinamide (CCMSLIB00005436501; dashed circle).[30]
(B) MS/MS patterns of ions of calyculin A (**3**, *m/z* 1009.5549) and 21-methoxy calyculinamide (**13**) (*m/z* 1041.5785) were very similar in both sample groups as well as in the database, suggesting similar parent ion structures.

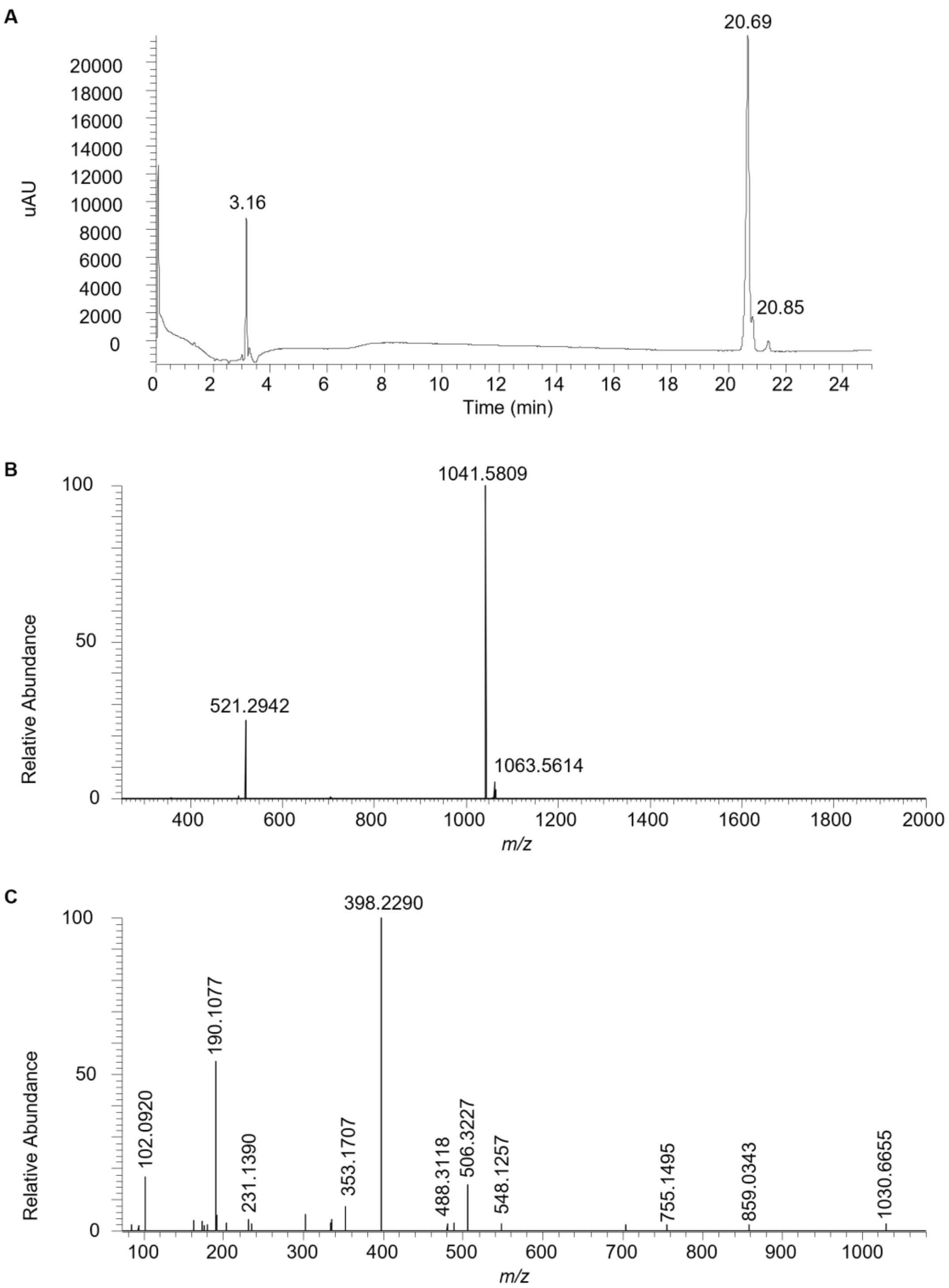

**Figure S43. HPLC-HESI-HRMS data of 21-methoxy calyculinamide (13).**

(A) UV trace of 21-methoxy calyculinamide (**13**) (λ 364 nm).

(B) MS data of 21-methoxy calyculinamide (**13**) (*m/z* 1041.5809 $[M+H]^+$, *m/z* 1063.5614 $[M+Na]^+$, *m/z* 521.2942 $[M+2H]^{2+}$).

(C) MS/MS fragmentation patterns of 21-methoxy calyculinamide (**13**) (normalized collision energy: stepped 27, 30).

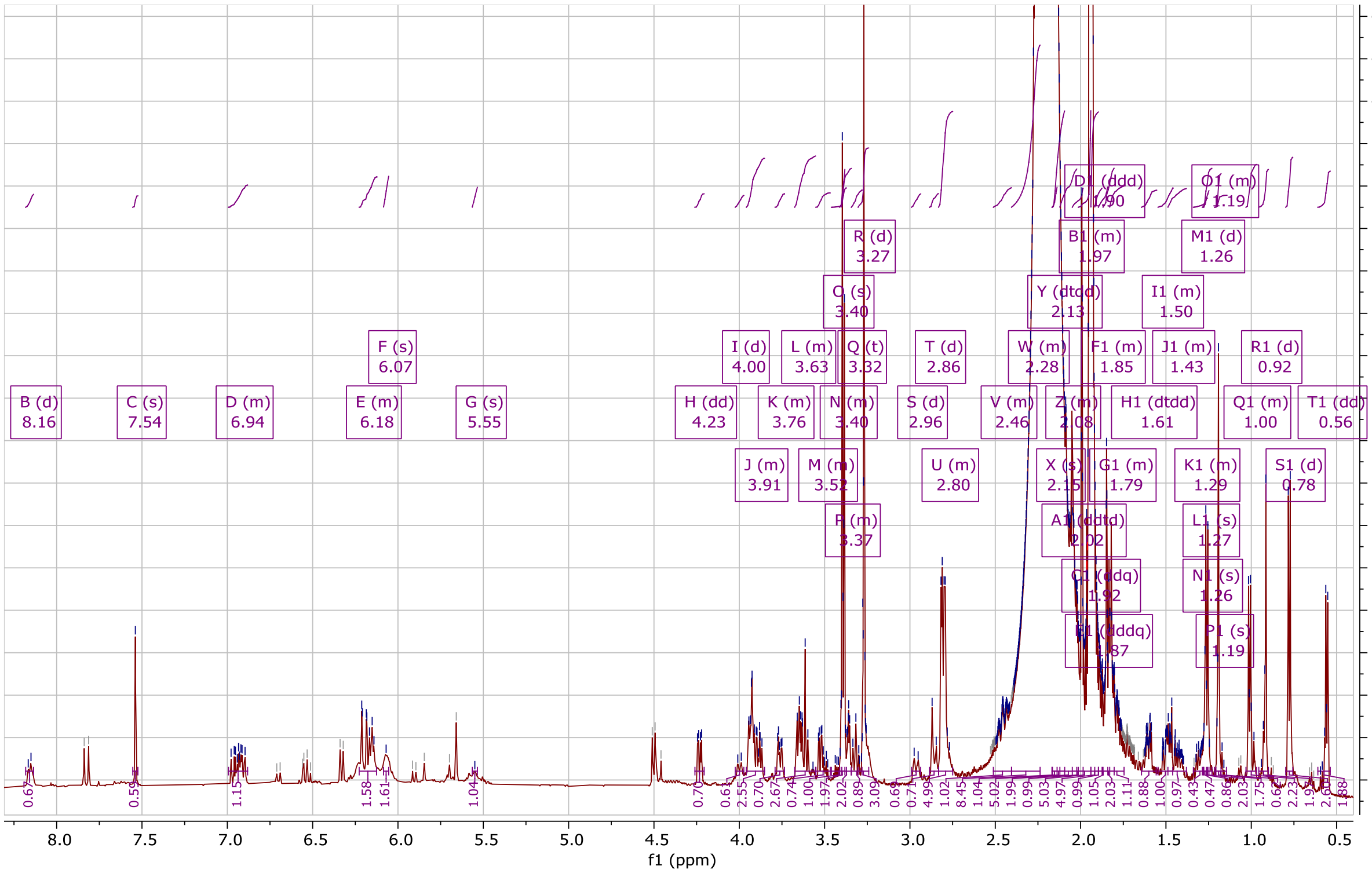

**Figure S44.** [1]**H NMR spectrum of 21-methoxy calyculinamide (13) in acetonitrile-*d*$_3$.**

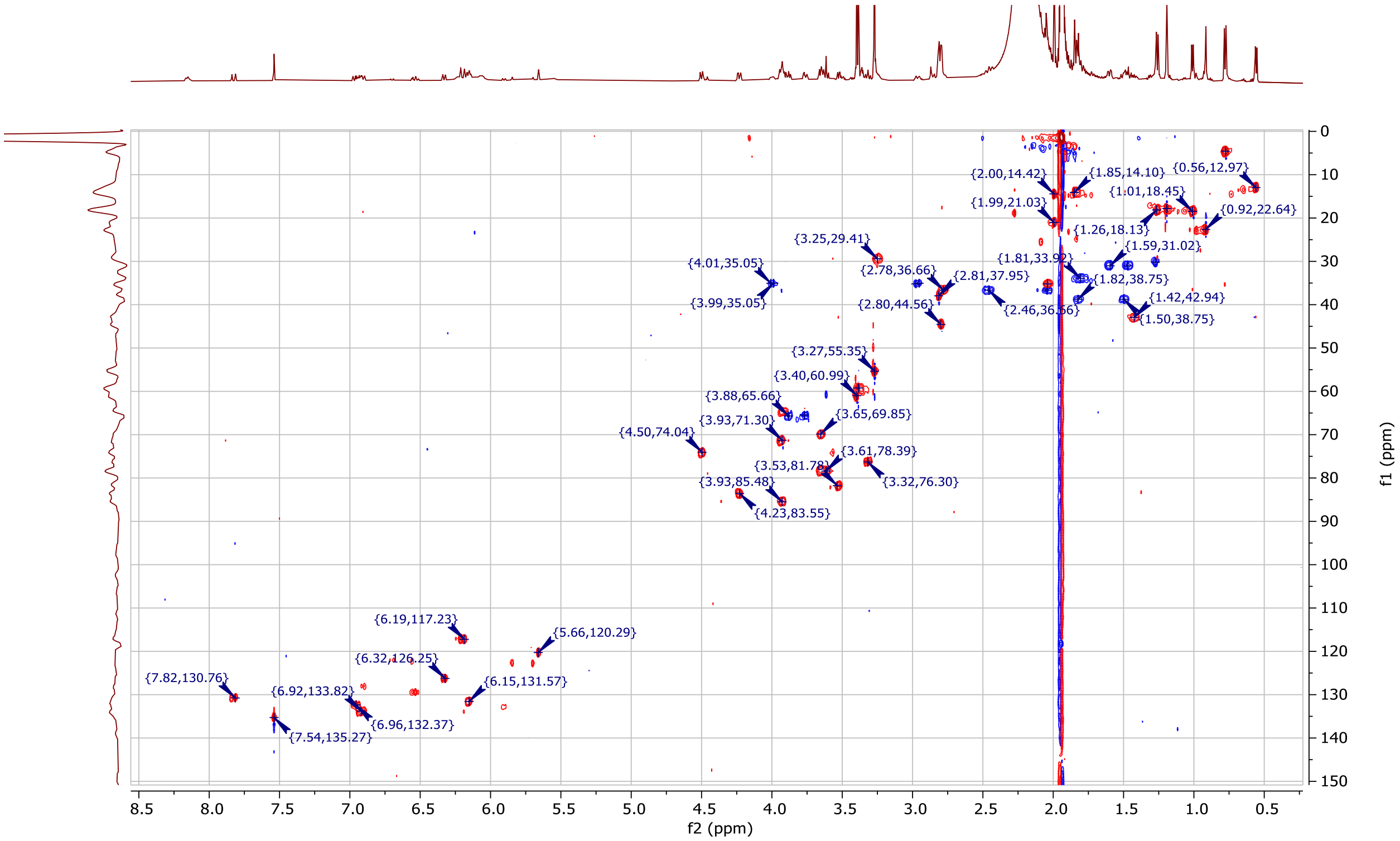

**Figure S45. HSQC spectrum of 21-methoxy calyculinamide (13) in acetonitrile-*d*$_3$.**

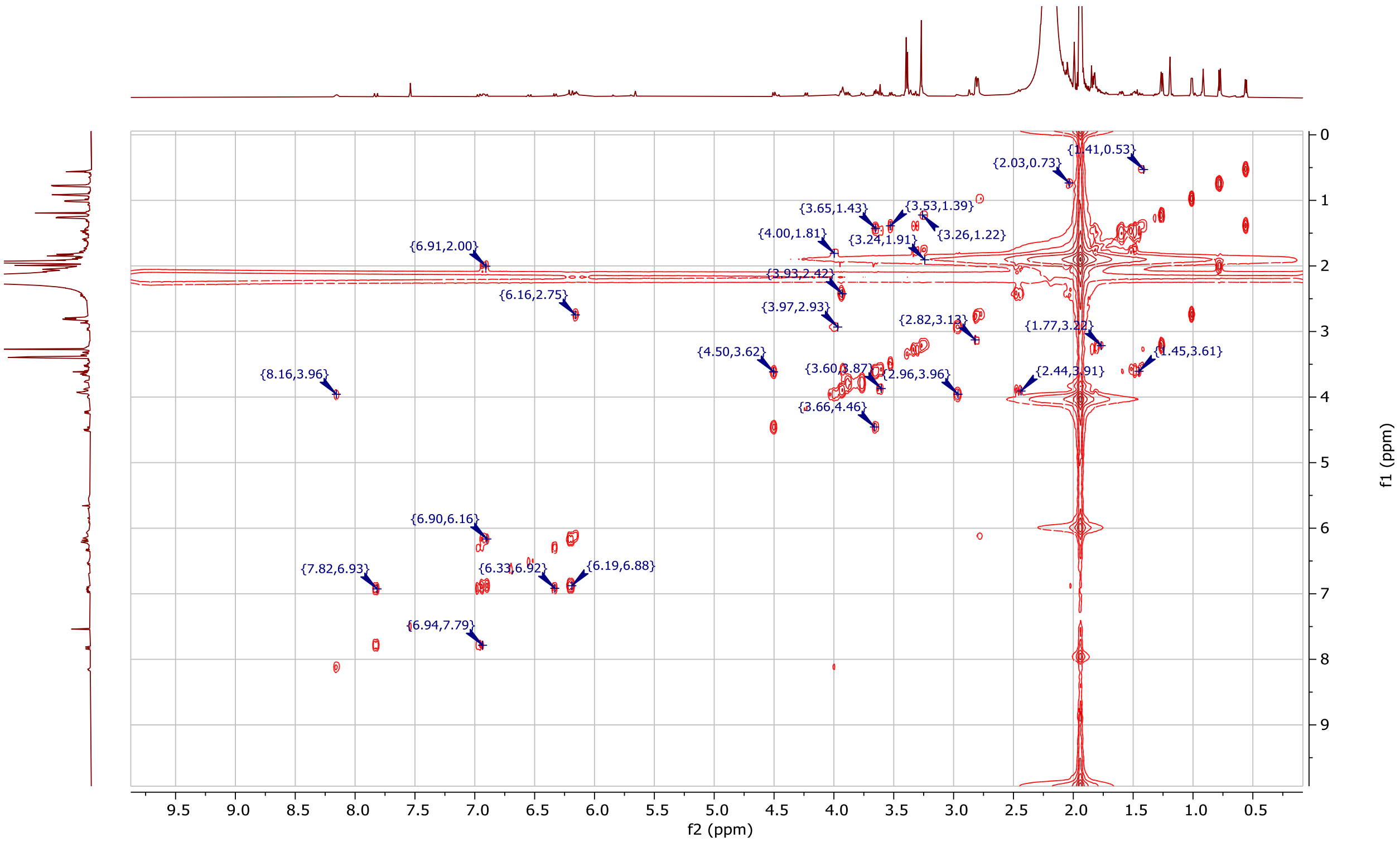

**Figure S46. Magnitude COSY spectrum of 21-methoxy calyculinamide (13) in acetonitrile-$d_3$**

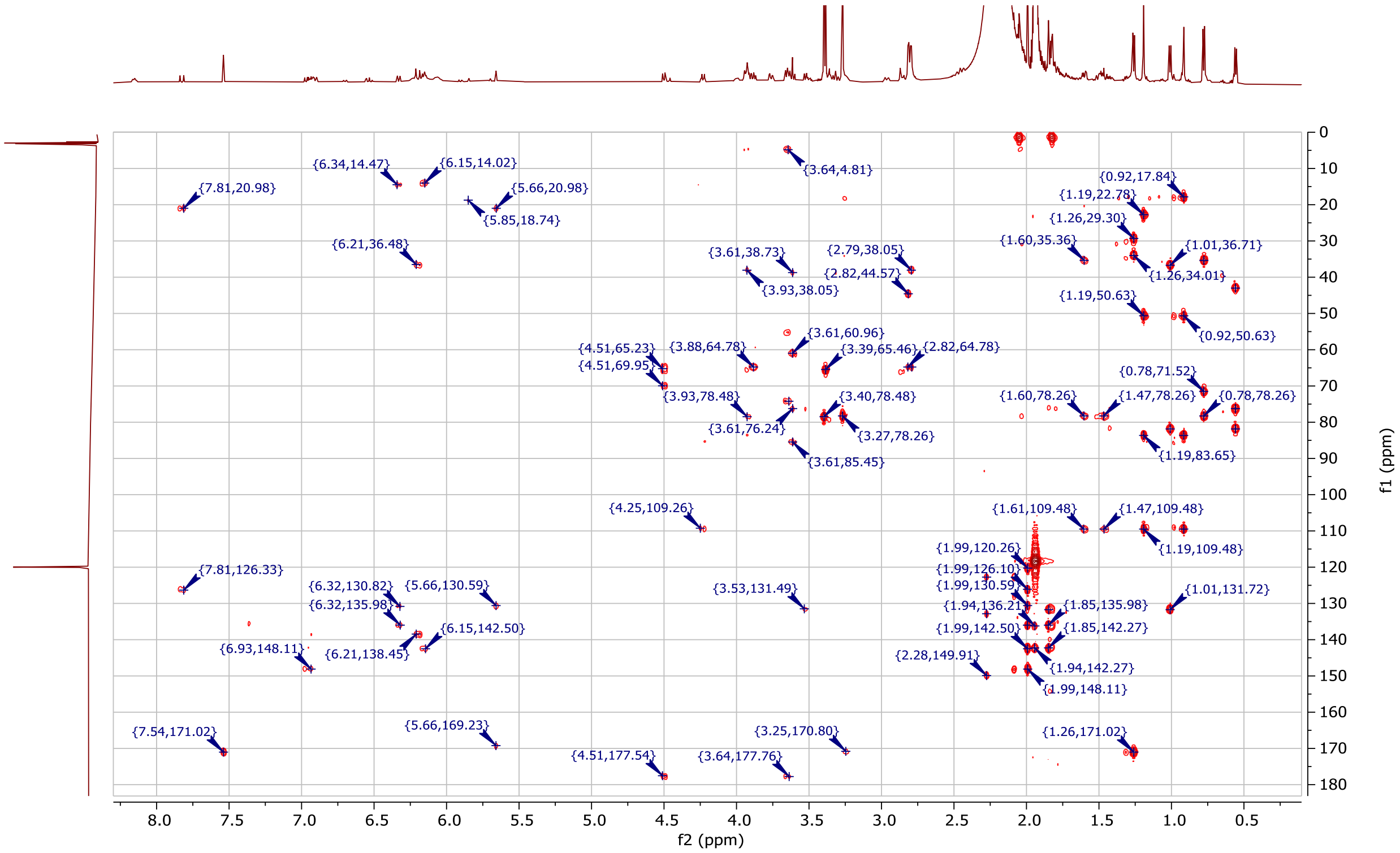

**Figure S47. HMBC spectrum of 21-methoxy calyculinamide (13) in acetonitrile-$d_3$.**

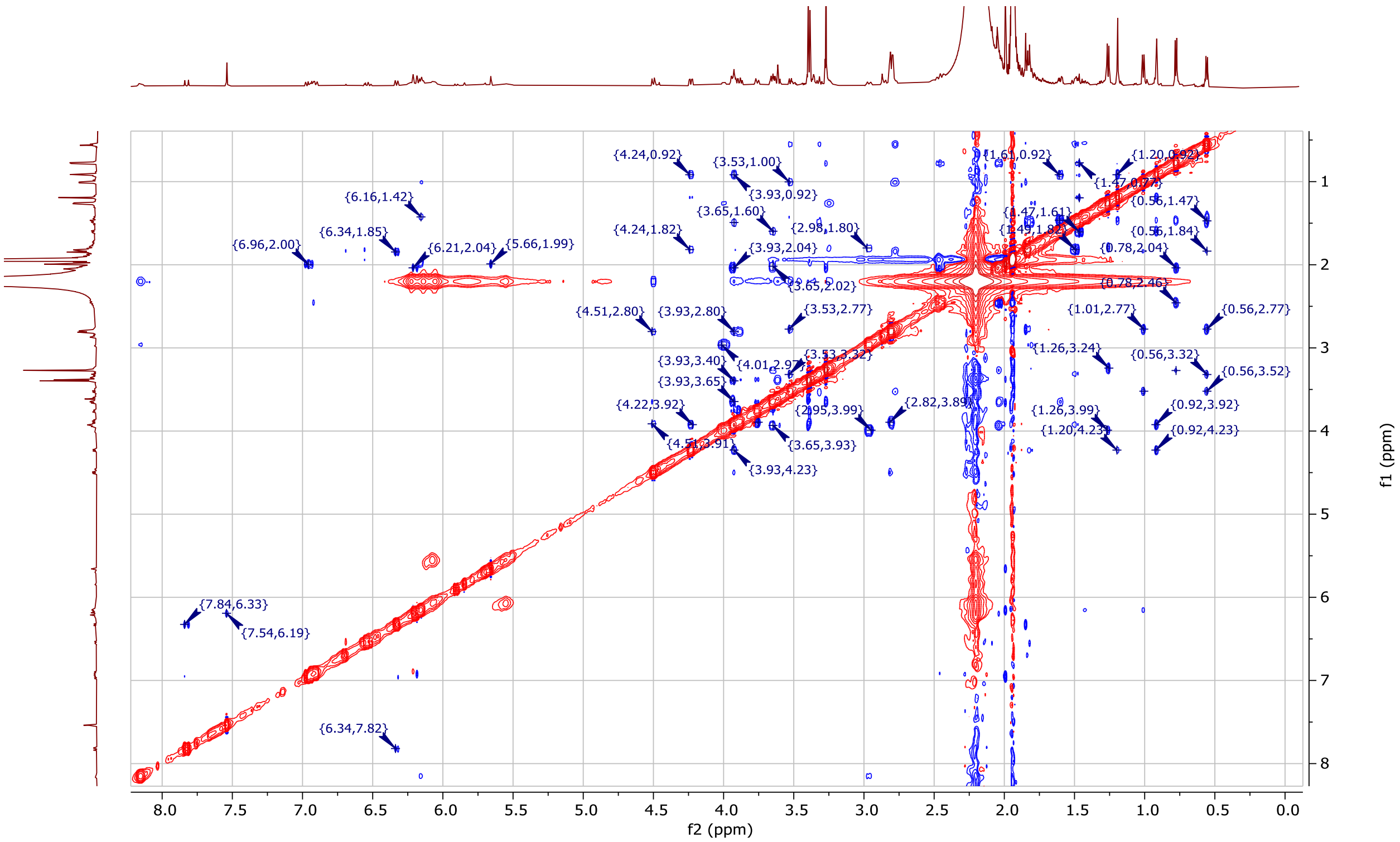

**Figure S48. NOESY spectrum of 21-methoxy calyculinamide (13) in acetonitrile-$d_3$.**

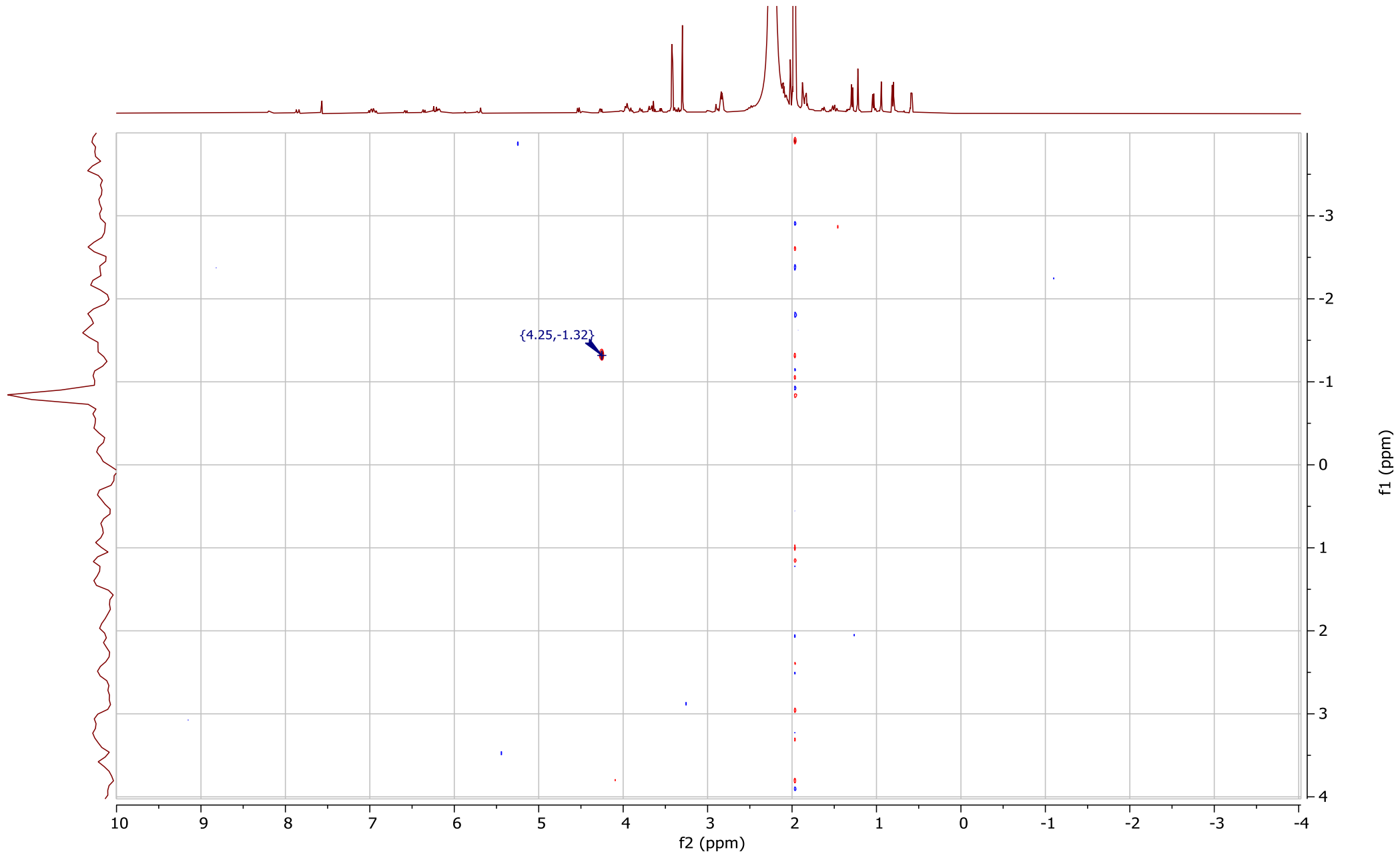

**Figure S49. $^{1}$H-$^{31}$Ph COSY spectrum of 21-methoxy calyculinamide (13) in acetonitrile-$d_3$.**

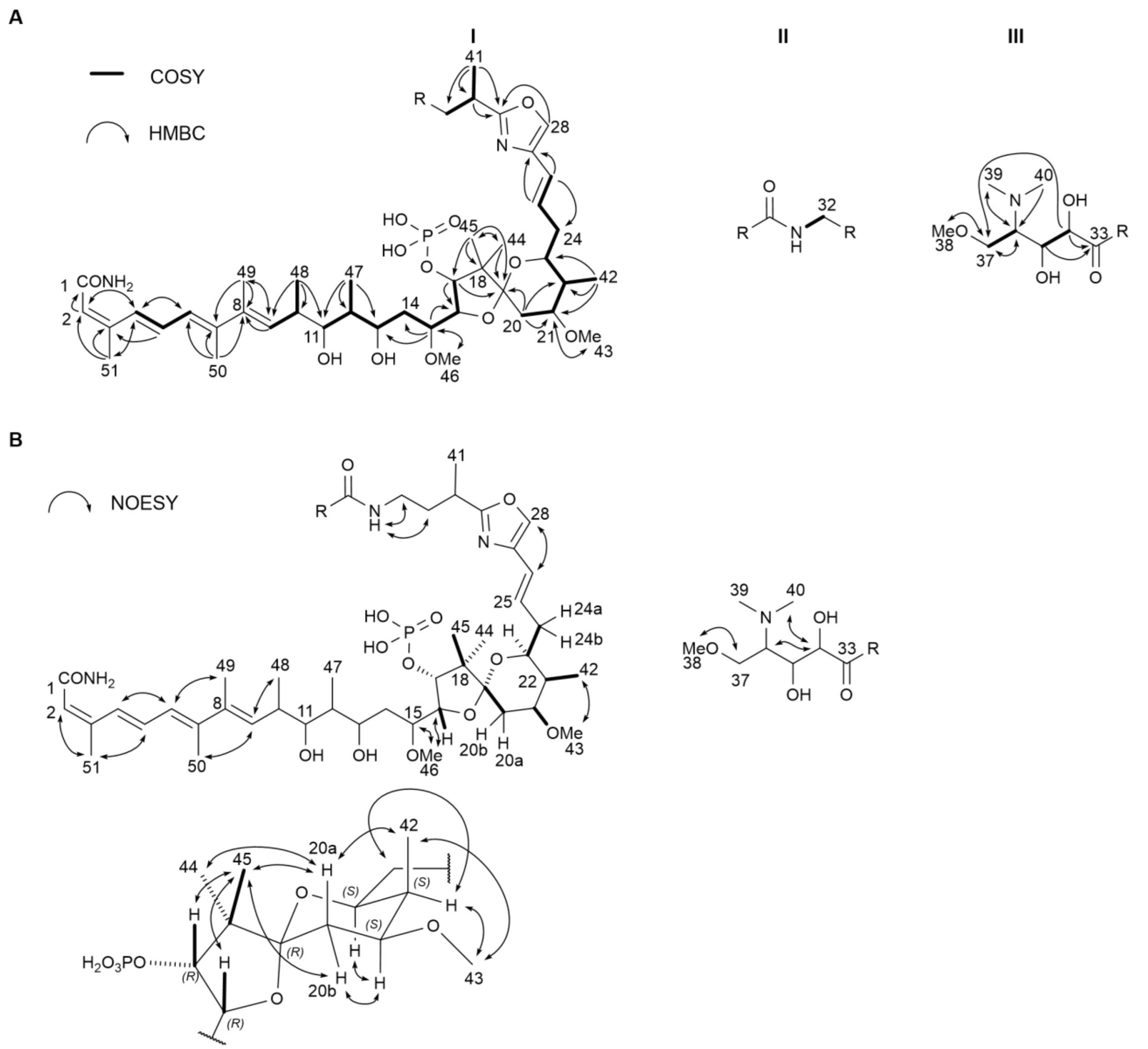

**Figure S50. Structure elucidation of 21-methoxy calyculinamide (13) in acetonitrile-$d_3$.**

(A) COSY and HMBC correlations of **13**.
(B) NOESY correlations of **13**.

**Table S8. NMR Data ($^1$H 600 MHz, $^{13}$C 150 MHz) for 21-methoxy calyculinamide (13) in acetonitrile-$d_3$.**

| No. | $^{13}$C | $^1$H (multiplicity) | $^{31}$P |
|---|---|---|---|
| 1 | 169.2, C | - | |
| 2 | 120.3, CH | 5.66 (s) | |
| 3 | 148.1, C | - | |
| 4 | 130.8, CH | 7.82 (d), $J$ = 15.4 Hz | |
| 5 | 132.4, CH | 6.96 (d), $J$ = 15.3, 11,1 Hz | |
| 6 | 126.3, CH | 6.32 (d), $J$ = 11.1 Hz | |
| 7 | 136.0, C | - | |
| 8 | 142.5, C | - | |
| 9 | 131.6, CH | 6.15 (m) | |
| 10 | 36.7, CH | 2.78 (m) | |
| 11 | 81.8, CH | 3.53 (dd), $J$ = 9.4, 2.5 Hz | |
| 12 | 42.9, CH | 1.42 (m) | |
| 13 | 76.3, CH | 3.32 (m) | |
| 14 | 38.8, $CH_2$ | 1.50/1.82 (m) | |
| 15 | 78.4, CH | 3.61 (m) | |
| 16 | 85.5, CH | 3.93 (m) | |
| 17 | 83.5, CH | 4.23 (dd), $J$ = 10.4, 3.9 Hz | |
| 18 | 50.6, C | - | |
| 19 | 109.5, C | - | |
| 20a | 31.0, $CH_2$ | 1.47 (m) | |
| 20b | | 1.59 (ddd), $J$ = 12.3, 4.5 Hz | |
| 21 | 78.4, CH | 3.65 (m) | |
| 22 | 35.2, CH | 2.04 (m) | |
| 23 | 71.3, CH | 3.93 (m) | |
| 24a | 36.7, $CH_2$ | 2.04 (m) | |
| 24b | | 2.46 (m) | |
| 25 | 133.8, CH | 6.92 (m) | |
| 26 | 117.2, CH | 6.19 (dd), $J$ = 16.0, 2.0 Hz | |
| 27 | 138.5, C | - | |
| 28 | 135.3, CH | 7.54 (s) | |
| 29 | 171.1, C | - | |
| 30 | 29.4, CH | 3.25 (m) | |
| 31 | 33.9, $CH_2$ | 1.81 (m) | |
| 32a | 35.1, $CH_2$ | 2.96 (d), $J$ = 14.0 Hz | |
| 32b | | 4.00 (m) | |
| 33 | 177.6, C | - | |
| 34 | 74.0, CH | 4.50 (d), $J$ = 10.0 Hz | |
| 35 | 69.9, CH | 3.65 (m) | |
| 36 | 64.7, CH | 3.93 (m) | |
| 37a | 65.5, CH | 3.77 (dd), $J$ = 12.2, 3.0 Hz | |
| 37b | | 3.88 (m) | |
| 38 | 59.2, $CH_3$ | 3.39 (s) | |
| 39 | 38.0, $CH_3$ | 2.81 (d), *J* = 4.2 Hz | |
| 40 | 44.6, $CH_3$ | 2.80 (d), *J* = 4.2 Hz | |
| 41 | 18.1, $CH_3$ | 1.26 (d), *J* = 6.9 Hz | |
| 42 | 4.6, $CH_3$ | 0.78 (d), *J* = 6.8 Hz | |
| 43 | 55.4, $CH_3$ | 3.27 (s) | |
| 44 | 17.8, $CH_3$ | 1.19 (s) | |
| 45 | 22.6, $CH_3$ | 0.92 (s) | |
| 46 | 61.0, $CH_3$ | 3.40 (s) | |
| 47 | 13.0, $CH_3$ | 0.56 (d), *J* = 6.7 Hz | |
| 48 | 18.5, $CH_3$ | 1.01 (d), *J* = 7.0 Hz | |
| 49 | 14.1, $CH_3$ | 1.85 (s) | |
| 50 | 14.6, $CH_3$ | 1.99 (s) | |
| 51 | 21.0, $CH_3$ | 1.99 (s) | |
| NH | - | 8.15 (m) | |
| P | | | -1.32 |

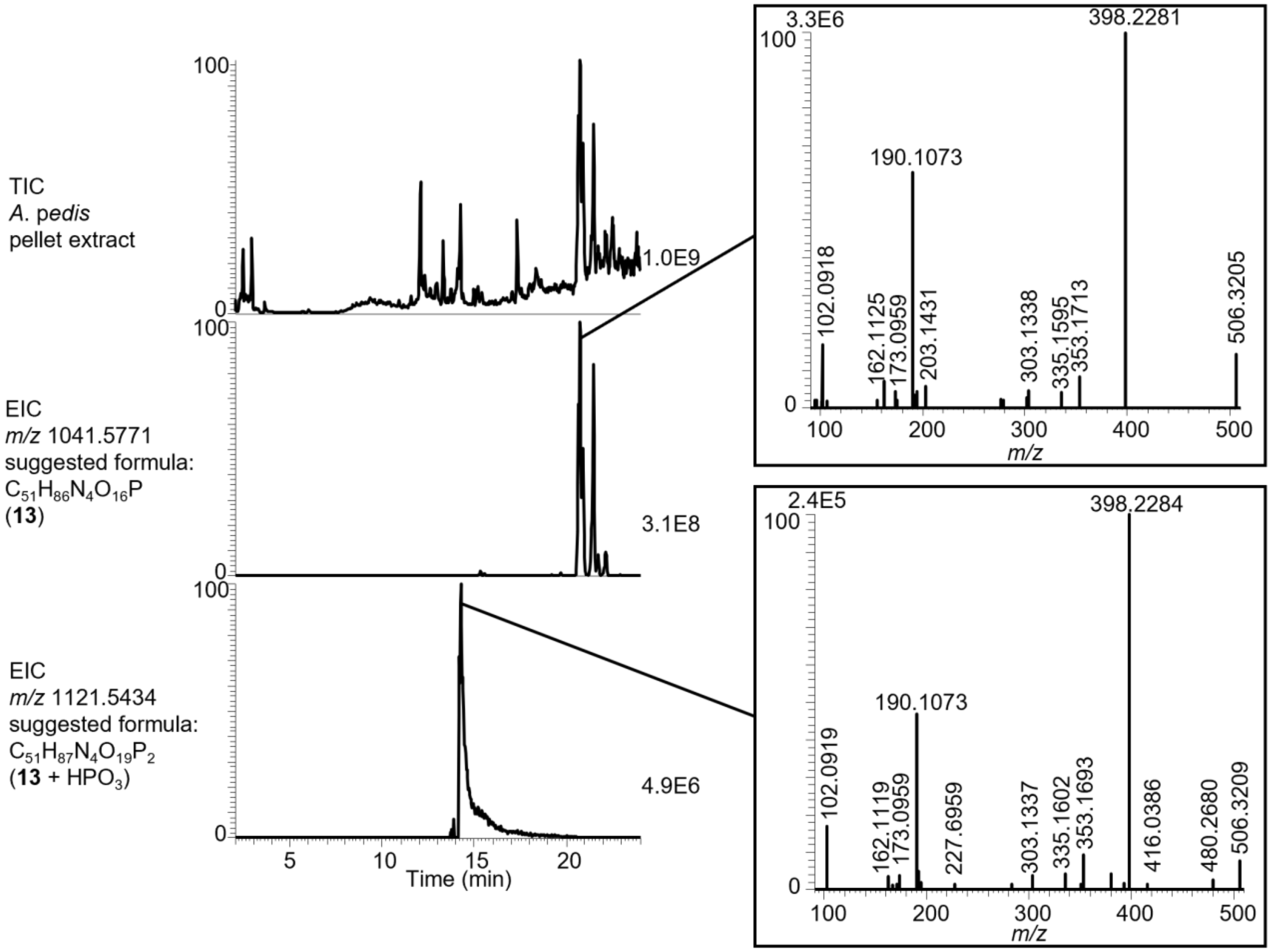

**Figure S51. Detection of 21-methoxy calyculinamide (13) and an ion corresponding to a putative phosphorylated congener in *A. pedis* cell pellets.**

The ion at *m/z* 1121.5434 was only detected in the cell pellet and not the supernatant. MS/MS fragmentation patterns indicate a similar structure to (**13**), with the difference of a phosphate group. Shown are the total ion chromatogram (TIC) and extracted ion chromatograms (EICs).

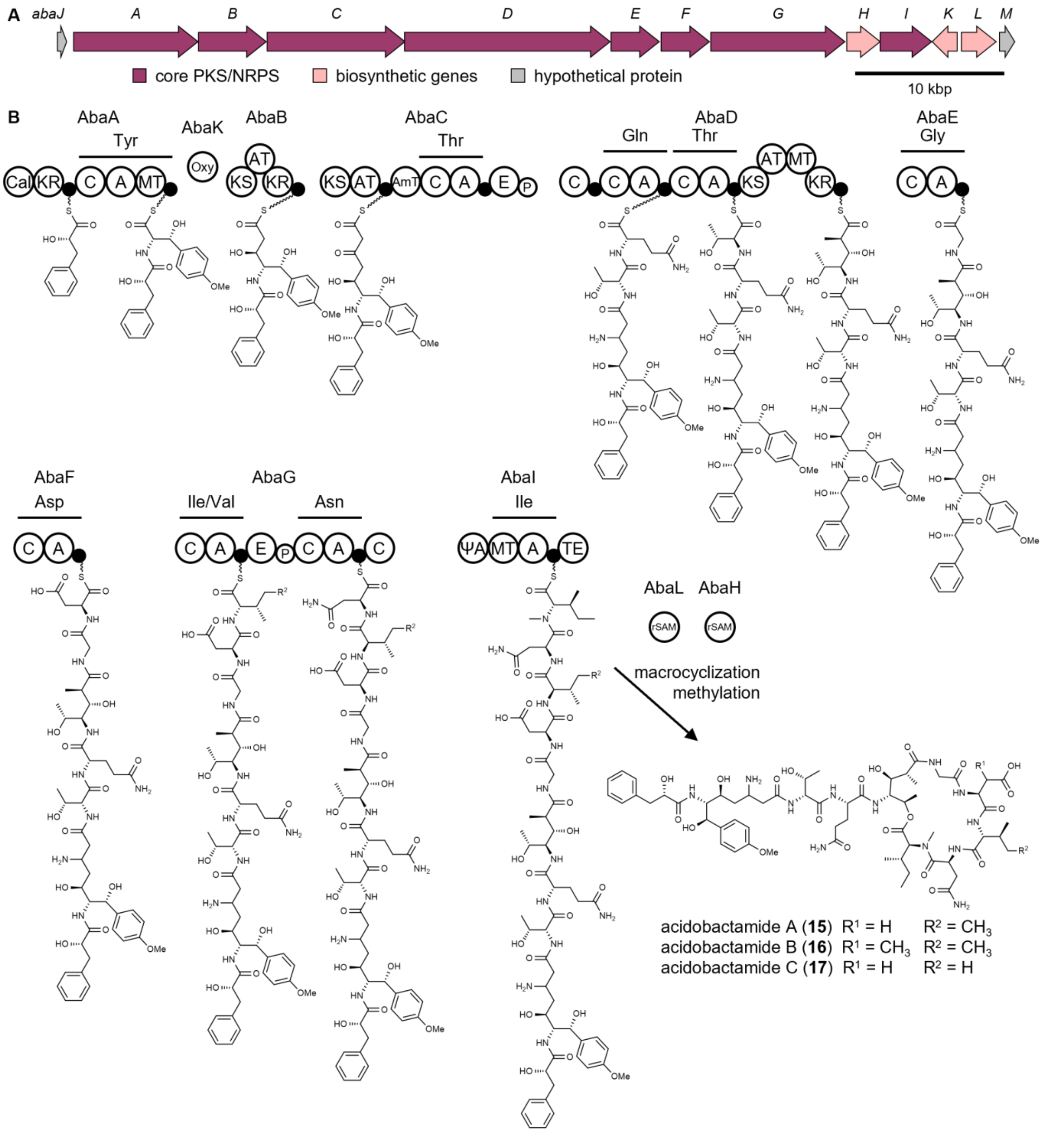

**Figure S52. Putative biosynthetic gene cluster for acidobactamides in *A. pedis* and biosynthetic model.**

(A) Genomic map of the aba PKS-NRPS hybrid BGC in *A. pedis*.
(B) Model of acidobactamide (**15**-**17**) biosynthesis. Acidobactamide production was putatively linked to the *aba* BGC based on predicted gene functions. The timing of tyrosine hydroxylation and C-methylation is speculative. AMT: aminotransferase; Cal: Co-enzyme A ligase; Oxy: amino acid β-hydroxylase; P: TIGR01720 domain; ΨA: pseudo A-domain; rSAM: radical *S*-adenosyl methionine methyltransferase.

**Table S9. ORFs of the *aba* genes and their putative functions.**

| ORF, GenBank accession number | Protein size [aa] | Proposed function | Closest homolog in GenBank, Source organism based on E value | Identity [%] | GenBank accession number |
|---|---|---|---|---|---|
| AbaJ, WP_207856589.1 | 209 | hypothetical protein | *Alteromonas* sp. ASW11-36 | 40 | WP_289365696.1 |
| AbaA, WP_207856588.1 | 2770 | NRPS-PKS (CAL, KR, PCP, C, A, MT, PCP) | *Hyella patelloides*. | 37 | WP_144865674.1 |
| AbaB, WP_207856587.1 | 1549 | PKS (KS, AT, KR, PCP) | *Coleofasciculus* sp. FACHB-1120 | 40 | WP_190816235.1 |
| AbaC, WP_207856586.1 | 3034 | NRPS-PKS (KS, AT, PCP, AmT, C, A, PCP, E) | *Methylocapsa palsarum* | 39 | WP_091679403.1 |
| AbaD, WP_207856585.1 | 4568 | NRPS-PKS (C, PCP, C, A, PCP, C, A, PCP, KS, AT, MT, KR, PCP) | Planctomycetota bacterium | 36 | MCP4261673.1 |
| AbaE, WP_207856584.1 | 1138 | NRPS (C, A, PCP) | Desulfobacterales bacterium | 43 | MCP4687842.1 |
| AbaF, WP_207856583.1 | 1073 | NRPS (C, A, PCP) | *Coleofasciculus* sp. FACHB-129 | 33 | WP_190677450.1 |
| AbaG, WP_207856582.1 | 3016 | NRPS (C, A, PCP, E, P, C, A, PCP, C) | *Candidatus* Thiomargarita nelsonii | 41 | TGO03755.1 |
| AbaH, WP_207856581.1 | 741 | radical SAM methyltransferase | Unassigned bacterium | 58 | MCP5047450.1 |
| AbaI, WP_207856580.1 | 1132 | NRPS (ΨA, MT, A, PCP, TE) | Thiotrichales bacterium HSG14 | 47 | MDM8560642.1 |
| AbaK, WP_207856579.1\| | 529 | amino acid β-hydroxylase | Acidobacteriota bacterium | 64 | MDJ0841517.1 |
| AbaL, WP_207856578.1 | 773 | radical SAM methyltransferase | *Candidatus* Scalindua sp. | 46 | MBT6052422.1 |
| AbaM, WP_207856577.1 | 340 | hypothetical protein | Desulfobacterales bacterium | 37 | MBF0101919.1 |

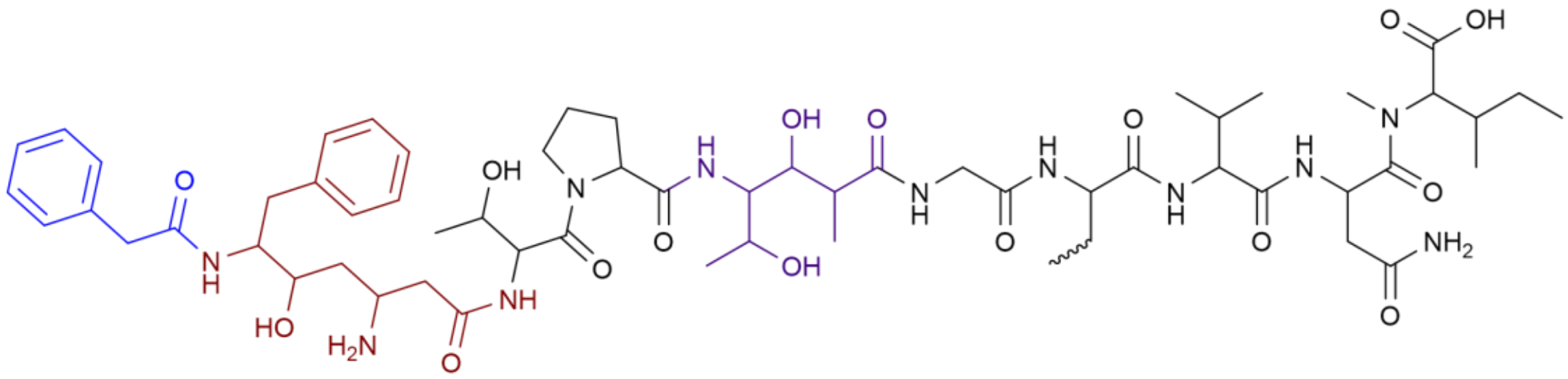

PRISM based product prediction of the *aba* BGC (**19**)

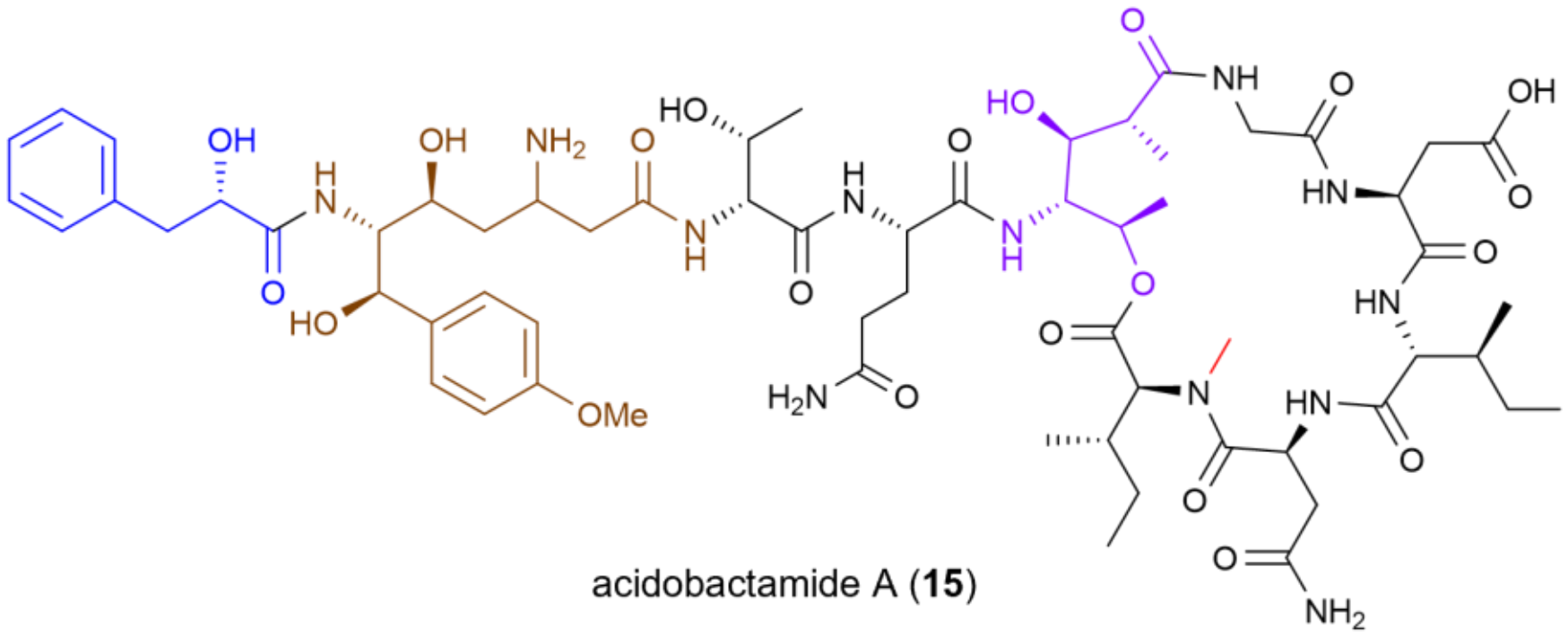

acidobactamide A (**15**)

**Figure S53. PRISM based product prediction of the *aba* BGC.**

PRISM predicted Hppa, Ddmpha, and Admha residues highlighted in blue, orange, and purple, respectively (**19**). The predicted moieties were confirmed in the isolated acidobactamides.

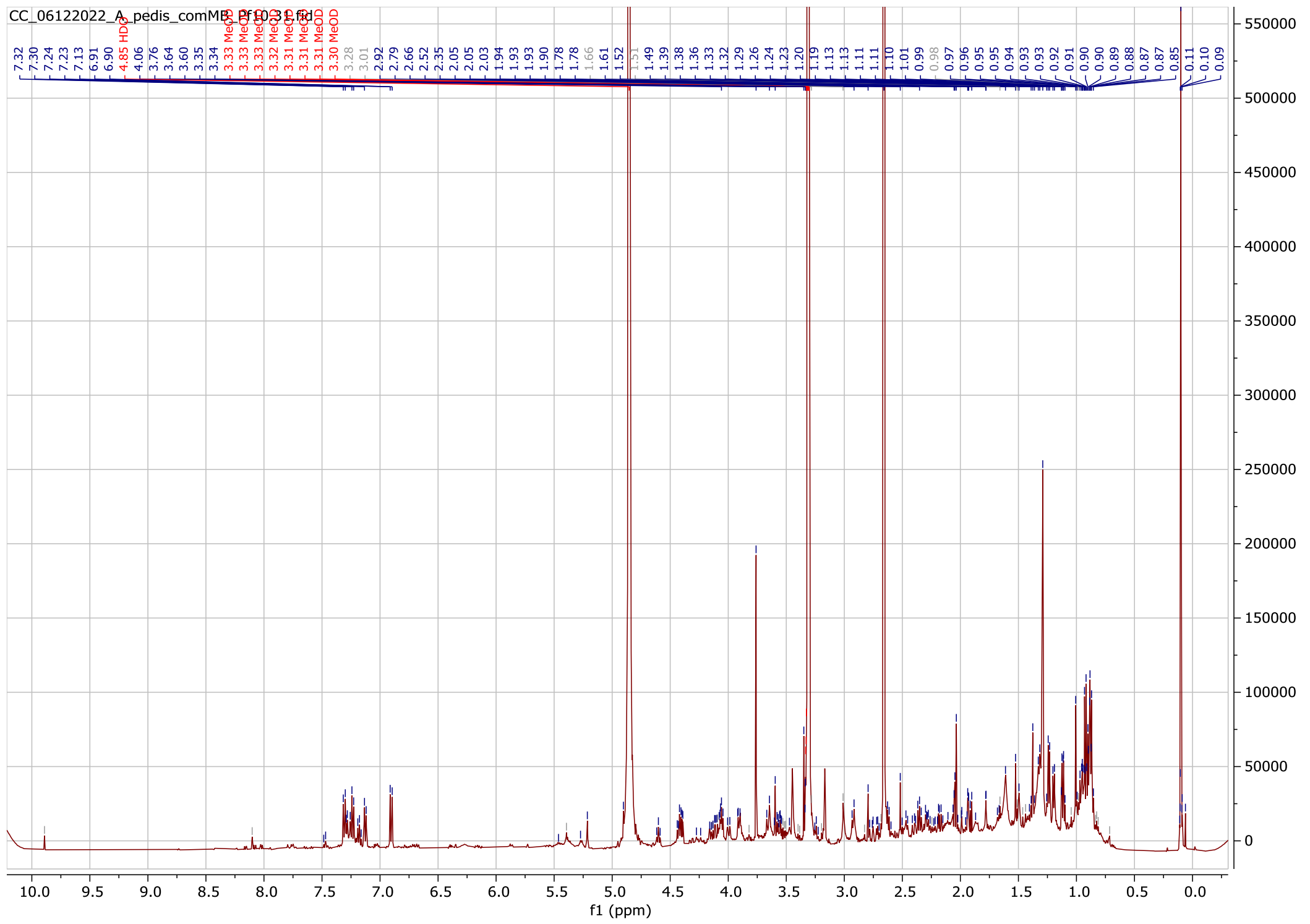

**Figure S54. $^1$H NMR spectrum of fraction F12 in methanol-*$d_4$*.**

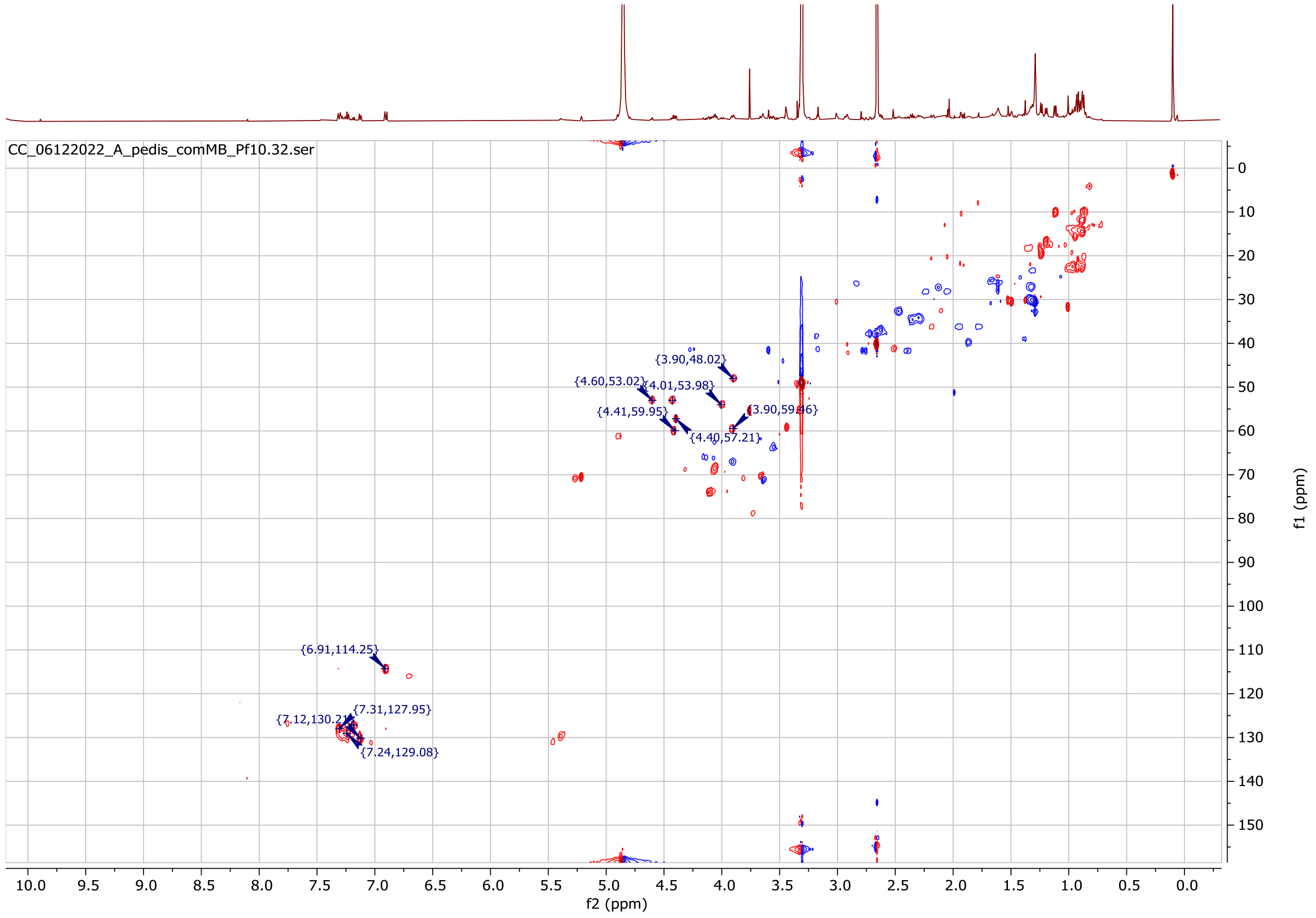

**Figure S55. HSQC spectrum of fraction F12 in methanol-*$d_4$*.**

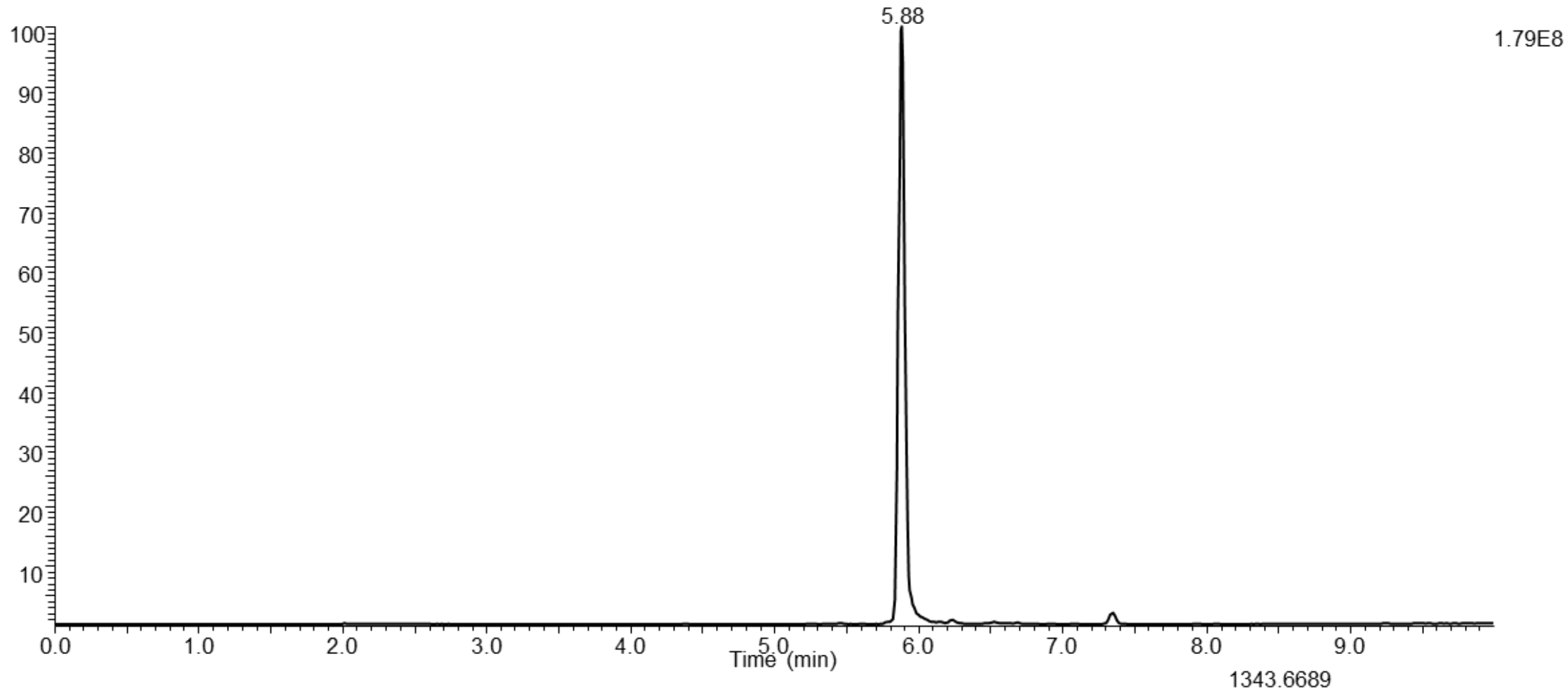

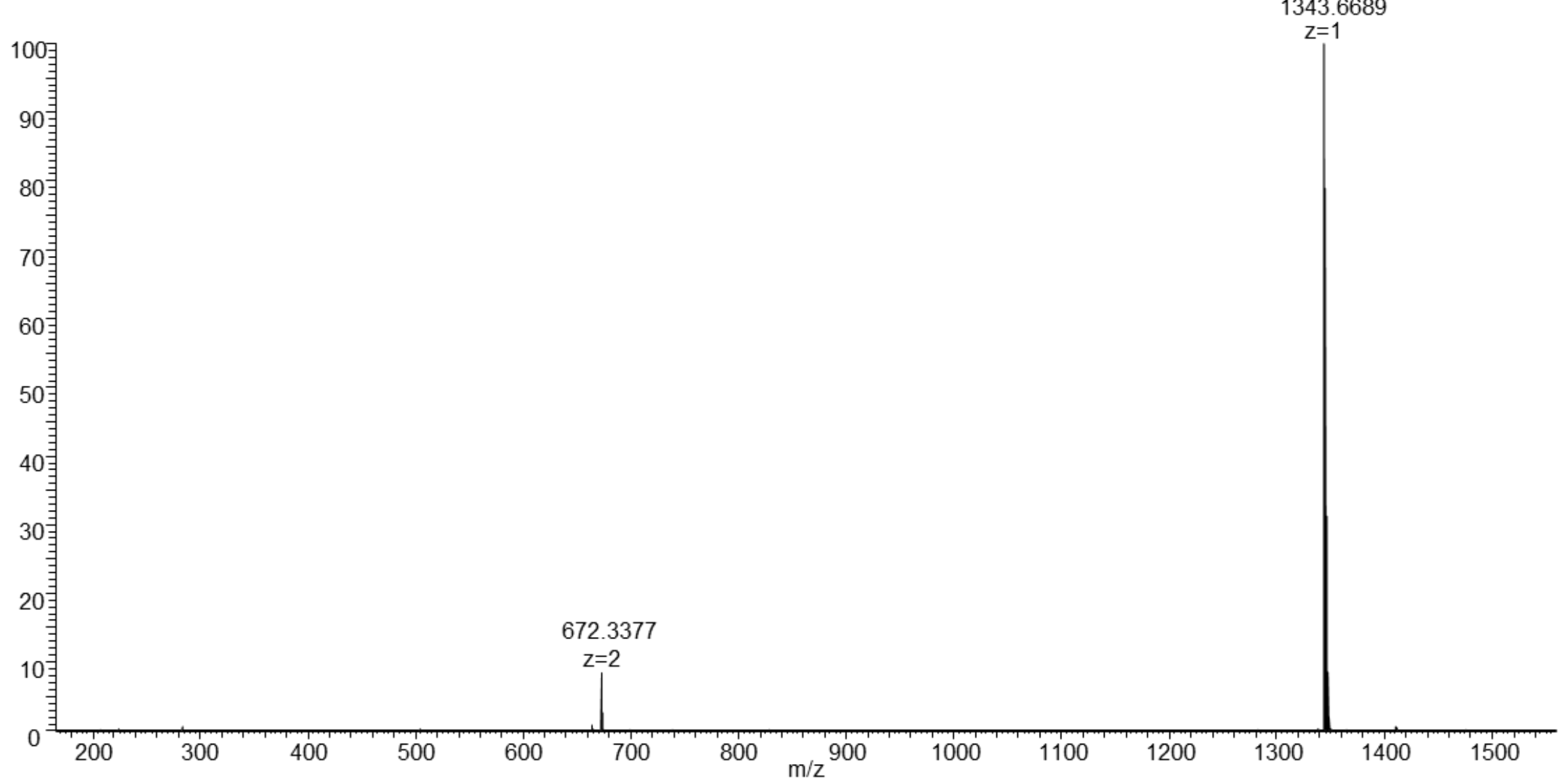

**Figure S56. HPLC-HESI-HRMS data of acidobactamide A (15)**

Top: Total ion chromatogram of isolated acidobactamide A (**15**).
Bottom: MS data of acidobactamide A (**15**) (m/z 1343.6689 $[M+H]^+$).

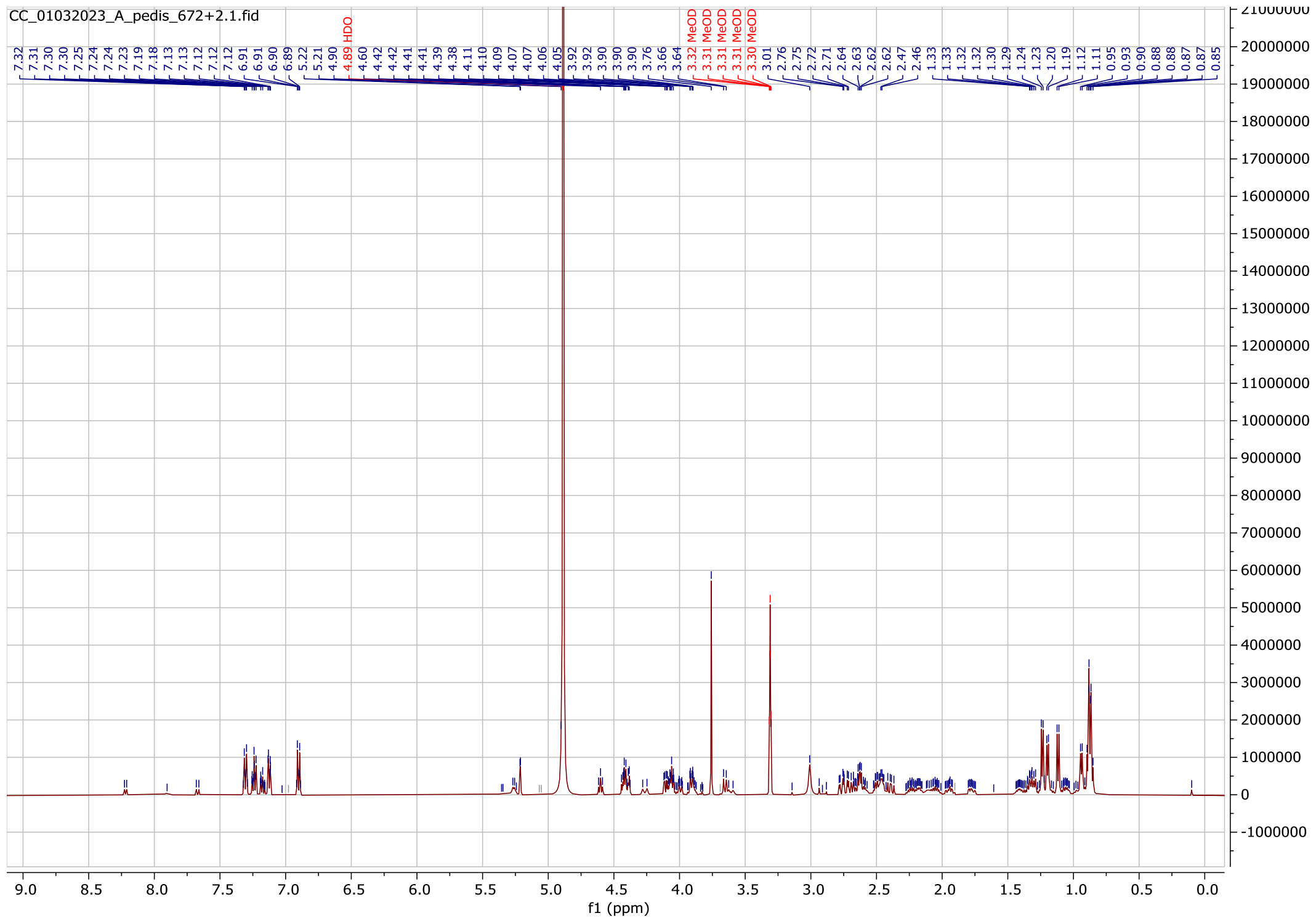

**Figure S57. [1]H NMR spectrum of acidobactamide A (15) in methanol-$d_4$.**

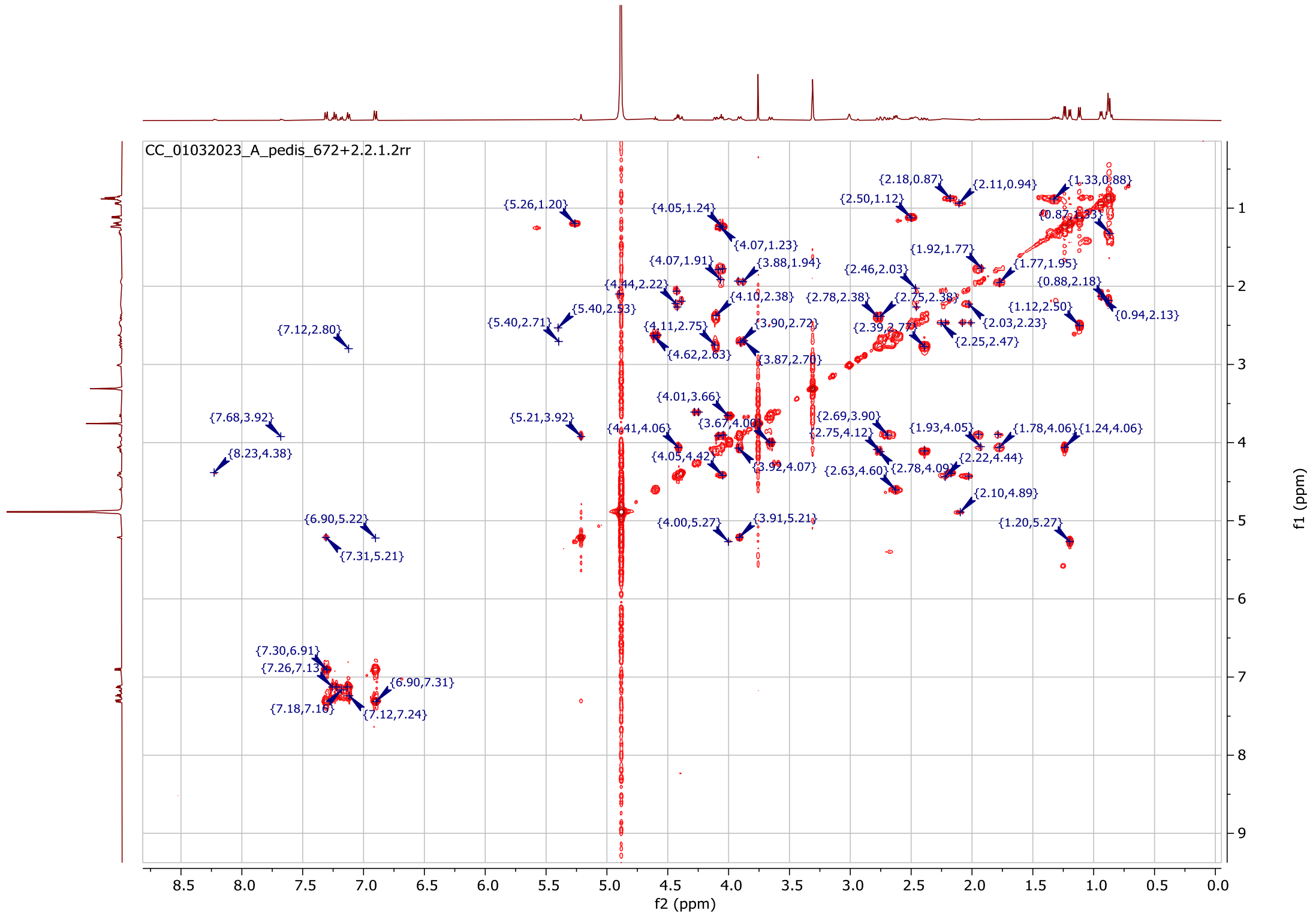

**Figure S58. COSY spectrum of acidobactamide A (15) in methanol-$d_4$.**

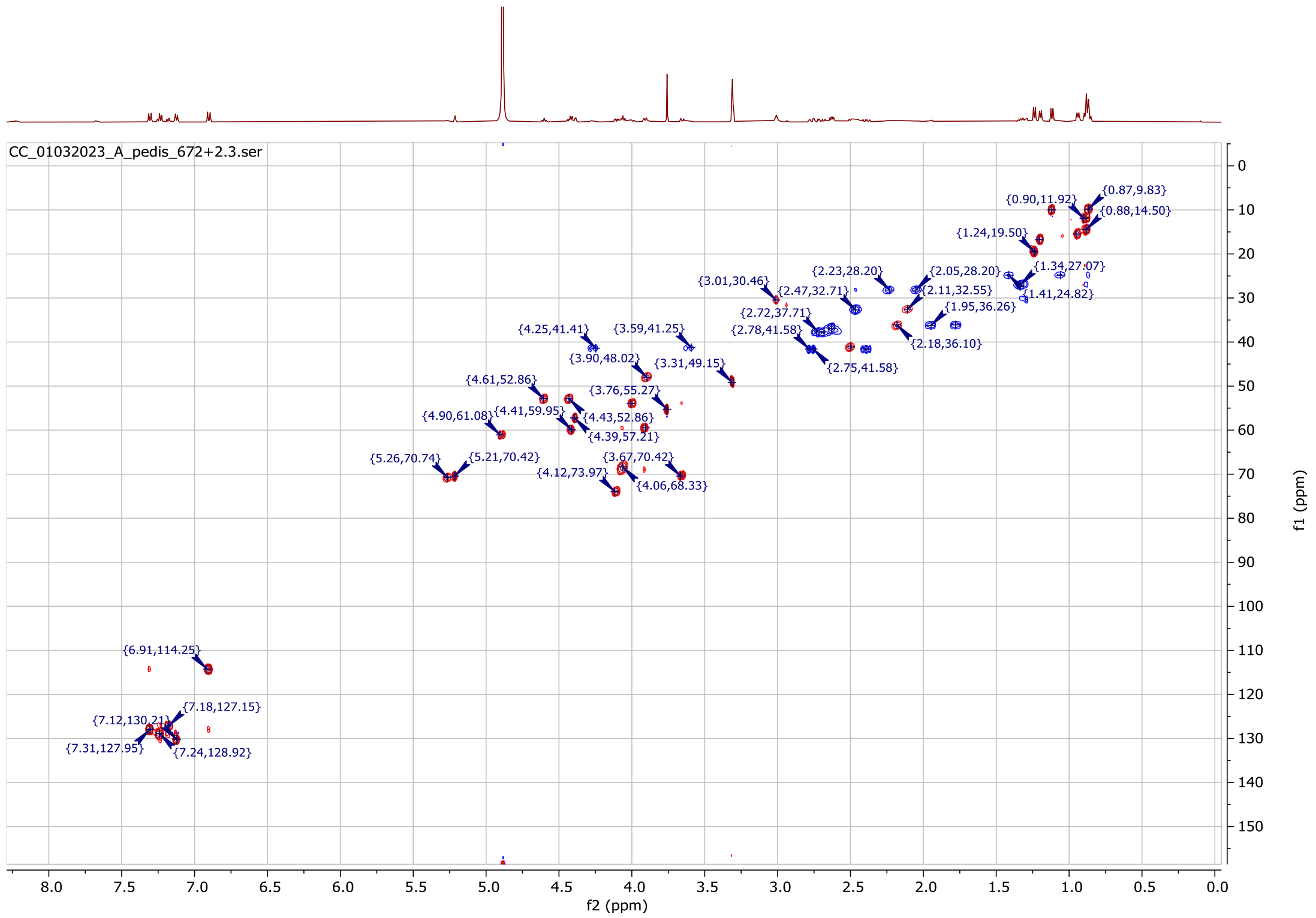

**Figure S59. HSQC spectrum of acidobactamide A (15) in methanol-$d_4$.**

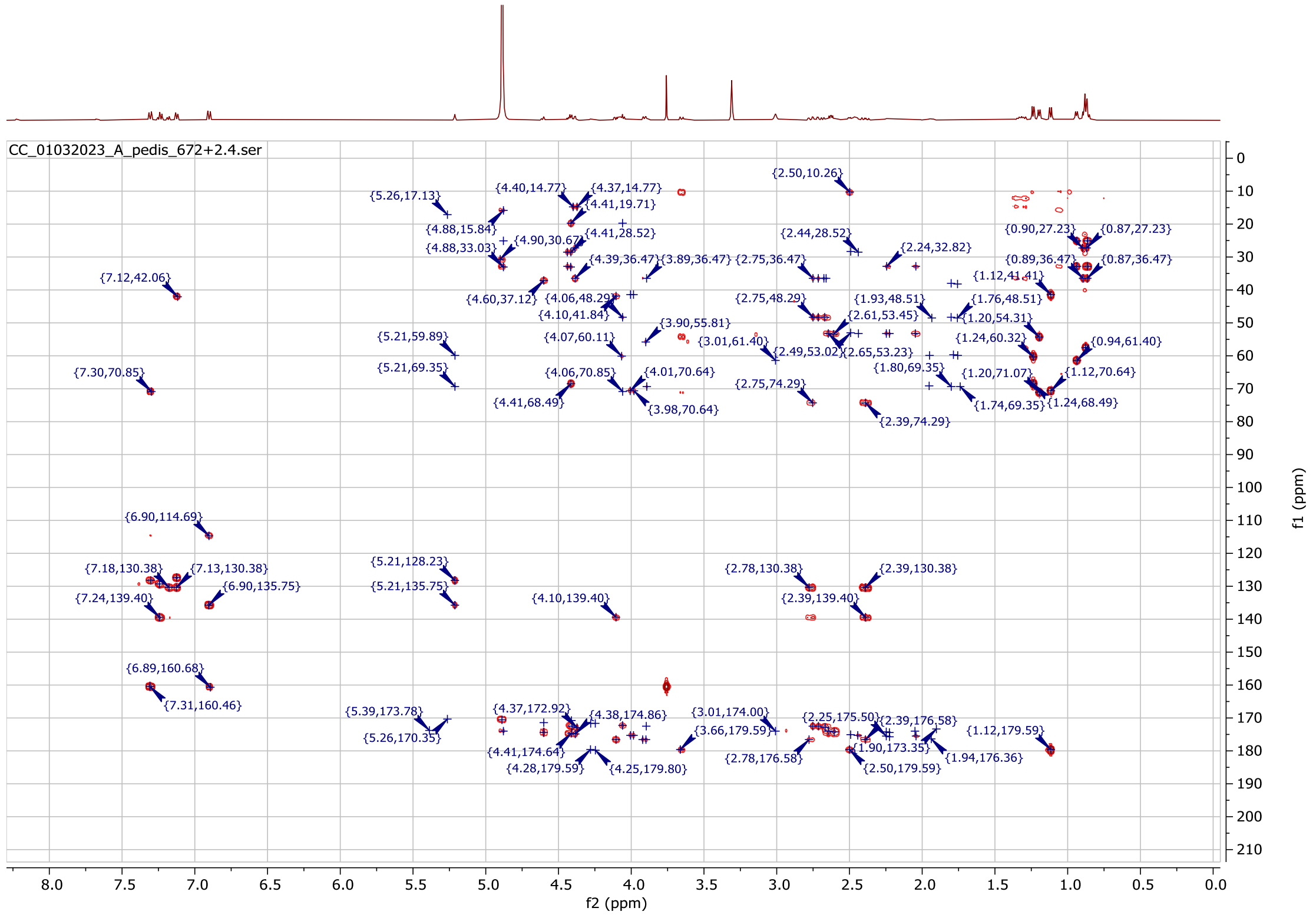

**Figure S60. HMBC spectrum of acidobactamide A (15) in methanol-$d_4$.**

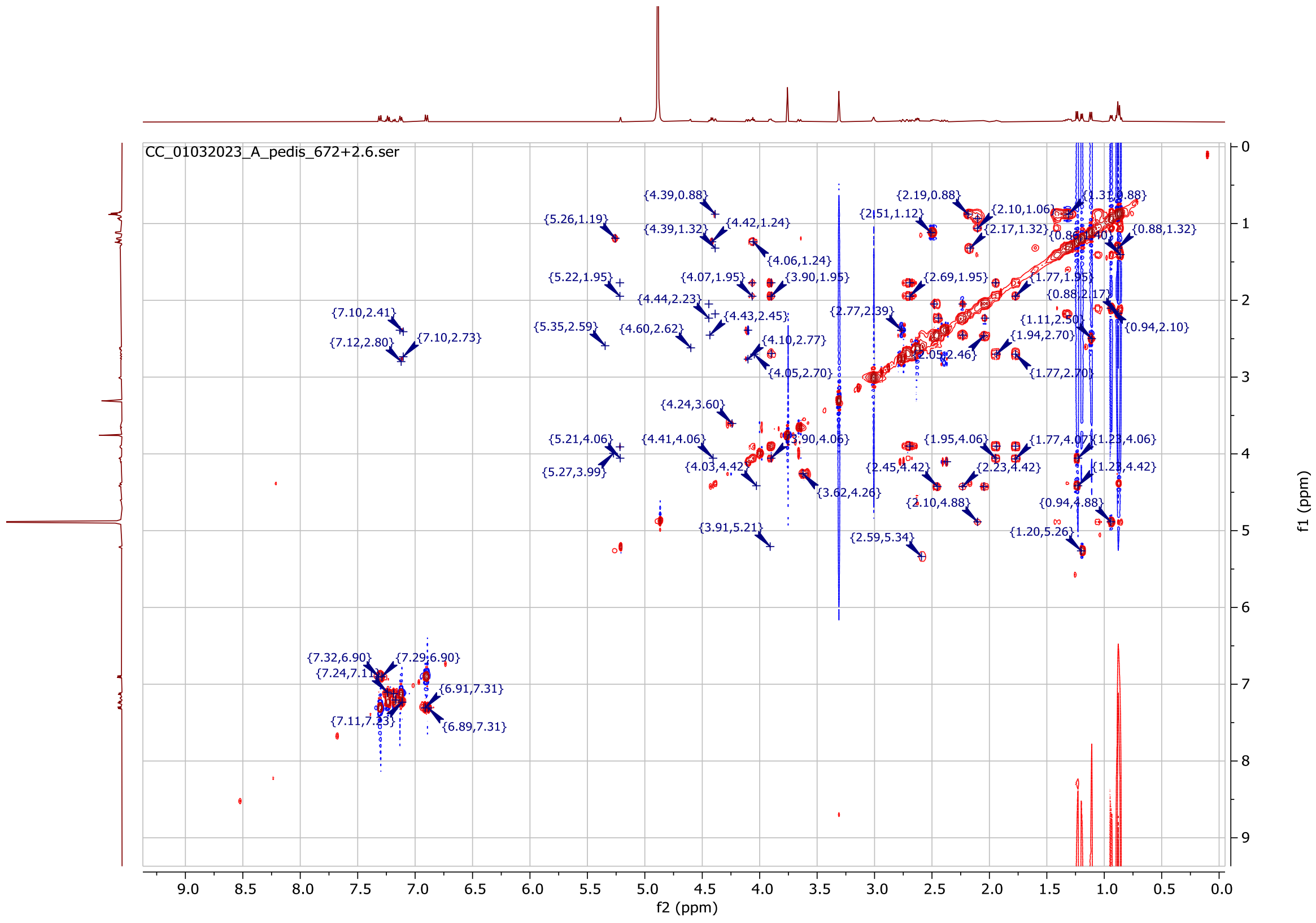

**Figure S61. TOCSY spectrum of acidobactamide A (15) in methanol-$d_4$.**

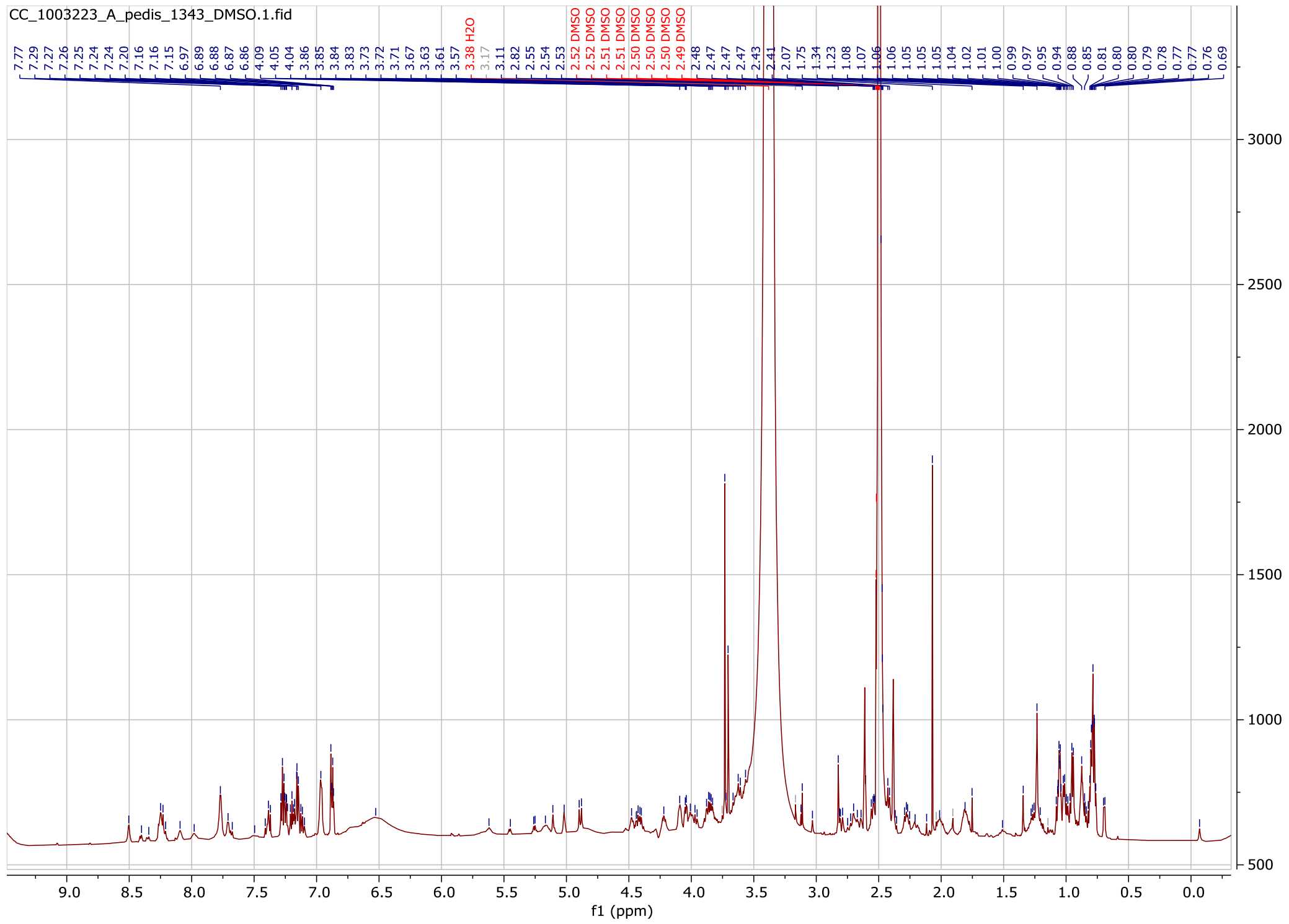

**Figure S62. $^{1}$H NMR spectrum of acidobactamide A (15) in DMSO-$d_6$.**

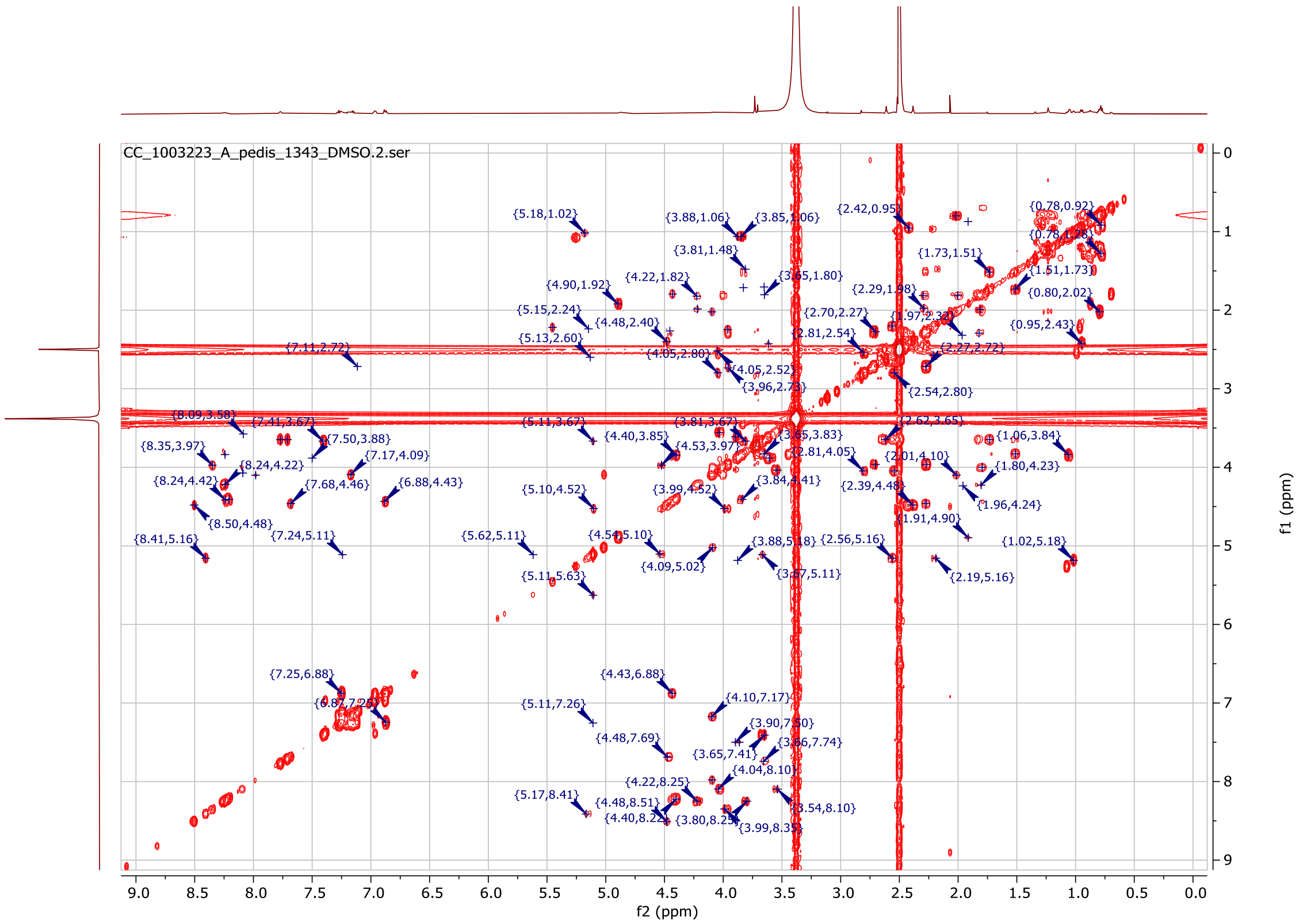

**Figure S63. COSY spectrum of acidobactamide A (15) in DMSO-$d_6$.**

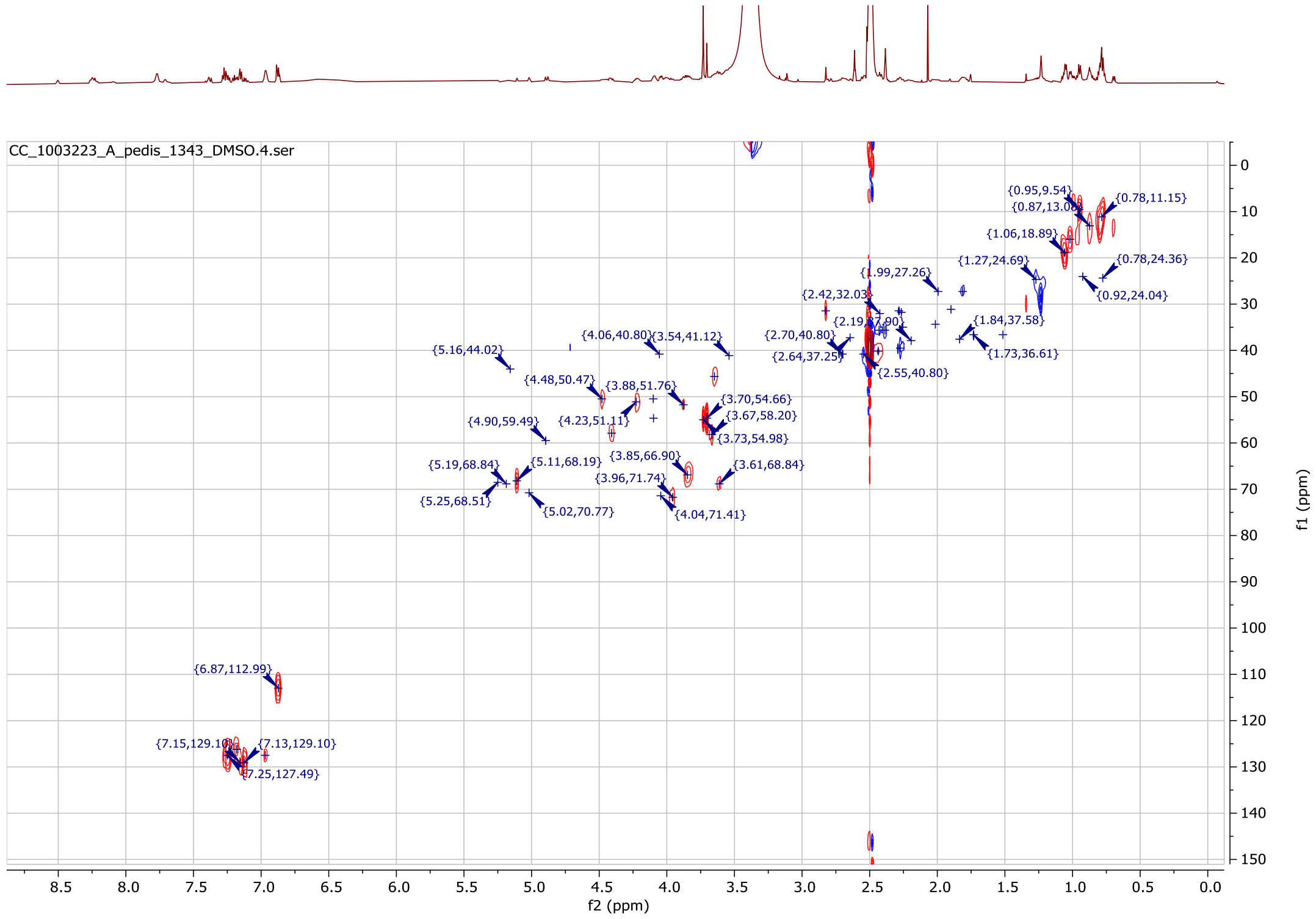

**Figure S64. HSQC spectrum of acidobactamide A (15) in DMSO-$d_6$.**

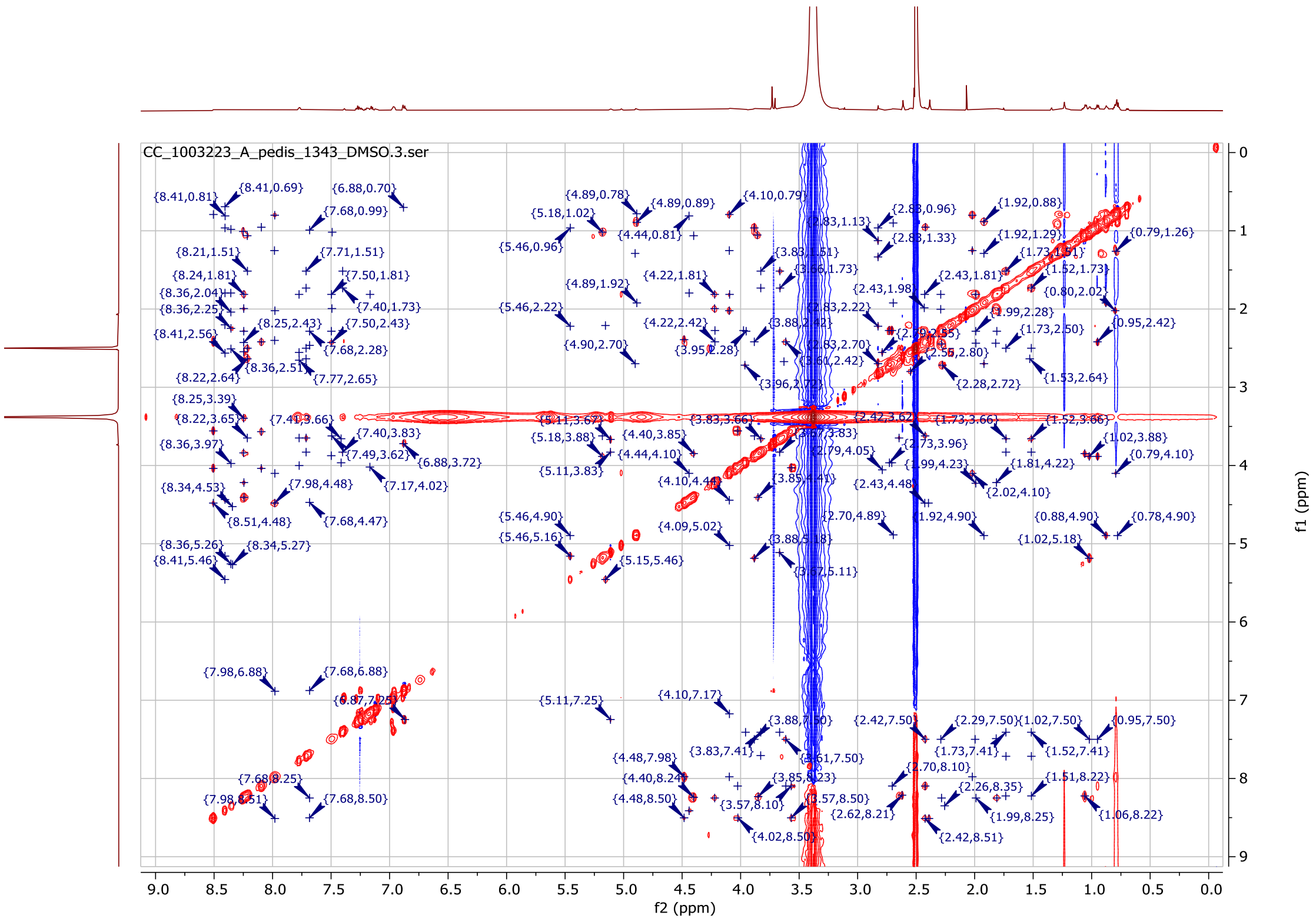

**Figure S65. NOESY spectrum of acidobactamide A (15) in DMSO-$d_6$.**

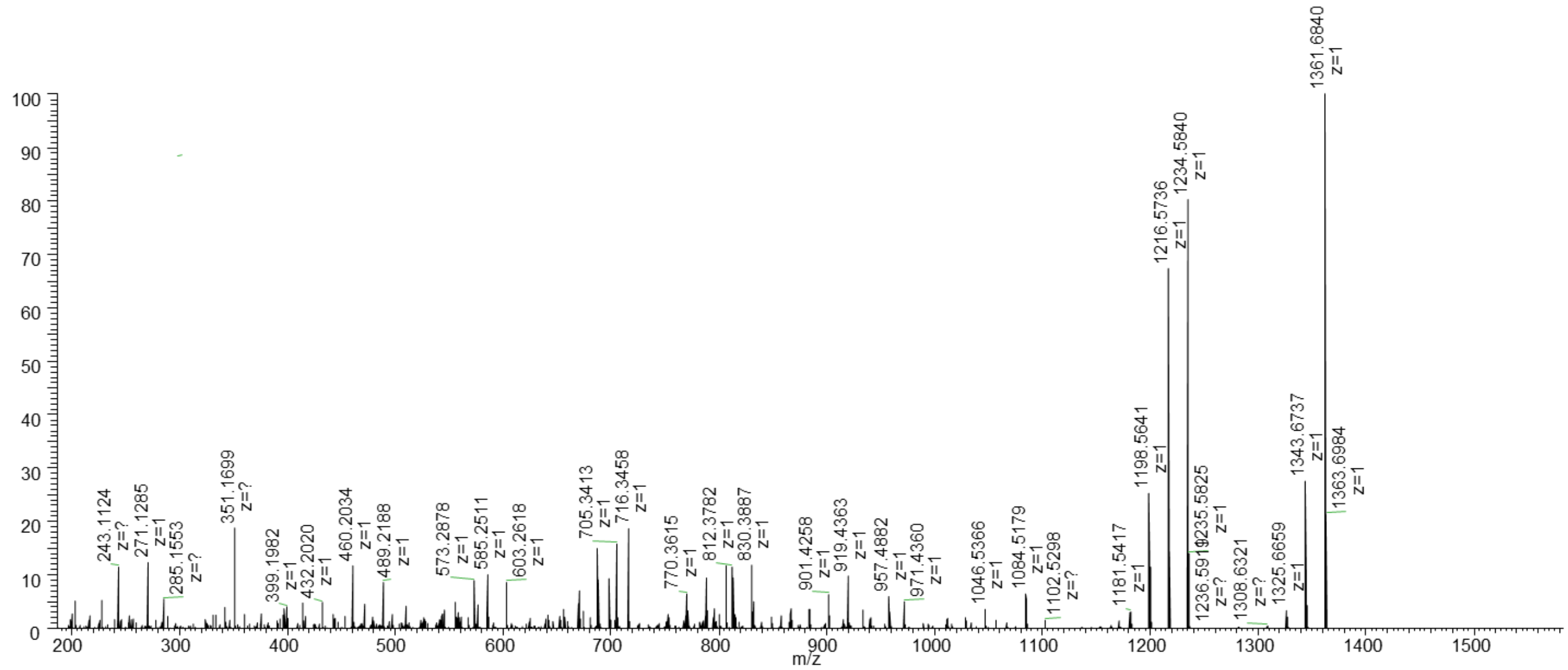

**Figure S66. HESI-HRMS MS/MS data of acidobactamide A NaOH hydrolysate (15).**

MS/MS data of acidobactamide A (**15**) NaOH hydrolysate (*m/z* 1361.6854 [M+H]$^+$).

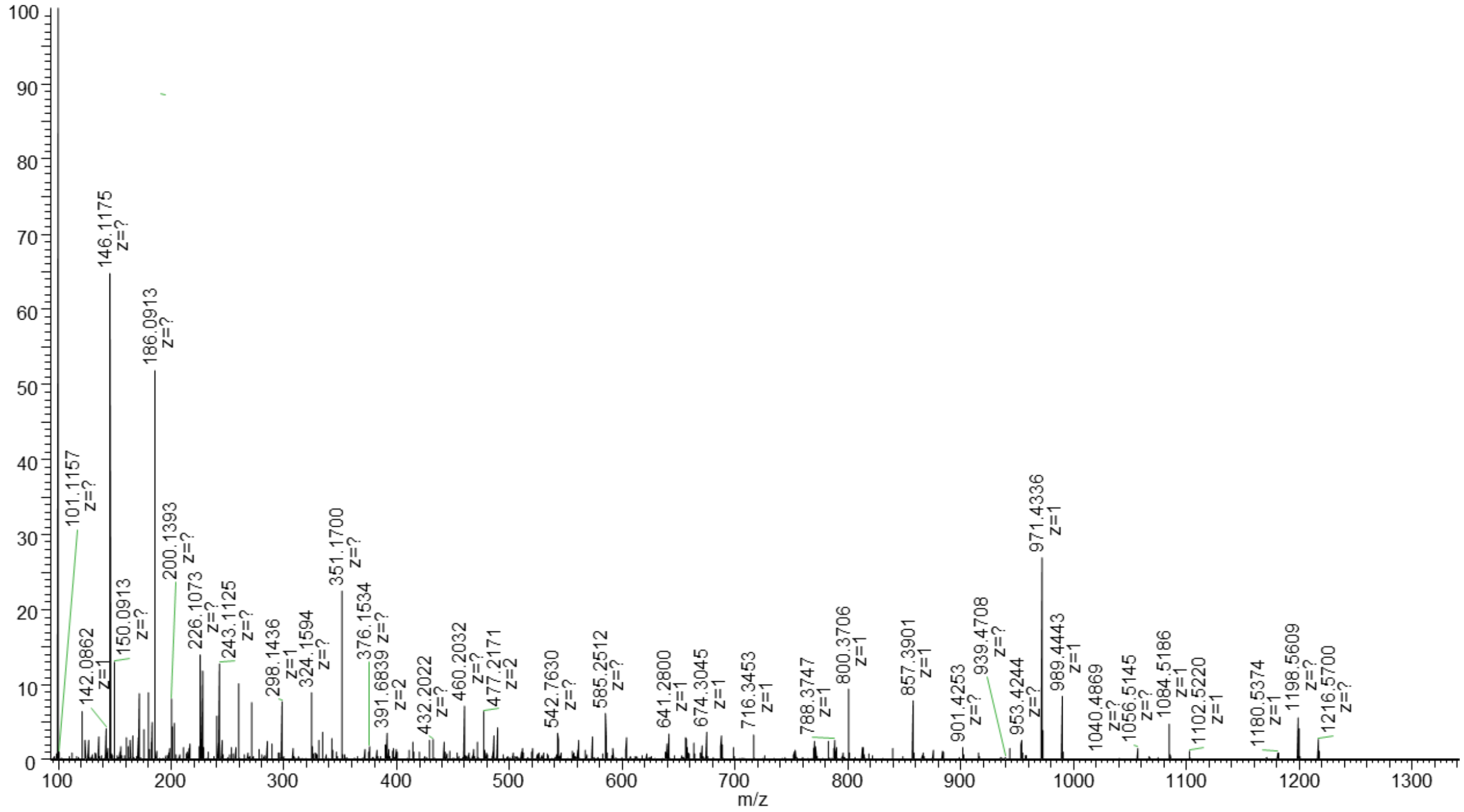

**Figure S67. HESI-HRMS MS/MS data of acidobactamide A NaOH hydrolysate (15).**

MS/MS data of acidobactamide A (**15**) NaOH hydrolysate (*m/z* 681.3454 [M+H]$^{2+}$).

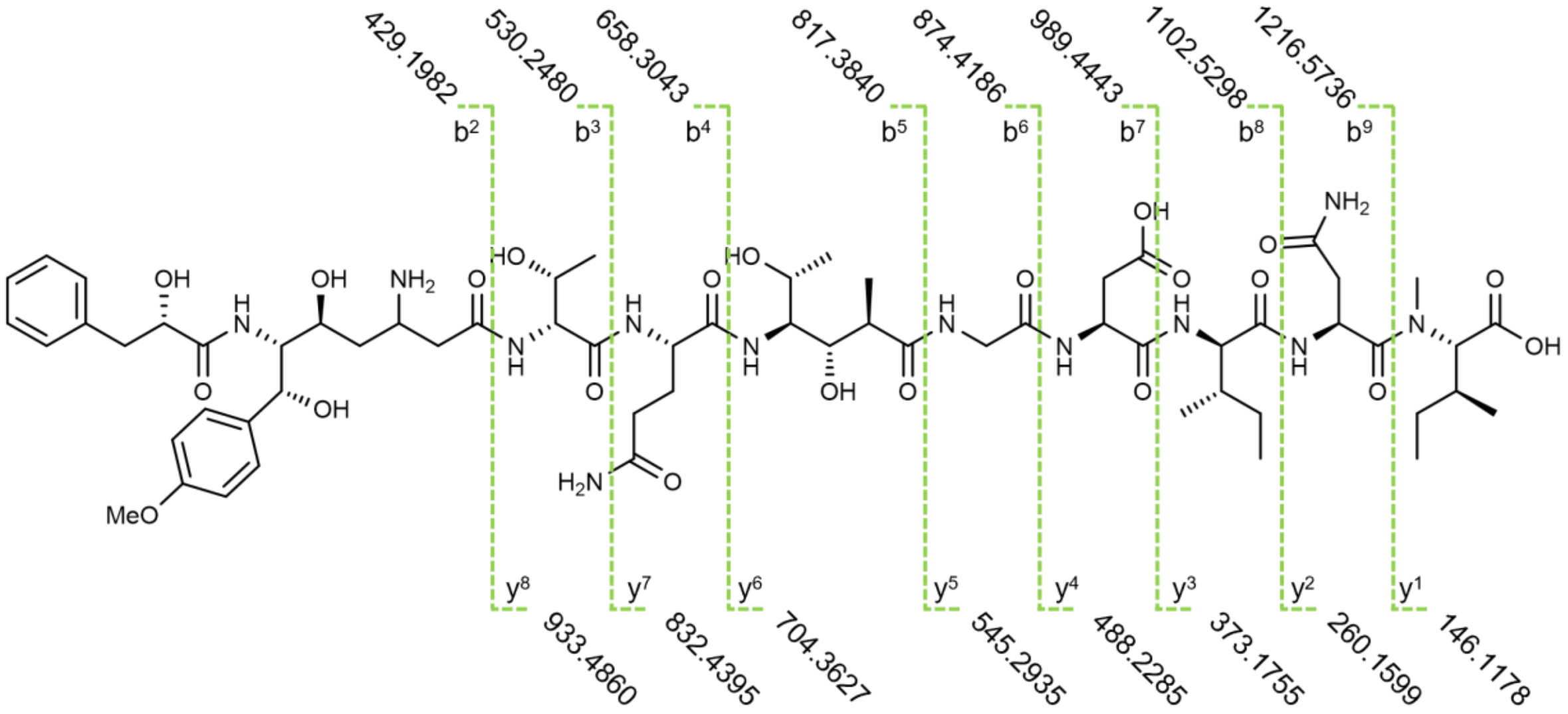

**Figure S68. MS/MS fragmentation of acidobactamide A (15) NaOH hydrolysate.**

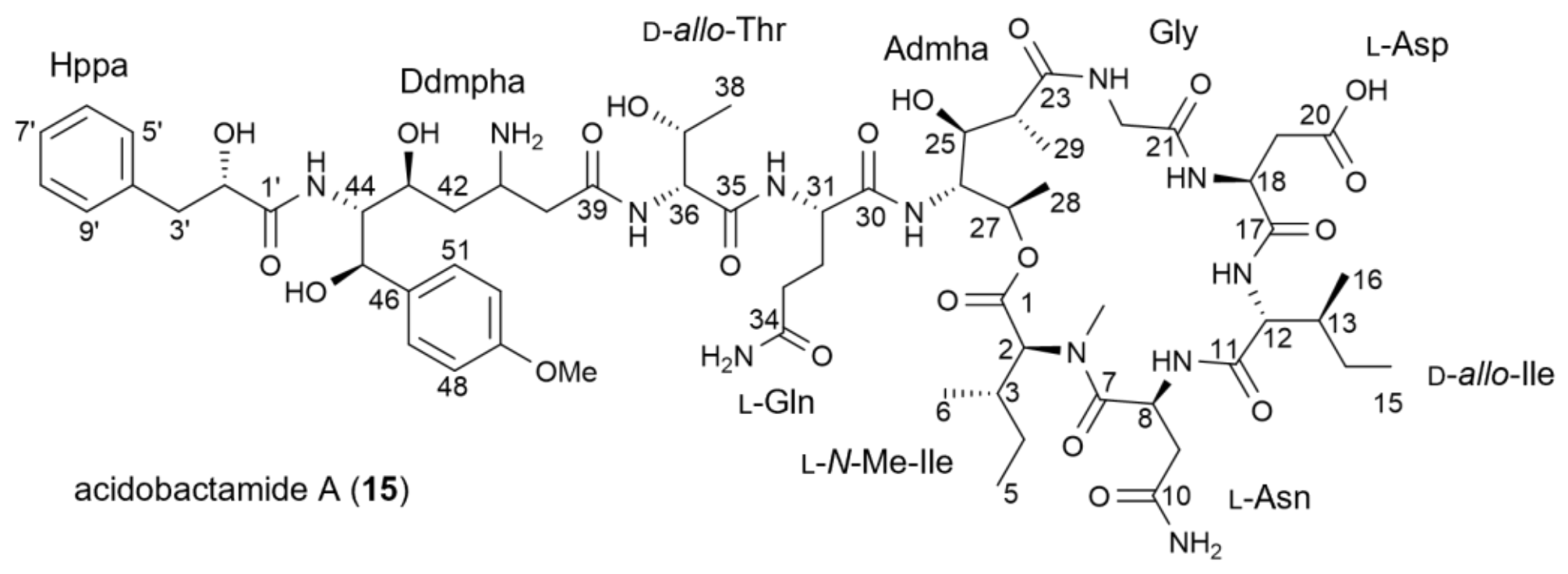

**Figure S69. Chemical structure of acidobactamide A (15).**

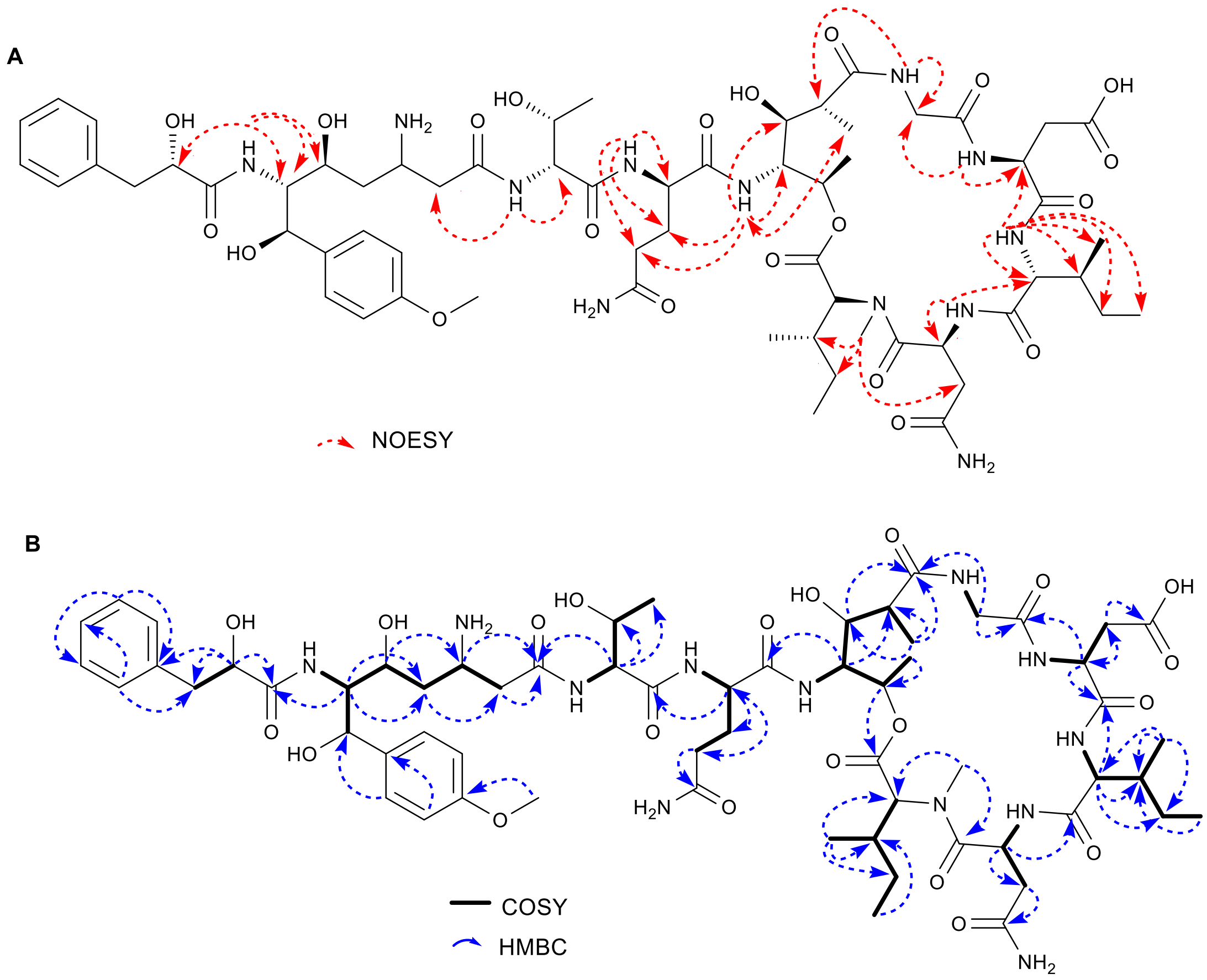

**Figure S70. Structure elucidation of acidobactamide A (15).**

(A) NOESY correlations of **15**.
(B) COSY and HMBC correlations of **15**.

**Table S10. NMR data ($^1$H 500 MHz, $^{13}$C 125 MHz) for acidobactamide A (15) in methanol-$d_4$.**

| | | Methanol-$d_4$ | | DMSO-$d_6$ | |
|---|---|---|---|---|---|
| NO | Amino acids | $^{13}$C/HSQC | $^1$H (multiplicity) | $^{13}$C/HSQC | $^1$H (multiplicity) |
| 1 | *N*Me-Ile | 170.6, C | | | |
| 2 | | 61.1, CH | 4.89 (m) | 59.5, CH | 4.90 (m) |
| 3 | | 32.6, CH | 2.11 (m) | 31.1, CH | 1.90 (m) |
| 4 | | 24.8, $CH_2$ | 1.06 (m), 1.42 (m) | 24.7, $CH_2$ | 0.89 (m), 1.25 (m) |
| 5 | | 9.8, $CH_3$ | 0.86 (m) | 11.1, $CH_3$ | 0.78 (m) |
| 6 | | 15.5, $CH_3$ | 0.94 (d), $J$ = 6.4 Hz | 13.1, $CH_3$ | 0.87 (m) |
| N-Me | | 30.5, $CH_3$ | 3.01 (s) | 31.5, $CH_3$ | 2.82 (m) |
| 7 | Asn | 173.4, C | | | |
| 8 | | 46.7 CH | 5.40 (m) | 44.0, CH | 5.16 (m) |
| 9 | | 38.0, $CH_2$ | 2.52 (m), 2.72 (m) | 37.6, $CH_2$ | 2.21 (m), 2.59 (m) |
| 10 | | 172.4, C | | | |
| NH | | | | | 8.41 $J$ = 7.7 Hz |
| 11 | Ile | 172.9, C | | | |
| 12 | | 57.2, CH | 4.39 (m) | 54.7, CH | 4.10 (m) |
| 33 | | 36.1, CH | 2.18 (m) | 34.4, CH | 2.01 (m) |
| 14 | | 27.1, $CH_2$ | 1.32 (m),0.88 (m) | 24.7, $CH_2$ | 0.92 (m), 1.27 (m) |
| 15 | | 11.9, $CH_3$ | 0.88 (m) | 11.2, $CH_3$ | 0.79 (m) |
| 16 | | 14.5, $CH_3$ | 0.88 (m) | 13.1, $CH_3$ | 0.79 (m) |
| NH | | | | | 7.98 (m) |
| 17 | Asp | 174.4, C | | | |
| 18 | | 52.9, CH | 4.60 (t), $J$ = 7.4 Hz | 50.5, CH | 4.48 (m) |
| 19 | | 36.9, $CH_2$ | 2.63 (m) | 35.9, $CH_2$ | 2.27 (m), 2.40 (m) |
| 20 | | 174.2, C | | | |
| NH | | | | | 8.50 (m) |
| 21 | Gly | 172.1, C | | | |
| 22 | | 41.3, $CH_2$ | 3.60 (d), $J$ = 16.3 Hz<br>4.24 (d), $J$ = 16.3 Hz | 41.1, $CH_2$ | 3.54 (m), 4.06 (m) |
| NH | | | | | 8.10 (m) |
| 23 | Admha | 179.8, C | | | |
| 24 | | 41.1, CH | 2.50 (m) | 40.2, CH | 2.44 (m) |
| 25 | | 70.4, CH | 3.65 (br d) $J$ = 10.0 Hz | 68.8, CH | 3.61 (m) |
| 26 | | 54.0, CH | 3.99 (br d), $J$ = 10.0 Hz | 51.8, CH | 3.88 (m) |
| 27 | | 70.7, CH | 5.27 (m), | 68.8, CH | 5.18 (m) |
| 28 | | 16.8, $CH_3$ | 1.20 (d), $J$ = 6.3 Hz | 15.9, $CH_3$ | 1.02 (d), $J$ = 6.5 Hz |
| 29 | | 10.0, $CH_3$ | 1.12 (d), $J$ = 7.1 Hz | 9.5, $CH_3$ | 0.95 (d), $J$ = 7.0 Hz |
| NH | | | | | 7.50 (m) |
| 30 | Gln | 175.2, C | | | |
| 31 | | 52.9, CH | 4.43 (m) | 51.1, CH | 4.23 (m) |
| 32 | | 28.2, $CH_2$ | 2.05 (m), 2.23 (m) | 27.3, $CH_2$ | 1.81 (m), 1.99 (m) |
| 33 | | 32.7, $CH_2$ | 2.47 (m) | 31.7, $CH_2$ | 2.27 (m), 2.42 (m) |
| 34 | | 174.0, C | | | |
| NH | | | | | 8.24 (m) |
| 35 | Thr | 172.3, C | | | |
| 36 | | 60.0, CH | 4.42 (m) | 57.8, CH | 4.41 (m) |
| 37 | | 68.3, CH | 4.07 (m) | 66.9, CH | 3.85 (m) |
| 38 | | 19.5, $CH_3$ | 1.24, (d) *J* = 6.3 Hz | 18.9, $CH_3$ | 1.06 (m) |
| NH | | | | | 8.22 (m) |
| 39 | Ddmpha | 172.4, C | | | |
| 40 | | 37.7, $CH_2$ | 2.72 (m) | 37.6, $CH_2$ | 2.50 (m)<br>2.64 (m) |
| 41 | | 48.0, CH | 3.90 (m) | 45.6, CH | 3.65 (m) |
| 42 | | 36.3, $CH_2$ | 1.78 (ddd), *J* = 3.8, 8.2, 15.1 Hz<br>1.95 (ddd), *J* = 3.4, 8.6, 15.1 Hz | 36.6, $CH_2$ | 1.51 (m)<br>1.73 (m) |
| 43 | | 68.3, CH | 4.07, (m) | 66.9, CH | 3.85 (m) |
| 44 | | 59.5, CH | 3.91 (m) | 58.2, CH | 3.67 (m) |
| 45 | | 70.4, CH | 5.21 (m) | 68.2, CH | 5.11 (m) |
| 46 | | 135.7, C | | | |
| 47/51 | | 128.0, C | 7.31 (d), *J* = 8.8 Hz | 127.5, CH | 6.87 (m), |
| 48/50 | | 114.3, CH | 6.90 (d), *J* = 8.8 Hz | 113.0, CH | 6.87 (d), *J* = 8.8 Hz |
| 49 | | 160.7, C | | | |
| $OCH_3$ | | 55.3, $CH_3$ | 3.76 (s) | 54.7, $CH_2$ | 3.70 (s) |
| 44-NH | | | | | 7.41 (m) |
| 1' | Hppa | 176.6, C | | | |
| 2' | | 73.9, CH | 4.10 (dd), *J* = 3.6, 9.4 Hz | 71.7, CH | 3.96 (m) |
| 3' | | 41.6, $CH_2$ | 2.40 (dd), *J* = 9.4, 14.0 Hz<br>2.76 (dd), *J* = 3.6, 14.0 Hz | 40.8, $CH_2$ | 2.55 (m)<br>2.70 (m) |
| 4' | | 139.4, C | | | |
| 5'/9' | | 130.2, CH | 7.12 (d), *J* = 7.4 Hz | 129.1, CH | 7.13 (m) |
| 6'/8' | | 128.9, CH | 7.24 (t), *J* = 7.5 Hz | 129.2, CH | 7.15 (m) |
| 7' | | 127.2, CH | 7.18 (t), *J* = 7.4 Hz | 126.2, CH | 7.18 (m) |
| 2'-OH | | | | | 8.35 (d) *J* = 10.9 Hz |

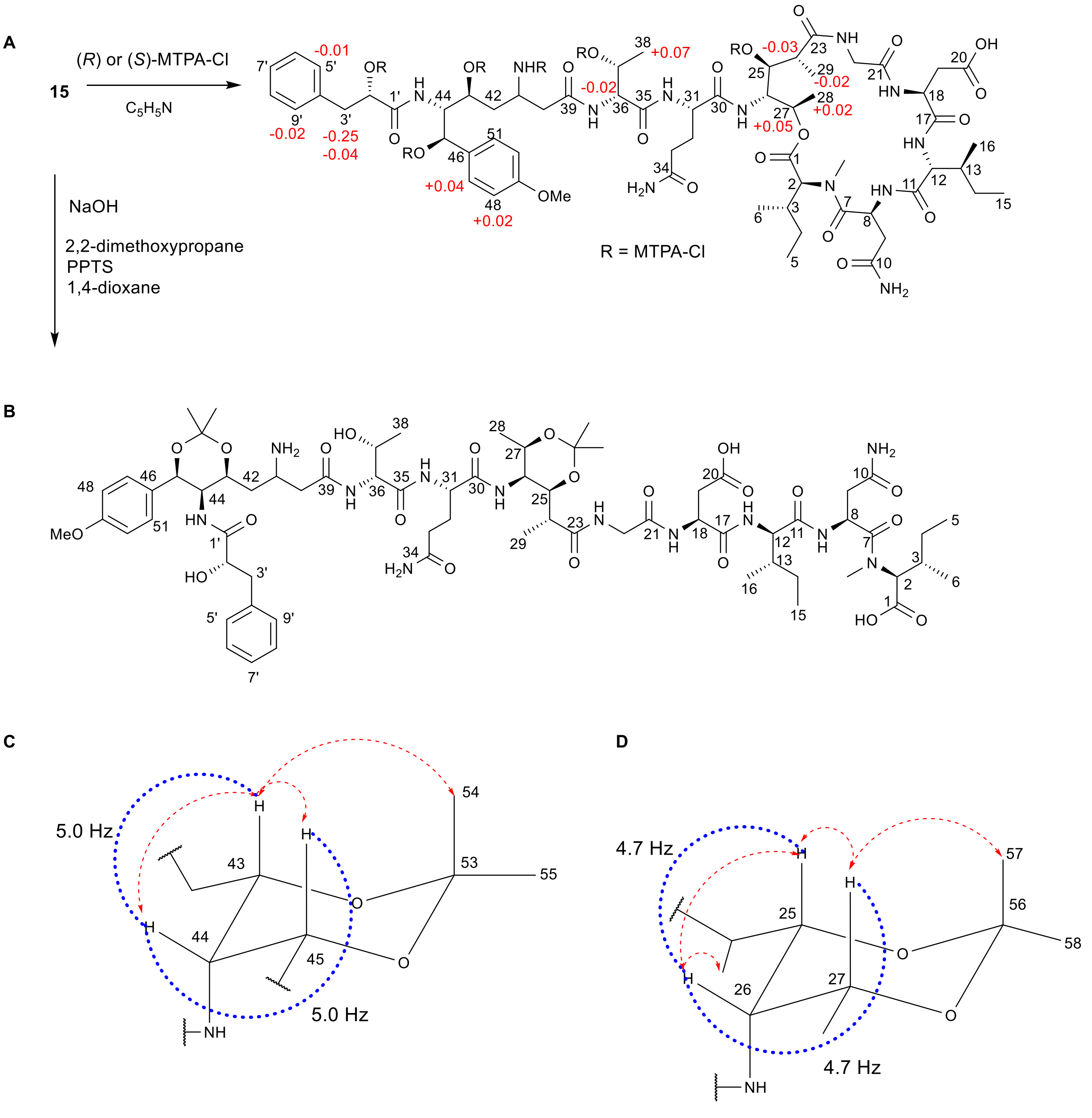

**Figure S71. Determination of the relative and absolute configuration of acidobactamide A (15) by NOESY correlations and the modified Mosher method.**

**Table S11. NMR Data ($^1$H 500 MHz, $^{13}$C 125 MHz) for 20 in methanol-$d_4$.**

| NO | Amino acids | $^{13}$C/HSQC | $^1$H (multiplicity) |
|---|---|---|---|
| 1 | *N*-Me-Ile | 174.9, C | |
| 2 | | 65.8, CH | 4.65 (m) |
| 3 | | 34.7, CH | 1.92 (m) |
| 4 | | 26.1, $CH_2$ | 0.95 (m), 1.39 (m) |
| 5 | | 11.3, $CH_3$ | 0.84 (m) |
| 6 | | 16.7, $CH_3$ | 0.97 (d), *J* = 6.3 Hz |
| N-Me | | 31.8, $CH_3$ | 3.11 (s) |
| 7 | Asn | 173.1, C | |
| 8 | | 48.5, CH | 5.20 (dd), *J* = 1.6, 6.3 Hz |
| 9 | | 37.7, $CH_2$ | 2.63 (m), 2.79 (m) |
| 10 | | 173.1, C | |
| 11 | Ile | 173.3, C | |
| 12 | | 57.7, CH | 4.44 (m) |
| 33 | | 37.6, CH | 2.03 (m) |
| 14 | | 27.1, $CH_2$ | 1.32 (m),0.90 (m) |
| 15 | | 11.9, $CH_3$ | 0.91 (m) |
| 16 | | 15.0, $CH_3$ | 0.86 (m) |
| 17 | Asp | 173.5, C | |
| 18 | | 51.9, CH | 4.67 (m) |
| 19 | | 37.2, $CH_2$ | 2.71 (m), 2.80 (m) |
| 20 | | 175.0, C | |
| 21 | Gly | 172.0, C | |
| 22 | | 43.7, $CH_2$ | 3.90 (d), *J* = 11.9 Hz<br>3.97 (d), *J* = 11.9 Hz |
| 23 | Admha | 179.6, C | |
| 24 | | 43.5, CH | 2.61 (m) |
| 25 | | 71.9, CH | 4.08 (m) |
| 26 | | 55.4, CH | 4.26 (m) |
| 27 | | 65.9, CH | 4.26 (m), |
| 28 | | 11.0, $CH_3$ | 1.17 (d), *J* = 7.3 Hz |
| 29 | | 20.3, $CH_3$ | 1.13 (d), *J* = 6.4 Hz |
| 30 | Gln | 176.5, C | |
| 31 | | 55.6, CH | 4.31 (m) |
| 32 | | 29.8, $CH_2$ | 1.98 (m), 2.23 (m) |
| 33 | | 33.5, $CH_2$ | 2.33 (m) |
| 34 | | 175.7, C | |
| 35 | Thr | 171.8, C | |
| 36 | | 60.4, CH | 4.36 (m) |
| 37 | | 68.2, CH | 4.09 (m) |
| 38 | | 19.7, $CH_3$ | 1.24, (d) *J* = 6.3 Hz |
| 39 | Ddmpha | 172.4, C | |
| 40 | | 37.7, $CH_2$ | 2.72 (m) |
| 41 | | 47.9, CH | 3.72 (m) |
| 42 | | 37.7, $CH_2$ | 1.78 (m), 1.95 (m) |
| 43 | | 70.9, CH | 3.96, (m) |
| 44 | | 55.6, CH | 4.31 (m) |
| 45 | | 72.0, CH | 5.26 (t), *J* = 5.0 Hz |
| 46 | | 130.6, C | |
| 47/51 | | 128.4, C | 7.27 (d), *J* = 8.5 Hz |
| 48/50 | | 114.4, CH | 6.91 (d), *J* = 8.5 Hz |
| 52 | | 160.4, C | |
| 53 | | 103.3, C | |
| 54 | | 24.3, $CH_3$ | 1.58 (s) |
| 55 | | 24.3, $CH_3$ | 1.46 (s) |
| $OCH_3$ | | 55.4, $CH_3$ | 3.75 (s) |
| 56 | | 103.1, C | |
| 57 | | 24.3, $CH_3$ | 1.57 (s) |
| 58 | | 24.5, $CH_3$ | 1.46 (s) |
| 1' | Hppa | 173.5, C | |
| 2' | | 73.6, CH | 3.95, (m) |
| 3' | | 41.4, $CH_2$ | 2.49 (m), 1.77 (m) |
| 4' | | 139.8, C | |
| 5'/9' | | 130.2, CH | 7.14 (d), $J$ = 7.4 Hz |
| 6'/8' | | 128.9, CH | 7.21 (t), $J$ = 7.5 Hz |
| 7' | | 127.0, CH | 7.15 (t), $J$ = 7.4 Hz |
| 2'-OH | | | |

**Table S12. NMR Data ($^{1}$H 600 MHz) for acidobactamide A (15) *S* and *R* MTPA ester in methanol-$d_4$.**

| | | *S*-MTPA ester | *R*-MTPA ester | Δ$\delta$(=$\delta_S$−$\delta_R$) |
|---|---|---|---|---|
| NO | Amino acids | $^{1}$H (multiplicity) | | |
| 1 | *N*-Me-Ile | | | +0.04 |
| 2 | | 4.92 (m) | 4.88 (m) | |
| 3 | | 2.11 (m) | 2.11 (m) | |
| 4 | | 1.06 (m), 1.42 (m) | 1.06 (m), 1.42 (m) | |
| 5 | | 0.87 (m) | 0.89 (m) | |
| 6 | | 0.95 (d), *J* = 6.4 Hz | 0.94 (d), *J* = 6.4 Hz | |
| N-Me | | 3.05 (s) | 3.05 (s) | |
| 7 | Asn | | | |
| 8 | | 5.34 (m) | 5.34 (m) | |
| 9 | | 2.53 (m), 2.72 (m) | 2.53 (m), 2.72 (m) | |
| 10 | | | | |
| 11 | Ile | | | |
| 12 | | 4.39 (m) | 4.39 (m) | |
| 33 | | 2.18 (m) | 2.18 (m) | |
| 14 | | 1.31 (m), 0.88 (m) | 1.31 (m),0.88 (m) | |
| 15 | | 0.88 (m) | 0.88 (m) | |
| 16 | | 0.89 (m) | 0.89 (m) | |
| 17 | Asp | | | |
| 18 | | 4.63 (t), *J* = 7.8 Hz | 4.63 (t), *J* = 7.8 Hz | |
| 19 | | 2.64 (m) | 2.63 (m) | |
| 20 | | | | |
| 21 | Gly | | | |
| 22 | | 3.62 (d), *J* = 16.0 Hz | 3.62 (d), *J* = 16.0 Hz | -0.001 |
| | | 4.27 (d), *J* = 16.0 Hz | 4.27 (d), *J* = 16.0 Hz | -0.002 |
| 23 | Admha | | | |
| 24 | | 2.54 (m) | 2.60 (m) | -0.06 |
| 25 | | 3.6 (m) | 3.62 (t) | 0.03 |
| 26 | | 5.40 (m) | 5.40 (m) | |
| 27 | | 4.10 (m) | 4.07 (t) | 0.05 |
| 28 | | 1.20 (d), *J* = 6.8 Hz | 1.18 (d), *J* = 6.6 Hz | 0.02 |
| 29 | | 1.12 (m) | 1.13 (d), *J* = 6.8 Hz | -0.01 |
| 30 | Gln | | | |
| 31 | | 4.42 (m) | 4.42 (m) | |
| 32 | | 2.04 (m), 2.24 (m) | 2.04 (m), 2.24 (m) | |
| 33 | | 2.47 (m) | 2.47 (m) | |
| 34 | | | | |
| 35 | Thr | | | |
| 36 | | 4.19 (m) | 4.17 (m) | -0.02 |
| 37 | | 5.52 (m) | 5.52 (m) | |
| 38 | | 1.28 (m) | 1.21 (m) | 0.07 |
| 39 | ddmpha | | | |
| 40 | | 2.82 (m) | 2.79 (m) | 0.03 |
| 41 | | 5.13 (m) | 5.11 (m) | 0.02 |
| 42 | | 1.59 (m) | 1.69 (m) | -0.1 |
| | | 2.01 (m) | 2.05 (m) | 0.01 |
| 43 | | 5.34 (m) | 5.34 (m) | |
| 44 | | 3.91 (m) | 3.83 (m) | 0.08 |
| 45 | | 6.70 (m) | 6.70 (m) | |
| 46 | | | | |
| 47/51 | | 7.32 (d), *J* = 8.9 Hz | 7.29 (d), *J* = 8.9 Hz | 0.03 |
| 48/50 | | 6.91 (d), *J* = 8.9 Hz | 6.89 (d), *J* = 8.9 Hz | 0.02 |
| 49 | | | | |
| $OCH_3$ | | 3.75 (s) | 3.75 (s) | |
| 1' | Hppa | | | |
| 2' | | 5.30 (m) | 5.30 (m) | |
| 3' | | 2.31 (m) | 2.57 (m) | -0.25 |
| | | 2.84 (m) | 2.88 (m) | -0.04 |
| 4' | | | | |
| 5'/9' | | 7.12 (d), *J* = 7.5 Hz | 7.13 (d), *J* = 7.4 Hz | -0.01 |
| 6'/8' | | 7.23 (t), *J* = 7.5 Hz | 7.25 (t), *J* = 7.4 Hz | -0.02 |
| 7' | | 7.17 (t), *J* = 7.5 Hz | 7.17 (t), *J* = 7.4 Hz | |
| 2'-OH | | | | |

Note: important $\Delta\delta(=\delta_S-\delta_R)$ are in blue

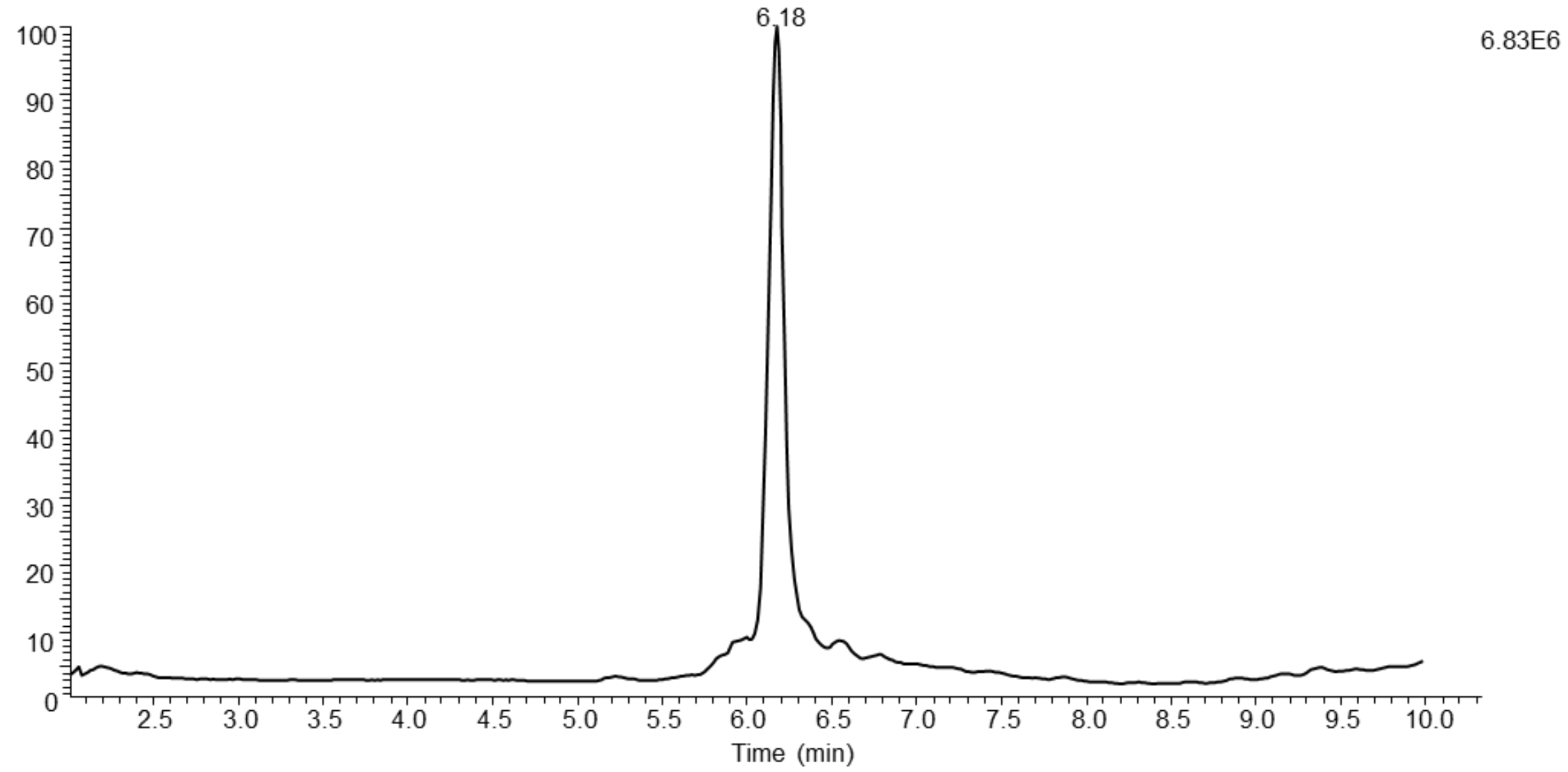

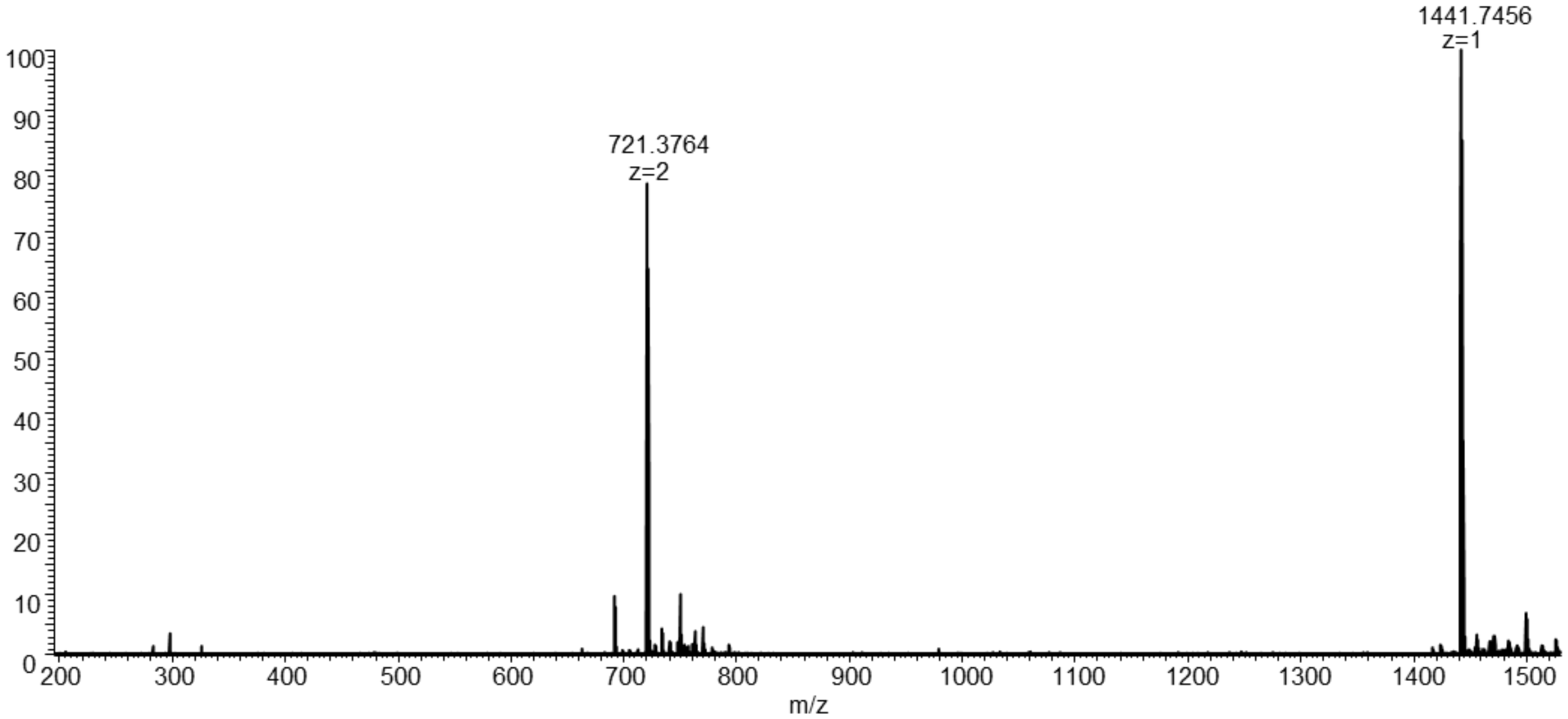

**Figure S72. HPLC-HESI-HRMS data of 20.**
Top: Total ion chromatogram of **20**.
Bottom: MS data of **20** (*m/z* 1441.7456 $[M+H]^+$).

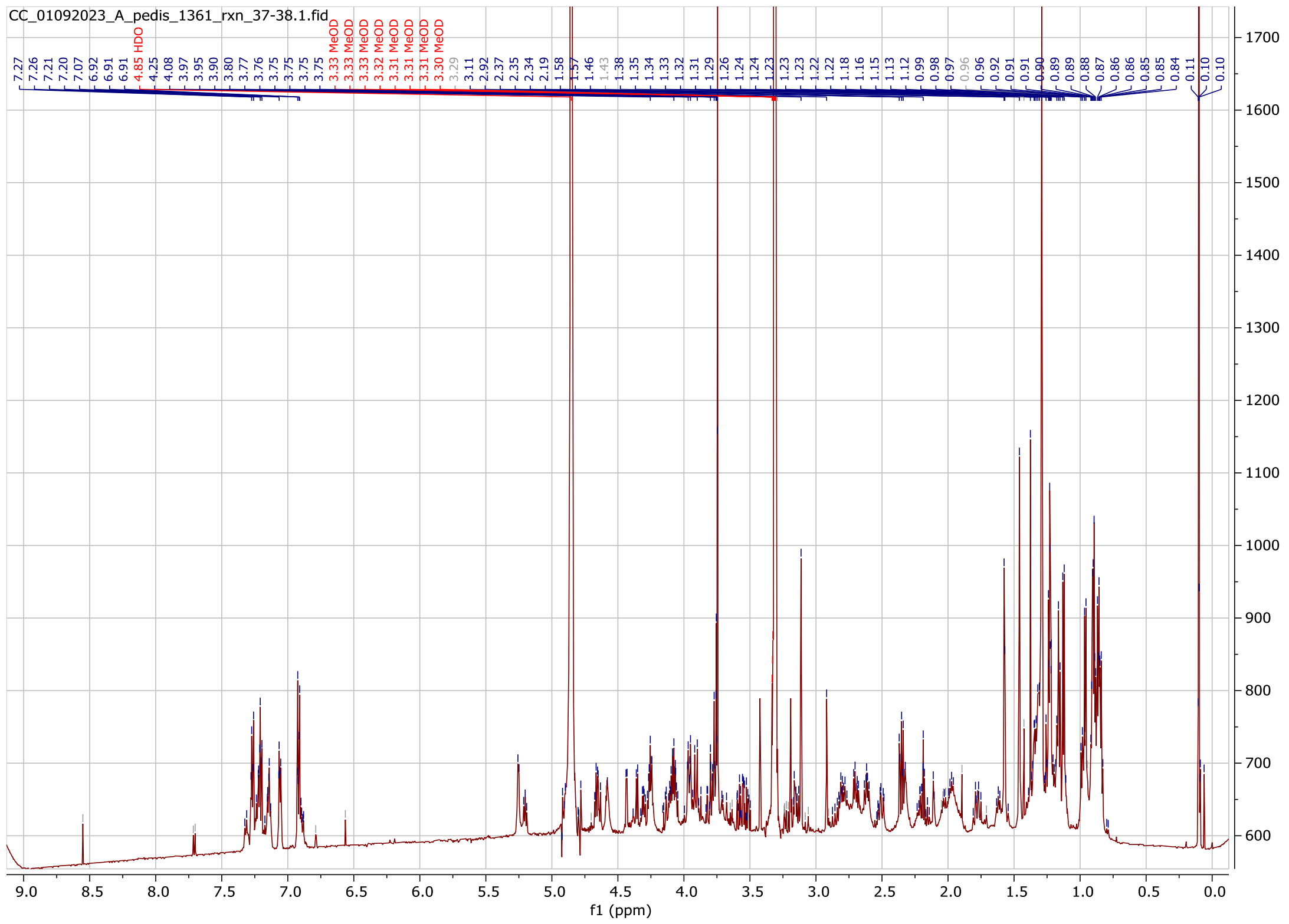

Figure S73. $^{1}H$ NMR spectrum of 20 in methanol-$d_4$.

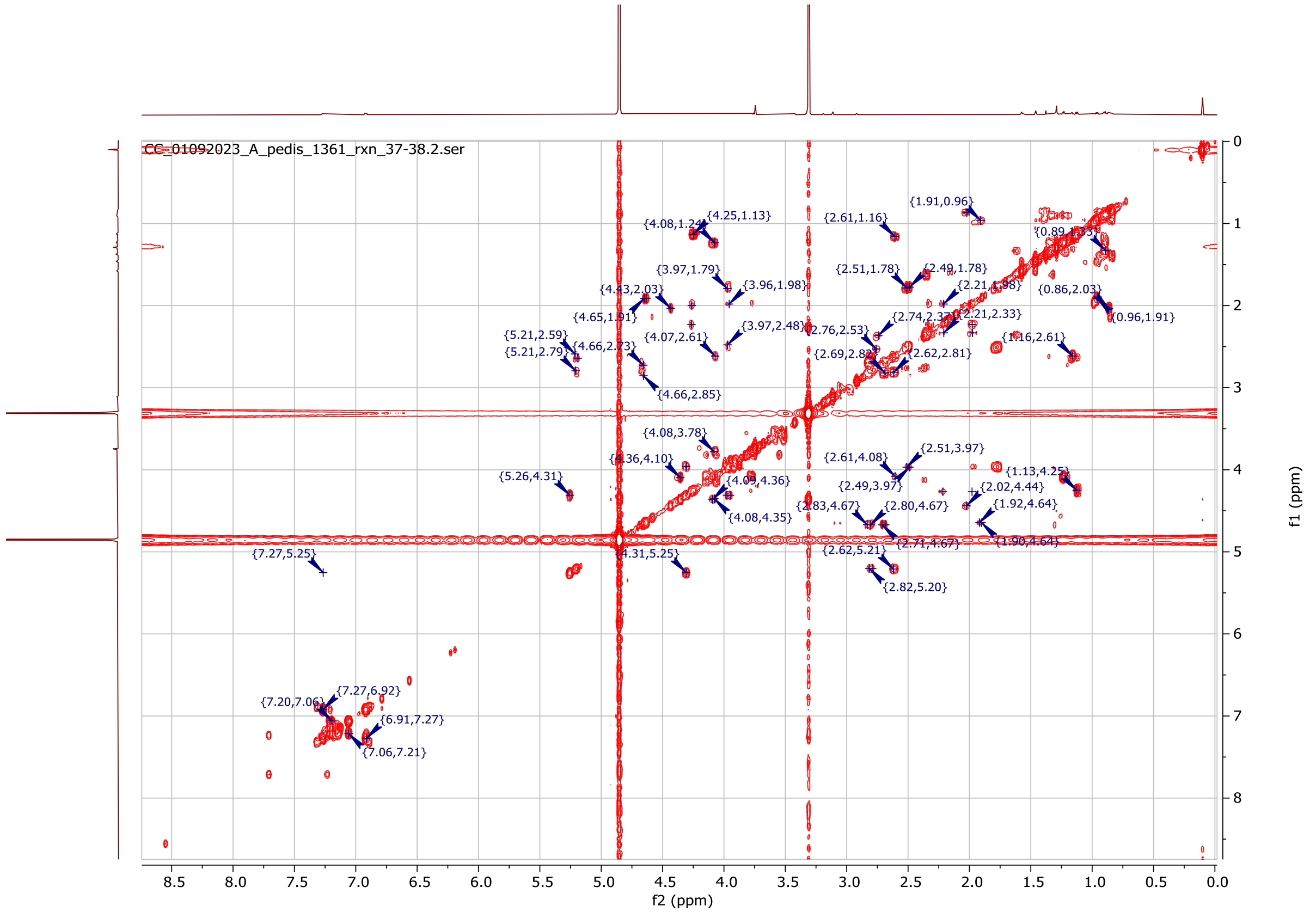

Figure S74. COSY spectrum of 20 in methanol-$d_4$.

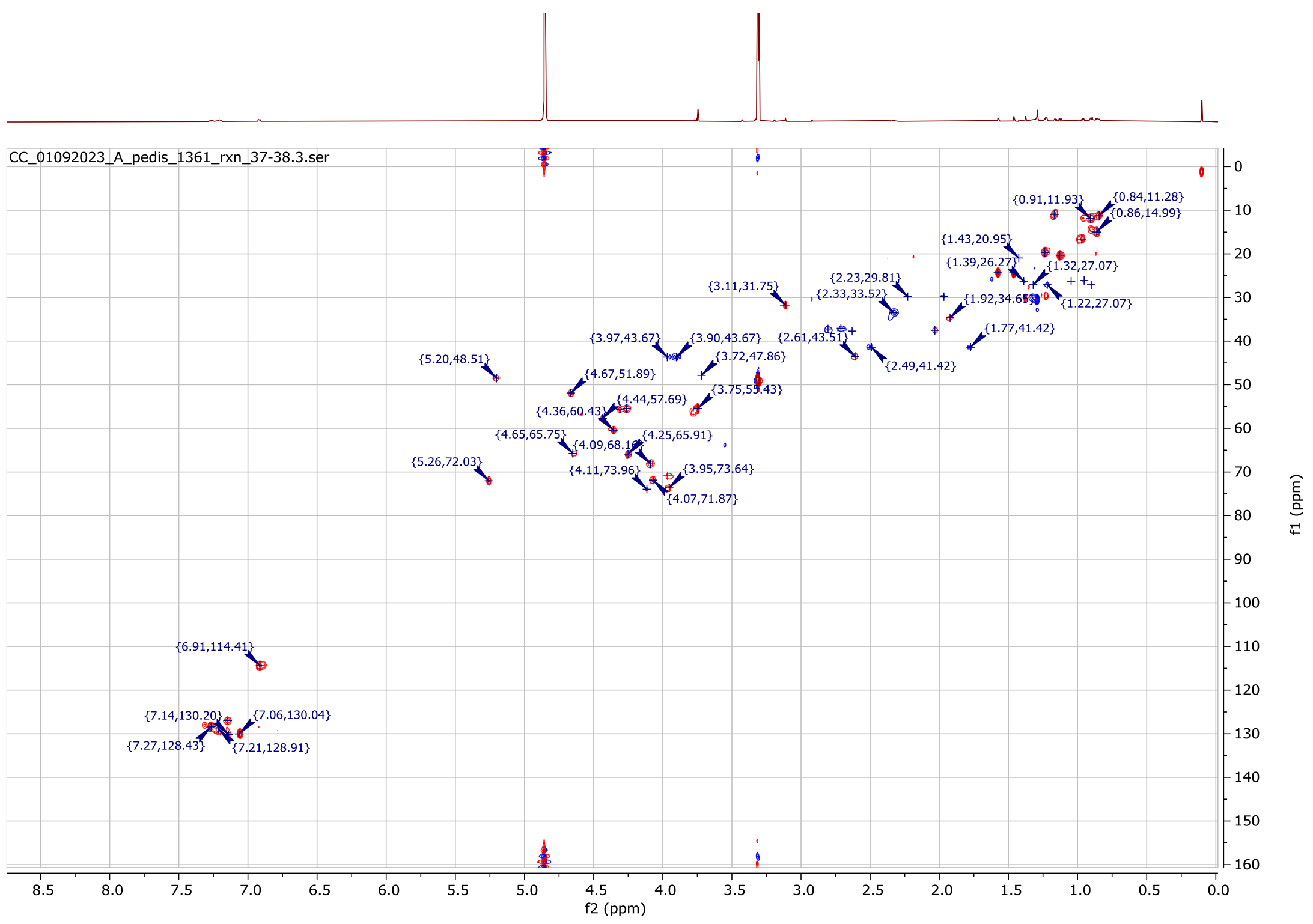

**Figure S75. HSQC spectrum of 20 in methanol-$d_4$.**

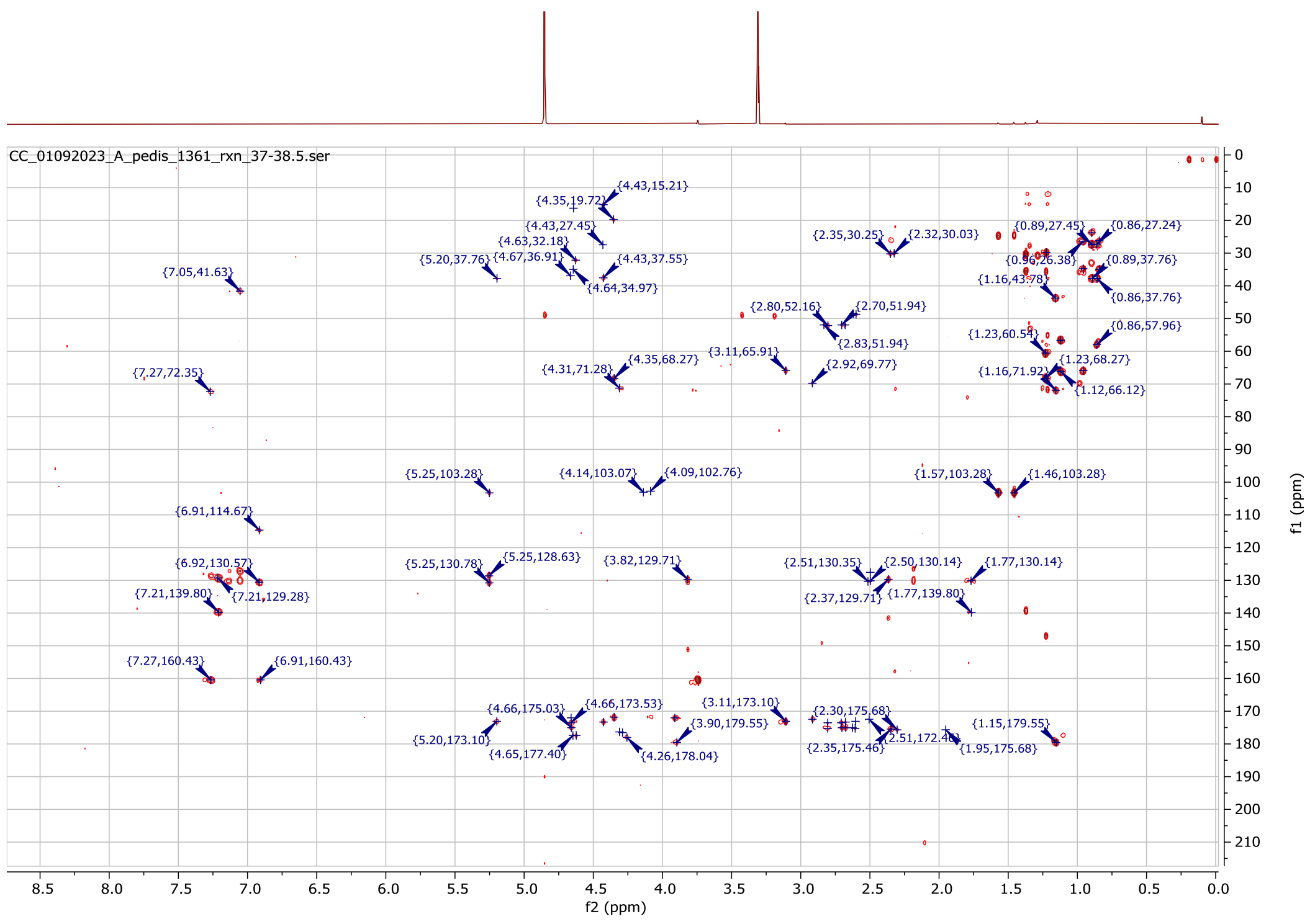

**Figure S76. HMBC spectrum of 20 in methanol-$d_4$.**

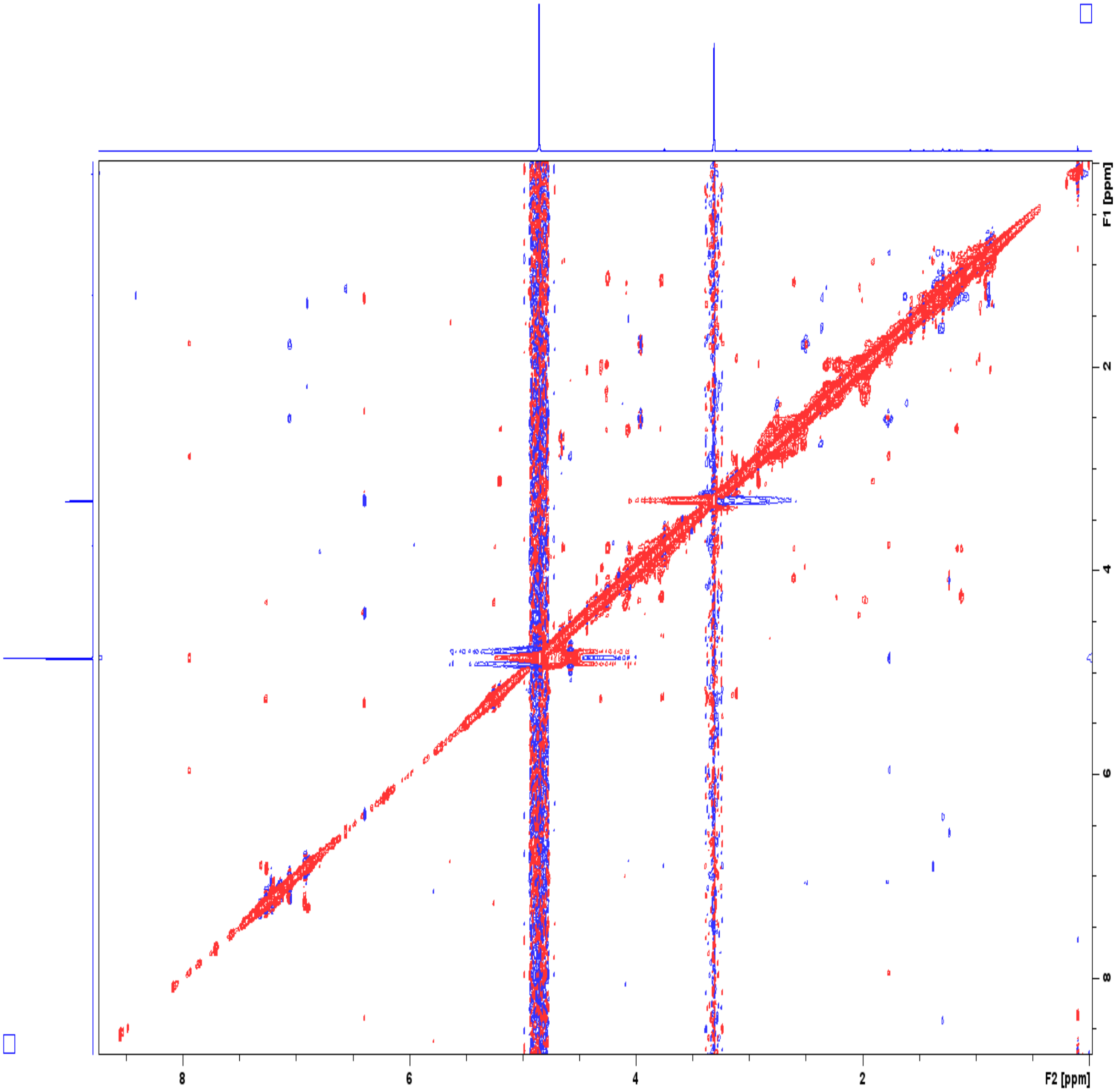

**Figure S77. NOESY spectrum of 20 in methanol-$d_4$.**

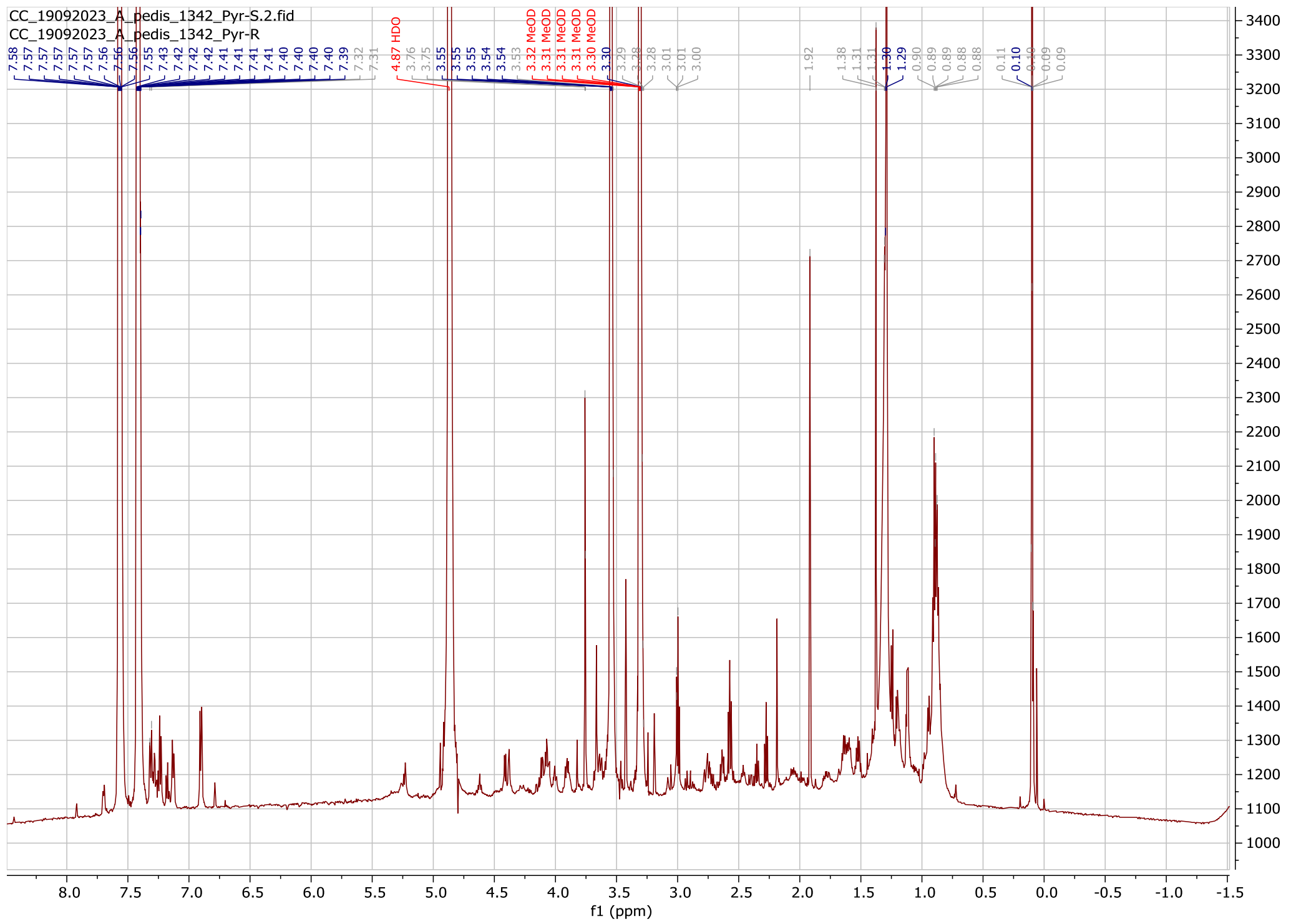

**Figure S78. $^1$H NMR spectrum of acidobactamide A (15) *R*-MTPA ester in methanol-$d_4$.**

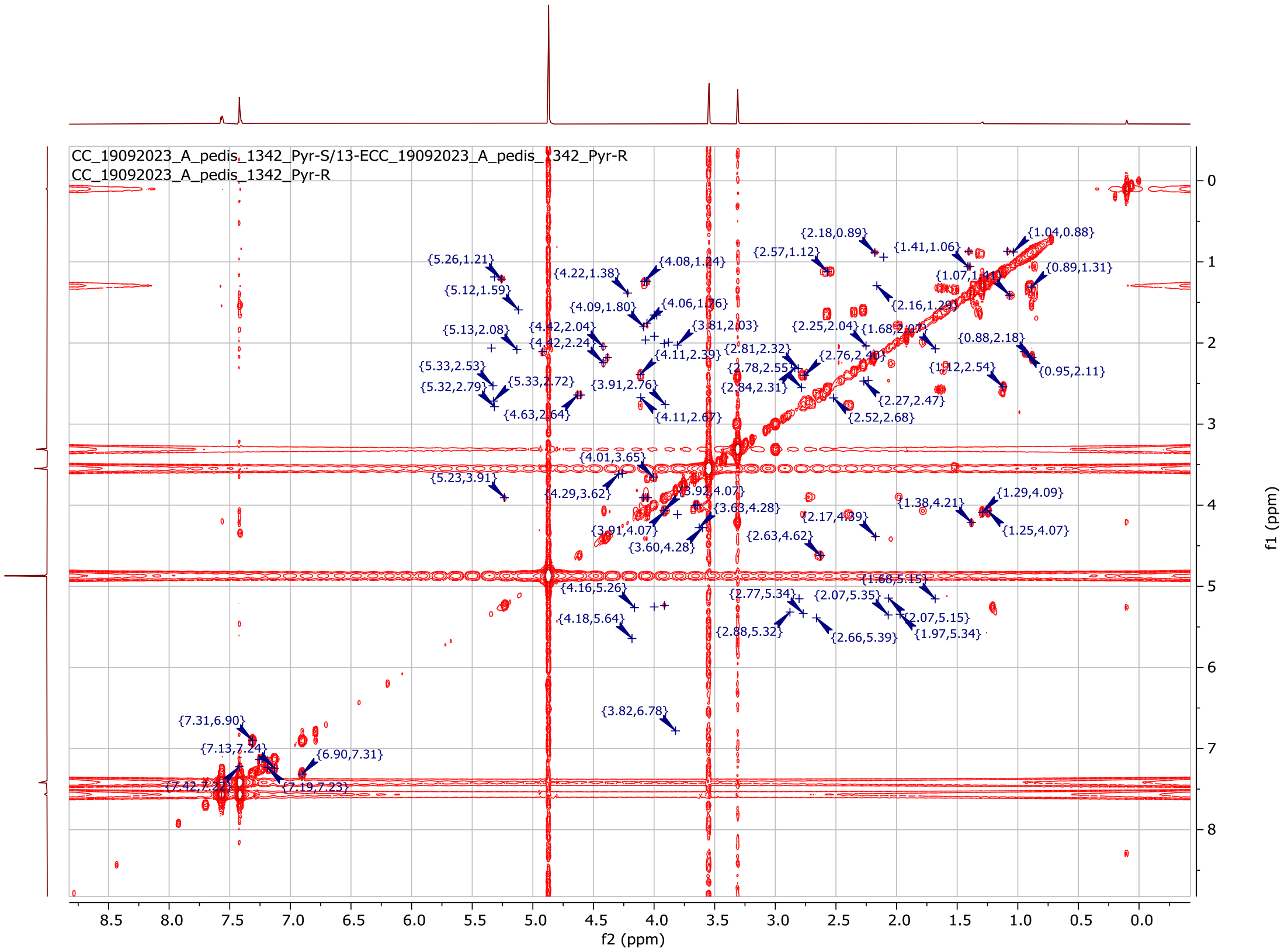

**Figure S79. COSY NMR spectrum of acidobactamide A (15) *R*-MTPA ester in methanol-$d_4$.**

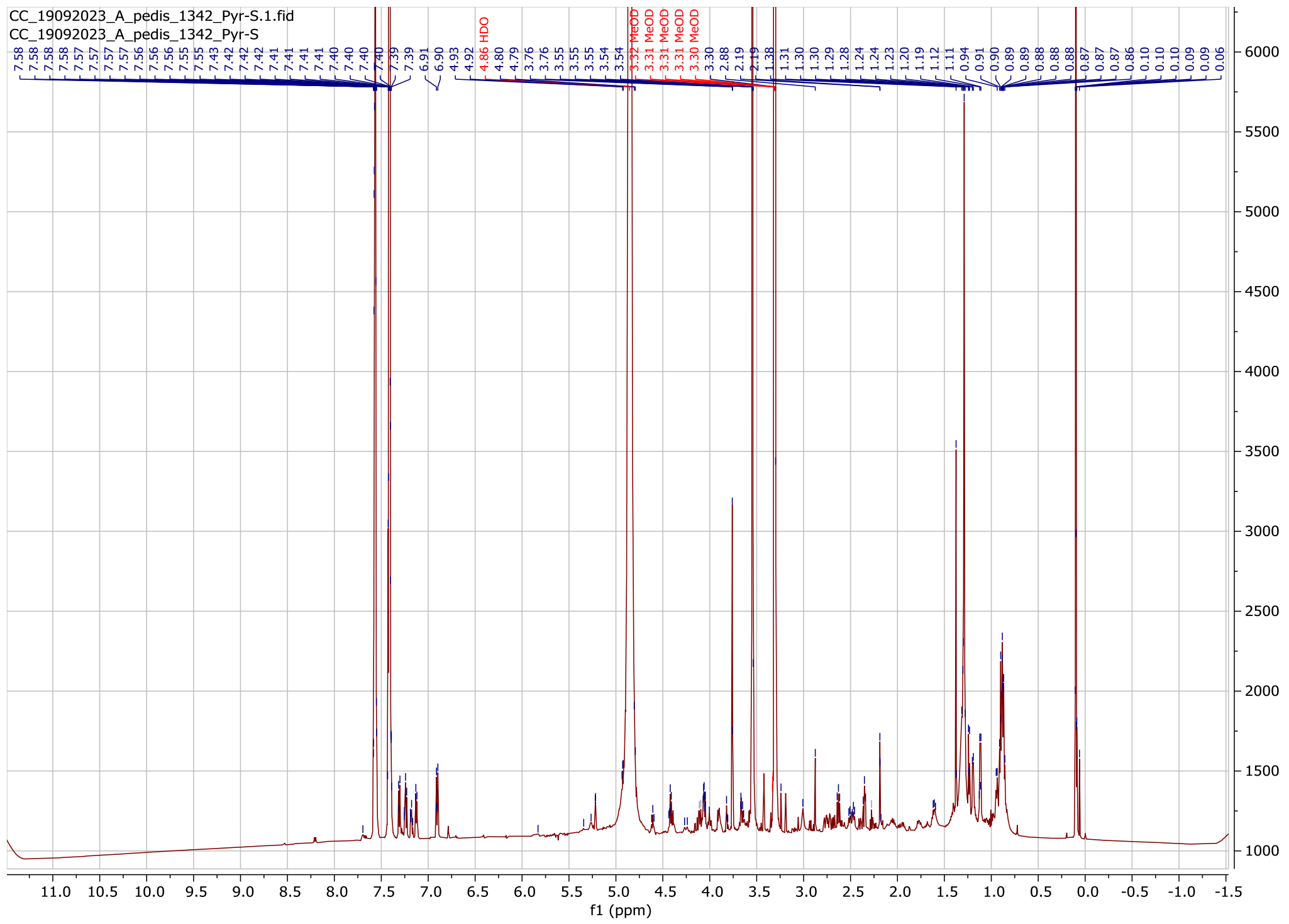

**Figure S80. $^{1}$H NMR spectrum of acidobactamide A (15) *S*-MTPA ester in methanol-$d_4$.**

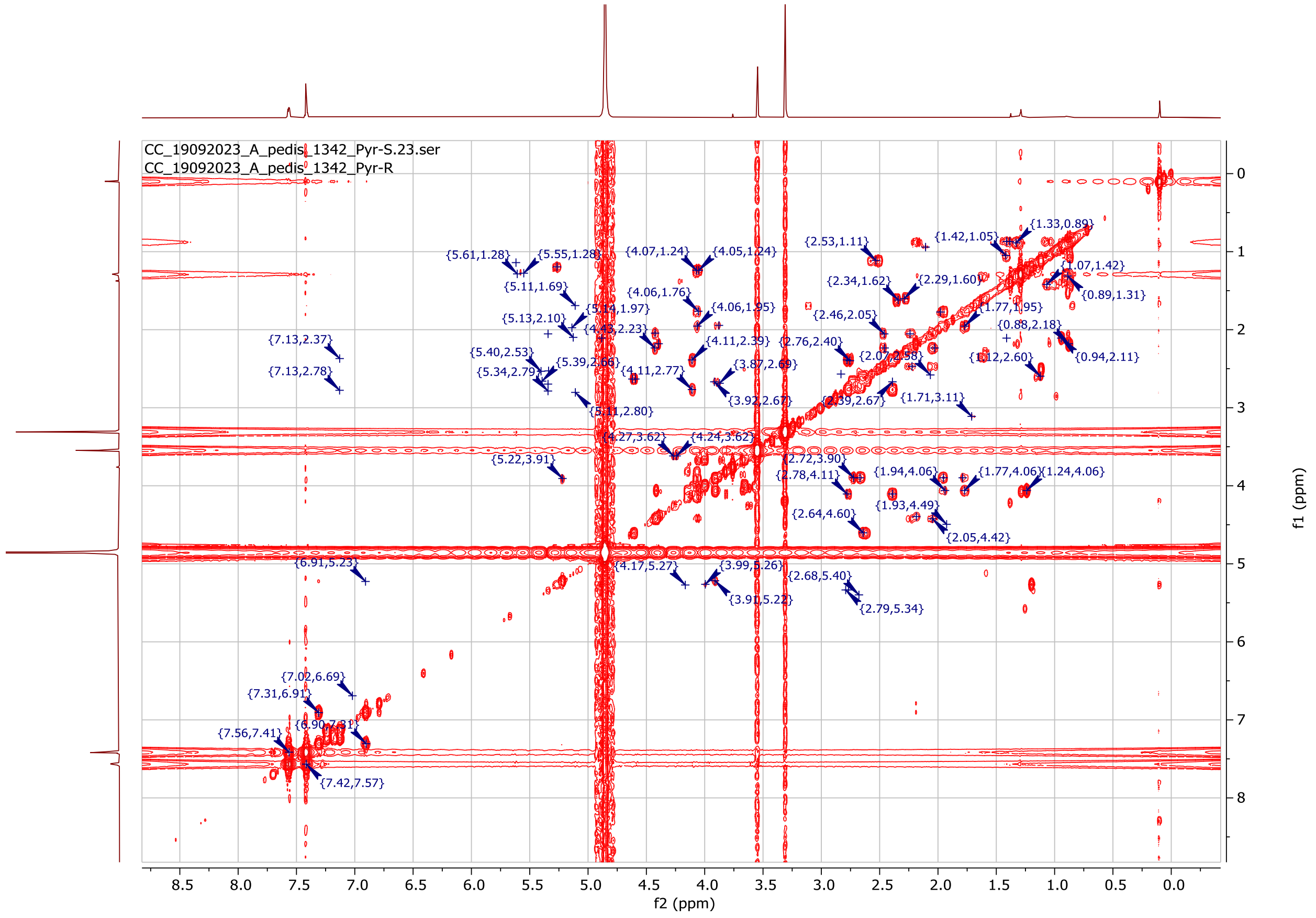

**Figure S81. COSY spectrum of acidobactamide A (15) *S*-MTPA ester in methanol-$d_4$.**

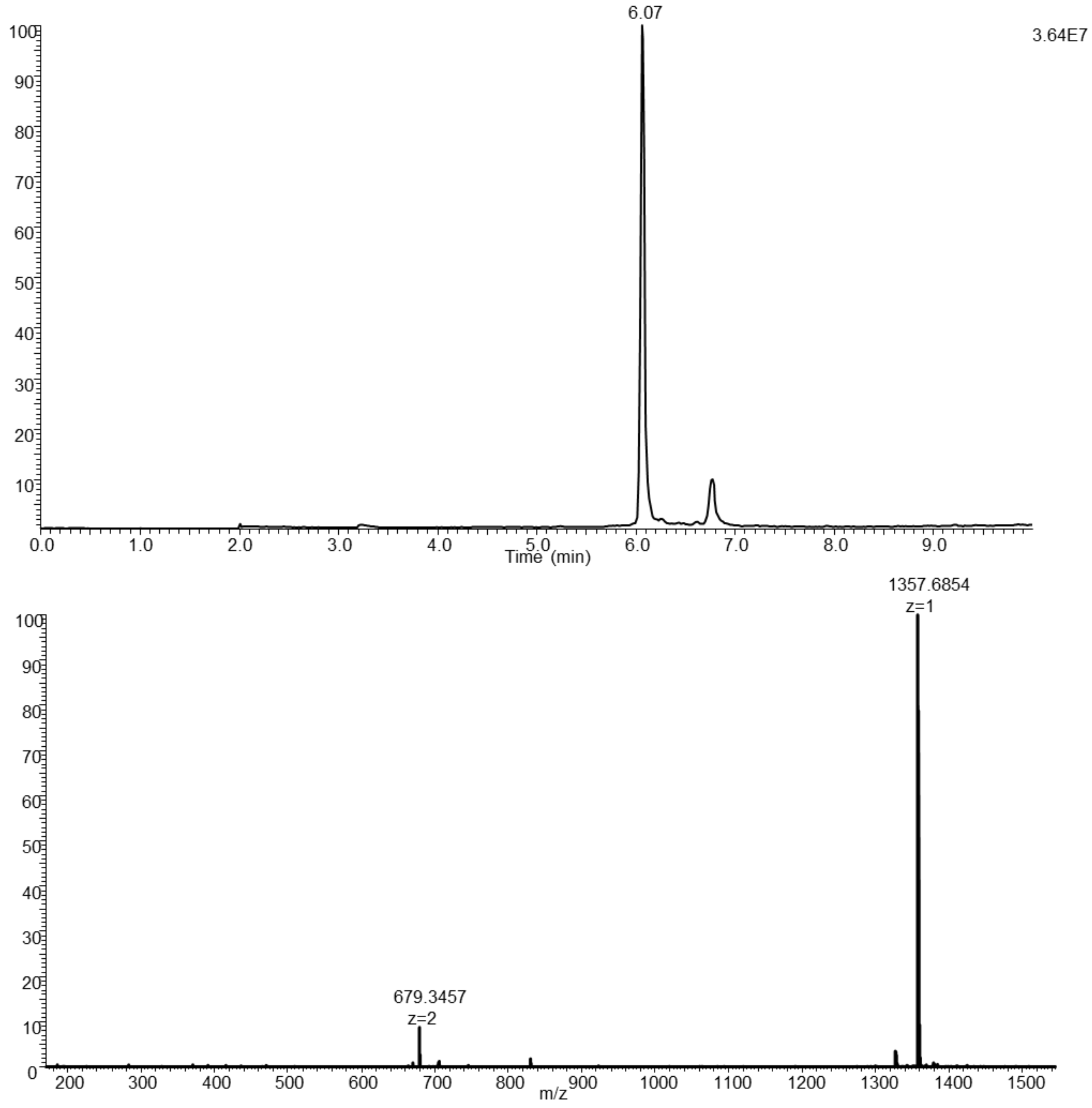

**Figure S82. HPLC-HESI-HRMS data of acidobactamide B (16).**

Top: Total ion chromatogram of isolated acidobactamide B (**16**).
Bottom: MS data of acidobactamide B (**16**) (*m/z* 1368.6854 $[M+H]^+$).

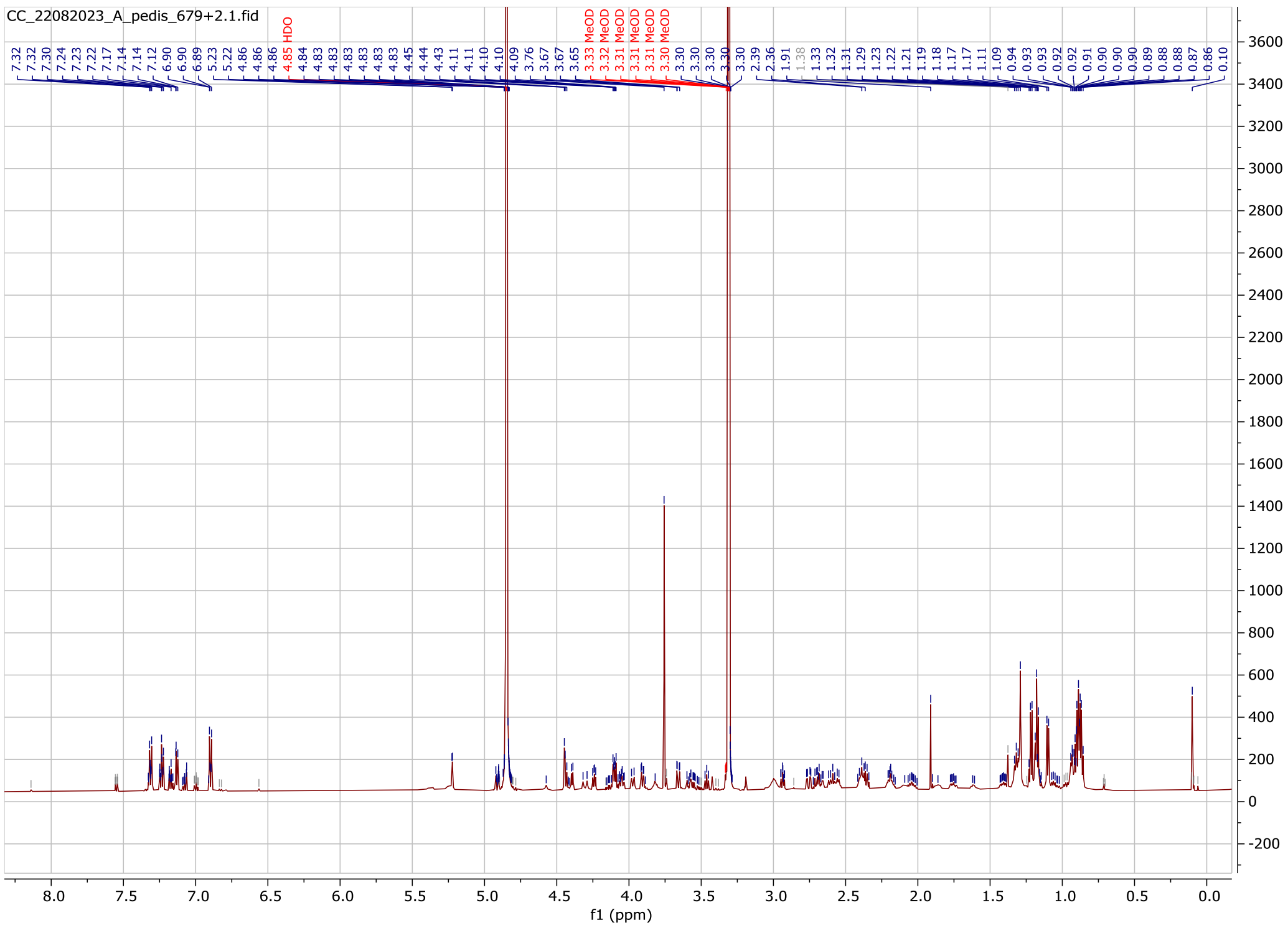

**Figure S83. $^1$H NMR spectrum of acidobactamide B (16) in methanol-$d_4$.**

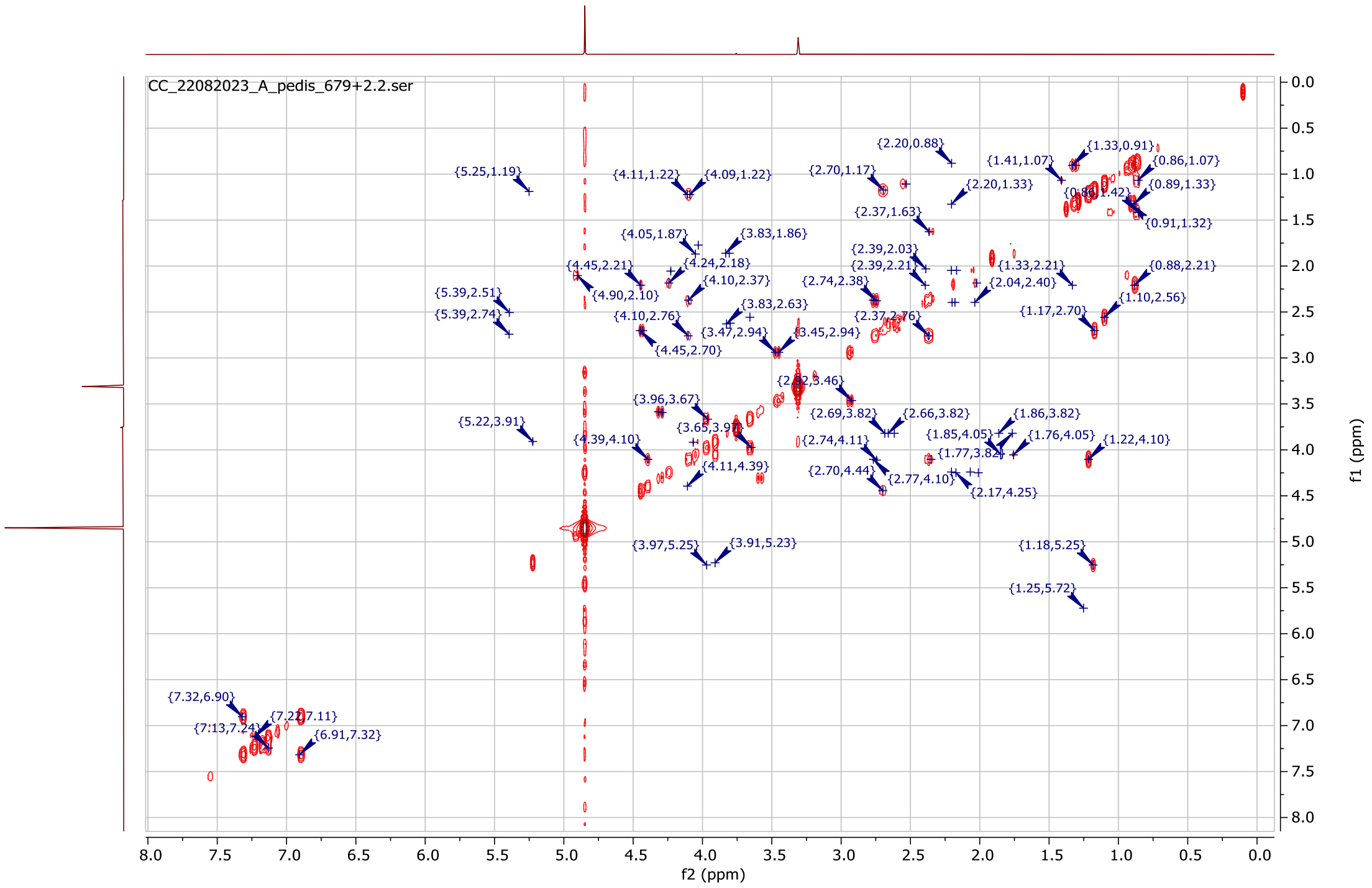

**Figure S84. COSY spectrum of acidobactamide B (16) in methanol-$d_4$.**

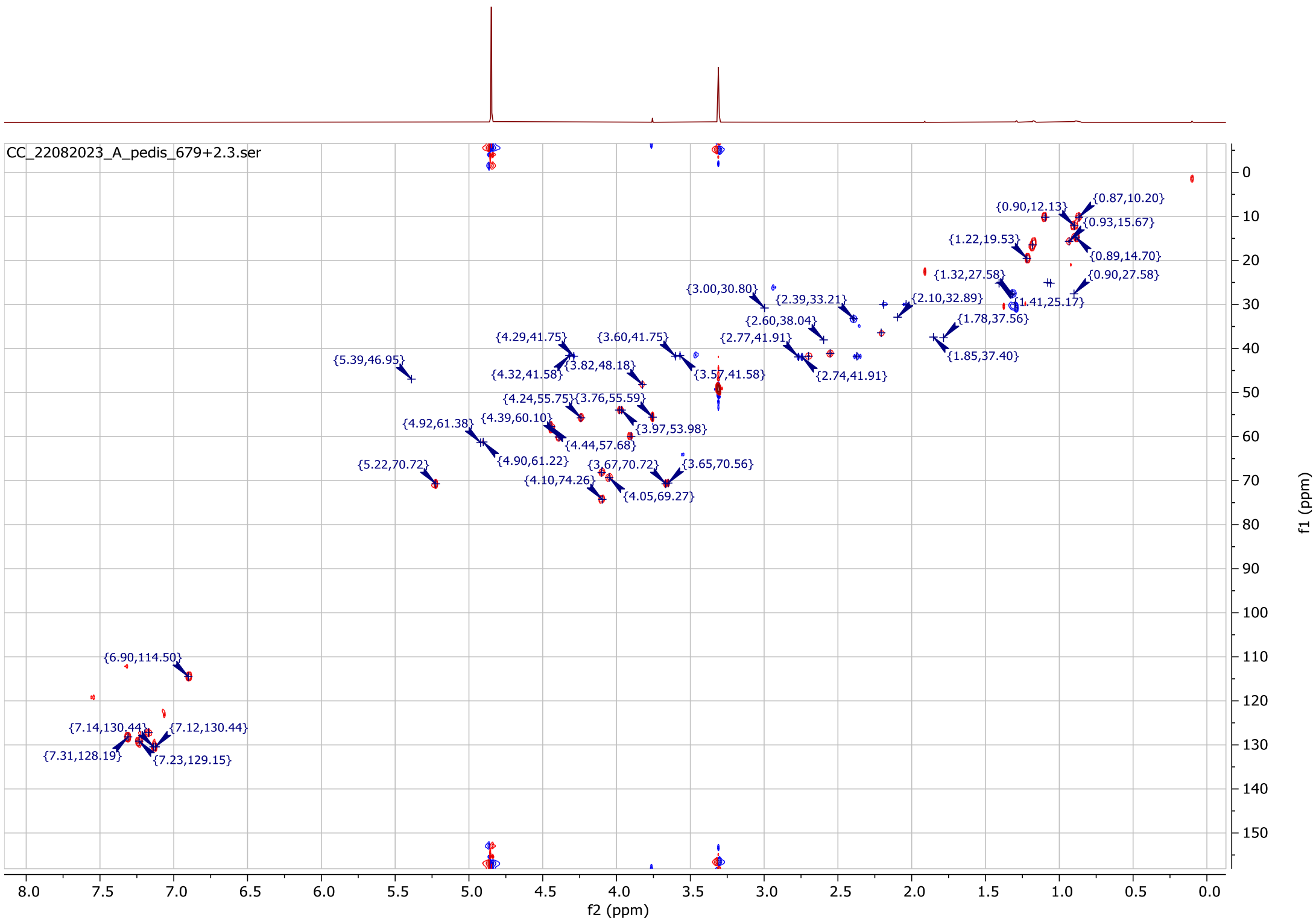

**Figure S85. HSQC spectrum of acidobactamide B (16) in methanol-$d_4$.**

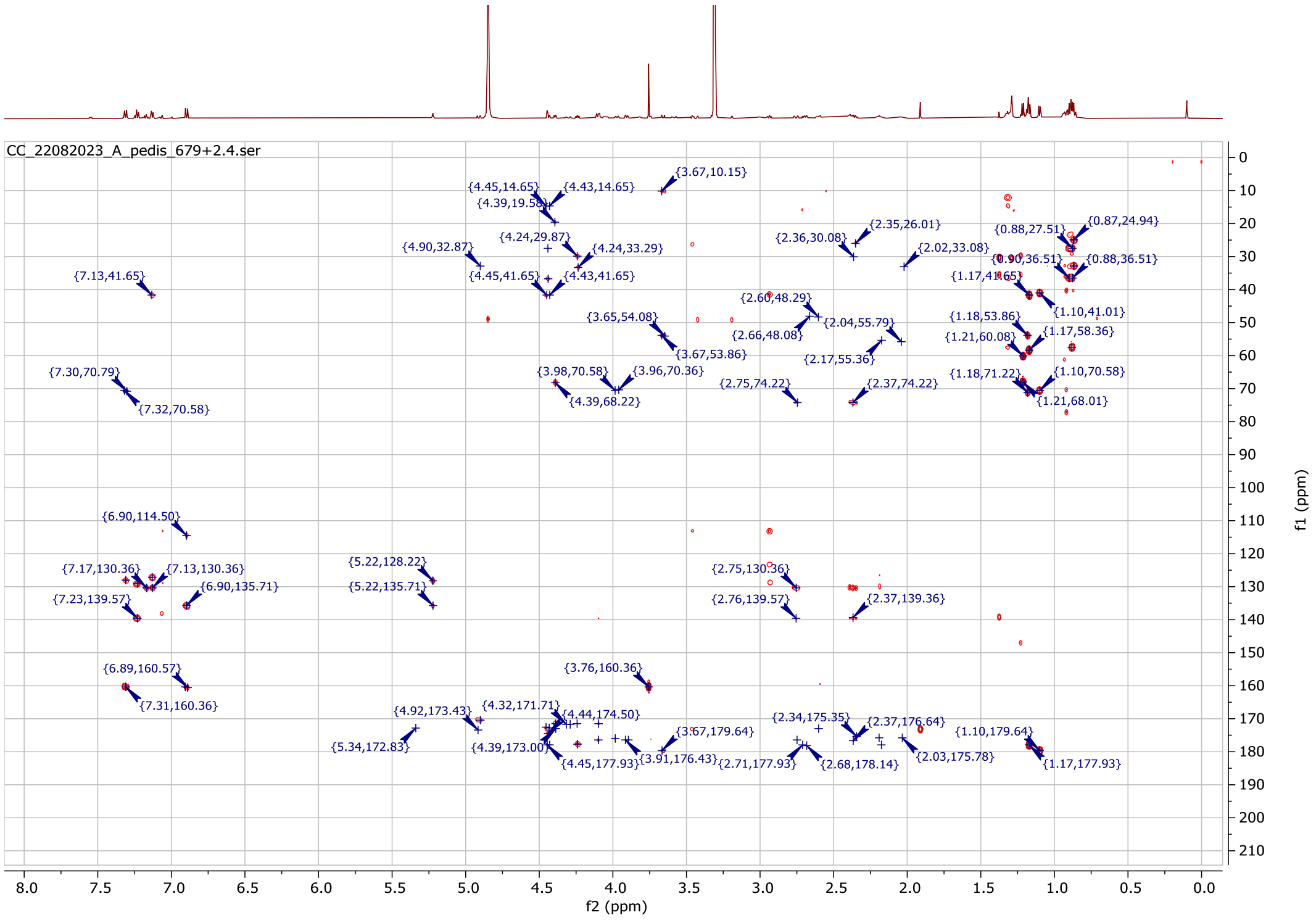

**Figure S86. HMBC spectrum of acidobactamide B (16) in methanol-$d_4$.**

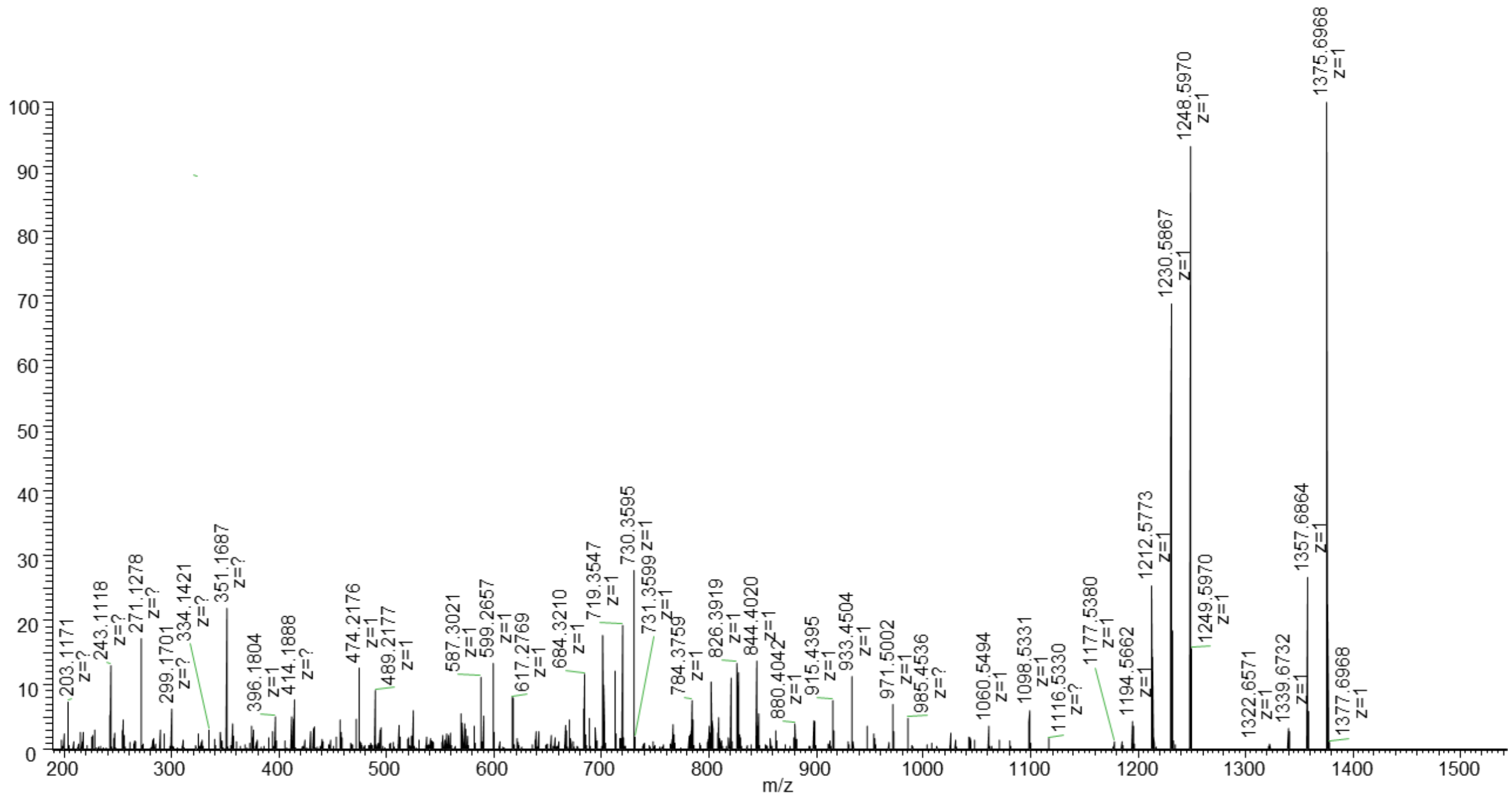

**Figure S87. HESI-HRMS MS/MS data of acidobactamide B NaOH hydrolysate (16).**

MS/MS data of acidobactamide B (**16**) NaOH hydrolysate (*m/z* 1375.6969 [M+H]$^{+}$).

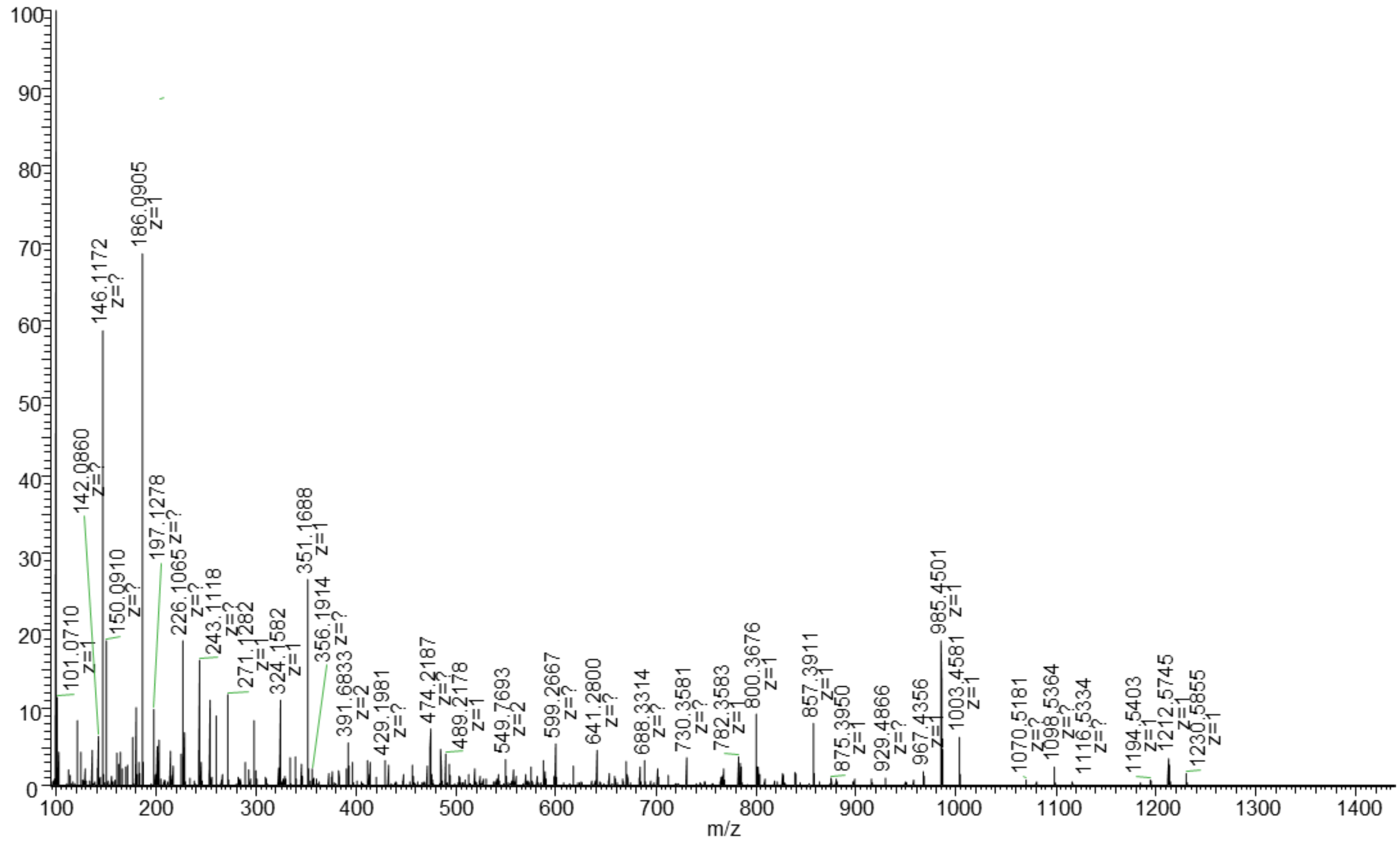

**Figure S88. HESI-HRMS MS/MS data of acidobactamide B (16) partial hydrolysis product.**

MS/MS data of acidobactamide B (**16**) NaOH hydrolysate (*m/z* 688.3523 [M+H]$^{2+}$).

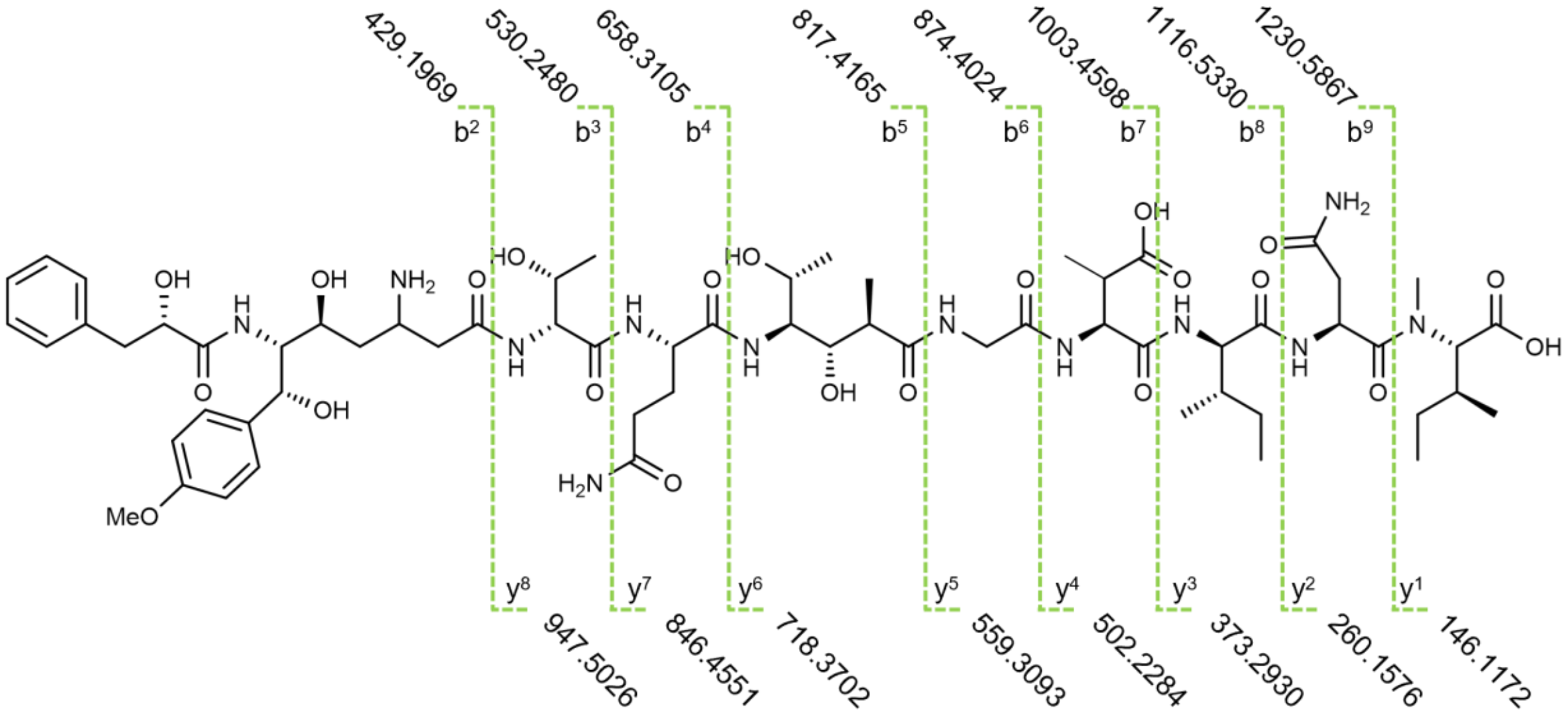

**Figure S89. MS/MS fragmentation of acidobactamide B (16) NaOH hydrolysate.**

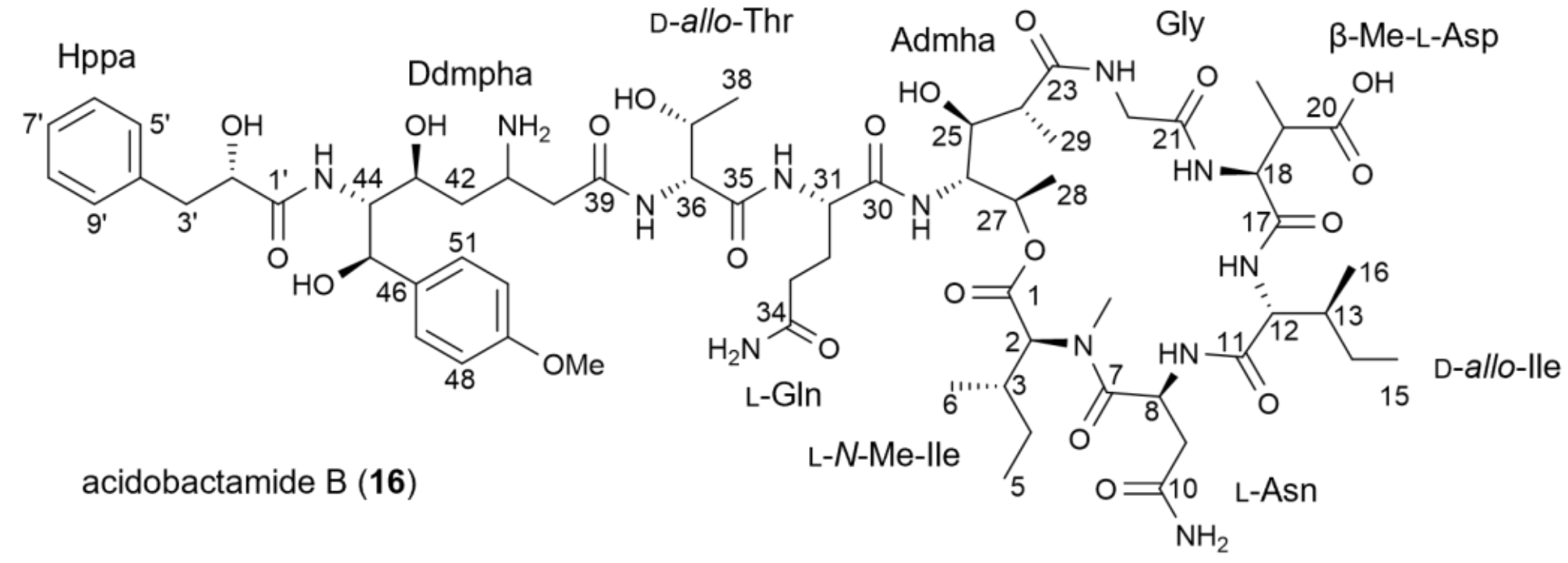

acidobactamide B (**16**)

**Figure S90. Chemical structure of acidobactamide B (16).**

**Table S13. NMR Data ($^1$H 600 MHz, $^{13}$C 150 MHz) for acidobactamide B (16) in methanol-$d_4$.**

| No. | Amino acids | $^{13}$C/HSQC | $^1$H (multiplicity) |
|---|---|---|---|
| 1 | *N*Me-Ile | 170.4, C | |
| 2 | | 61.2, CH | 4.91 (d), *J* = 11.6 Hz |
| 3 | | 32.0, CH | 2.10 (m) |
| 4 | | 25.1, $CH_2$ | 1.07 (m), 1.41 (m) |
| 5 | | 10.2, $CH_3$ | 0.87 (m) |
| 6 | | 15.7, $CH_3$ | 0.93 (d), *J* = 6.2 Hz |
| *N*Me | | 30.8, $CH_3$ | 3.00 (s) |
| 7 | Asn | 173.7, C | |
| 8 | | 46.9, CH | 5.39 (m) |
| 9 | | 38.0, $CH_2$ | 2.52 (m), 2.74 (m) |
| 10 | | 172.4, C | |
| 11 | Ile | 172.7, C | |
| 12 | | 57.7, CH | 4.44 (m) |
| 13 | | 36.4, CH | 2.21 (m) |
| 14 | | 27.6, $CH_2$ | 1.32 (m), 0.90 (m) |
| 15 | | 12.1, $CH_3$ | 0.90 (m) |
| 16 | | 14.7, $CH_3$ | 0.89 (m) |
| 17 | β-Me-Asp | 172.8, C | |
| 18 | | 58.0, CH | 4.43 (m) |
| 19 | | 41.8, CH | 2.70 (m) |
| 20 | | 16.5, $CH_3$ | 1.18 (m) |
| 21 | | 177.9, C | |
| 22 | Gly | 171.7, C | |
| 23 | | 41.8, $CH_2$ | 3.58 (d), *J* = 16.6 Hz<br>4.30 (d), *J* = 16.6 Hz |
| 24 | Admha | 179.6, C | |
| 25 | | 41.1, CH | 2.55 (m) |
| 26 | | 70.7, CH | 3.66 (dd) *J* = 1.4,10.3 HZ |
| 27 | | 54.0, CH | 3.97 (br d), *J* = 10.3 HZ |
| 28 | | 70.7, CH | 5.22 (m), |
| 29 | | 16.5, $CH_3$ | 1.18 (m) |
| 30 | | 10.2, $CH_3$ | 1.10 (d) *J* = 7.0 Hz |
| 31 | Gln | 177.7, C | |
| 32 | | 55.8, CH | 4.24 (dd), *J* = 5.2, 7.1 Hz |
| 33 | | 30.0, $CH_2$ | 2.04 (m), 2.19 (m) |
| 34 | | 33.2, $CH_2$ | 2.39 (m) |
| 35 | | 175.7, C | |
| 36 | Thr | 171.5, C | |
| 37 | | 60.1, CH | 4.39 (d), *J* = 6.5 Hz |
| 38 | | 68.1, CH | 4.10 (m) |
| 39 | | 19.5, $CH_3$ | 1.22, (d) *J* = 6.6 Hz |
| 40 | Ddmpha | 173.0, C | |
| 41 | | 38.0, $CH_2$ | 2.63 (m), 2.68 (m) |
| 42 | | 48.2, CH | 3.82 (m) |
| 43 | | 36.6, $CH_2$ | 1.77 (m)<br>1.87 (m) |
| 44 | | 69.3, CH | 4.05 (td), *J* = 3.0, 8.7 Hz |
| 45 | | 59.9, CH | 3.90 (m) |
| 46 | | 70.7, CH | 5.23 (m) |
| 47 | | 135.7, C | |
| 48/52 | | 128.2, C | 7.31 (d), *J* = 8.7 Hz |
| 49/51 | | 114.5, CH | 6.90 (d), *J* = 8.7 Hz |
| 50 | | 160.4, C | |
| OCH3 | | 55.6, $CH_3$ | 3.76 (s) |
| 1' | Hppa | 176.4, C | |
| 2' | | 74.3, CH | 4.10 (m) |
| 3' | | 41.9, $CH_2$ | 2.37 (dd), *J* = 9.3, 14.0 Hz |
| | | | 2.75 (dd), *J* = 3.5, 14.0 Hz |
| 4' | | 139.8, C | |
| 5'/9' | | 130.4, CH | 7.12 (d), *J* = 7.5 Hz |
| 6'/8' | | 129.2, CH | 7.23 (t), *J* = 7.5 Hz |
| 7' | | 127.2, CH | 7.17 (d), *J* = 7.5 Hz |

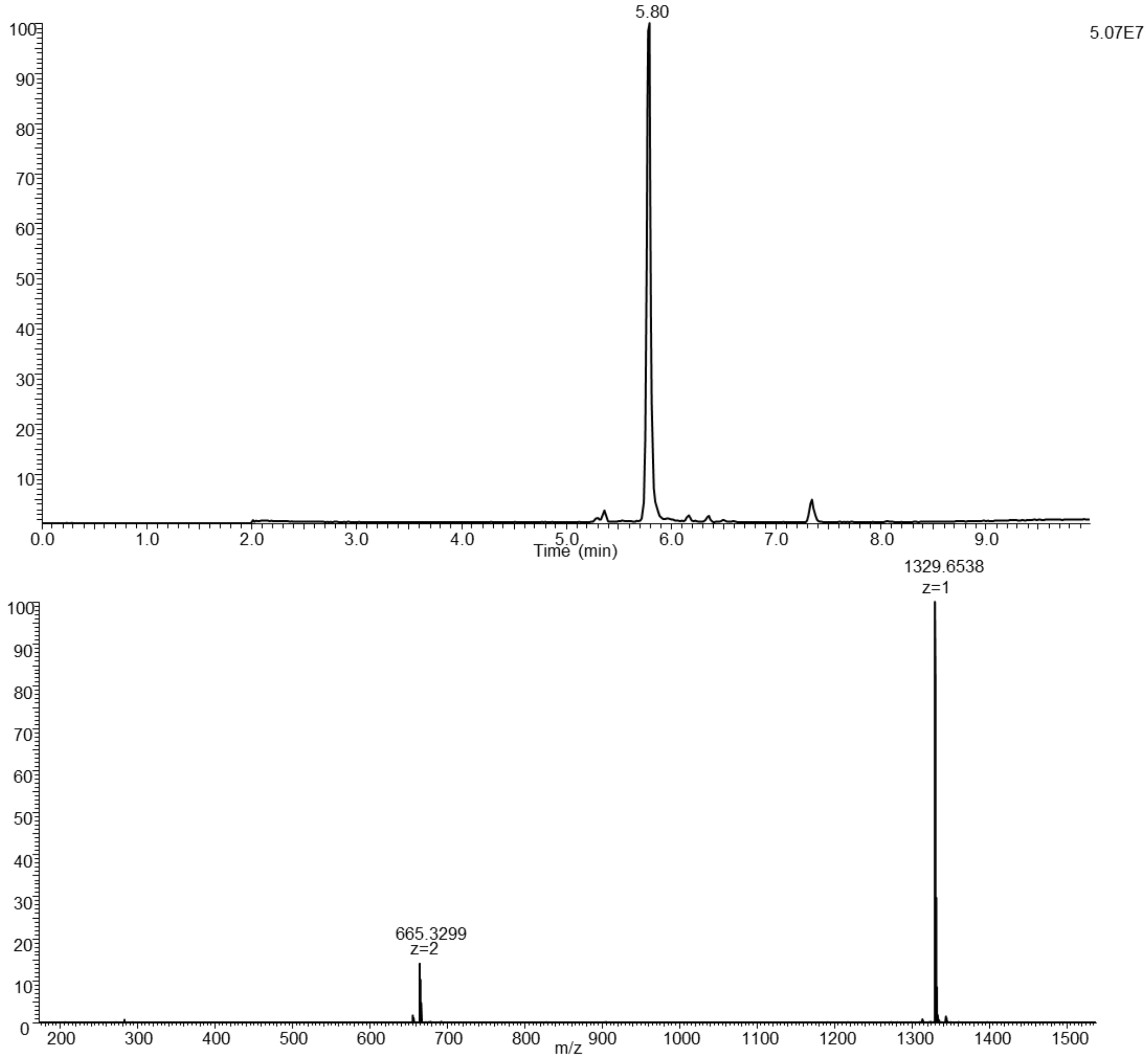

**Figure S91. HPLC-HESI-HRMS data of acidobactamide C (17).**

Top: Total ion chromatogram of isolated acidobactamide C (**17**).
Bottom: MS data of acidobactamide C (**17**) (m/z 1329.6538 $[M+H]^+$).

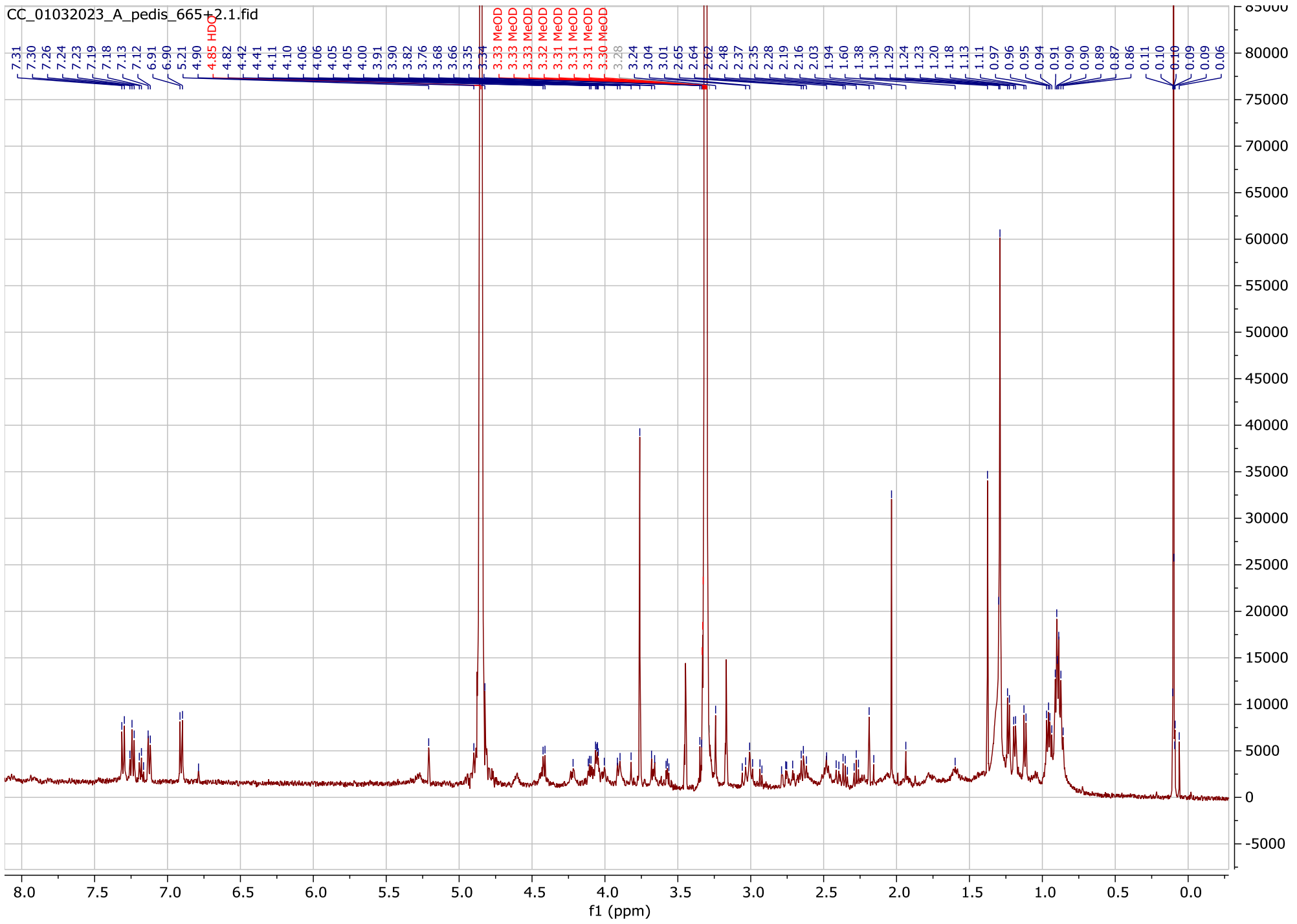

**Figure S92. [1]H NMR spectrum of acidobactamide C (17) in methanol-$d_4$.**

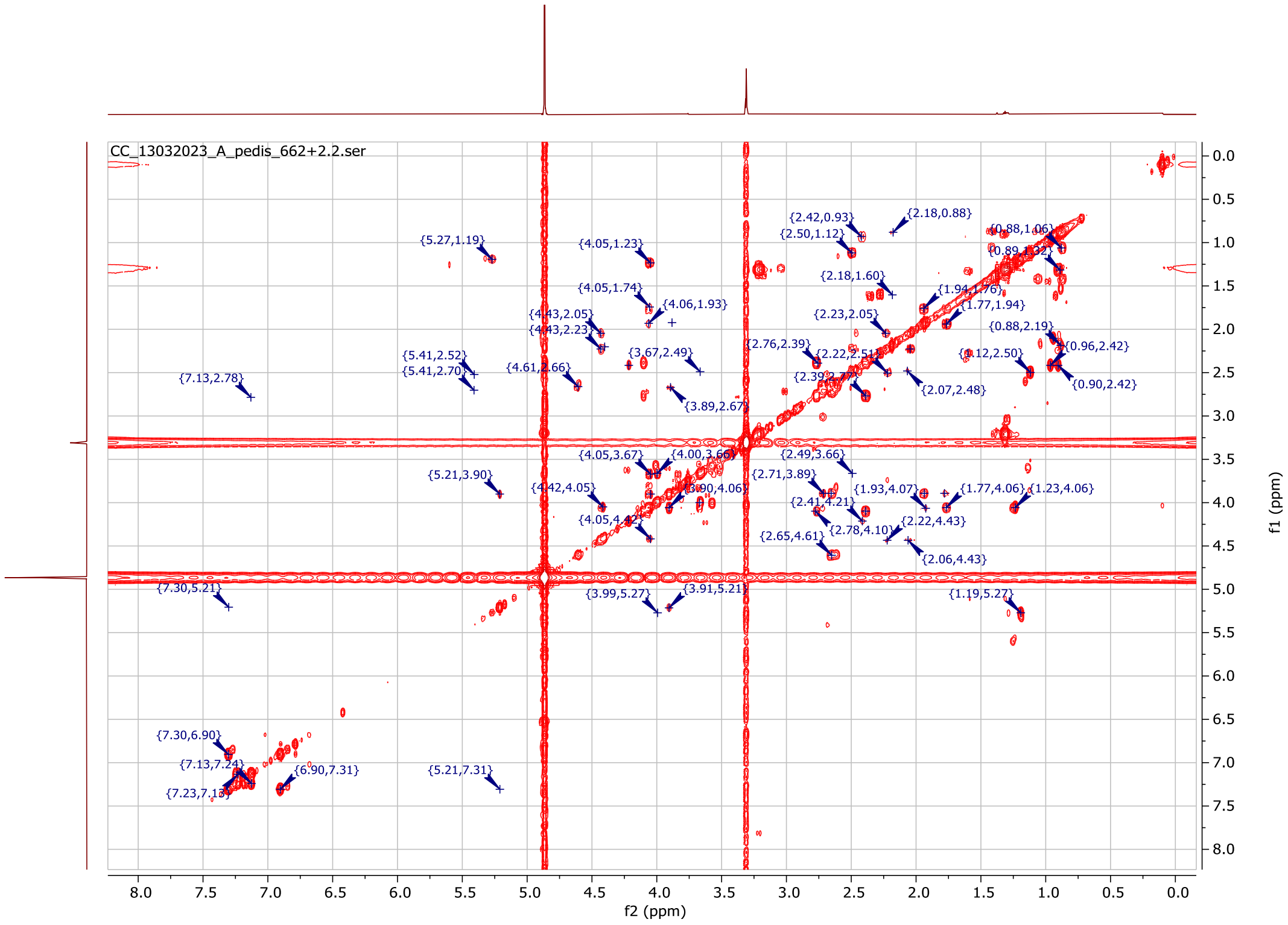

**Figure S93. COSY spectrum of acidobactamide C (17) in methanol-$d_4$.**

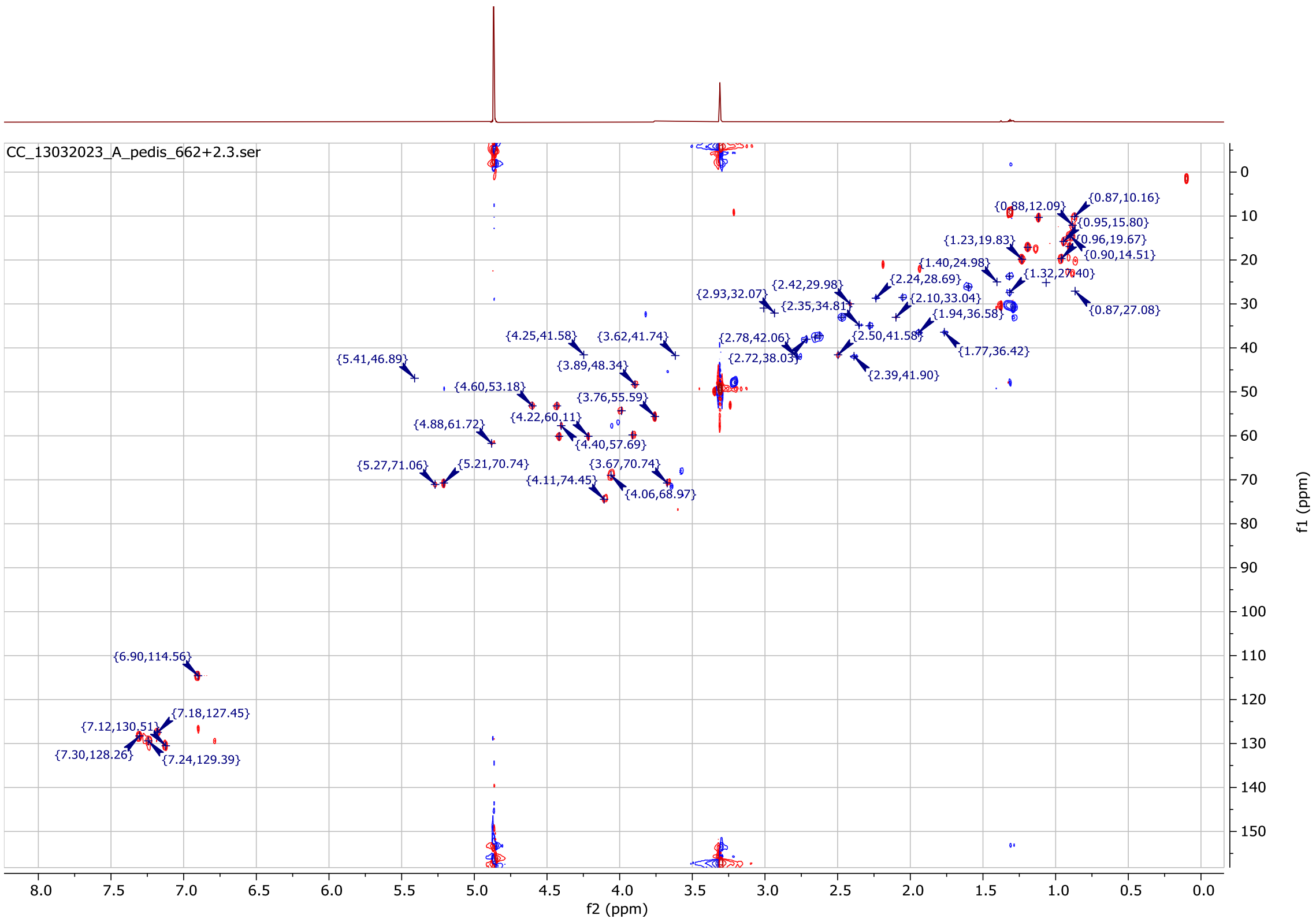

**Figure S94. HSQC spectrum of acidobactamide C (17) in methanol-$d_4$.**

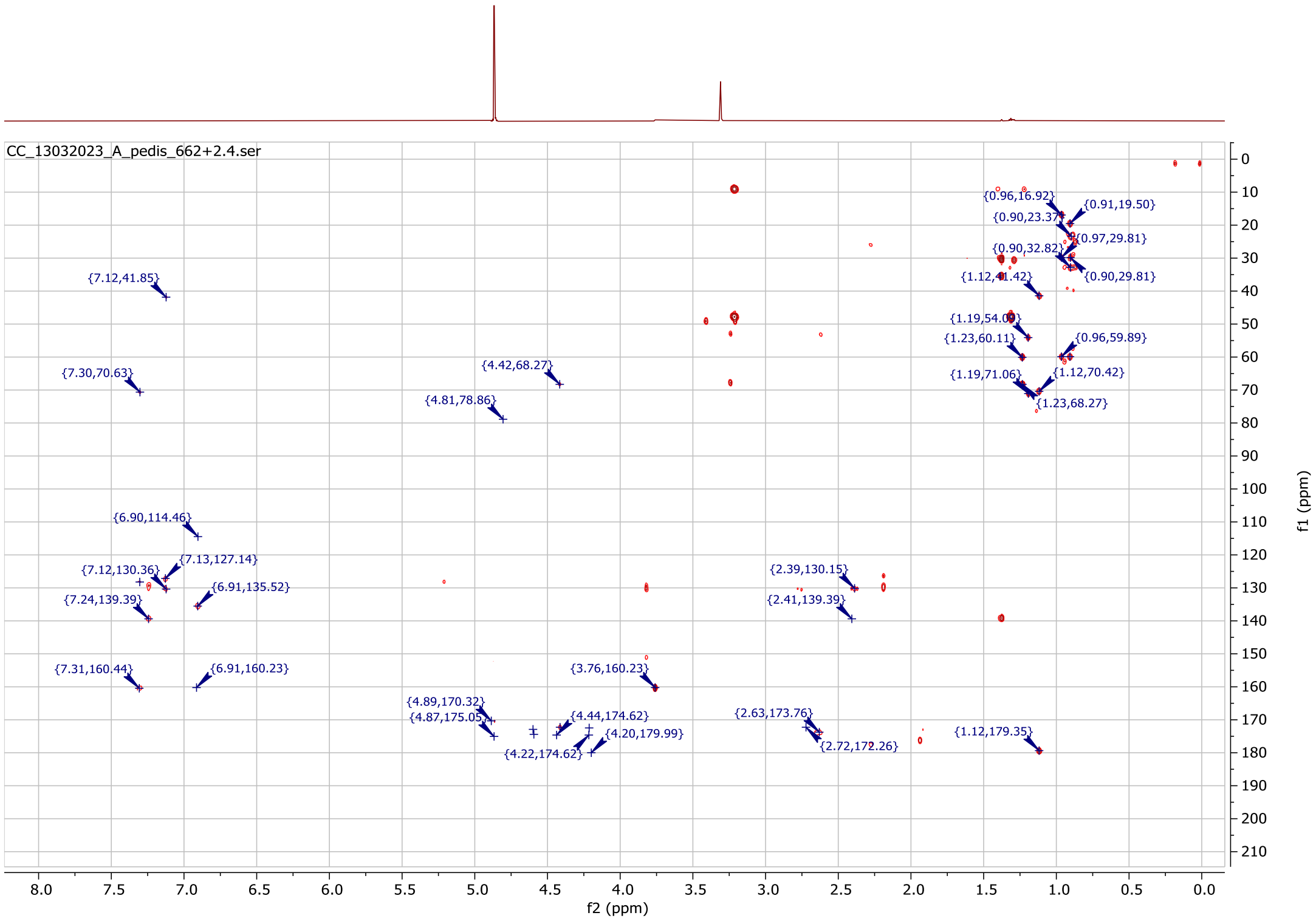

**Figure S95. HMBC spectrum of acidobactamide C (17) in methanol-$d_4$.**

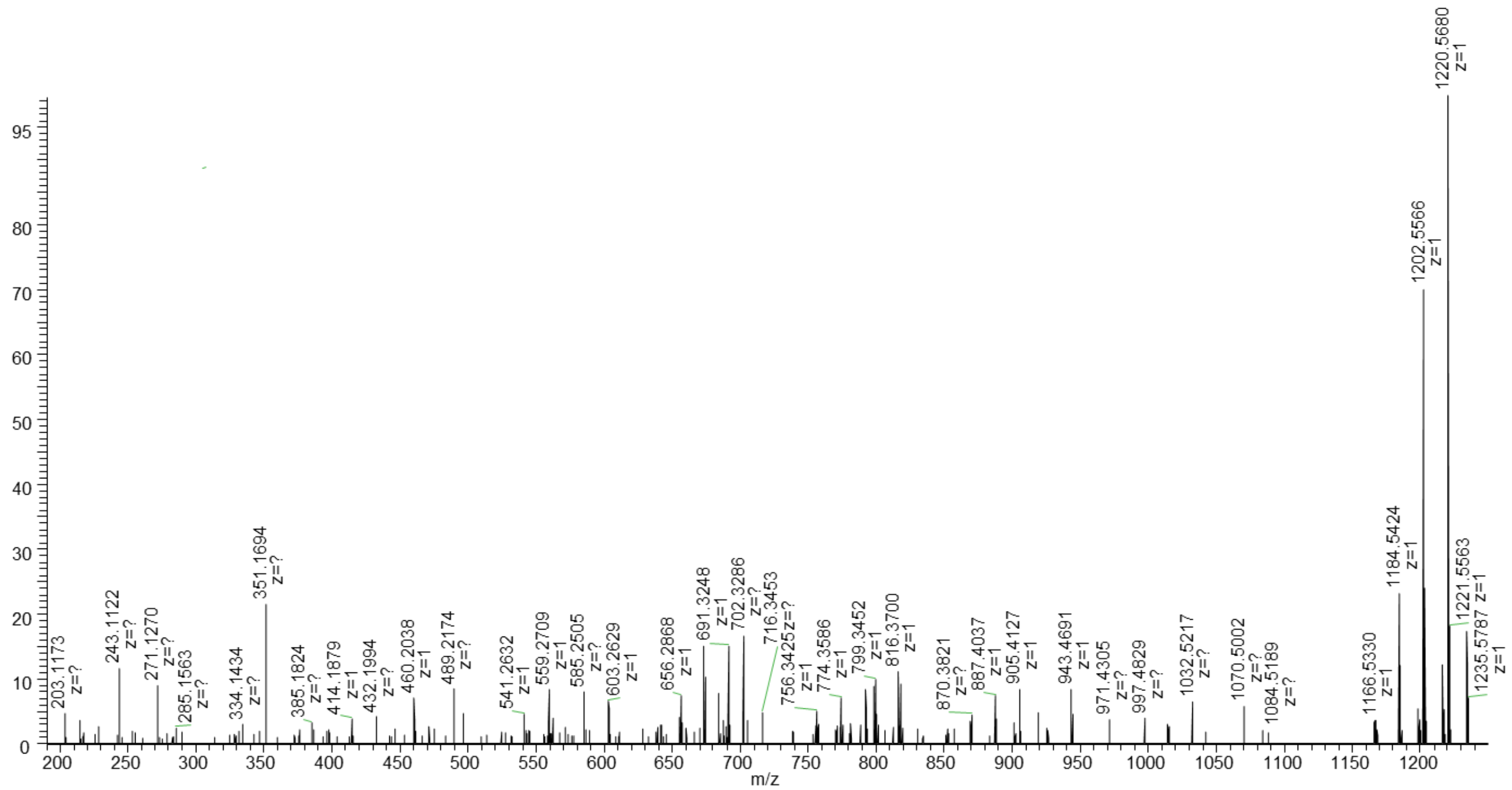

**Figure S96. HESI-HRMS MS/MS data of acidobactamide C NaOH hydrolysate (17).**

MS/MS data of acidobactamide C (**17**) NaOH hydrolysate (*m/z* 1347.6663 [M+H]$^{+}$).

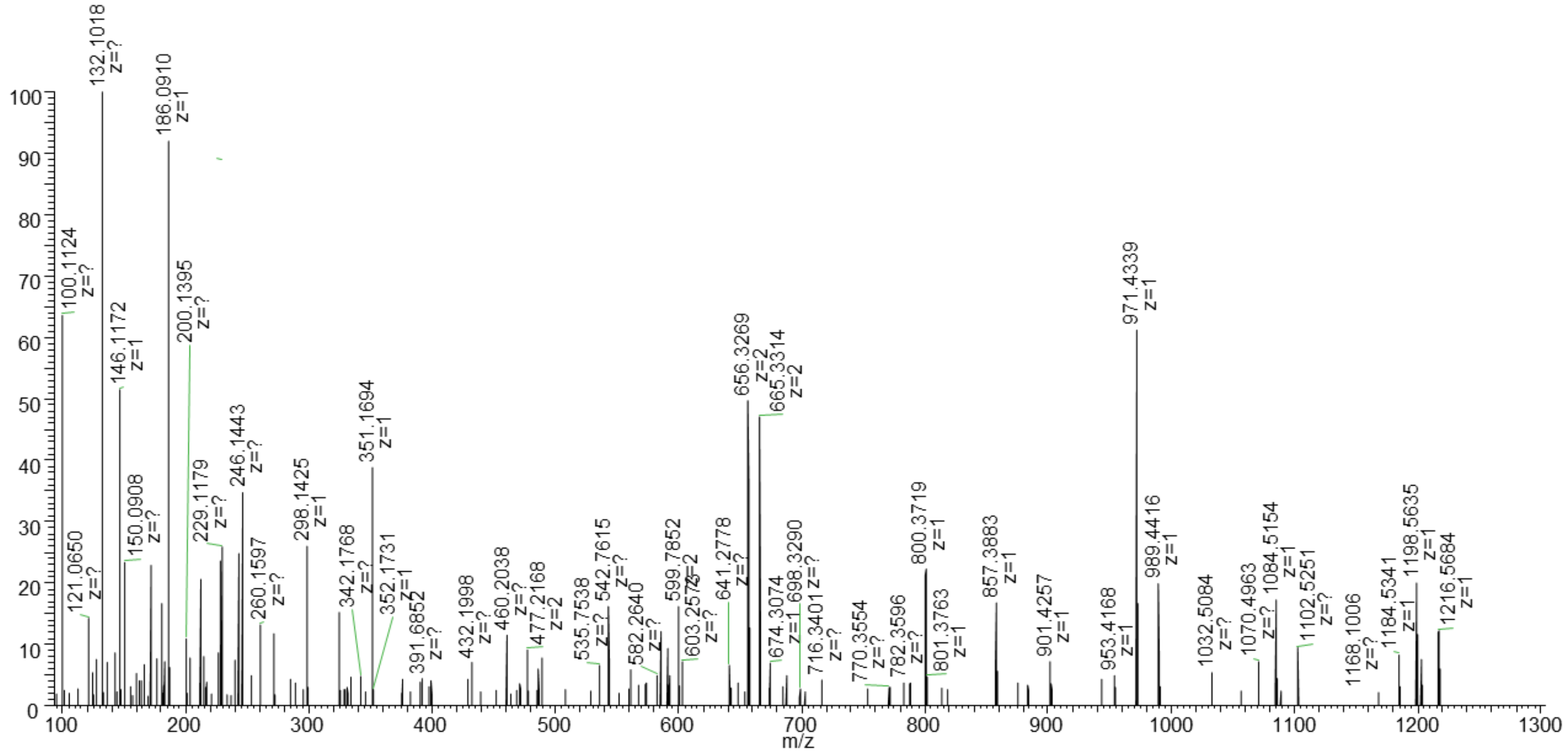

**Figure S97. HESI-HRMS MS/MS data of acidobactamide C NaOH hydrolysate (17).**

MS$^{2}$ data of acidobactamide C (**17**) NaOH hydrolysate (*m/z* 674.3367 [M+H]$^{2+}$).

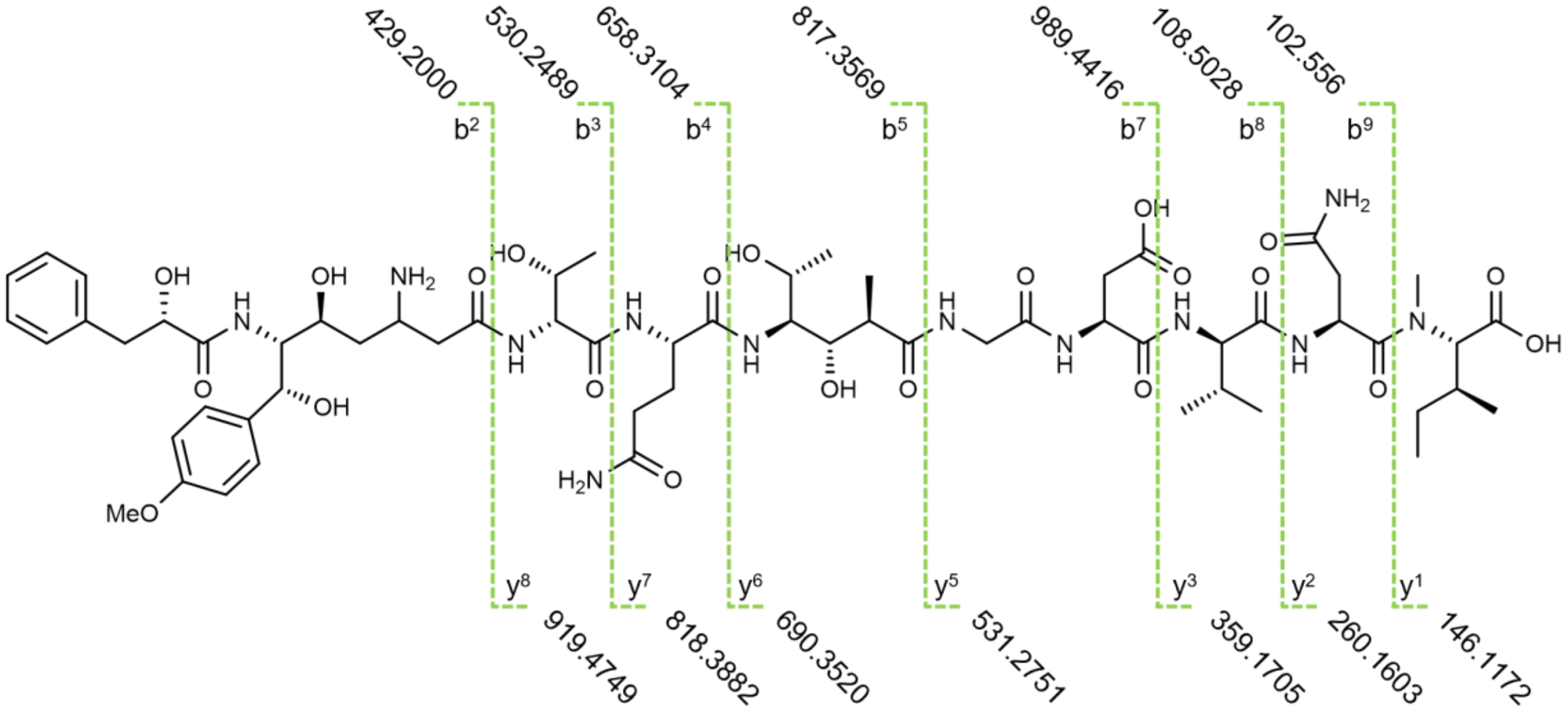

**Figure S98. MS/MS fragmentation of acidobactamide C (17) NaOH hydrolysate.**

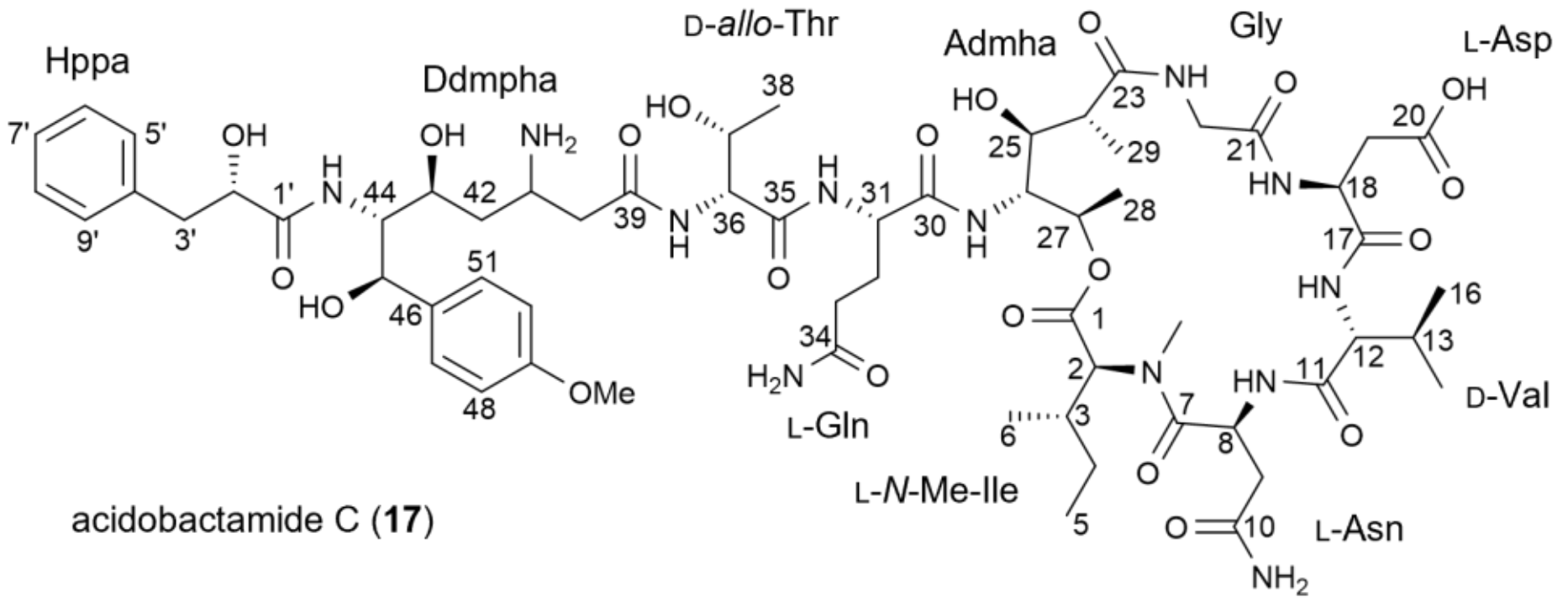

acidobactamide C (**17**)

**Figure S99. Chemical structure of acidobactamide C (17).**

**Table S14. NMR Data ($^{1}$H 500 MHz, $^{13}$C 125 MHz) for acidobactamide C (17) in methanol-$d_4$.**

| No. | Amino acids | $^{13}$C/HSQC | $^{1}$H (multiplicity) |
|---|---|---|---|
| 1 | *N*Me-Ile | 170.3, C | |
| 2 | | 61.7, CH | 4.88 (m) |
| 3 | | 33.0, CH | 2.10 (m) |
| 4 | | 25.0, $CH_2$ | 1.06 (m), 1.40 (m) |
| 5 | | 10.2, $CH_3$ | 0.87 (m) |
| 6 | | 15.8, $CH_3$ | 0.94 (d), $J$ = 6.2 Hz |
| *N*Me | | 30.8, $CH_3$ | 3.01 (s) |
| 7 | Asn | 173.7, C | |
| 8 | | 46.7, CH | 5.40 (m) |
| 9 | | 38.0, $CH_2$ | 2.55 (m), 2.72 (m) |
| 10 | | 172.4, C | |
| 11 | Val | 172.5, C | |
| 12 | | 60.1, CH | 4.21 (m) |
| 13 | | 29.9, CH | 2.41 (m) |
| 14 | | 16.9, $CH_3$ | 0.90 (m) |
| 15 | | 19.7, $CH_3$ | 0.96 (d), $J$ = 7.1 Hz |
| 16 | Asp | 174.4, C | |
| 17 | | 53.2, CH | 4.61 (t), $J$ = 7.2 Hz |
| 18 | | 37.2, $CH_2$ | 2.62 (m) |
| 19 | | 173.7, C | |
| 20 | Gly | 174.6, C | |
| 21 | | 41.7, $CH_2$ | 3.61 (d), $J$ = 16.1 Hz |
| | | | 4.24 (m) |
| 22 | Admha | 179.4, C | |
| 23 | | 41.6, CH | 2.49 (m) |
| 24 | | 70.7, CH | 3.67 (br d) $J$ = 10.3 HZ |
| 25 | | 54.3, CH | 3.99 (br d), $J$ = 10.2 HZ |
| 26 | | 71.1, CH | 5.27 (m), |
| 27 | | 17.1, $CH_3$ | 1.19 (d), $J$ = 6.5 Hz |
| 26 | | 10.3, $CH_3$ | 1.11 (d) $J$ = 7.3 Hz |
| 29 | Gln | 175.2, C | |
| 30 | | 53.2, CH | 4.43 (m) |
| 31 | | 28.7, $CH_2$ | 2.05 (m), 2.23 (m) |
| 32 | | 33.0, $CH_2$ | 2.47 (m) |
| 33 | | 173.9, C | |
| 34 | Thr | 172.3, C | |
| 35 | | 60.1, CH | 4.41 (m) |
| 36 | | 68.9, CH | 4.06 (m) |
| 37 | | 19.8, $CH_3$ | 1.23, (d) $J$ = 6.5 Hz |
| 38 | Ddmpha | 172.3, C | |
| 39 | | 38.0, $CH_2$ | 2.65 (m), 2.72 (m) |
| 40 | | 48.3 | 3.89 (m) |
| 41 | | 36.6, $CH_2$ | 1.76 (ddd), $J$ = 3.8, 8.1, 15.2 Hz |
| | | | 1.94 (ddd), $J$ = 3.5, 8.3, 15.1 Hz |
| 42 | | 68.9, CH | 4.06 (m) |
| 43 | | 59.9, CH | 3.91 (m) |
| 44 | | 70.7, CH | 5.21 (m) |
| 45 | | 135.5, C | |
| 46 | | 128.3, C | 7.30 (d), $J$ = 8.9 Hz |
| 47 | | 114.6, CH | 6.90 (d), $J$ = 8.9 Hz |
| 48 | | 160.2, C | |
| 49 | | 55.6, $CH_3$ | 3.75 (s) |
| 1' | Hppa | 176.5, C | |
| 2' | | 74.4, CH | 4.10 (), $J$ = 3.5, 9.4 Hz |

| | | |
|---|---|---|
| 3' | 42.1, $CH_2$ | 2.38 (dd), *J* = 9.4, 13.9 Hz<br>2.78 (dd), *J* = 3.5, 13.9 Hz |
| 4' | 139.4, C | |
| 5'/9' | 130.5, CH | 7.12 (d), *J* = 7.6 Hz |
| 6'/8' | 129.4, CH | 7.24 (t), *J* = 7.7 Hz |
| 7' | 127.5, CH | 7.18 (d), *J* = 7.7 Hz |

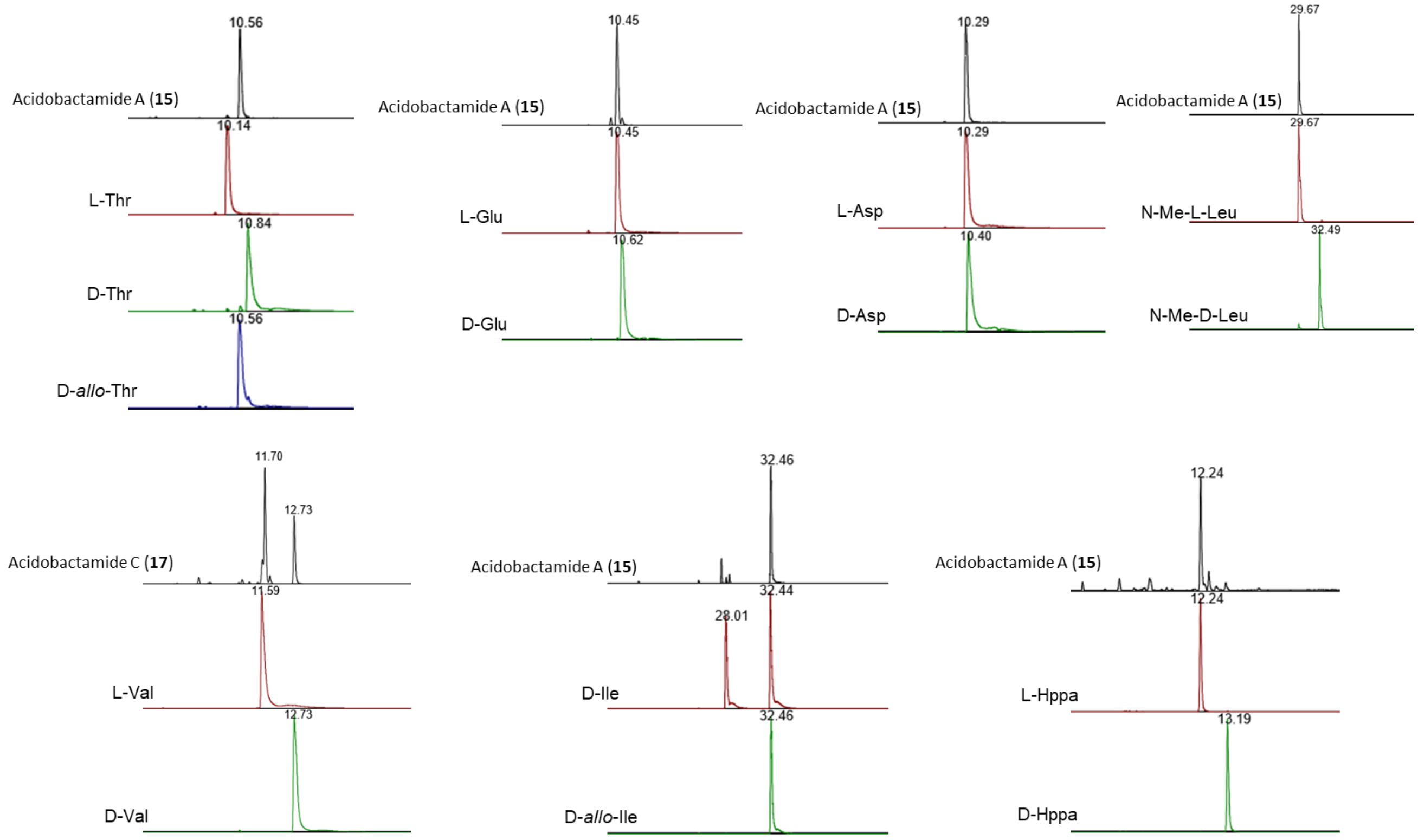

**Figure S100. Marfey's analysis of acidobactamide A (15) and C (17).**

Comparison of Extracted Ion Chromatograms (EICs) of the acidobactamide A (**15**) and C (**17**) hydrolysates to the EICs of the D- and L-amino acid standards.

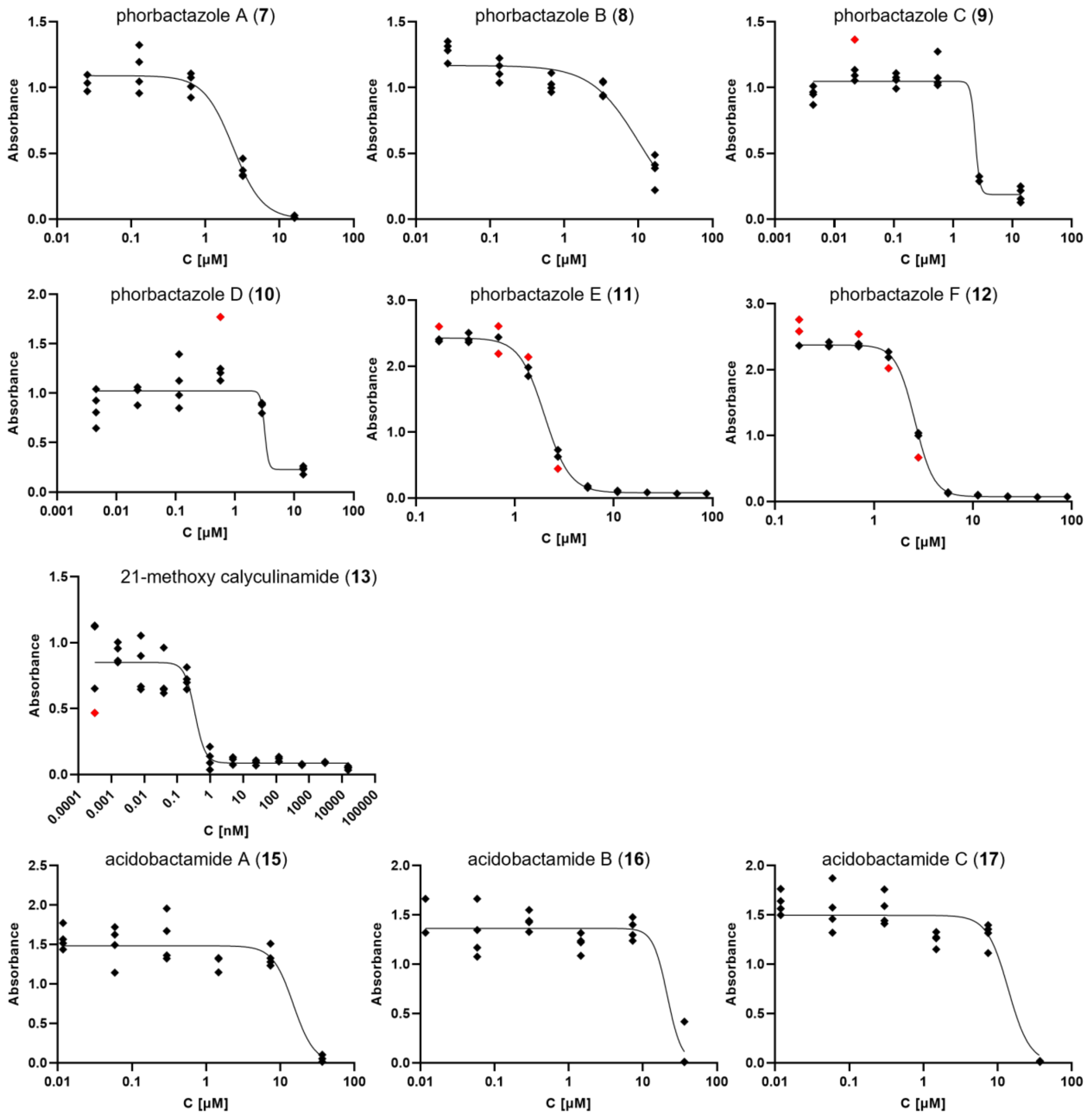

**Figure S101. Bioactivity of isolated compounds against HeLa human cervical cancer cells.**

Cells were treated with compounds **7**-**13**, **15**-**17** in triplicates or quadruplicates. From the data collected in a MTT assay after three days of incubation, the $IC_{50}$-values were calculated: $IC_{50,\,\mathbf{7}}$ = 2.4 µM, $R^2$ = 0.967; $IC_{50,\,\mathbf{8}}$ = 10.2 µM, $R^2$ = 0.895; $IC_{50,\,\mathbf{9}}$ = 2.4 µM, $R^2$ = 0.962; $IC_{50,\,\mathbf{10}}$ = 3.2 µM, $R^2$ = 0.798; $IC_{50,\,\mathbf{11}}$ = 2.0 µM, $R^2$ = 0.999; $IC_{50,\,\mathbf{12}}$ = 2.6 µM, $R^2$ = 0.999; $IC_{50,\,\mathbf{13}}$ = 0.3 nM, $R^2$ = 0.919; $IC_{50,\,\mathbf{15}}$ = 15.0 µM, $R^2$ = 0.889; $IC_{50,\,\mathbf{16}}$ = 21.4 µM, $R^2$ = 0.880; $IC_{50,\,\mathbf{17}}$ = 13.9 µM, $R^2$ = 0.913. A least squares regression using a four-parameter logistic function was used. The bottom and the $IC_{50}$ were constrained to "greater than 0". For the fit, outliers (red) were eliminated using a ROUT coefficient Q=1% in GraphPad Prism 9.2.0 (GraphPad Software).

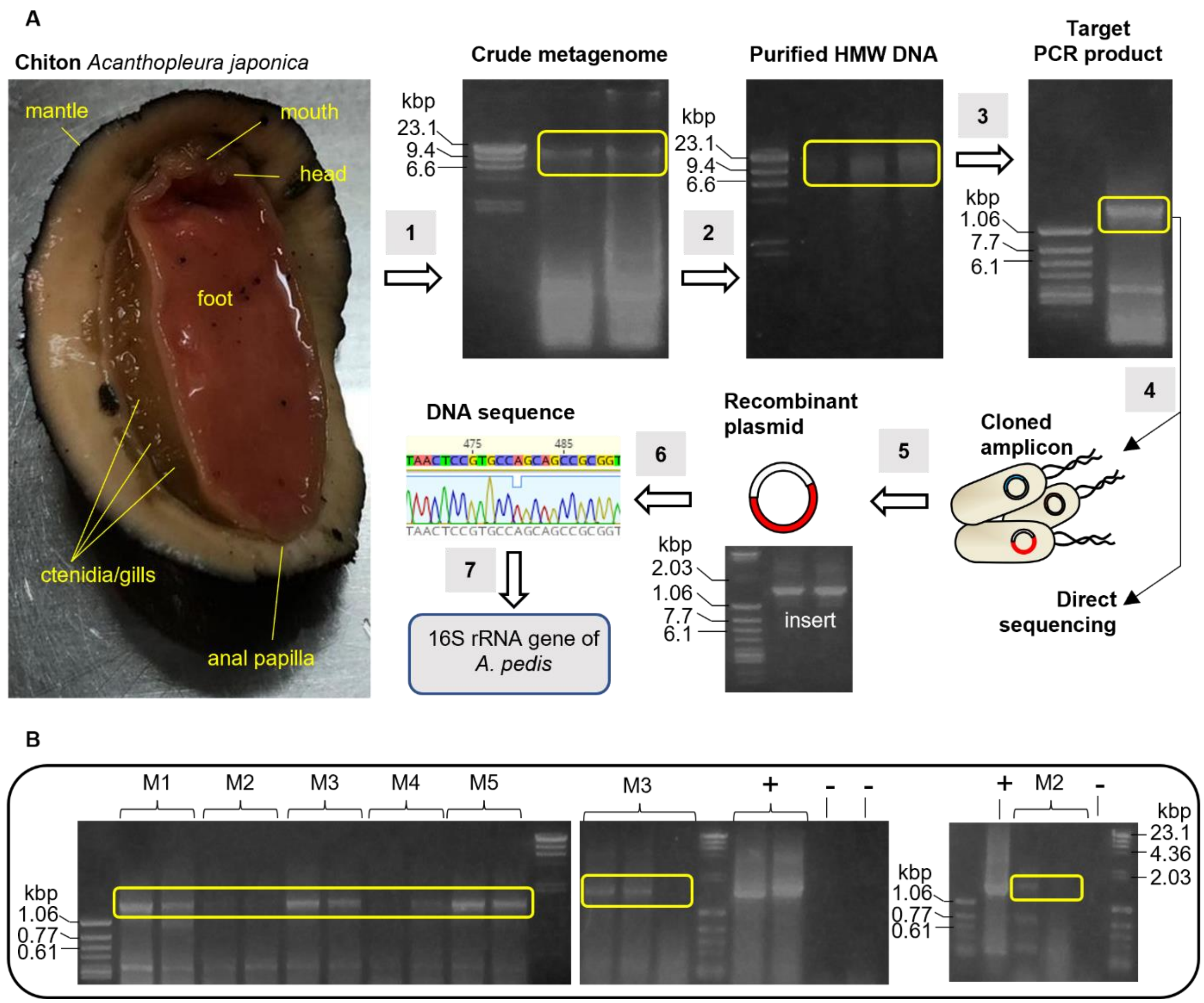

**Figure S102. Detection of the bacterium *A. pedis* in the chiton *A. japonica*.**

(A) Metagenomic DNA was initially extracted from the body of a collected chiton [step **1**], which was then purified by gel-extraction [step **2**]. The purified HMW metagenomic DNA was used as a PCR template to target the 16S rRNA gene of *A. pedis* [step **3**]. The target PCR product of 1.4 kb was cloned into *E. coli* DH5α [step **4**]. The resulting clones were screened by colony PCR using M13 primers [step **5**]. Recombinant plasmids were prepared from three positive clones and subjected to DNA sequencing [steps **6**]. The resulting DNA sequence was queried against the non-redundant nucleotide database using BLAST (blastn)[31] [step **7**].
(B) The DNA electrophoretic gels showing correct-sized amplicons of ~1.4 kb in five of the six different metagenome samples (M1-M5) [step **3** in **A**], but not in the negative controls. Metagenomes of the five chiton individuals used as the PCR templates were labelled as M1, M2, M3, M4, and M5. The HMW DNA marker: λ DNA digested with *Hind*III, and LMW DNA marker: ΦX174 virion DNA digested with *Hin*cII (Nippon Gene). Abbreviations: “+”, positive control (16S amplicon of *A. pedis*); “-“, negative control (without DNA template); HMW, high-molecular-weight; and LMW, low-molecular-weight.

**Table S15. The Minimum Inhibition Concentration (MIC) of compounds 7-10 and 15-17 was tested against a panel of bacteria.**

MIC tests were conducted in 96-well microtiter well plates in LB medium. The test organisms' cultures were adjusted $OD_{600nm}$ of 0.04. 100 µL of the test pathogen was aliquoted in all the rows except row A that was aliquoted with 186.7 µL. 13.3 µL aliquots of compounds **7-10** and **13** at 1.5 mg/mL in MeOH and the reference drugs (kanamycin 50 µ/mL in water) were pipetted on the first row (A) of the plate and mixed by repeated pipetting, before transferring 100 µL to the second row. A 1:1 serial dilution was done in the subsequent rows, and 100 µL discarded after the last row. For compounds **15-17**, 4 µL aliquots of the compounds at 2.5 mg/mL was used. Plates were incubated at 37 °C for 24 h. The lowest concentration of the drug preventing visible growth of the pathogen was taken as the MIC.

| | Minimum Inhibition Concentration (MIC) for compounds **7-10, 13** and **15-17** (µg/mL) | | | | | | | | |
|---|---|---|---|---|---|---|---|---|---|
| Test organisms | **7** | **8** | **9** | **10** | **13** | **15** | **16** | **17** | **Kanamycin (µg/mL)** |
| *Bacillus subtilis* 168 DMS 402 | 100 | - | - | - | - | - | - | - | < 0.04 |
| *Staphylococcus aureus* DMS 18827 MRSA | 100 | - | - | - | - | - | - | - | < 0.04 |
| *Pseudomonas stutzeri* LMG5838 | 100 | / | 100 | / | / | - | - | - | < 0.04 |
| *Escherichia coli* DSM 22664 | - | / | - | / | / | - | - | - | < 0.04 |
| *Klebsiella pneumoniae* DSM 25721 | - | - | - | - | - | - | - | - | < 0.04 |
| *Acinetobacter baumannii* DSM 30007 | 100 | 100 | 100 | 100 | 100 | - | - | - | < 0.039 |
| *Enterococcus faecalis* DSM 2570 | 100 | / | 100 | / | / | - | - | - | < 0.039 |

-: no activity at concentration ≤ 100 µg/mL for **7-10**, **13** and ≤ 50 µg/mL for **15-17**
/ not tested

**Table S16. Detection of the *A. pedis* 16S rRNA gene in metagenomes of different chiton specimens collected in Chiba.**

Experiments were conducted in the Sapporo laboratory where no *A. pedis* cultures were present.

| Chiton label | Corrected-size band | Sequence of *A. pedis* 16S rRNA gene |
|---|---|---|
| M1 | + | + |
| M2 | + | + |
| M3 | + | ND |
| M4 | + | - |
| M5 | + | - |
| M6 | - | - |

ND: not determined.

**Table S17. Sequences of *A. pedis* 16S rRNA gene amplicon obtained from the chiton metagenomes M1 and M2.**

Nucleotide sequence of 16S rRNA gene amplicon from M1 metagenome was present in the recombinant plasmid pM1a6. Nucleotide sequence of 16S rRNA gene amplicon from the M2 metagenome was present in the recombinant plasmid pMII.2.4. Binding sites of the forward primer APedis16S-F2 primer and the reverse primer 16SAPedis-R1 are underlined. One nucleotide (A base) highlighted in red is found as G base in the reference sequence.

| sample | sequence |
|---|---|
| **M1** | TACGCGTTGTTTTGGCTTCGGCTGAGGCAACGAGTGGCAGACGGGTGAGTAACGCGTGGG<br>CAATCTACCTTCGAGTGGGGGATACCATCCCGAAAGGGGTGTTAATACCGCATAACACCTTC<br>GAGCCTTTGGGCTTGACGGTCAAAGTTGGGGATCTGGGAAACCGGACCTGGTGCTTGAAGA<br>GGAGCTCGCGTCAGATTAGCTAGTTGGTGAGGTAACGGCTCACCAAGGCAACGATCTGTAG<br>CCGGCCTGAGAGGGTGATCGGCCACACTGGAACTGAGACACGGTCCAGACTCCTACGGGA<br>GGCAGCAGTGGGGAATTTTGCGCAATGGGGGAAACCCTGACGCAGCAACGCCGCGTGGGT<br>GATGAAGCATCTTGGTGTGTAAAGCCCTGTCGTTAGGGACTAAGGACGGTTGATTAAGAGTT<br>AATCGTCTTGAAGGTACCTGAAGAGGAAGCCCCGGCTAACTCCGTGCCAGCAGCCGCGGTA<br>ATACGGA**A**GGGGCAAGCGTTATTCGGAATTACTGGGCGTAAAGGGCGCGTAGGCGGCCTG<br>GTCAGTGGGAAGTGAAAGCCCTCGGCTCAACCGAGGAATAGCTTCCCATACTGCCAAGCTA<br>GAGTATGGGAGAGGGAAGTGGAATATCCGGTGTAGCGGTGAAATGCGTAGAGATCGGATG<br>GAACACCAGTGGCGAAGGCGACTTCCTGGACCATCACTGACGCTGATGCGCGAAAGCGTG<br>GGGAGCAAACAGGATTAGATACCCTGGTAGTCCACGCCCTAAACGATGAACACTTTGTGGTA<br>CGGGTATCGACCCCTGTACTGCAGGAGCTAACGCATTAAGTGTTCCGCCTGGGGAGTACGG<br>TCGCAAGGCTGAAACTCAAAGGAATTGACGGGGGCCCGCACAAGCGGTGGAGCATGTGGT<br>TTAATTCGAAGCAACGCGCAGAACCTTACCTGGGCTTAAACTGCAGTGGACGGTACCAGAG<br>ATGGTGCTTTTCCTTCGGGAACTGCTGTAGAGGTGCTGCATGGCTGTCGTCAGCTCGTGTC<br>GTGAGATGTTGGGTTAAGTCCCGCAACGAGCGCAACCCCTGCTTCTAGTTGCTAACAGGTTA<br>AGCTGAGCACTCTAGAGGGACTGCCTGGGCAACCAGGAGGAAGGCGGGGATGACGTCAAG<br>TCCTCATGGCCCTTATGTCCAGGGCTACACACGTGCTACAATGGGCGGTACAGAGCGCAGC<br>GAACTCGCGAGAGTAAGCAAATCGCACAAAGCCGTCCTCAGTTCGGATCGCAGTCTGCAAC<br>TCGACTGCGTGAAGCTGGAATCGCTAGTAATCGGAGATCAGCACGCTCCGGTGAATACGTT<br>CCCGGGCCTTGTACACACCGCCCGTCACATCACGAAAGCTGGTTGCACCTGAAAACGGTGG<br>GCTAACCCCTTGGGGAGGTAGCTGTTTACGGTGTGATTGGTG |
| **M2** | TACGCGTTGTTTTGGCTTCGGCTGAGGCAACGAGTGGCAGACGGGTGAGTAACGCGTGGG<br>CAATCTACCTTCGAGTGGGGGATACCATCCCGAAAGGGGTGTTAATACCGCATAACACCTTC<br>GAGCCTTTGGGCTTGACGGTCAAAGTTGGGGATCTGGGAAACCGGACCTGGTGCTTGAAGA<br>GGAGCTCGCGTCAGATTAGCTAGTTGGTGAGGTAACGGCTCACCAAGGCAACGATCTGTAG<br>CCGGCCTGAGAGGGTGATCGGCCACACTGGAACTGAGACACGGTCCAGACTCCTACGGGA<br>GGCAGCAGTGGGGAATTTTGCGCAATGGGGGAAACCCTGACGCAGCAACGCCGCGTGGGT<br>GATGAAGCATCTTGGTGTGTAAAGCCCTGTCGTTAGGGACTAAGGACGGTTGATTAAGAGTT<br>AATCGTCTTGAAGGTACCTGAAGAGGAAGCCCCGGCTAACTCCGTGCCAGCAGCCGCGGTA<br>ATACGGAGGGGGCAAGCGTTATTCGGAATTACTGGGCGTAAAGGGCGCGTAGGCGGCCTG<br>GTCAGTGGGAAGTGAAAGCCCTCGGCTCAACCGAGGAATAGCTTCCCATACTGCCAAGCTA<br>GAGTATGGGAGAGGGAAGTGGAATATCCGGTGTAGCGGTGAAATGCGTAGAGATCGGATG<br>GAACACCAGTGGCGAAGGCGACTTCCTGGACCATCACTGACGCTGATGCGCGAAAGCGTG<br>GGGAGCAAACAGGATTAGATACCCTGGTAGTCCACGCCCTAAACGATGAACACTTTGTGGTA<br>CGGGTATCGACCCCTGTACTGCAGGAGCTAACGCATTAAGTGTTCCGCCTGGGGAGTACGG<br>TCGCAAGGCTGAAACTCAAAGGAATTGACGGGGGCCCGCACAAGCGGTGGAGCATGTGGT<br>TTAATTCGAAGCAACGCGCAGAACCTTACCTGGGCTTAAACTGCAGTGGACGGTACCAGAG<br>ATGGTGCTTTTCCTTCGGGAACTGCTGTAGAGGTGCTGCATGGCTGTCGTCAGCTCGTGTC<br>GTGAGATGTTGGGTTAAGTCCCGCAACGAGCGCAACCCCTGCTTCTAGTTGCTAACAGGTTA<br>AGCTGAGCACTCTAGAGGGACTGCCTGGGCAACCAGGAGGAAGGCGGGGATGACGTCAAG<br>TCCTCATGGCCCTTATGTCCAGGGCTACACACGTGCTACAATGGGCGGTACAGAGCGCAGC<br>GAACTCGCGAGAGTAAGCAAATCGCACAAAGCCGTCCTCAGTTCGGATCGCAGTCTGCAAC<br>TCGACTGCGTGAAGCTGGAATCGCTAGTAATCGGAGATCAGCACGCTCCGGTGAATACGTT<br>CCCGGGCCTTGTACACACCGCCCGTCACATCACGAAAGCTGGTTGCACCTGAAAACGGTGG<br>GCTAACCCCTTGGGGAGGTAGCTGTTTACGGTGTGATTGGTG |

**Table S18. Closest similarity hits to the sequences of the *A. pedis* 16S rRNA gene amplicon obtained from the chiton metagenomes M1 and M2.**

90% identity are shown and sorted according to sequence coverage.

| Organism | Isolate/clone/ strain | Accession number | Coverage (%) | Identity (%) | Source (location) |
|---|---|---|---|---|---|
| Uncultured *Holophagales* | bacterium clone OTU20 | MT858037.1 | 100 | 100 | gut microbiota of *Seriola rivoliana* (Mexico) |
| *Acanthopleuribacter pedis* | NBRC 101209 | NR_041599.1 | 100 | 100 | *Acanthopleura japonica* (Japan) |
| *Sulfidibacter corallicola* | M133 | CP071793.1 | 99 | 92.24 | *Porites lutea* (China) |
| Uncultured bacterium | MM11 13B87 | MG601316.1 | 97 | 90.98 | Sinkhole (environmental sample) |
| Uncultured bacterium | SGUS495 | FJ202764.1 | 97 | 90.88 | *Montastraea faveolata* |
| Uncultured bacterium | APC-3439-J3E10 | KF616733.1 | 97 | 90.70 | Hydrate ridge (OR, USA) |
| Uncultured bacterium | SHFG595 | FJ203188.1 | 97 | 90.55 | *Montastraea faveolata*-diseased tissue (environmental sample) |
| Uncultured bacterium | SGUS817 | FJ202144.1 | 97 | 90.06 | *Montastraea faveolata* kept in aquarium for 23 days |
| Uncultured bacterium | BC_B2_5b | EU622297.1 | 97 | 90.04 | Magneto-FISH captured genomic DNA associated with ANME-2c archaea (Eel River Basin methane seep 520 m depth) |
| Bacterium | N2yML4 | EF629834.1 | 95 | 90.32 | *Mycale laxissima* aquaculture (isolated at Conch Reef, Key Largo, Florida) |
| Acidobacterium | DG1540 | KC295408.1 | 95 | 90.01 | *Emiliania huxleyi* RCC 1214 |
| Uncultured bacterium | F9P413000_O19 | HQ673552.1 | 87 | 90.09 | (Northeast subarctic Pacific Ocean, Station P4, 1300m depth) |
| Uncultured bacterium | AP61 | JQ347353.1 | 83 | 91.10 | *Acropora pruinosa* (Da Ya Bay, China) |
| Uncultured bacterium | husub-g5 | AB573135.1 | 74 | 90.37 | surface sediment sample (Hiroshima, Hiroshima Bay) |
| Uncultured bacterium | A10pH75c | JQ178582.1 | 66 | 96.77 | crustose coralline algae (Great Barrier Reef)) |

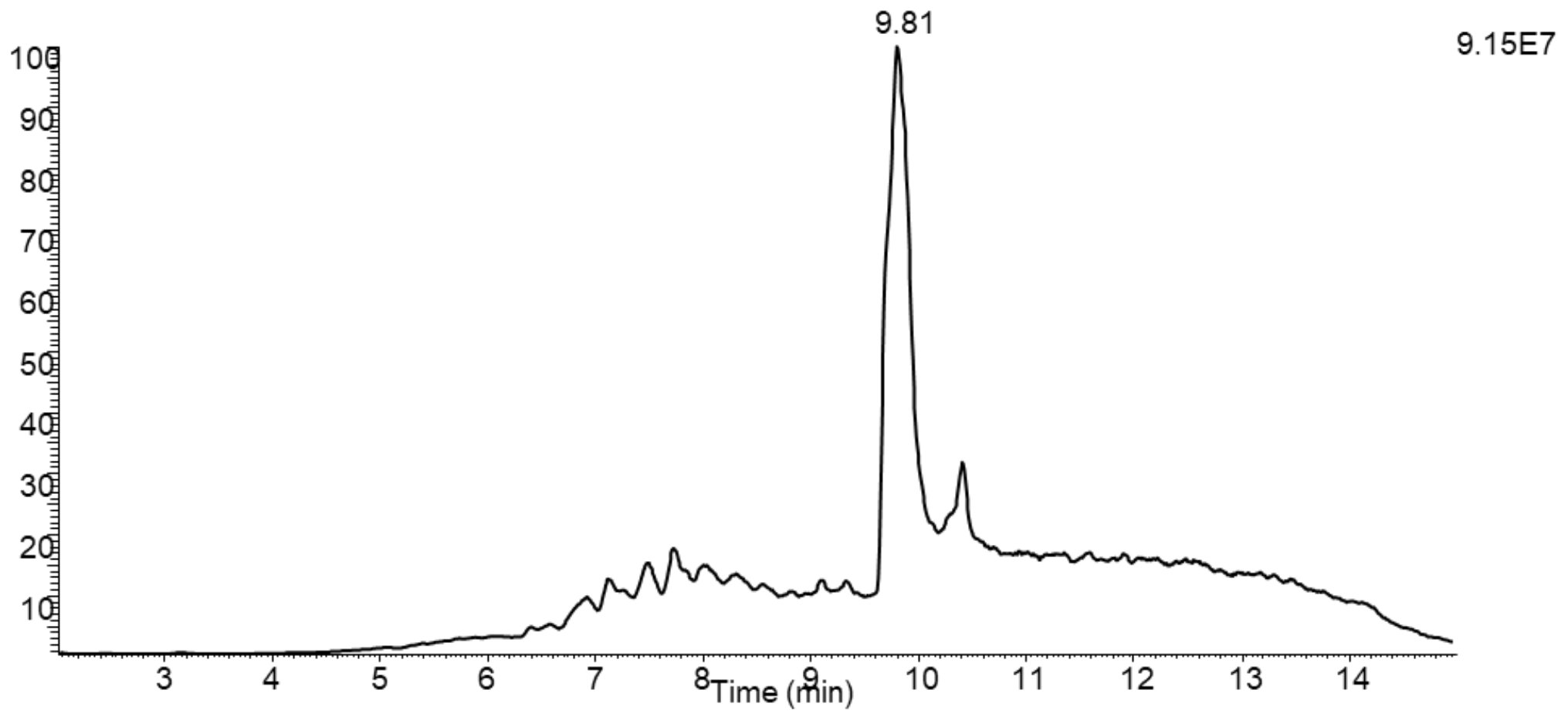

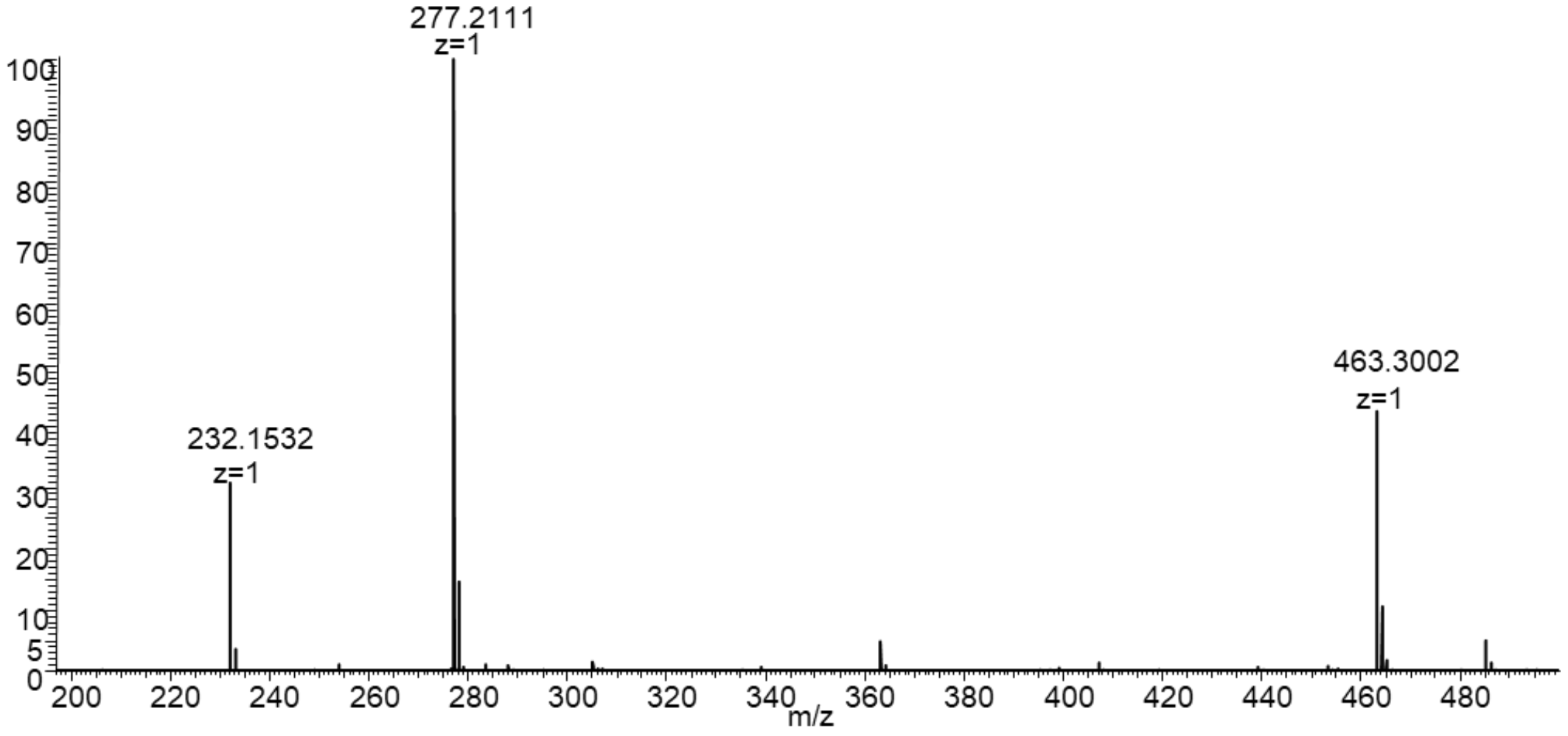

**Figure S103. HPLC-HESI-HRMS data of Boc-L-Ile-OH.**
Top: Total ion chromatogram of isolated Boc-D-Ile-OH
Bottom: MS data of Boc-D-Ile-OH (m/z 232.1532 $[M+H]^+$).

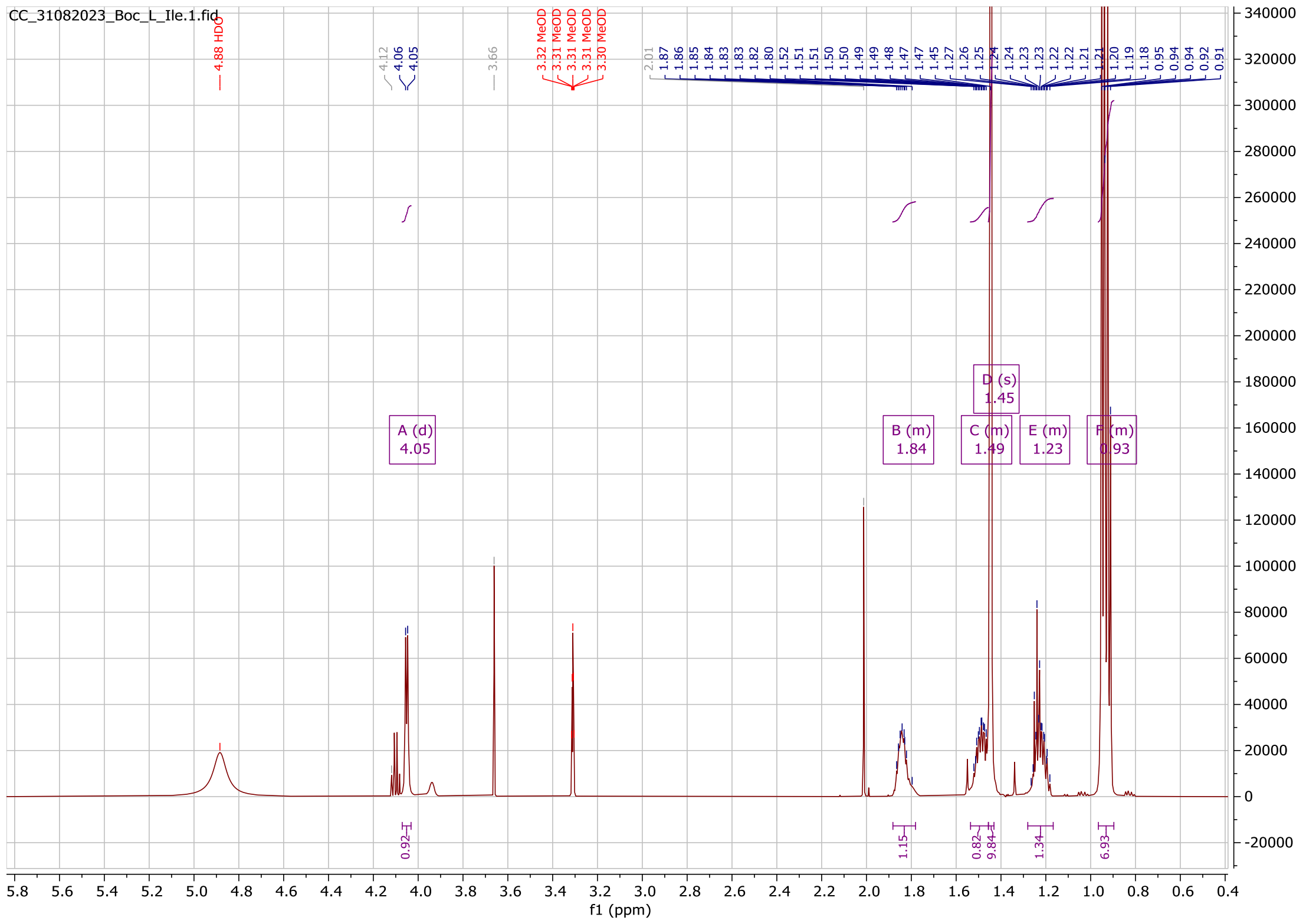

**Figure S104. $^{1}$H NMR spectrum of Boc-L-Ile-OH in methanol-$d_4$.**

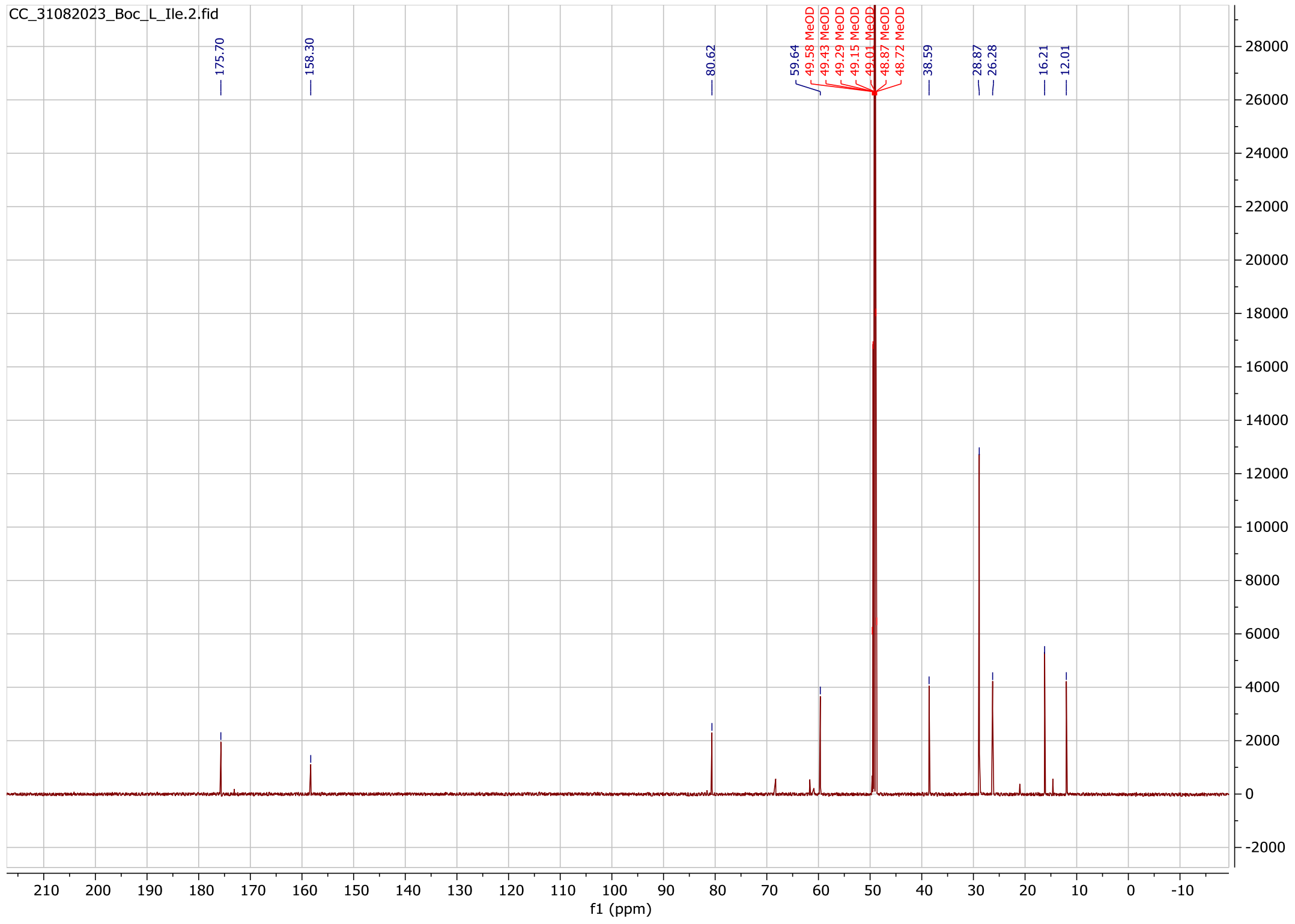

**Figure S105. $^{13}$C NMR spectrum of Boc-L-Ile-OH in methanol-$d_4$.**

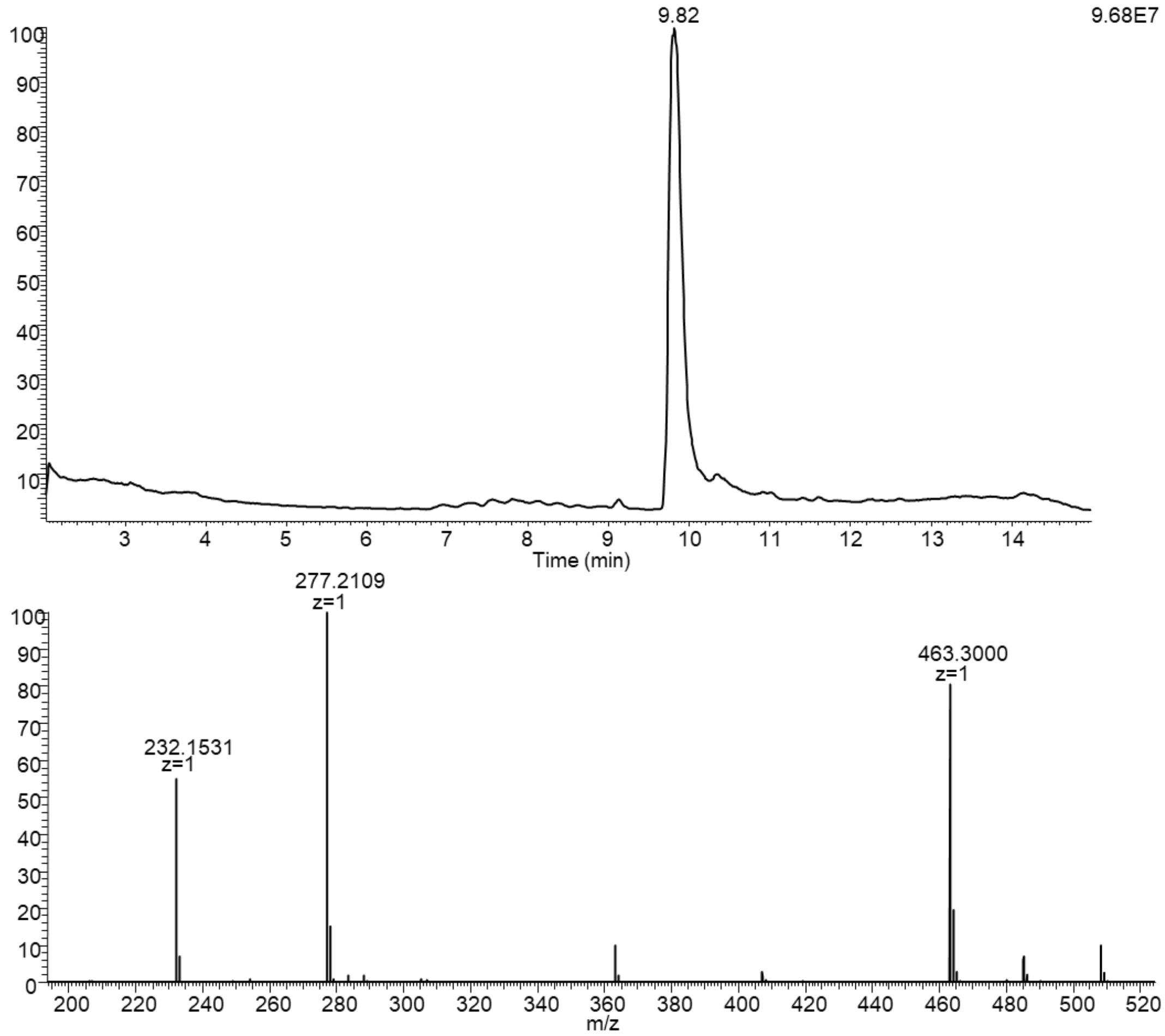

**Figure S106. HPLC-HESI-HRMS data of Boc-D-Ile-OH.**

Top: Total ion chromatogram of isolated Boc-D-Ile-OH
Bottom: MS data of Boc-D-Ile-OH (m/z 232.1531 $[M+H]^+$).

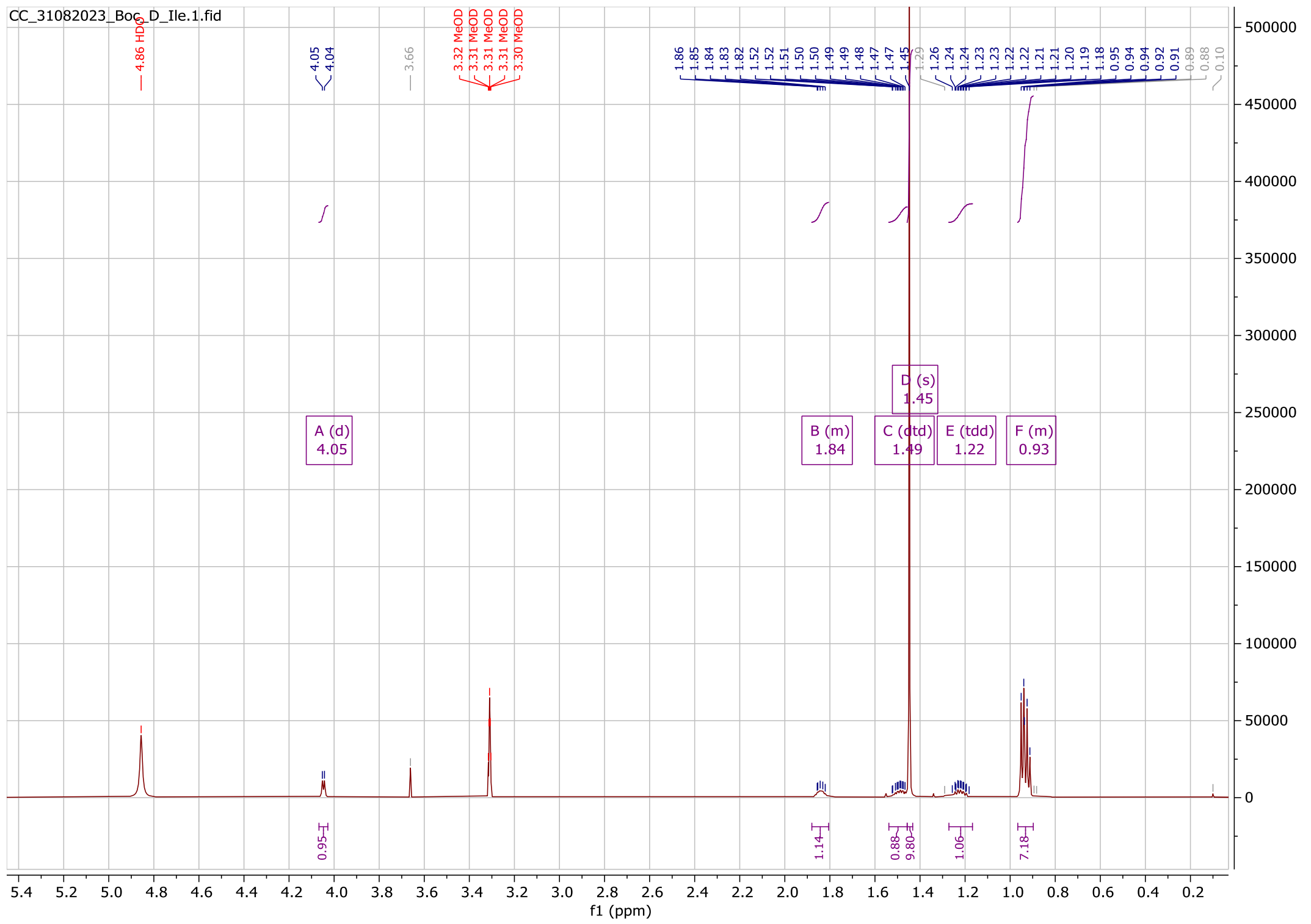

**Figure S107. $^{1}$H NMR spectrum of Boc-D-Ile-OH in methanol-$d_4$.**

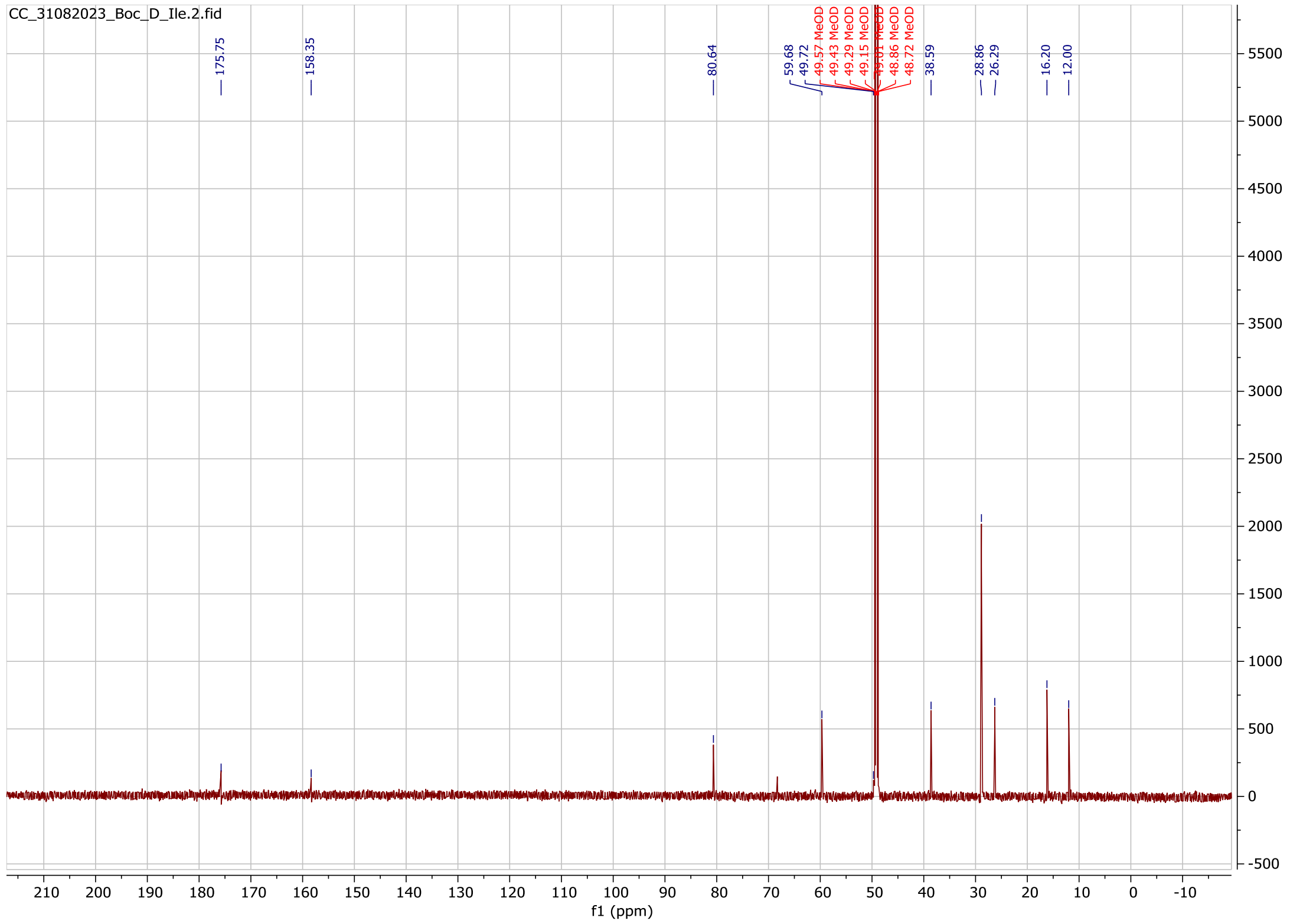

**Figure S108. $^{13}$C NMR spectrum of Boc-D-Ile-OH in methanol-$d_4$.**

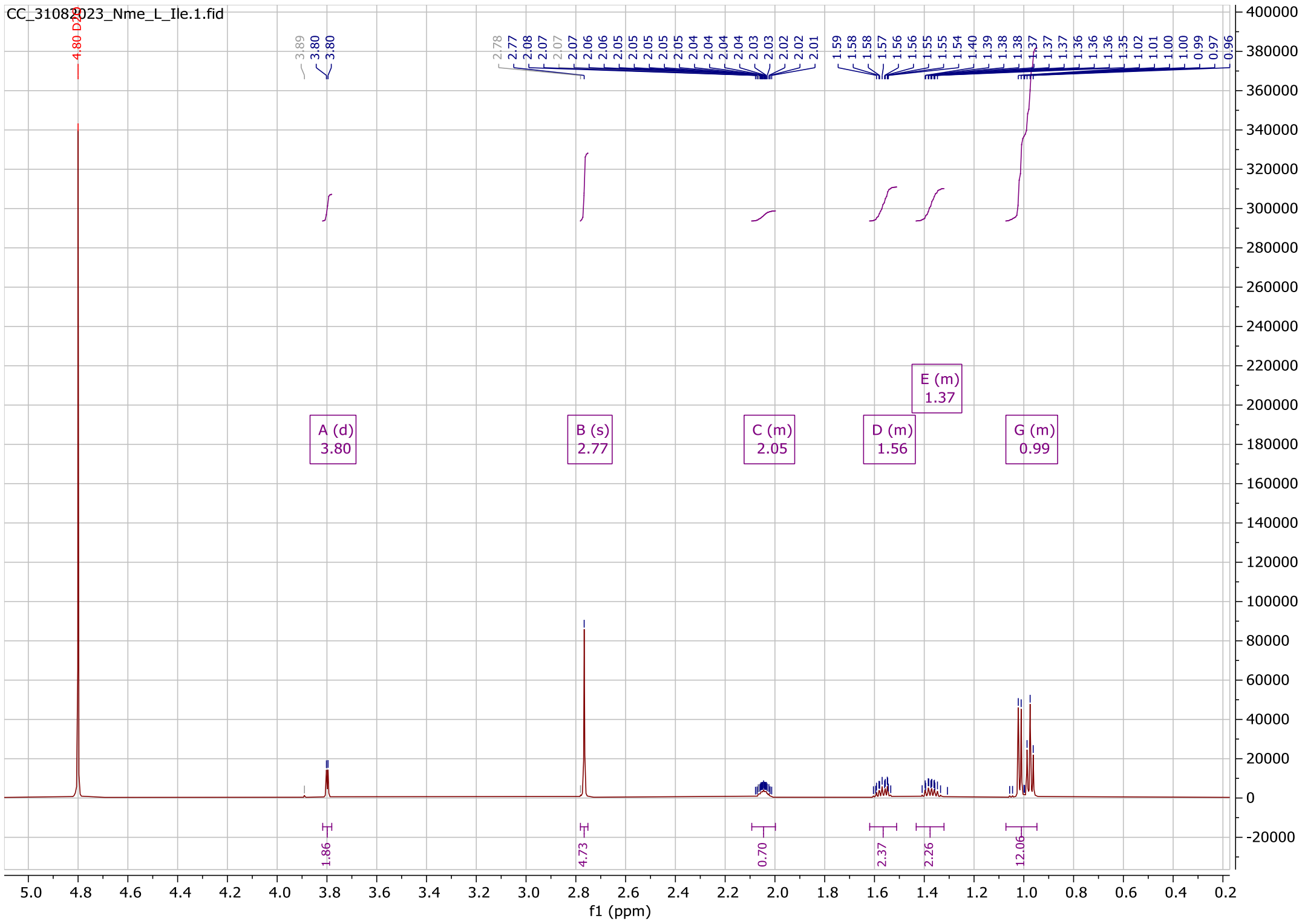

**Figure S109. $^{1}H$ NMR spectrum of *N*-Me-L-Ile in $D_2O$.**

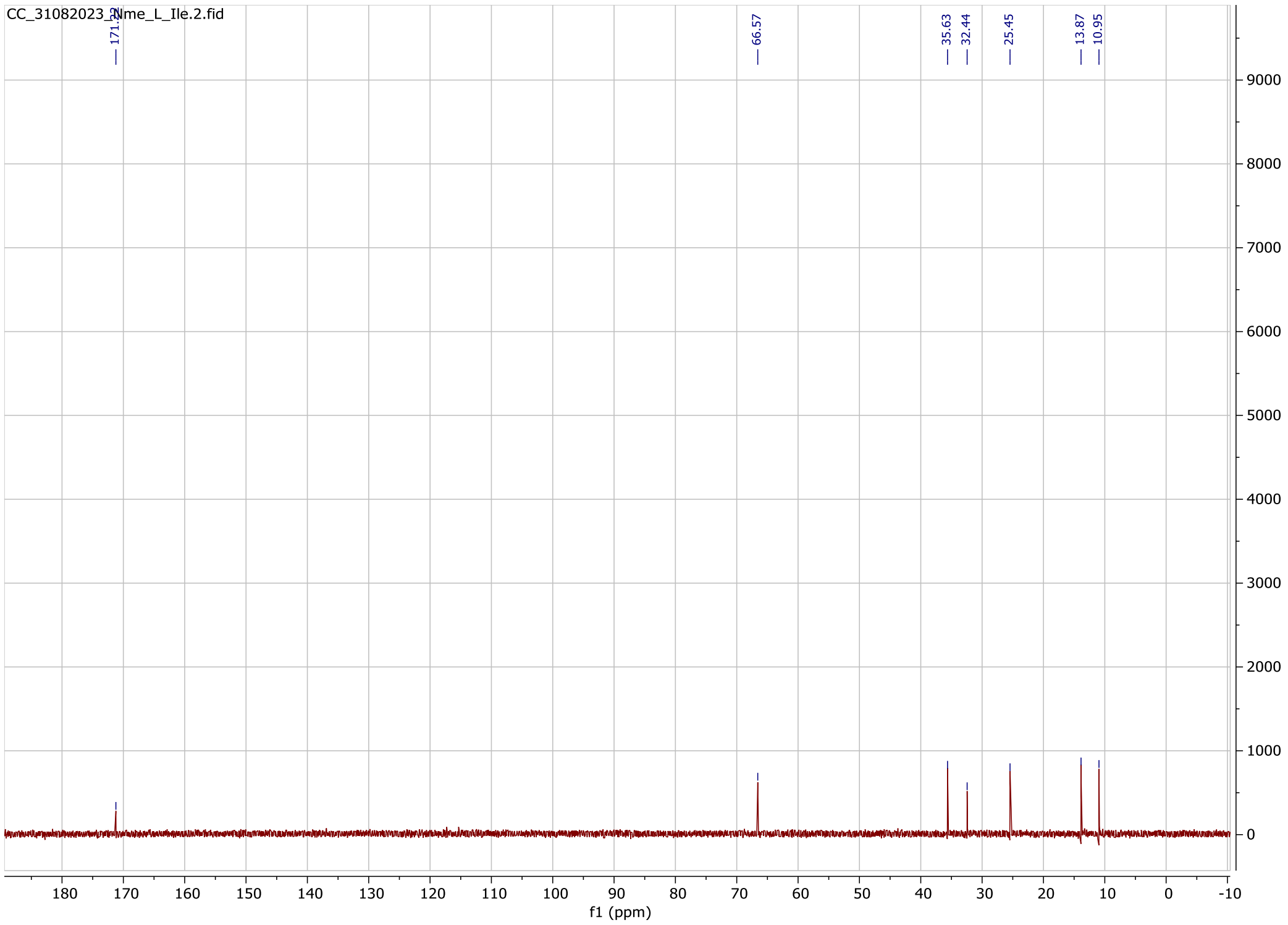

**Figure S110. $^{13}C$ NMR spectrum of *N*-Me-L-Ile in $D_2O$.**

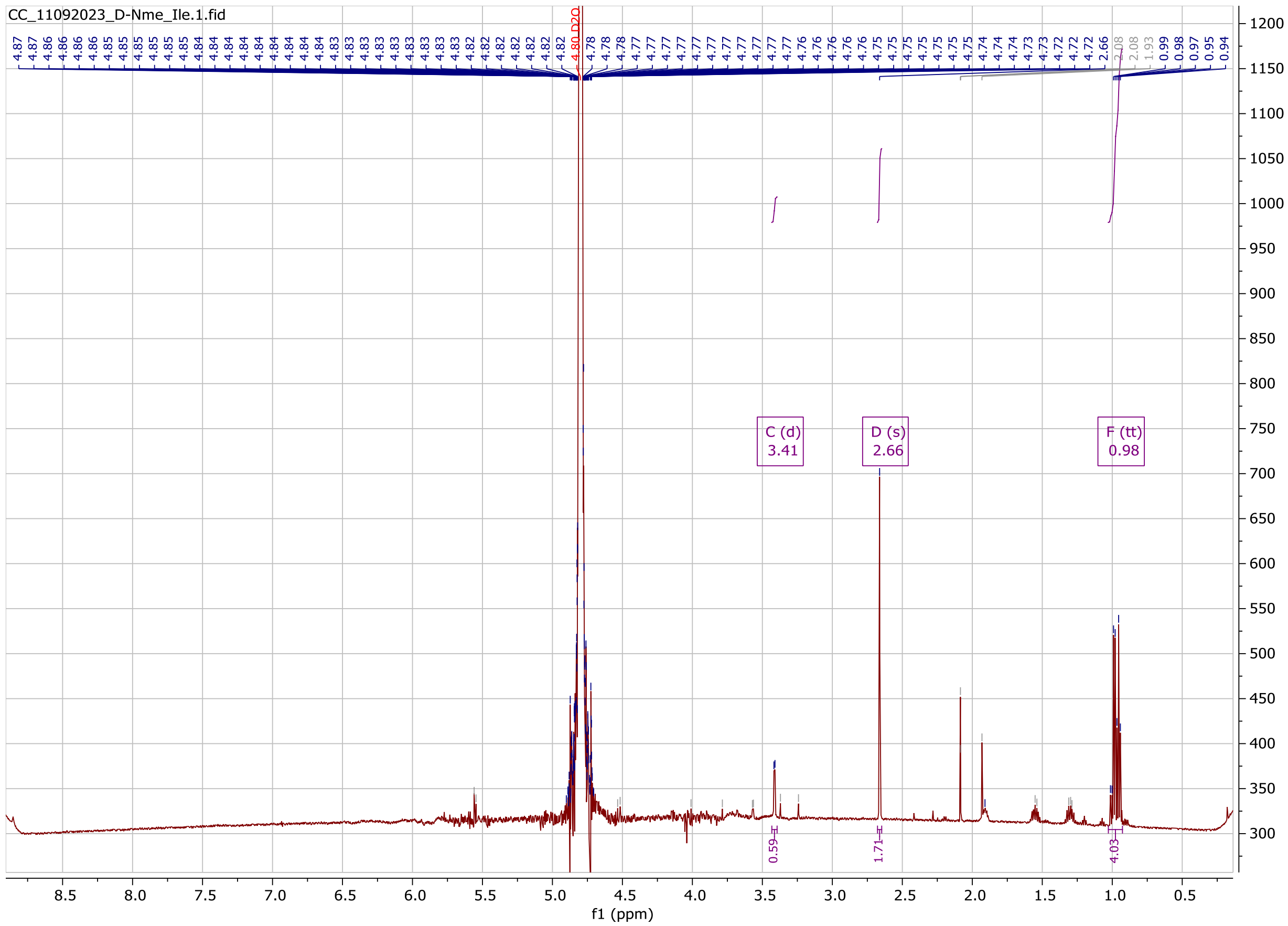

**Figure S111.** $^1$**H NMR spectrum of *N*-Me-D-Ile in $D_2O$.**

**Supplemental references**